\documentclass[a4paper, 11pt]{article}

\renewcommand{\baselinestretch}{1.2}
\usepackage{graphicx}
\usepackage{subcaption}
\usepackage{dcolumn}
\usepackage{bm}
\usepackage{multirow}

\usepackage{centernot}

\usepackage[normalem]{ulem}
\usepackage{cancel}
\usepackage[nosort]{cite}

\usepackage{amsmath}
\numberwithin{equation}{section}
\allowdisplaybreaks

\usepackage{amsthm}
\usepackage{amstext}
\usepackage{amssymb}
\usepackage{mathrsfs}
\usepackage{amsfonts}
\usepackage{amsbsy} 
\usepackage{tensor}
\usepackage{physics}
\usepackage{mathtools}

\usepackage{wasysym}

\usepackage{latexsym}
\usepackage[american]{babel}
\usepackage{bbm}
\usepackage[backref=page, colorlinks,citecolor=pastelblue,linkcolor=pastelblue,urlcolor=pastelblue,hyperindex]{hyperref}

\newcommand*{\figref}[2][]{%
  \hyperref[{#2}]{%
    \ref*{#2}%
    \ifx\\#1\\%
    \else
      \,#1%
    \fi
  }%
}

\usepackage[all,matrix,cmtip]{xy}

\usepackage{csquotes}
\MakeOuterQuote{"}

\usepackage{footmisc}

\usepackage[dvipsnames]{xcolor}

\definecolor{myblue}{HTML}{407fbf}
\definecolor{mygreen}{HTML}{38a25f}
\definecolor{myred}{HTML}{c6453c}
\definecolor{mypurple}{HTML}{9864b2}
\colorlet{mybluemute}{myblue!75!white}

\definecolor{red}{rgb}{1,0,0}
\definecolor{blue}{rgb}{0,0,1}
\definecolor{dblue}{rgb}{0,0,0.4}
\definecolor{green}{rgb}{0,1,0}
\definecolor{black}{rgb}{0,0,0}
\definecolor{white}{rgb}{1,1,1}
\definecolor{niceBlue}{RGB}{20,10,237}
\definecolor{pastelblue}{RGB}{20,93,160}

\definecolor{brn}{rgb}{.8,.4,.0}
\definecolor{redo}{rgb}{1,.5,.0}
\definecolor{ddgrn}{rgb}{0,0.4,0}
\definecolor{dgrn}{rgb}{0,0.55,0}
\definecolor{dbl}{rgb}{0,0,0.5}

\usepackage[bbgreekl]{mathbbol}

\newcommand{\Z}{\mathbb{Z}}
\newcommand{\C}{\mathbb{C}}
\newcommand{\R}{\mathbb{R}}

\newcommand{\G}{\mathbb{G}}

\newcommand{\p}[1]{\prime\,}

\renewcommand{\t}[1]{\widetilde{#1}} 
\newcommand{\h}[1]{\hat{#1}} 

\newcommand{\ii}{\hspace{1pt}\mathrm{i}\hspace{1pt}}
\newcommand{\ee}{\hspace{1pt}\mathrm{e}}

\renewcommand{\dd}{\hspace{1pt}\mathrm{d}}

\newcommand{\<}{\langle}
\renewcommand{\>}{\rangle}

\renewcommand{\Im}{{\rm Im}}

\newcommand{\pp}{\partial}

\newcommand{\bpm}{\begin{pmatrix}}
\newcommand{\epm}{\end{pmatrix}}
\newcommand{\bmm}{\begin{matrix}}
\newcommand{\emm}{\end{matrix}}

\newcommand{\cA}{\mathcal{A}} 

\newcommand{\cC}{ {\cal C} }

\newcommand{\cG}{ {\cal G} } 
\newcommand{\cH}{ {\cal H} }

\newcommand{\cL}{ {\cal L} } 
 
\newcommand{\cN}{ {\cal N} }

\newcommand{\cR}{ {\cal R} }

\newcommand{\cW}{\mathcal{W}} 
\newcommand{\cX}{ {\cal X} } 
 
\newcommand{\cZ}{ {\cal Z} } 

\usepackage{euscript}

\newcommand\scrH         {\mathscr{H}}

\newcommand{\al}{\alpha} 
\newcommand{\bt}{\beta} 
\newcommand{\del}{\delta} 
\newcommand{\Del}{\Delta} 
\newcommand{\eps}{\epsilon} 
 
\newcommand{\ga}{\gamma} 
\newcommand{\Ga}{\Gamma} 
\newcommand{\ka}{\kappa} 
\newcommand{\la}{\lambda} 
\newcommand{\La}{\Lambda} 
\newcommand{\om}{\omega} 
\newcommand{\Om}{\Omega}

\newcommand{\si}{\sigma}

\newcommand{\Hom}{\mathrm{Hom}}

\DeclareMathOperator{\Aut}{Aut}

\DeclareMathOperator{\Rep}{\mathsf{Rep}}
\renewcommand{\Vec}{\mathsf{Vec}}

\newcommand{\ssb}{\overset{\mathrm{ssb}}{\longrightarrow}}

\newcommand{\Uone}{\mathrm{U}(1)}

\usepackage{tikz}
\usepackage{tikz-cd}
\usetikzlibrary{arrows}
\usetikzlibrary{intersections}
\usetikzlibrary{shapes.geometric}
\usetikzlibrary{decorations.pathmorphing, patterns,shapes}
\usetikzlibrary{decorations.markings}
\usetikzlibrary{calc}

\usepackage{enumerate}

\usepackage{hhline}

\usepackage{pifont}

\definecolor{anom}{HTML}{B1002C}
\definecolor{spt}{HTML}{0057B8}
\definecolor{topCell}{HTML}{f2f2f2}
\definecolor{site1}{HTML}{b3cde3}
\definecolor{site2}{HTML}{ccebc5}
\definecolor{site3}{HTML}{decbe4}
\definecolor{link1}{HTML}{ffffcc}
\definecolor{link2}{HTML}{fbb4ae}
\definecolor{link3}{HTML}{fed9a6}

\usepackage{xpatch}

\makeatletter
\patchcmd{\@ssect@ltx}
    {\addcontentsline{toc}{#1}{\protect\numberline{}#8}}
    {}
    {}
    {}
\makeatother

\DeclareMathOperator{\Irr}{Irr}
\DeclareMathOperator{\Map}{Map}

\usepackage{scalerel}
\usepackage{xstring}
\newsavebox{\trivertexbox}

\newcommand{\trivertex}[1]{%
    \sbox{\trivertexbox}{\kern0em\tikz{
        \path (-1.3,-{sqrt(3)/2-0.3}) rectangle (1.3,{sqrt(3)/2+0.3});
        \coordinate (0) at (0,0);
        \coordinate (e) at (1,0);
        \coordinate (w) at (-1,0);
        \coordinate (nw) at (-0.5,{sqrt(3)/2});
        \coordinate (ne) at (0.5,{sqrt(3)/2});
        \coordinate (sw) at (-0.5,-{sqrt(3)/2});
        \coordinate (se) at (0.5,-{sqrt(3)/2});
        \draw[line width=5pt] (e) -- (w);
        \draw[line width=5pt] (nw) -- (se);
        \draw[line width=5pt] (ne) -- (sw);
        \filldraw[gray] (#1) circle (10pt);
    }\kern0em}%
    \scalerel*{\usebox{\trivertexbox}}{[]}%
}

\tikzset{
    baseline={([yshift=-.5ex]current bounding box.center)},
    every picture
}

\newsavebox{\tricornerbox}
\newcommand{\tricorner}[2][up]{%
    \sbox{\tricornerbox}{\kern0em\tikz{
        \path (0,-1.3) -- (0,1.3);
        \IfEqCase{#1}{%
            {up}{%
                \coordinate (apex)  at (0,{sqrt(3)/2});
                \coordinate (left)  at (-1,{-sqrt(3)/2});
                \coordinate (right) at (1,{-sqrt(3)/2});
            }%
            {down}{%
                \coordinate (apex)  at (0,{-sqrt(3)/2});
                \coordinate (left)  at (-1,{sqrt(3)/2});
                \coordinate (right) at (1,{sqrt(3)/2});
            }%
        }%
        \coordinate (0) at (0,0);
        \draw[line width=5pt] (apex) -- (left) -- (right) -- cycle;
        \ifx\relax#2\relax\else
            \filldraw[gray] (#2) circle (12pt);
        \fi
    }\kern0em}%
    \scalerel*{\usebox{\tricornerbox}}{[]}%
}

\newsavebox{\triarrowbox}
\sbox{\triarrowbox}{\kern0em\tikz{
    \path (-1.68,-0.18) rectangle (0.18,1.05);
    \draw[line width=2.5pt]
        (-1.5, 0.866025) -- (-1, 0) -- (0, 0) -- (-0.5, 0.866025) -- cycle;
    \draw[line width=2.5pt]
        (-1, 0) -- (-0.5, 0.866025);
    \filldraw[gray] (-1,0) circle (6pt);
}\kern0em}

\newsavebox{\trilinkbox}
\newcommand{\trilink}[1]{%
    \sbox{\trilinkbox}{\tikz[line width=0.5pt, scale=0.22]{
        \IfEqCase{#1}{%
            {a}{
                \draw (0, 0.866025) -- (0.5, 0) -- (1.5, 0) -- (1, 0.866025) -- cycle;
                \draw (0.5, 0) -- (1, 0.866025);
                \filldraw[gray] (0.5,0) circle (5pt);
                \filldraw[gray] (1, 0.866025) circle (5pt);
            }%
            {b}{
                \draw (0, 0) -- (1, 0) -- (1.5, 0.866025) -- (0.5, 0.866025) -- cycle;
                \draw (1, 0) -- (0.5, 0.866025);
                \filldraw[gray] (1,0) circle (5pt);
                \filldraw[gray] (0.5, 0.866025) circle (5pt);
            }%
            {c}{
                \draw (0.5, 0.866025) -- (0, 0) -- (0.5, -0.866025) -- (1, 0) -- cycle;
                \draw (0, 0) -- (1, 0);
                \filldraw[gray] (0,0) circle (5pt);
                \filldraw[gray] (1, 0) circle (5pt);
            }%
        }%
    }}%
    \mathbin{\usebox{\trilinkbox}}%
}

\newsavebox{\triedgebox}
\newcommand{\triedge}[2][up]{%
    \sbox{\triedgebox}{\kern0em\tikz{
        \IfEqCase{#1}{%
            {up}{
                \coordinate (C) at (0.5,{sqrt(3)/2});
                \coordinate (A) at (0,0);
                \coordinate (B) at (1,0);
            }%
            {down}{
                \coordinate (C) at (0.5,0);             \coordinate (A) at (0,{sqrt(3)/2});
                \coordinate (B) at (1,{sqrt(3)/2});
            }%
        }%
        \draw[line width=2pt] (A) -- (B) -- (C) -- cycle;
        \IfEqCase{#2}{%
            {s}{\filldraw[gray] ($(A)!0.5!(B)$) circle (5pt);}%
            {ne}{\filldraw[gray] ($(C)!0.5!(A)$) circle (5pt);}%
            {nw}{\filldraw[gray] ($(B)!0.5!(C)$) circle (5pt);}%
        }%
        \path (0.5,-0.23) -- (0.5,1.1);
    }\kern0em}%
    \mathbin{\scalerel*{\usebox{\triedgebox}}{[]}}%
}

\newcommand{\trilabelpad}{0.7} 

\newsavebox{\trilabelbox}
\newcommand{\trilabel}[3][up]{%
    \sbox{\trilabelbox}{\tikz[baseline=(0.base),scale=0.1]{
        \pgfmathsetmacro{\p}{\trilabelpad}
        \IfEqCase{#1}{%
            {up}{%
                \coordinate (apex)  at (0,{sqrt(3)/2});
                \coordinate (left)  at (-1,{-sqrt(3)/2});
                \coordinate (right) at (1,{-sqrt(3)/2});
                \coordinate (apex-label)  at (0,{sqrt(3)/2+\p});
                \coordinate (left-label)  at ({-1-\p},{-sqrt(3)/2-\p});
                \coordinate (right-label) at ({1+\p},{-sqrt(3)/2-\p});
            }%
            {down}{%
                \coordinate (apex)  at (0,{-sqrt(3)/2});
                \coordinate (left)  at (-1,{sqrt(3)/2});
                \coordinate (right) at (1,{sqrt(3)/2});
                \coordinate (apex-label)  at (0,{-sqrt(3)/2-\p});
                \coordinate (left-label)  at ({-1-\p},{sqrt(3)/2+\p});
                \coordinate (right-label) at ({1+\p},{sqrt(3)/2+\p});
            }%
        }%
        \node[inner sep=0pt, outer sep=0pt] (0) at (0,0) {\vphantom{X}};
        \coordinate (0-label) at (0,0);
        \draw[line width=0.4pt] (apex) -- (left) -- (right) -- cycle;
        \node[inner sep=0pt, outer sep=0pt] at (#2-label) {\tiny$\smash{#3}$};

    }}%
    \mathbin{\usebox{\trilabelbox}}%
}

\definecolor{defectblue}{RGB}{143,159,197}
\definecolor{junctionorange}{RGB}{242,112,70}

\begin{document}

\thispagestyle{empty}
\fontsize{12pt}{20pt}
\hfill 
\vspace{13mm}
\begin{center}
{\huge Lattice 2-group symmetries:\\\vspace{6pt}operators, defects, and gauging}
\\[13mm]
{\large Lucas Z. Brito$^{\,\G}$ and Salvatore D. Pace$^{\,\rho,\, [\bt]}$
}

\bigskip
{\it 
$^\G$ Department of Physics, Harvard University, Cambridge, MA, USA \\
$^{\rho}$ Department of Physics, Massachusetts Institute of Technology, Cambridge, MA, USA \\
$^{[\bt]}$ School of Natural Sciences, Institute for Advanced Study, Princeton, NJ, USA \\ [.6em]}

\bigskip
\today
\end{center}

\bigskip

\begin{abstract}
\noindent

We construct and study lattice realizations of finite 2-group symmetries in ${2+1}$d quantum lattice systems with finite-dimensional tensor-product Hilbert spaces.
We focus on two broad classes of 2-groups with 0-form symmetry group $G$ and 1-form symmetry group $A$: split 2-groups with trivial Postnikov class ${[\beta]\in\mathcal{H}^3(G,A)}$, and central 2-groups with trivial action ${\rho\colon G\to\text{Aut}(A)}$. 
In both cases, we construct symmetry operators on the full tensor-product Hilbert space that become 2-group symmetry operators when restricted to the topological subspace of the lattice $A$ 1-form symmetry.
While the lattice split 2-group symmetry operators are onsite, the lattice central 2-group symmetry operators are not, and can only be made onsite after introducing ancillae.
We extensively explore various manifestations of $\rho$ and $[\beta]$ for these lattice 2-group symmetry operators and demonstrate their agreement with expectations from quantum field theory.
These manifestations arise in the transformation of operators carrying symmetry charge, the structure of lattice 2-group symmetry defects, and the dual fusion 2-category symmetries obtained by gauging the lattice 2-group symmetries.
We further propose families of local symmetric Hamiltonians for both classes of lattice 2-group symmetries and identify exactly solvable limits lying in phases with spontaneous 2-group symmetry breaking and nontrivial symmetry-enriched topological order.
In one such limit, the gauged Hamiltonians are exactly solvable lattice realizations of the corresponding 2-group gauge theories, whose ground-state degeneracies we calculate.

\end{abstract}

\vfill

\newpage

\pagenumbering{arabic}
\setcounter{page}{1}
\setcounter{footnote}{0}

{\renewcommand{\baselinestretch}{.88} \parskip=0pt
\setcounter{tocdepth}{3}
\tableofcontents}

\vspace{20pt}
\hrule width\textwidth height .8pt
\vspace{13pt}

\section{Introduction}\label{sec:introduction}

Quantum lattice models (QLMs) and quantum field theories (QFTs) are two foundational frameworks of modern theoretical physics, providing complementary descriptions of quantum systems with many degrees of freedom. 
From the perspective of quantum field theory, a QLM can provide a microscopic realization of a QFT, placing it in a mathematically controlled setting amenable to beyond-perturbative study. 
Conversely, from the perspective of quantum lattice models, a QFT captures the universal long-distance and low-energy behavior of a QLM, providing powerful tools for characterizing its phases, phase transitions, and emergent phenomena. 
Understanding the precise relationship between these frameworks remains a longstanding problem of both conceptual and practical importance.
One particularly fruitful approach to this problem is to study two fundamental structures shared by QLMs and QFTs: symmetries and their anomalies.

This program of relating the structural aspects of QLMs and QFTs has been substantially enriched by the advent of generalized symmetries. In the modern formulation, a global symmetry of a relativistic QFT is characterized by a collection of topological defects.
This viewpoint leads naturally to both higher-form and non-invertible symmetries.
An $n$-form symmetry is generated by topological defects of codimension ${n+1}$ in spacetime~\cite{GW14125148}; for ${n>0}$, it is referred to as a higher-form symmetry. A non-invertible symmetry, by contrast, is characterized by topological defects whose fusion is not group-like and may produce a direct sum of defects~\cite{BT170402330,T171209542,CLS180204445,TW191202817,KLW200514178}.
Topological defects of varying codimension and invertibility commonly coexist, and together with their junctions and higher junctions, naturally organize into a
higher-categorical structure. 
See Refs.~\cite{M220403045, CDI220509545, S230518296, S230800747} for reviews on generalized symmetries.

In ${1+1}$d, the relationship between generalized symmetries in relativistic QFTs and QLMs has been studied systematically for symmetries described by unitary fusion categories. 
Every unitary fusion category symmetry can be realized in an anyon-chain QLM~\cite{FTL0612341, AMF160107185, BG170102800, LDO211209091, BBS240505964, JSW241008884}. The Hilbert spaces of these models, however, are generally constrained and do not factorize into tensor products of onsite Hilbert spaces.
The situation is more subtle for QLMs with tensor-product Hilbert spaces. 
When every local Hilbert space is finite-dimensional, only restricted classes of unitary fusion categories can be realized as symmetries~\cite{EJ250705185, I260212053, JY260309949, WIS260515194}. 
By contrast, if the local Hilbert spaces are allowed to be infinite-dimensional, every unitary fusion category symmetry can be realized~\cite{BJ260521327}.

In ${2+1}$d, the relationship between generalized symmetries in relativistic QFTs and QLMs is far less systematically understood.  
There are generalizations of anyon chains to ${2+1}$d called fusion surface models, which provide lattice realizations of symmetries described by unitary fusion 2-categories~\cite{IO230505774, EF240804006, E250114722, IHT250609177}.
The mathematical notion of a fusion $2$-category was introduced in Ref.~\cite{DR181211933}, and a classification of (multi-)fusion $2$-categories was recently obtained in Ref.~\cite{DHJ241105907}.
However, much like anyon chains in ${1+1}$d, these models generally possess constrained Hilbert spaces that do not factorize into tensor products of onsite Hilbert spaces.
For QLMs with tensor-product Hilbert spaces, however, little is known about which fusion $2$-category symmetries from QFT admit such lattice realizations.

An important additional feature of ${2+1}$d systems is the possibility of higher-form symmetries.\footnote{Throughout this work, we do not consider $d$-form symmetries in ${d+1}$-dimensional systems.}
A higher-form symmetry operator is necessarily a topological operator. An operator $T_C$ supported on a spatial locus $C$ is a topological operator if and only if it is invariant under local deformations of $C$: ${T_C=T_{\t{C}}}$ whenever $C$ and $\t{C}$ are homologous.

Although a tensor product Hilbert space supports operators whose support lies on positive-codimension loci, it does not support topological operators.
Indeed, if ${T_C}$ and ${T_{\t{C}}}$ have disjoint support, they are necessarily different operators on a tensor-product Hilbert space.
Consequently, a tensor product Hilbert space does not support higher-form symmetry operators.

A higher-form symmetry can still be realized in a QLM with tensor-product Hilbert space in the following sense.
Consider a collection of symmetry operators $\{T_C\}$ satisfying ${T_C = B_{C,\t{C}}\, T_{\t{C}}}$ for $\t{C}$ homologous to $C$. 
While a generic $T_C$ is not a topological operator on the full Hilbert space, it is on the subspace where every ${B_{C,\t{C}} = 1}$.
If $\{T_C\}$ are higher-form symmetry operators on this subspace, they can provide a lattice realization of the higher-form symmetry.
When they do, we refer to $\{T_C\}$ acting on the full tensor-product Hilbert space as lattice higher-form symmetry operators.\footnote{What we call a lattice higher-form symmetry is sometimes called a non-topological higher-form symmetry, a faithful higher-form symmetry, or a non-relativistic higher-form symmetry~\cite{S190910544, QRH201002254, OPH230104706}.
It is also sometimes just called higher-form symmetry despite its symmetry operators not being topological.
They commonly appear in QLMs with tensor-product Hilbert space, but can also arise in non-relativistic QFTs~\cite{S190910544}.} 
Thus, a lattice realization of a higher-form symmetry consists of both the microscopic operators $\{T_C\}$ and the local constraints under which their dependence on the precise shape of $C$ disappears. Such lattice higher-form symmetries in QLMs have been explored from several complementary directions~\cite{
Y150803468,
KSK180505367,
W181202517,
QRH201002254,
OPH230104706,
PW230105261,
SNH230404792,
HNK230507063,
TC230703180,
EHN231006701,
LLM231016839,
XRK231116235,
XPK240200127,
CSS240513105,
LXP250217572,
HKP250410569,
PAL250702036,
FKCR250912304,
FCH251023701,
LTL260120935,
HPC260512601
}.

Let us illustrate this using a finite triangular lattice $\La$ with a qubit on each link ${ij}$ and tensor product Hilbert space ${\bigotimes_{ij\in\La_1}\C^2}$. 
Consider a $\Z_2$-valued lattice 1-cycle ${\Upsilon = \sum_{ij\in\La_1}\Upsilon_{ij} [ij]}$, where ${\Upsilon_{ij}\in \{0,1\}}$ satisfies the 1-cycle condition ${(\pp\Upsilon)_k = \sum_{ij\,\mid\, k\in ij}\Upsilon_{ij} = 0\bmod 2}$ for each site $k$. Equivalently, an even number of links with ${\Upsilon_{ij}=1}$ meet at every site. 
Define
\begin{equation}
    W^{(\Upsilon)} = \prod_{ij\in\La_1} Z_{ij}^{\Upsilon_{ij}}.
\end{equation}
These operators are not topological on the full Hilbert space. However, defining
\begin{equation}
    B_{ijk} = Z_{jk}Z_{ik}Z_{ij} \equiv
    \begin{tikzpicture}[decoration={markings, mark=at position 0.55 with {\arrow{>}}}, scale=1.9]
    \coordinate (c) at (0,0);
    \coordinate (e) at (1,0);
    \coordinate (w) at (-1,0);
    \coordinate (nw) at (-0.5,{sqrt(3)/2});
    \coordinate (ne) at (0.5,{sqrt(3)/2});
    \coordinate (sw) at (-0.5,-{sqrt(3)/2});
    \coordinate (se) at (0.5,-{sqrt(3)/2});
    %
    \coordinate (t1) at ($(c)!0.333!(e) + (c)!0.333!(ne) - (c)$);   
    \coordinate (t2) at ($(c)!0.333!(ne) + (c)!0.333!(nw) - (c)$);  
    \coordinate (t3) at ($(c)!0.333!(nw) + (c)!0.333!(w) - (c)$);   
    \coordinate (t4) at ($(c)!0.333!(w) + (c)!0.333!(sw) - (c)$);   
    \coordinate (t5) at ($(c)!0.333!(sw) + (c)!0.333!(se) - (c)$);  
    \coordinate (t6) at ($(c)!0.333!(se) + (c)!0.333!(e) - (c)$);   
    %
    %
    \draw[postaction=decorate, color=lightgray] (w) node[color=gray, left] {\footnotesize $i$} -- node[anchor=mid, color=black] {$Z_{ij}$} (c);
    \draw[postaction=decorate, color=lightgray] (c) node[color=gray, right] {\footnotesize $j$} -- node[anchor=mid, color=black] {$Z_{jk}$} (nw);
    \draw[postaction=decorate, color=lightgray] (w) -- node[anchor=mid, color=black] {$Z_{ik}$} (nw) node[color=gray, above] {\footnotesize $k$};
\end{tikzpicture},
\end{equation}
each $W^{(\Upsilon)}$ satisfies ${W^{(\Upsilon)} = \prod_{ijk\in D}B_{ijk}\,W^{(\t{\Upsilon})}}$ for ${\t{\Upsilon} = \Upsilon - \pp D}$ and, therefore, is a topological operator on the subspace ${\scrH\mid_{B_{ijk} = 1}}$.
In fact, $\{W^{(\Upsilon)}\}$ are $\Z_2$ 1-form symmetry operators on this subspace. Therefore, $\{W^{(\Upsilon)}\}$ are lattice $\Z_2$ 1-form symmetry operators on the full Hilbert space.

In this paper, we explore the interplay between finite invertible 0-form and lattice 1-form symmetry in QLMs with tensor-product Hilbert space. 
In relativistic QFT, invertible 0-form and 1-form symmetries form a categorical structure called a 2-group, and the associated combined symmetry is called a 2-group symmetry~\cite{KT13094721, S150804770, T171209542, CI180204790, BH180309336}.
2-group symmetries appear in a wide range of QFTs and play an important role in characterizing symmetry-enriched topological phases~\cite{BBCW14104540, FV151101502, BC170609464}.
Guided by this structure, we will construct and study lattice realizations of 2-group symmetries. That is, 0-form and lattice 1-form symmetry operators that become 2-group symmetry operators when restricted to a subspace in which the lattice 1-form symmetry operators are topological.
Although particular examples of nontrivial $n$-group symmetries in QLMs have appeared previously~\cite{CET200805652, BCH220807367, BCHK221111764, DT230101259,  BHK231105674, OE260402856, LSS260406307}, a systematic treatment has remained limited in QLMs with tensor-product Hilbert space. 
Here, we study general finite lattice 2-group symmetries and perform a comprehensive analysis of their symmetry operators, symmetry defects, and gauging.
In the remainder of this introduction, we will review 2-group symmetries in relativistic QFT and then provide a self-contained summary of our
main results.

\subsection{Review of 2-group symmetry}\label{sec:review-of-2group-symmetry}

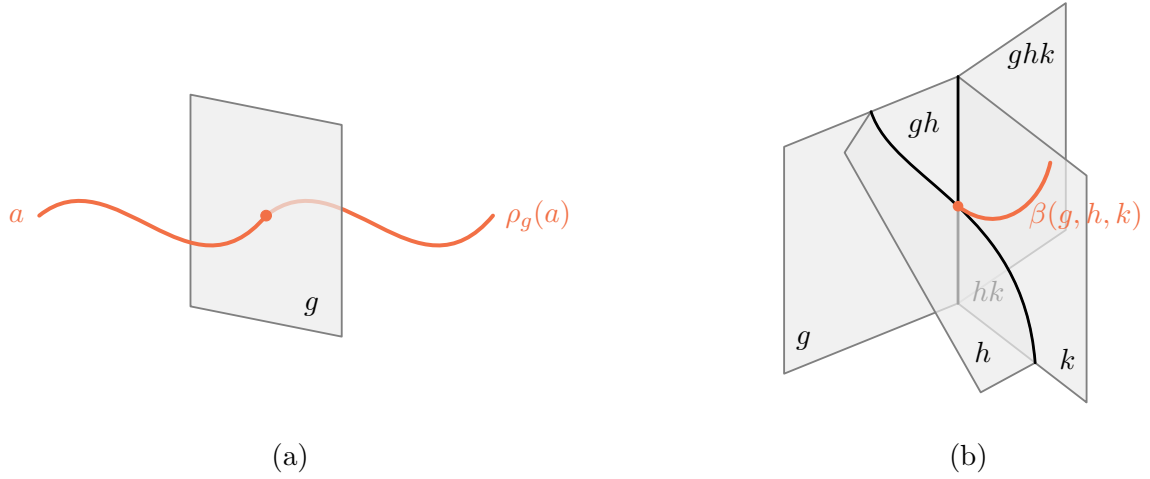
\begin{figure}
    \centering
    \begin{tikzpicture}[
        x=1cm, y=1cm,
        line cap=round,
        line join=round,
        every node/.append style={text=black},
        sheet/.style={
            draw=gray,
            line width=.7pt,
            fill=gray!15,
            fill opacity=.7
        },
        junction/.style={
            black,
            line width=1.05pt
        },
    ]

    \begin{scope}[local bounding box=panelA, shift={(-1.3,-1.4)}, scale=2]
        \draw[color=junctionorange, line width=1.5] (0.5, 0.6) .. controls (1, 1) and (1.5,0) .. (2,0.6);
        \path[sheet] (0,0) -- (1,-0.2) -- (1, 1.2) -- (0, 1.4) -- cycle;
        \draw[color=junctionorange, line width=1.5] (-1, 0.6) .. controls (-0.5, 1) and (0,0) .. (0.5,0.6);
        \filldraw[color=junctionorange] (0.5,0.6) circle (1pt);
        \node () at (0.8, 0) {$g$};
        \node[color=junctionorange] () at (-1.15, 0.6) {$a$};
        \node[color=junctionorange] () at (2.3, 0.6) {$\rho_g(a)$};
    \end{scope}

    \begin{scope}[local bounding box=panelB, shift={(9,0)}, scale=0.5]

        \coordinate (O) at (-0.3,-0.15);

        %
        \coordinate (Jghk) at (-0.3, 3.28);
        \coordinate (JgHK) at (-0.3,-2.72);

        \coordinate (G1) at (-4.90, 1.42);
        \coordinate (R1) at (2.55, 5.22);
        \coordinate (K1) at (3.10, 0.66);
        \coordinate (K2) at ($(K1) - (Jghk) + (JgHK)$);
        \coordinate (R2) at ($(R1) - (Jghk) + (JgHK)$);

        \coordinate (Jgh) at ($0.5*(Jghk) - 0.5*(G1) + (G1)$);
        \coordinate (H1)  at ($0.2*(K1) - 0.2*(G1) + (G1)$);
        \coordinate (G2)  at ($(G1) - (Jghk) + (JgHK)$);
        \coordinate (H2)  at ($0.65*(K2) - 0.65*(G2) + (G2)$);
        \coordinate (Jhk) at ($0.4*(JgHK) - 0.4*(K2) + (K2)$);

        \coordinate (C0) at (1.42, -0.2);
        \coordinate (C4) at (-2,0.5);

        \path[sheet]
            (Jghk) -- (Jgh) -- (G1) -- (G2) -- (JgHK) -- cycle;

        \path[sheet]
            (O) -- (Jghk) -- (R1) -- (R2) -- (JgHK) -- cycle;

        \path[sheet]
            (Jghk) -- (K1) -- (K2) -- (Jhk) -- (JgHK) -- cycle;

        \draw[junction] (Jghk) -- (O);
        \draw[junction] (JgHK) -- (O);

        \node at (0.5, -2.4) {$hk$};

        \path[sheet]
            (Jgh) -- (H1) -- (H2) -- (Jhk) .. controls (C0) and (C4) .. (Jgh);

        \draw[junction] (Jhk) .. controls (C0) and (C4) .. (Jgh);

        \node at (-4.38,-3.7) {$g$};
        \node at (0.36,-4) {$h$};
        \node at (2.6,-4.2) {$k$};

        \node at (-1.2, 2) {$gh$};
        \node at ( 1.64, 3.83) {$ghk$};

        \coordinate (C5) at (1.9,-0.1);
        \coordinate (c) at (O);
        \draw[junctionorange, line width=1.5pt]
            (c) .. controls (0.82,-1) and (C5) .. (2.14,1.);

        \fill[junctionorange] (c) circle (4pt);

        \node[text=junctionorange, anchor=west] at (1.3,-0.4) {$\bt(g,h,k)$};
    \end{scope}

    \coordinate (labelline) at (0,-3.4);
    \node at (panelA.south |- labelline) {(a)};
    \node at (panelB.south |- labelline) {(b)};

    \end{tikzpicture}
    \caption{Defect interpretation in ${2+1}$d of the action $\rho$ and Postnikov class $[\bt]$ of the 2-group~\eqref{2GrpSymData}.
    (a) The action $\rho$ causes an ${a\in A}$ topological defect line to become a $\rho_g(a)$ topological defect line when penetrating a ${g\in G}$ topological defect surface. (b) The Postnikov class $[\bt]$ determines how a topological defect line can end on a 0-dimensional junction of three $G$ topological defect surfaces.}
    \label{fig:2group-data}
\end{figure}

A 2-group is a categorified group that naturally describes the interplay between invertible 0-form and 1-form symmetries in relativistic QFT. More precisely, a 2-group $\G$ is a monoidal groupoid in which every object is invertible up to isomorphism.\footnote{Equivalently, delooping a monoidal groupoid $\G$ produces a one-object 2-groupoid $B\G$. The term 2-group is sometimes used to refer to either $\G$ or $B\G$.} 
Up to equivalence, a 2-group is specified by four pieces of data:
\begin{equation}\label{2GrpSymData}
    \G
    \equiv
    (G,A,\rho,[\bt]).
\end{equation}
Here, ${G = \pi_0(\G)}$ is the group of isomorphism classes of objects and ${A = \pi_1(\G)}$ is the automorphism group of the monoidal identity. The group $A$ is necessarily abelian. 
We take both $G$ and $A$ to be finite throughout this work. 
The group homomorphism
\begin{equation}
\begin{aligned}
    \rho\colon G &\to \Aut(A)\\
    g &\mapsto \rho_g
\end{aligned}
\end{equation}
specifies an action of $G$ on $A$, and
\begin{equation}
    [\bt]\in \cH^3_\rho(G,A)
\end{equation}
is the Postnikov class. 
When $[\bt]$ is trivial, we call $\G$ a split 2-group. When $\rho$ is trivial, we call $\G$ a central 2-group. 
See~\cite{BL0307200,KT13094721} for further mathematical introductions to 2-groups, and the Appendix of~\cite{PZB231008554} for general $n$-groups.

For a relativistic QFT with 2-group symmetry $\G$, the data in Eq.~\eqref{2GrpSymData} admit a direct interpretation in terms of topological defects~\cite{BH180309336}. The codimension-1 defects are labeled by elements ${g\in G}$ and fuse according to the group law of $G$, thereby generating a $G$ 0-form symmetry. Likewise, the codimension-2 defects are labeled by elements ${a\in A}$ and fuse according to the abelian group law of ${A}$, thereby generating an $A$ 1-form symmetry. The interplay between these two symmetries is encoded by $\rho$ and $[\bt]$.
The action $\rho$ determines how a 1-form symmetry defect transforms when transported through a 0-form symmetry defect. On the other hand, a cocycle representative $\bt$ of the Postnikov class controls the associator for the fusion of 0-form symmetry defects and allows a 1-form symmetry defect to end at the corresponding junction. These two manifestations are illustrated in Fig.~\ref{fig:2group-data}. An equivalent formulation uses background $\G$ gauge fields, as reviewed in Appendix~\ref{app:double-from-path-integral}. In this description, the action $\rho$ and a cocycle representative $\bt$ of the Postnikov class appear in the gauge-transformations and generalized flatness conditions, thereby coupling the $G$ and $A$ background 1-form and 2-form fields, respectively.

A finite 2-group symmetry may carry an 't Hooft anomaly. When the anomaly vanishes, the symmetry can be gauged. Gauging it in a trivial gapped phase produces a topological finite 2-group gauge theory~\cite{MP0608484,thorngrenThesis,DT180210104}.
In ${2+1}$d, the dual symmetry after gauging is described by the fusion 2-category ${2\Rep(\G)}$ of 2-representations of $\G$~\cite{BSW220805973, BBFP220805993}.
This fusion 2-category symmetry is the electric symmetry of $\G$ gauge theory. It also appears as the magnetic symmetry of certain ${2+1}$d nonlinear sigma models~\cite{CT230700939, P230805730, PZB231008554}.

\subsection{Summary}\label{Sec:summary}

\begin{figure}[t!]
    \centering
 \begin{tikzpicture}[scale=2,
    decoration={markings, mark=at position 0.55 with {\arrow{>}}},
    d/.style={postaction=decorate},
    dot/.style={lightgray!50!white, opacity=1}
]

    \def\Nx{3}
    \def\Ny{3}

    \pgfmathsetmacro{\a}{1}
    \pgfmathsetmacro{\b}{0.8660254}  

    \foreach \i in {0,...,\Ny} {
        \foreach \j in {0,...,\Nx} {

            \coordinate (v_\i_\j)
                at (\j*\a + 0.5*\i*\a, \i*\b);

        }
    }

    \foreach \i in {0,...,\Ny} {
        \foreach \j in {0,...,\numexpr\Nx-1} {

            \pgfmathtruncatemacro{\jp}{\j+1}

            \draw[d, color=gray] (v_\i_\j) -- (v_\i_\jp);

            \coordinate (c_h_\i_\j)
                at ($(v_\i_\j)!0.5!(v_\i_\jp)$);

        }
    }

    \foreach \i in {0,...,\numexpr\Ny-1} {
        \foreach \j in {0,...,\Nx} {

            \pgfmathtruncatemacro{\ip}{\i+1}

            \draw[d, color=gray] (v_\i_\j) -- (v_\ip_\j);

            \coordinate (c_dr_\i_\j)
                at ($(v_\i_\j)!0.5!(v_\ip_\j)$);

        }
    }

    \foreach \i in {0,...,\numexpr\Ny-1} {
        \foreach \j in {1,...,\Nx} {

            \pgfmathtruncatemacro{\ip}{\i+1}
            \pgfmathtruncatemacro{\jm}{\j-1}

            \draw[d, color=gray] (v_\i_\j) -- (v_\ip_\jm);

            \coordinate (c_dl_\i_\j)
                at ($(v_\i_\j)!0.5!(v_\ip_\jm)$);

        }
    }

    \fill[color=lightgray!30] (v_0_1) -- (v_1_0) -- (v_0_0) -- cycle;
    \draw[thick, black, postaction=decorate] (v_0_0) -- (v_0_1); 
    \draw[thick, black, postaction=decorate] (v_0_0) -- (v_1_0); 
    \draw[thick, black, postaction=decorate] (v_0_1) -- (v_1_0); 

    \node (i) at (0, -0.1) {$i$};
    \node (i) at (1, -0.1) {$j$};
    \node (i) at (0.4, 0.9) {$k$};
    \node (2simp) at (-0.25, 0.5) {$ijk$};
    \node (1simp) at (0.3, -0.5) {$ij$};
    \node () at (1, -0.5) {$i<j<k$};

    \coordinate (blt_mid) at ($(v_0_1)!0.5!(v_1_0)$);
    \coordinate (blt_center) at ($(v_0_0)!0.6667!(blt_mid)$);

    \draw[->] (2simp) to[bend right=30] (blt_center);
    \draw[->] (1simp) to[bend right=30] (0.5, -0.1);

\end{tikzpicture}
    \caption{
    We consider quantum lattice systems whose spatial lattice is a triangular lattice. There is a corresponding simplicial complex whose branching structure is displayed here. Namely, the orientation of each link always points upwards and to the right.}
    \label{fig:lattice-branching-structure}
\end{figure}

We now summarize the main results of this work. We consider a finite triangular lattice $\La$ triangulating a spatial torus, equipped with the branching structure shown in Fig.~\ref{fig:lattice-branching-structure}. We place a $G$-qudit on each site and an $A$-qudit on each link, giving the tensor-product Hilbert space
\begin{equation}
    \scrH
    =
    \bigotimes_{i\in\La_0}\C[G]
    \otimes
    \bigotimes_{ij\in\La_1}\C[A].
\end{equation}
A computational basis state is denoted by
${\ket{\{g_i\},\{a_{ij}\}}}$.
We make use of the standard group qudit operators, which we review in App.~\ref{app:group-qudit-review}. 

We consider lattice realizations of split 2-group symmetries in Sec.~\ref{sec:split-2group-symmetry}, and of central 2-group symmetries with nontrivial Postnikov class in Sec.~\ref{sec:central-2group-symmetry}.
Both the split and central constructions contain the same lattice $A$ 1-form symmetry operators,
\begin{equation}\label{eq:summary-1-form-symmetry-ops}
    T^{(\ga)}
    =
    \prod_{ij\in\La_1}
    X_{ij}^{(\ga_{ij})},
    \qquad
    \ga\in Z_1(\La^\vee;A),
\end{equation}
where ${\ga = \sum_{ij\in\La_1} \ga_{ij}[ij]^\vee}$ is an $A$-valued 1-cycle on the dual lattice $\La^\vee$. 
The $A$-qudit operator $X^{(\la)}$ satisfies ${X^{(\la)}\ket{a} = \ket{a+\la}}$.
These operators are defined on the full tensor-product Hilbert space but become topological only after restricting to the subspace $\scrH_{\mathrm{top}}$ defined in Eq.~\eqref{HtopDefSplit2GrpSection}. On this subspace, the $A$-qudit configurations obey the local redundancy ${a_{ij} \sim a_{ij}+\la_j-\la_i}$, and $T^{(\ga)}$ becomes a topological operator that depends only on the homology class ${[\ga]\in H_1(\La^\vee;A)}$.

\noindent\textbf{Split 2-groups.}
In Sec.~\ref{sec:split-2group-symmetry}, we consider the case in which the Postnikov class is trivial. The 0-form symmetry part of this lattice split 2-group symmetry is implemented by the onsite operators
\begin{equation}\label{eq:split-sym-ops-summary}
    U^{(h)}
    =
    \prod_{i\in\La_0}
    \overrightarrow{X}^{(h)}_i
    \prod_{ij\in\La_1}
    P_{ij}^{(\rho_h)},
    \qquad
    h\in G.
\end{equation}
The $G$-qudit operator $\overrightarrow{X}^{(h)}$ satisfies ${\overrightarrow{X}^{(h)}\ket{g} = \ket{hg}}$, whereas the $A$-qudit operator $P^{(\rho_h)}$ for ${h\in G}$ satisfies ${P^{(\rho_h)}\ket{a} = \ket{\rho_h(a)}}$.
The operators~\eqref{eq:split-sym-ops-summary} are $G$ symmetry operators: ${U^{(g)}U^{(h)}=U^{(gh)}}$. Their interplay with the lattice $A$ 1-form symmetry operator~\eqref{eq:summary-1-form-symmetry-ops} is
\begin{equation}\label{eq:summary-split-interplay}
    U^{(h)}
    T^{(\ga)}
    U^{(h)\dagger}
    =
    T^{(\rho_h\triangleright\ga)},
    \qquad
    \rho_h\triangleright\ga
    :=
    \sum_{ij\in\La_1}
    \rho_h(\ga_{ij})[ij]^\vee.
\end{equation}
At the level of symmetry defects, we show in Sec.~\ref{sec:split-2group-sym-defects} that transporting an ${a\in A}$ 1-form symmetry defect through an ${h\in G}$ 0-form symmetry defect changes its label by ${a\mapsto \rho_h(a)}$, reproducing the defect-crossing relation shown in Fig.~\figref[a]{fig:2group-data}. This and~\eqref{eq:summary-split-interplay} demonstrate that the symmetry operators~\eqref{eq:summary-1-form-symmetry-ops} and~\eqref{eq:split-sym-ops-summary} realize a split 2-group symmetry on the topological subspace $\scrH_\mathrm{top}$. Here, restricting to $\scrH_\mathrm{top}$ does not modify the 0-form symmetry of the lattice split 2-group symmetry but is necessary to make the lattice 1-form symmetry operators topological.

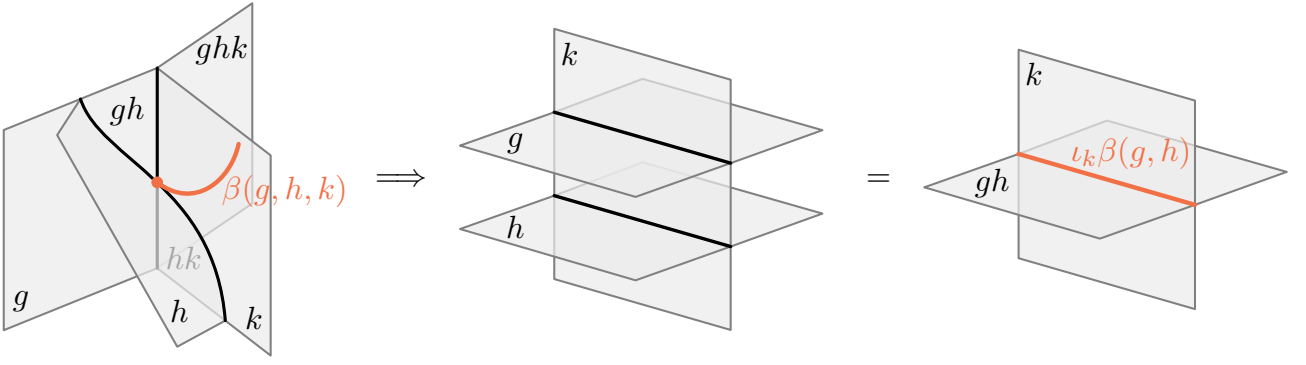
\begin{figure}[t!]
\centering

\resizebox{\textwidth}{!}{%

\begin{tikzpicture}[
    scale=.4,
    x=1cm,
    y=1cm,
    line cap=round,
    line join=round,
    every node/.style={text=black},
    sheet/.style={
        draw=gray,
        line width=.7pt,
        fill=gray!15,
        fill opacity=.7
    },
    junction/.style={
        black,
        line width=1.05pt
    }
]
    \coordinate (O) at (-0.3,-0.15);

    %
    \coordinate (Jghk) at (-0.3, 3.28);
    \coordinate (JgHK) at (-0.3,-2.72);
    
    \coordinate (G1) at (-4.90, 1.42);
    
    \coordinate (H2) at (-0,-5.4);
    
    \coordinate (R1) at (2.55, 5.22);
    
    \coordinate (K1) at (3.10, 0.66);
    \coordinate (K2) at ($(K1) - (Jghk) + (JgHK) $);
    \coordinate (R2) at ($(R1) - (Jghk) + (JgHK) $);
    
    \coordinate (Jgh) at ($0.5*(Jghk) - 0.5*(G1) + (G1)$);
    \coordinate (H1) at ($0.2*(K1) - 0.2*(G1) + (G1)$);
    \coordinate (G2) at ($(G1) - (Jghk) + (JgHK)$);
    \coordinate (H2) at ($0.65*(K2) - 0.65*(G2) + (G2)$);
    \coordinate (Jhk) at ($0.4*(JgHK) - 0.4*(K2) + (K2)$);


    \coordinate (C0) at (1.42, -0.2);
    \coordinate (C4) at (-2,0.5);

    \path[sheet]
        (Jghk) 
        -- (Jgh)
        -- (G1)
        -- (G2)
        -- (JgHK)
        -- cycle;

    \path[sheet]
        (O)
        -- (Jghk)
        -- (R1)
        -- (R2)
        -- (JgHK)
        -- cycle;


    \coordinate (C1) at (0.2, -3);
    \coordinate (C2) at (-3.3,-1.2);

    \path[sheet]
        (Jghk)
        -- (K1)
        -- (K2)
        -- (Jhk)
        -- (JgHK)
        -- cycle;

    \draw[junction] (Jghk) -- (O);
    \draw[junction] (JgHK) -- (O);
    
    \node at (0.5, -2.4) {$hk$};

    \path[sheet]
        (Jgh)
        -- (H1) -- (H2)
        -- (Jhk) .. controls (C0) and (C4) .. (Jgh);

    \draw[junction] (Jhk) .. controls (C0) and (C4) .. (Jgh);


    \node at (-4.38,-3.7) {$g$};
    \node at (0.36,-4) {$h$};
    \node at (2.6,-4.2) {$k$};

    \node at (-1.2, 2) {$gh$};
    \node at ( 1.64, 3.83) {$ghk$};


    \coordinate (C5) at (1.9,-0.1);
    \coordinate (c) at (O);
    \draw[
        junctionorange,
        line width=1.5pt
    ]
        (c)
        .. controls
            (0.82,-1)
            and (C5)
        .. (2.14,1.);

    \fill[junctionorange]
        (c) circle (5pt);

    \node[
        text=junctionorange,
        anchor=west
    ] at (1.3,-0.4)
        {$\bt(g,h,k)$};

\end{tikzpicture}
\hspace{2pt}\(\Longrightarrow\)\quad
\begin{tikzpicture}[
    scale=.5,
    x=1cm,
    y=1cm,
    line cap=round,
    line join=round,
    every node/.style={text=black}
]
    \coordinate (A) at (-2.10,0.58);
    \coordinate (C) at (2.13,-0.64);

    \filldraw[
        draw=gray,
        line width=.7pt,
        fill=gray!15,
        fill opacity=.7
    ]
        (A)
        -- (0.02,1.38)
        -- (4.33,0.15)
        -- (C)
        -- cycle;

    \filldraw[
        draw=gray,
        line width=.7pt,
        fill=gray!15,
        fill opacity=.7
    ]
        ($(A)+(0,2)$)
        -- (0.02,3.38)
        -- (4.33,2.15)
        -- ($(C)+(0,2)$)
        -- cycle;

    \filldraw[
        draw=gray,
        line width=.7pt,
        fill=gray!15,
        fill opacity=.7
    ]
        (-2.10,4.58)
        -- (2.13,3.36)
        -- (2.13,-2.64)
        -- (-2.10,-1.42)
        -- cycle;

    \filldraw[
        draw=gray,
        line width=.7pt,
        fill=gray!15,
        fill opacity=.7
    ]
        (-4.36,-0.22)
        -- (A)
        -- (C)
        -- (-0.15,-1.45)
        -- cycle;

    \filldraw[
        draw=gray,
        line width=.7pt,
        fill=gray!15,
        fill opacity=.7
    ]
        (-4.36,1.78)
        -- ($(A)+(0,2)$)
        -- ($(C)+(0,2)$)
        -- (-0.15,0.55)
        -- cycle;

    \draw[black,line width=1.15pt]
        (A) -- (C);

    \draw[black,line width=1.15pt]
        ($(A)+(0,2)$) -- ($(C)+(0,2)$);

    \node at (-3.03,1.83) {$g$};
    \node at (-3.03,-.17) {$h$};
    \node at (-1.73,3.98) {$k$};
\end{tikzpicture}
\quad\(=\)\quad
\begin{tikzpicture}[
    scale=.5,
    x=1cm,
    y=1cm,
    line cap=round,
    line join=round,
    every node/.style={text=black}
]
    \coordinate (A) at (-2.10,0.58);
    \coordinate (C) at (2.13,-0.64);

    \filldraw[
        draw=gray,
        line width=.7pt,
        fill=gray!15,
        fill opacity=.7
    ]
        ($(A)+(0,.5)$)
        -- (0.02,1.88)
        -- (4.33,.65)
        -- ($(C)+(0,.5)$)
        -- cycle;

    \filldraw[
        draw=gray,
        line width=.7pt,
        fill=gray!15,
        fill opacity=.7
    ]
        (-2.10,3.58)
        -- (2.13,2.36)
        -- (2.13,-2.64)
        -- (-2.10,-1.42)
        -- cycle;

    \filldraw[
        draw=gray,
        line width=.7pt,
        fill=gray!15,
        fill opacity=.7
    ]
        (-4.36,.28)
        -- ($(A)+(0,.5)$)
        -- ($(C)+(0,.5)$)
        -- (-0.15,-.95)
        -- cycle;

    \draw[
        junctionorange,
        line width=1.5pt
    ]
        ($(A)+(0,.5)$) -- ($(C)+(0,.5)$);

    \node at (-2.73,.38) {$gh$};
    \node at (-1.73,3)   {$k$};

    \node[text=junctionorange]
        at (.6,1.1)
        {$\iota_k\bt(g,h)$};
\end{tikzpicture}%
}

\caption{The Postnikov class $[\bt]$ of a 2-group symmetry modifies the 0-form symmetry operator algebra in the presence of a 0-form symmetry defect. 
In the $k$-defect sector, multiplying ${C_G(k)\leq G}$ symmetry operators creates a 1-form symmetry operator according to the slant product ${\iota_k\bt}$.}
\label{fig:beta-junction-resolution}
\end{figure}

\noindent\textbf{Central 2-groups.}
In Sec.~\ref{sec:central-2group-symmetry}, we instead take the $G$-action on $A$ to be trivial and allow a nontrivial Postnikov class. We introduce the 0-form symmetry operators
\begin{equation}\label{eq:central-sym-ops-summary}
    U^{(h,[f])}
    =
    \sum_{\{g_i\},\{a_{ij}\}}
    \ketbra{
        \{hg_i\},
        \left\{
            a_{ij}
            +\bt(h,g_i,g_i^{-1}g_j)
            -f(g_i)+f(g_j)
        \right\}
    }{
        \{g_i\},\{a_{ij}\}
    },
\end{equation}
where ${\bt\in\cZ^3(G,A)}$ and
\begin{equation}
    h\in G,
    \qquad
    [f]\in\Map(G,A)/A_{\mathrm{const}},
\end{equation}
with ${A_{\mathrm{const}}\leq\Map(G,A)}$ the subgroup of constant maps. The group ${\Map(G,A)/A_{\mathrm{const}}\cong A^{|G|-1}}$. 
The operators~\eqref{eq:central-sym-ops-summary} are not onsite. We show that they cannot be made onsite using only a QCA, but can be using finite-dimensional ancillae and a finite-depth circuit.
The operators
\begin{equation}
    S_f
    \equiv
    U^{(1,[f])}
    =
    \sum_{\{g_i\},\{a_{ij}\}}
    \ketbra{
        \{g_i\},
        \{a_{ij}-f(g_i)+f(g_j)\}
    }{
        \{g_i\},\{a_{ij}\}
    }
\end{equation}
furnish a faithful representation of ${\Map(G,A)/A_{\mathrm{const}}}$. The operators ${U^{(h)}\equiv U^{(h,[0])}}$, on the other hand, do not furnish a faithful representation of $G$. Instead,
\begin{equation}
    U^{(g)}U^{(h)}
    =
    U^{(gh)}S_{b_{g,h}},
    \qquad
    b_{g,h}(k)
    =
    \bt(g,h,k).
\end{equation}
We show that the operators~\eqref{eq:central-sym-ops-summary} furnish a faithful representation of a finite group $\t{G}$ that fits into the extension
\begin{equation}\label{eq:summary-tilde-G-extension}
    1
    \longrightarrow
    \Map(G,A)/A_{\mathrm{const}}
    \longrightarrow
    \t{G}
    \overset{p}{\longrightarrow}
    G
    \longrightarrow
    1.
\end{equation}
On the topological subspace
$\scrH_{\mathrm{top}}$, however, every $S_f$ acts trivially:
\begin{equation}
    S_f\big|_{\scrH_{\mathrm{top}}}
    =
    1.
\end{equation}
The operators~\eqref{eq:central-sym-ops-summary} are then $G$ symmetry operators on $\scrH_\mathrm{top}$, and restricting to $\scrH_\mathrm{top}$ implements the quotient ${p\colon \t{G}\to G}$ in~\eqref{eq:summary-tilde-G-extension}. It is interesting to note that, unlike for the split case, for the central 2-group lattice operators, restricting to $\scrH_\mathrm{top}$ also modifies the 0-form symmetry operators in addition to making the lattice 1-form symmetry operators topological.

We show that the symmetry operators~\eqref{eq:summary-1-form-symmetry-ops} and~\eqref{eq:central-sym-ops-summary} on $\scrH_\mathrm{top}$ are central 2-group symmetry operators with Postnikov class $[\bt]$. In particular, we show that these symmetry operators on $\scrH_\mathrm{top}$ realize the following two manifestations of the Postnikov class.
In Sec.~\ref{sec:obstructionToSubSym}, we show that $[\bt]$ is an obstruction to the 0-form symmetry being a subsymmetry: operators carrying $A$ 1-form symmetry charge transform under the 0-form symmetry. Namely, a loop operator of charge ${\chi\in\h{A}\equiv \Hom(A,\Uone)}$ transforms by a so-called anomaly changing operator for the ${1+1}$d $G$ symmetry anomaly $[\chi\circ\bt]\in\cH^3(G,\Uone)$, as expected for 2-group symmetry in QFT~\cite{HJS260804248}.
Then, in Sec.~\ref{sec:central-2group-sym-defects}, we show that in the presence of a ${(k,[s])\in \t{G}}$ symmetry defect, the remaining 0-form symmetry operators on $\scrH_\mathrm{top}$ satisfy the 0-form symmetry group law up to a 1-form symmetry operator appearing on the defect, which is determined by the slant product $\iota_k\bt$ (up to 2-coboundaries associated with operator redefinitions).
See Fig.~\ref{fig:beta-junction-resolution}. As we show in App.~\ref{sec:defectSectorFusionApp}, this is expected from a central 2-group symmetry in QFT.

\noindent\textbf{Symmetric Hamiltonians.} For both the split and central lattice 2-group symmetry operators, we construct families of symmetric local Hamiltonians. Among these are the Hamiltonians
\begin{equation}
    H_{\G\text{-Potts}}^{(\rho)}
    \qquad\text{and}\qquad
    H_{\G\text{-Potts}}^{(\bt)}
\end{equation}
in Eqs.~\eqref{eq:split-2group-pottsmodel} and~\eqref{eq:central-2group-pottsmodel}. The former commutes with the
split lattice 2-group operators ${U^{(h)}}$ and ${T^{(\ga)}}$, while the latter commutes with the central lattice 2-group operators ${U^{(h,[f])}}$ and ${T^{(\ga)}}$. 
These Hamiltonians are generalizations of the Potts model to 2-groups $\G$. 
We show that these $\G$-Potts models have various exactly solvable points that realize spontaneous symmetry breaking phases of the lattice 2-group symmetries. This includes a trivial $\G$ symmetric phase, as well as a phase in which the entire lattice 2-group symmetry is spontaneously broken.

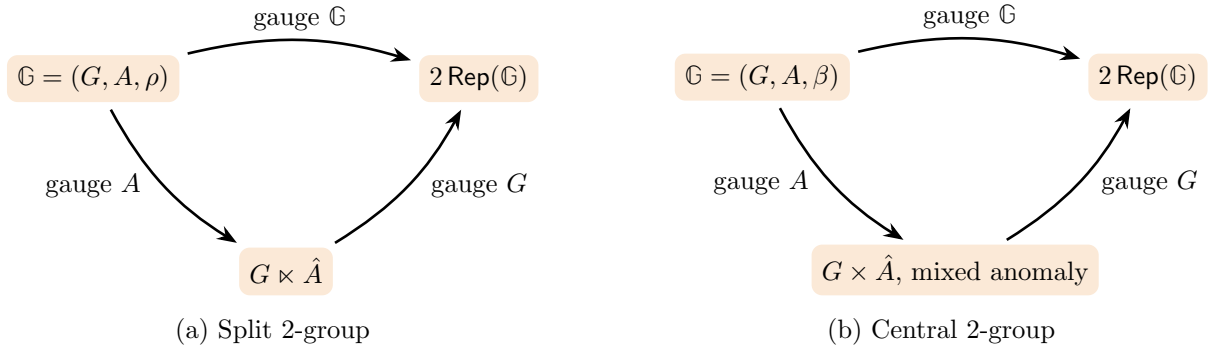
\begin{figure}
    \centering
    \tikzset{
        arr/.style={->, >={Stealth}, line width=1pt,
                    shorten >=3pt, shorten <=3pt},
        vtx/.style={fill=orange!70!lightgray!20, draw=none, rounded corners,
                    inner sep=0.35em, align=center},
    }
    \begin{subfigure}{0.48\textwidth}
        \centering
        \begin{tikzpicture}[
            baseline={([yshift=-.5ex]current bounding box.center)},
            scale=0.85,
            every node/.append style={font=\rmfamily\small},
            ]
            \node[vtx] (topleft)  at (-3, 3) {$\G=(G, A, \rho)$};
            \node[vtx] (topright) at ( 3, 3) {$2\Rep(\G)$};
            \node[vtx] (bot)      at ( 0, 0) {$G\ltimes \h{A}$};

            \draw[arr, bend left=15]  (topleft) to node[midway, above]            {gauge $\G$} (topright);
            \draw[arr, bend right=15] (topleft) to node[midway, left,  xshift=-1ex] {gauge $A$}         (bot);
            \draw[arr, bend right=15] (bot)     to node[midway, right, xshift=1ex]  {gauge $G$}         (topright);
        \end{tikzpicture}
        \caption{Split 2-group}
        \label{fig:2group-split}
    \end{subfigure}
    \hfill
    \begin{subfigure}{0.48\textwidth}
        \centering
        \begin{tikzpicture}[
            baseline={([yshift=-.5ex]current bounding box.center)},
            scale=0.85,
            every node/.append style={font=\rmfamily\small},
            ]
            \node[vtx] (topleft)  at (-3, 3) {$\G=(G, A, \bt)$};
            \node[vtx] (topright) at ( 3, 3) {$2\Rep(\G)$};
            \node[vtx] (bot)      at ( 0, 0) {$G\times \h{A}$, mixed anomaly};

            \draw[arr, bend left=15]  (topleft) to node[midway, above]            {gauge $\G$} (topright);
            \draw[arr, bend right=15] (topleft) to node[midway, left,  xshift=-1ex] {gauge $A$}         (bot);
            \draw[arr, bend right=15] (bot)     to node[midway, right, xshift=1ex]  {gauge $G$}         (topright);
        \end{tikzpicture}
        \caption{Central 2-group}
        \label{fig:2group-central}
    \end{subfigure}
    \caption{Gauging web of 2-group symmetry operators in field theory. In the split 2-group case, gauging the 1-form $A$-symmetry leads to a dual semidirect product 0-form $G\ltimes \hat{A}$-symmetry for $\hat{A}=\Hom(A,\text{U}(1))$. In the central 2-group case, gauging the 1-form $A$-symmetry leads to a direct product 0-form $G\times\hat{A}$-symmetry with mixed anomaly depending on $[\beta]$. Gauging the 0-form $G$ symmetry in both cases leads to a dual $2\Rep \mathbb{G}$ symmetry.}
    \label{fig:2group-gauging-web}
\end{figure}

\noindent\textbf{Gauging and dual symmetries.}
In Secs.~\ref{sec:split-gauging} and~\ref{sec:central-gauging}, we gauge the lattice 2-group symmetries using a two-step gauging procedure. 
In both cases, we first gauge the lattice $A$ 1-form symmetry. In
the split case, this produces the dual 0-form symmetry
\begin{equation}
    G\ltimes\h{A},
\end{equation}
where the $G$-action on $\h{A}$ is given by ${\rho_g\triangleright\chi = \chi\circ\rho_{g^{-1}}}$.
In the central case, the gauging procedure for the lattice $A$ 1-form symmetry also trivializes each $S_f$, causing the $\t{G}$ 0-form symmetry to become a $G$ 0-form symmetry. The full dual 0-form symmetry is 
\begin{equation}
    G\times\h{A}.
\end{equation}
We show it has a mixed anomaly determined by $[\bt]$. Gauging the remaining $G$ 0-form symmetry in both cases produces a dual symmetry which is a lattice realization of the fusion 2-category
\begin{equation}
    2\Rep(\G).
\end{equation}
In particular, there is a lattice $\Rep(G)$ 1-form symmetry operator, and in its topological subspace the dual symmetry operators realize the $2\Rep(\G)$ symmetry.
We discuss the 0-form part of these lattice operators in detail in App.~\ref{app:non-invertible-minimal-coupling}.
In the Hilbert space where all lattice 1-form symmetry operators are topological, these gauging procedures yield the gauging web displayed in Fig.~\ref{fig:2group-gauging-web}, which agrees with the expectations of gauging split and central 2-group symmetries in QFT.

\noindent\textbf{Lattice 2-group gauge theory.}
We also minimally couple the $\G$-Potts Hamiltonians through this gauging procedure to find Hamiltonian realizations of 2-group lattice gauge theory. At convenient fixed points in their deconfined phases, they become
the split and central 2-group quantum-double Hamiltonians
\begin{equation}
    H_{\G}^{(\rho)}
    \qquad\text{and}\qquad
    H_{\G}^{(\bt)}
\end{equation}
defined in Eqs.~\eqref{eq:split-quantum-double} and~\eqref{eq:central-quantum-double}. These are exactly solvable and generalizations of Kitaev's quantum double model for 2-groups. As shown in App.~\ref{app:double-from-path-integral}, their ground-state subspaces are naturally identified with the Hilbert spaces of untwisted finite split and central 2-group gauge theories, respectively. In both the split and central cases, we find closed-form expressions for their ground state degeneracies on the spacial sphere and torus.

\noindent\textbf{Examples.}
Finally, in Secs.~\ref{sec:split-example} and~\ref{sec:central-example}, we illustrate the general constructions in two simple nontrivial, self-contained examples. For the split case, we take ${G=\Z_2}$ and ${A=\Z_n}$, with the nontrivial element of $\Z_2$ acting by charge conjugation: ${\rho_{-1}(\la)=-\la}$. For the central example, we take ${G=A=\Z_2}$ and choose the nontrivial type-I Postnikov cocycle which causes the microscopic 0-form symmetry group to be ${\t{G} = \Z_4}$. In both cases, using conventional quantum gates, we write down their respective lattice 2-group symmetry operators, Potts model Hamiltonians, $2\Rep(\G)$ symmetry operators, and 2-group quantum double models.
Together, these examples give concrete realizations of the distinct roles played by the action $\rho$ in a split 2-group and by the Postnikov class $[\bt]$ in a central 2-group.

\section{Split 2-group symmetry}\label{sec:split-2group-symmetry}

This section presents and explores symmetry operators for a split 2-group symmetry. By split 2-group, we mean a 2-group whose Postnikov class is trivial. The data specifying a split 2-group $\G$ is ${(G,A,\rho)}$. We assume that the 0-form symmetry group $G$ is finite, and that the 1-form symmetry group $A$ is finite abelian. 
(We use multiplicative notation for the $G$ group law, and additive notation for the $A$ group law.) There is an action of $G$ on $A$, which is described by the group homomorphism
\begin{equation}
\begin{aligned}
    \rho\colon G &\to \Aut(A)\\
    g &\mapsto \rho_g,
\end{aligned}
\end{equation}
where $\Aut(A)$ is the automorphism group of $A$. We note that the $G$-action on $A$ induces a $G$-action on ${\h{A}\equiv \Hom(A,\Uone)}$ given by
\begin{equation}\label{rhoActonRep}
    \rho_{g} \triangleright \chi = \chi\circ\rho_{g^{-1}}, \qquad \chi\in\h{A}.
\end{equation}

\subsection{Symmetry operators}\label{sec:split-2group-sym-ops}

Consider a ${2+1}$d quantum lattice system. We take the underlying two-dimensional space to be a torus, and the spatial lattice to be a triangular lattice $\La$ with finitely many sites. More precisely, $\La$ is a simplicial complex whose sites form a triangular lattice. We denote a general 0-, 1-, and 2-simplex of $\La$ by $i$, $ij$, and $ijk$ with ${i<j<k}$ and use the branching structure shown in Fig.~\ref{fig:lattice-branching-structure}. Furthermore, we will refer to 0-simplices as sites, 1-simplices as links, and 2-simplices as plaquettes.
We denote by $\La_n$ the set of all $n$-simplices of $\La$.
On each lattice site ${i\in\La_0}$ resides a $G$-qudit ${\scrH_i = \mathrm{span}_\C(\ket{g}\mid g\in G)}$, and on each link ${ij\in\La_1}$ resides an $A$-qudit ${\scrH_{ij} = \mathrm{span}_\C(\ket{a}\mid a\in A)}$.
The total Hilbert space of the system admits the tensor product decomposition
\begin{equation}\label{GsitesAlinksHilb1}
    \scrH = 
    \bigotimes_{i\in\La_0} \C[G]
    \otimes 
    \bigotimes_{ij\in\La_1} \C[A].
\end{equation}
It admits a natural basis formed by the states ${\bigotimes_{i\in\La_0} \ket{g_i}\bigotimes_{ij\in\La_1} \ket{a_{ij}}\equiv \ket{\{g_i\}, \{a_{ij}\}}}$.

We first present unitary operators on~\eqref{GsitesAlinksHilb1} that are lattice split 2-group ${\G = (G,A,\rho)}$ symmetry operators.

The $G$ 0-form symmetry operators are ${\{U^{(h)}\}_{h\in G}}$ with
\begin{equation}\label{eq:split-2group-G-sym-op}
    U^{(h)} = \prod_{i\in\La_0} \overrightarrow{X}_i^{(h)}
    \prod_{ij\in\La_1}
    P_{ij}^{(\rho_h)}.
\end{equation}
The generalized Pauli operator $\overrightarrow{X}_i^{(h)}$ acts on $G$-qudits as ${\overrightarrow{X}^{(h)}\ket{g} = \ket{hg}}$ whereas $P_{ij}^{(\rho_h)}$ acts on $A$-qudits as ${P^{(\rho_h)}\ket{a} = \ket{\rho_h(a)}}$. (See App.~\ref{app:group-qudit-review} for a review of these group-valued qudit operators.)
The $G$ group law
\begin{equation}
    U^{(h_1)}U^{(h_2)} = U^{(h_1h_2)}
\end{equation}
follows from ${\overrightarrow{X}^{(h_1)}\overrightarrow{X}^{(h_2)} = \overrightarrow{X}^{(h_1h_2)}}$ and ${P^{(\rho_{h_1})}P^{(\rho_{h_2})} = P^{(\rho_{h_1 h_2})}}$.

The lattice $A$ 1-form symmetry operators are ${\{T^{(\ga)}\}_{\ga\in Z_1(\La^\vee; A)}}$ with\footnote{While the 1-form symmetry operator~\eqref{eq:split-2group-A-sym-op} includes a product over all links, it will act nontrivially only on 1-cycles on the dual lattice since $\ga$ is a 1-cycle.}
\begin{equation}\label{eq:split-2group-A-sym-op}
    T^{(\ga)} = \prod_{ij\in\La_1} X^{(\ga_{ij})}_{ij}.
\end{equation}
Here, ${\ga \equiv \sum_{ij\in\La_1} \ga_{ij}\, [ij]^\vee}$ is an $A$-valued $1$-cycle of the dual lattice $\La^\vee$.
Namely, each ${\ga_{ij}\in A}$ and ${[ij]^\vee}$ denotes the link dual to $[ij]$. 
The group $Z_1(\La^\vee; A)$ is the group of all $A$-valued $1$-cycles on $\La^\vee$. Its group law is ${\ga_1 + \ga_2 \equiv \sum_{ij\in\La_1} ((\ga_1)_{ij} + (\ga_2)_{ij})\, [ij]^\vee}$.
It follows from ${X^{(a_1)}X^{(a_2)} = X^{(a_1+a_2)}}$ that the lattice 1-form symmetry operators $T^{(\ga)}$ satisfy the $Z_1(\La^\vee; A)$ group law: 
\begin{equation}\label{latticeA1FSgrouplaw}
    T^{(\ga_1)}T^{(\ga_2)} = T^{(\ga_1+\ga_2)}.
\end{equation}

These lattice $G$ 0-form and $A$ 1-form symmetry operators have a nontrivial interplay and satisfy
\begin{equation}\label{split2grpOperatorAction}
    U^{(h)}\, T^{(\ga)}\, U^{(h)^\dag} = T^{(\rho_h\triangleright\, \ga)} ,
\end{equation}
where ${\rho_h\triangleright\ga \equiv \sum_{ij} \rho_h(\ga_{ij})\, [ij]^\vee}$. This is precisely the operator algebra for the split 2-group symmetry $\G$.
Therefore,~\eqref{eq:split-2group-G-sym-op} and~\eqref{eq:split-2group-A-sym-op} are split 2-group symmetry operators.
These $\G$ symmetry operators are onsite.

The lattice 1-form symmetry operators~\eqref{eq:split-2group-A-sym-op} are not topological operators on the full Hilbert space. They are topological operators generating a 1-form symmetry in the subspace
\begin{equation}\label{HtopDefSplit2GrpSection}
    \scrH_\text{top} \!=\! \left\{\ket{\psi}\!\in\!\scrH \mid A_i^{(\la)}\ket{\psi} = \ket{\psi}\!~\forall~i\in\La_0,~\la\in A\right\},
    \quad
    (A_i^{(\la)} =
X_{\trivertex{w}}^{(\la)}
X_{\trivertex{sw}}^{(\la)}
X_{\trivertex{se}}^{(\la)}
X_{\trivertex{e}}^{(-\la)}
X_{\trivertex{ne}}^{(-\la)} X_{\trivertex{nw}}^{(-\la)})
\end{equation}
where the symbols $\trivertex{w}$, $\trivertex{sw}$, etc., denote the corresponding links emanating from the site $i$.
A convenient basis of $\scrH_\text{top}$ is given by the vectors
\begin{equation}\label{Htopbasisvectors}
    \ket{\left\{g_i\right\}, \left\{[a_{ij}]\right\}} = \sum_{\la\in C^0(\La,A)} \ket{\left\{g_i\right\}, \left\{a_{ij}+\la_j - \la_i\right\}},
\end{equation}
where ${[a_{ij}] = [a_{ij} + \la_j - \la_i]}$ for each ${\la\in C^0(\La,A)}$.
The 1-form symmetry operators satisfy
\begin{equation}
    T^{(\ga)}\ket{\left\{g_i\right\}, \left\{[a_{ij}]\right\}} =  T^{(\ga + \pp \la^\vee)}\ket{\left\{g_i\right\}, \left\{[a_{ij}]\right\}},
\end{equation}
for every $A$-valued 1-boundary ${\pp\la^\vee\in B_1(\La^\vee; A)}$ of the dual lattice.
Therefore, in $\scrH_\text{top}$, the 1-form symmetry operators $T^{(\ga_1)}$ and $T^{(\ga_2)}$ are equivalent whenever $\ga_1$ and $\ga_2$ are homologous.
Importantly, we note that the 2-group action~\eqref{split2grpOperatorAction} applies both when $T^{(\ga)}$ is topological and when it is not. (As we discuss in Sec.~\ref{sec:central-2group-symmetry}, this is not true for nonsplit 2-groups.)

\subsubsection{Split 2-group Ising model}\label{sec:split-2group-ising-model}

There are infinitely many local Hamiltonians that commute with the split 2-group symmetry operators~\eqref{eq:split-2group-G-sym-op} and~\eqref{eq:split-2group-A-sym-op}. For example, consider the local Hamiltonian\footnote{The first term of~\eqref{eq:split-2group-isingmodel} satisfies ${\tr[Z_i^{(\Ga)} Z_{j}^{(\bar{\Ga})}] = \sum_{\{g_k\}} \tr_\Ga(g_ig_j^{-1})\ketbra{\{g_k\}}{\{g_k\}}}$.}
\begin{align}\label{eq:split-2group-isingmodel}
    H_{\G\text{-Ising}}^{(\rho)} = 
    -\frac12\bigg(& \sum_{ij\in\La_1} \sum_{\Ga\in\Irr G} J^{(\Ga)}_G \tr[Z_i^{(\Ga)} Z_{j}^{(\bar{\Ga})}]
    + \sum_{i\in \La_0} \sum_{g\in G}  h^{\!(g)}_G\, \overleftarrow{X}_{i}^{(g)}
    \\
    &
    +  \sum_{ijk\in\La_2} \sum_{\chi\in \h{A}}  J_A^{(\chi)} B_{ijk}^{(\rho_{g_i}\triangleright \chi)}
    + \sum_{ij\in \La_1} \sum_{\la\in A} h^{\!(\la)}_A X^{(\rho_{g_i}(\la))}_{ij}
    + \sum_{i\in\La_0} \frac{1}{|A|}  \sum_{\la \in A}  A_i^{(\la)}\bigg)
    +
    \mathrm{H.c.}
    ,\nonumber
\end{align}
where ${J^{(\Ga)}_G, h^{\!(g)}_G, J^{(\chi)}_A, h^{\!(\la)}_A \in \R}$,
 $\bar{\Ga}$ satisfies ${\bar{\Ga}(g) = \Ga(g^{-1})}$,
and the plaquette operator 
\begin{equation}
    B_{ijk}^{(\chi)} =
Z_{ij}^{(\chi)} 
Z_{jk}^{(\chi)}
Z_{ik}^{(\chi)\,\dag}.
\end{equation}
This Hamiltonian commutes with the split 2-group symmetry operators, and we will refer to it as the split $\G$-Ising model.
The first two terms acting only on the $G$-qudits are the $G$-Ising model Hamiltonian.
The third and fourth terms are generalizations of the $A$-Higgs model Hamiltonian with
\begin{align}
    X_{ij}^{(\rho_{g_i}(\la))} &= \sum_{\{g_{l}\}, \{a_{l k}\}} \ketbra{\{g_{l}\}, \{a_{l k} + \del_{ij,l k}\,\rho_{g_i}(\la)\}}{\{g_{l}\}, \{a_{l k}\}},\\
    Z_{ij}^{(\rho_{g_i}\triangleright \chi)} &= \sum_{\{g_{l}\}, \{a_{l k}\}} 
    \chi(\rho_{g_i^{-1}}(a_{ij}))
    \ketbra{\{g_{l}\}, \{a_{l k}\}}{\{g_{l}\}, \{a_{l k}\}}.
\end{align}
The $\rho_{g_i}$s appearing in these operators are necessary for them to commute with the $G$ 0-form symmetry operators $\{U^{(h)}\}_{h\in G}$ for generic coupling constants.
Lastly, the fifth term enforces the low-energy subspace of $H_{\G~\mathrm{Ising}}$ to lie within the topological subspace~\eqref{HtopDefSplit2GrpSection} of the $A$ 1-form symmetry operator.

The $\G$ Ising model Hamiltonian~\eqref{eq:split-2group-isingmodel} has a rich phase diagram. For instance, consider the case where each ${J^{(\Ga)}_G = d_\Ga/|G|}$ (with $d_\Ga$ the dimension of the irrep $\Ga$), ${h^{\!(g)}_G = h_G/|G|}$, ${J^{(\chi)}_A = 1/|A|}$, and ${h^{\!(\la)}_A = h_A/|A|}$. In this limit, the Hamiltonian can be simplified to
\begin{equation}\label{eq:split-2group-pottsmodel}
\begin{aligned}
    H_{\G\text{-Potts}}^{(\rho)} = 
    &- \frac1{|G|} \bigg( \sum_{ij\in\La_1} \sum_{\Ga\in\Irr G} d_\Ga \tr[Z_i^{(\Ga)} Z_{j}^{(\bar{\Ga})}]
    + h_G \, \sum_{i\in \La_0} \sum_{g\in G}  \, \overleftarrow{X}_{i}^{(g)}
    \bigg)
    \\
    &
    - \frac1{|A|} \bigg( \sum_{ijk\in\La_2} \sum_{\chi\in \h{A}}   B_{ijk}^{(\chi)}
    + \sum_{i\in\La_0}   \sum_{\la \in A}  A_i^{(\la)}
    + h_A \sum_{ij\in \La_1} \sum_{\la\in A} X^{(\la)}_{ij}
    \bigg)
    .
\end{aligned}
\end{equation}
We call this the split $\G$-Potts model.
Note that it is independent of $\rho$ and the $G$- and $A$-qudits have decoupled. 
The first two terms are the $G$-Potts Hamiltonian, while the last three terms are the $A$ quantum-double Hamiltonian deformed by a string-tension term.
It has four exactly-solvable points, which lie in the following gapped $\G$-SSB phases:
\setlength{\tabcolsep}{8pt} \renewcommand{\arraystretch}{1.5} 
\begin{center}
\begin{tabular}{c|c|c} 
  \quad Parameters \quad   &  \quad  $\G$-SSB pattern \quad  & \quad  ground-state degeneracy\quad  \\ 
\hhline{=|=|=}
$h_G = h_A = 0$  & $\G\ssb 1$ & $|H^0(\La,G)| \, |H^1(\La,A)|$  \\
\hline 
$h_G = 0, h_A \to\infty$  & $\G\ssb A^{(1)}$ &  $|H^0(\La,G)|$ \\
\hline 
$h_G \to\infty, h_A =0$  & $\G\ssb G^{(0)}$ &  $|H^1(\La,A)|$ \\
\hline 
$h_G, h_A \to\infty $  & $\G\ssb \G$ &  1 \\
\end{tabular}
\end{center}
\renewcommand{\arraystretch}{1}
Since we assume that $\La$ triangulates a torus, ${|H^0(\La,G)| = |G|}$ and ${|H^1(\La,A)| = |A|^2}$.
The phase with SSB pattern ${\G\ssb G^{(0)}}$ is an SET. In this phase, the 0-form $G$ symmetry operators~\eqref{eq:split-2group-G-sym-op} provide a microscopic realization of the anyon automorphism of a $G$-SET~\cite{BBCW14104540}.

\subsection{Symmetry defects}\label{sec:split-2group-sym-defects}

The lattice 0-form and 1-form symmetry operators~\eqref{eq:split-2group-G-sym-op} and \eqref{eq:split-2group-A-sym-op} have corresponding codimension-1 and -2 symmetry defects, respectively.
Since both symmetry operators are onsite, it is straightforward to insert their symmetry defects using their truncated symmetry operators.
(See~\cite{S230805151} for a systematic discussion of inserting symmetry defects in ${1+1}$d, which straightforwardly generalizes to ${2+1}$d.)

The symmetry operator action~\eqref{split2grpOperatorAction} of $G^{(0)}$ on $A^{(1)}$ implies an action at the level of symmetry defects.
Suppose there is a ${\la\in A}$ 1-form symmetry defect at ${ijk\in\La_2}$.
A unitary operator that moves this symmetry defect from the plaquette $ijk$ to the plaquette $lmn$ is
\begin{equation}\label{1-formsymDefectMoveOp}
    \prod_{ij\in \La_1} X_{ij}^{(C_{ij}^\vee)} 
\end{equation}
where $C^\vee$ is any $A$-valued 1-chain on $\La^\vee$ satisfying ${\pp C^\vee = \la\,[lmn]^\vee - \la\, [ijk]^\vee}$.
We now insert an ${h\in G}$ 0-form symmetry defect along the dual 1-cycle $\eta^\vee$. 
One way to do this is to put the lattice on $\R^2$ and consider the truncated operator 
\begin{equation}
    D_{\eta^\vee}^{(h)} = \prod_{i\text{ right of }\eta^\vee} \overrightarrow{X}_i^{(h)}
        P^{(\rho_h)}_{\trivertex{e}_i}
        P^{(\rho_h)}_{\trivertex{ne}_i}
        P^{(\rho_h)}_{\trivertex{se}_i}.
\end{equation}
The site and link qudits nontrivially acted upon by $D^{(h)}_{\eta^\vee}$ are indicated by the gray dots:
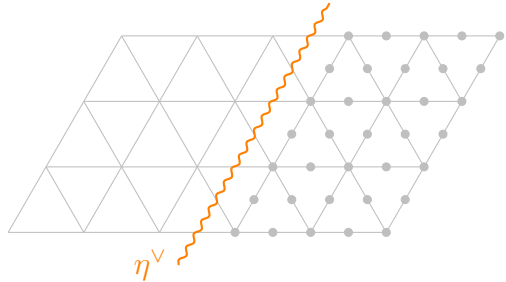
\begin{equation*}
\begin{tikzpicture}[scale=1,
    decoration={markings, mark=at position 0.55 with {\arrow{>}}},
    d/.style={postaction=decorate}
]

    \def\Nx{5}
    \def\Ny{3}

    \pgfmathsetmacro{\a}{1}
    \pgfmathsetmacro{\b}{0.8660254}  

    \foreach \i in {0,...,\Ny} {
        \foreach \j in {0,...,\Nx} {

            \coordinate (v_\i_\j)
                at (\j*\a + 0.5*\i*\a, \i*\b);

        }
    }

    \foreach \i in {0,...,\Ny} {
        \foreach \j in {0,...,\numexpr\Nx-1} {

            \pgfmathtruncatemacro{\jp}{\j+1}

            \draw[color=lightgray] (v_\i_\j) -- (v_\i_\jp);

            \coordinate (c_h_\i_\j)
                at ($(v_\i_\j)!0.5!(v_\i_\jp)$);

        }
    }

    \foreach \i in {0,...,\numexpr\Ny-1} {
        \foreach \j in {0,...,\Nx} {

            \pgfmathtruncatemacro{\ip}{\i+1}

            \draw[color=lightgray] (v_\i_\j) -- (v_\ip_\j);

            \coordinate (c_dr_\i_\j)
                at ($(v_\i_\j)!0.5!(v_\ip_\j)$);

        }
    }

    \foreach \i in {0,...,\numexpr\Ny-1} {
        \foreach \j in {1,...,\Nx} {

            \pgfmathtruncatemacro{\ip}{\i+1}
            \pgfmathtruncatemacro{\jm}{\j-1}

            \draw[color=lightgray] (v_\i_\j) -- (v_\ip_\jm);

            \coordinate (c_dl_\i_\j)
                at ($(v_\i_\j)!0.5!(v_\ip_\jm)$);

        }
    }
    \filldraw[lightgray] (c_h_0_3) circle (1.5pt);
    \filldraw[lightgray] (c_h_1_3) circle (1.5pt);
    \filldraw[lightgray] (c_h_2_3) circle (1.5pt);
    \filldraw[lightgray] (c_h_3_3) circle (1.5pt);

    \filldraw[lightgray] (c_h_0_4) circle (1.5pt);
    \filldraw[lightgray] (c_h_1_4) circle (1.5pt);
    \filldraw[lightgray] (c_h_2_4) circle (1.5pt);
    \filldraw[lightgray] (c_h_3_4) circle (1.5pt);
    
    \filldraw[lightgray] (c_dl_0_4) circle (1.5pt);
    \filldraw[lightgray] (c_dl_1_4) circle (1.5pt);
    \filldraw[lightgray] (c_dl_2_4) circle (1.5pt);

    \filldraw[lightgray] (c_dl_0_5) circle (1.5pt);
    \filldraw[lightgray] (c_dl_1_5) circle (1.5pt);
    \filldraw[lightgray] (c_dl_2_5) circle (1.5pt);

    \filldraw[lightgray] (c_dr_0_5) circle (1.5pt);
    \filldraw[lightgray] (c_dr_1_5) circle (1.5pt);
    \filldraw[lightgray] (c_dr_2_5) circle (1.5pt);

    \filldraw[lightgray] (c_dr_0_4) circle (1.5pt);
    \filldraw[lightgray] (c_dr_1_4) circle (1.5pt);
    \filldraw[lightgray] (c_dr_2_4) circle (1.5pt);

    \filldraw[lightgray] (c_dr_0_3) circle (1.5pt);
    \filldraw[lightgray] (c_dr_1_3) circle (1.5pt);
    \filldraw[lightgray] (c_dr_2_3) circle (1.5pt);

    \filldraw[lightgray] (v_0_3) circle (1.5pt);
    \filldraw[lightgray] (v_1_3) circle (1.5pt);
    \filldraw[lightgray] (v_2_3) circle (1.5pt);
    \filldraw[lightgray] (v_3_3) circle (1.5pt);

    \filldraw[lightgray] (v_0_4) circle (1.5pt);
    \filldraw[lightgray] (v_1_4) circle (1.5pt);
    \filldraw[lightgray] (v_2_4) circle (1.5pt);
    \filldraw[lightgray] (v_3_4) circle (1.5pt);

    \filldraw[lightgray] (v_0_5) circle (1.5pt);
    \filldraw[lightgray] (v_1_5) circle (1.5pt);
    \filldraw[lightgray] (v_2_5) circle (1.5pt);
    \filldraw[lightgray] (v_3_5) circle (1.5pt);
    
    \draw[orange, thick, decorate, decoration={snake, amplitude=.3mm, segment length=2mm}] ($(c_h_0_2)+0.5*(-0.5, -\b)$) node[left] {$\eta^\vee$} -- ($(c_h_3_2)+0.5*(0.5,\b)$);
\end{tikzpicture}
\end{equation*}
The 1-form symmetry defect movement operator~\eqref{1-formsymDefectMoveOp} in the presence of this 0-form symmetry defect becomes
\begin{equation}
    D_{\eta^\vee}^{(h)}
    \left(\prod_{ij\in \La_1} X_{ij}^{(C_{ij}^\vee)}\right)
    D_{\eta^\vee}^{(h)\dag}
    = 
    \prod_{ij\text{ left of }\eta^\vee} X_{ij}^{(C_{ij}^\vee)}
    \prod_{ij\text{ right of }\eta^\vee} X_{ij}^{(\rho_h(C_{ij}^\vee)\,)}.
\end{equation}
For example, the movement operator acting horizontally only near the 0-form symmetry defect may look like:
\begin{equation}\label{eq:crossing-action-1form-defect}
\begin{tikzpicture}[scale=1.5,
]

    \def\Nx{5}
    \def\Ny{1}

    \pgfmathsetmacro{\a}{1}
    \pgfmathsetmacro{\b}{0.8660254}  

    \foreach \i in {0,...,\Ny} {
        \foreach \j in {0,...,\Nx} {

            \coordinate (v_\i_\j)
                at (\j*\a + 0.5*\i*\a, \i*\b);

        }
    }

    \foreach \i in {0,...,\Ny} {
        \foreach \j in {0,...,\numexpr\Nx-1} {

            \pgfmathtruncatemacro{\jp}{\j+1}

            \draw[color=lightgray] (v_\i_\j) -- (v_\i_\jp);

            \coordinate (c_h_\i_\j)
                at ($(v_\i_\j)!0.5!(v_\i_\jp)$);

        }
    }

    \foreach \i in {0,...,\numexpr\Ny-1} {
        \foreach \j in {0,...,\Nx} {

            \pgfmathtruncatemacro{\ip}{\i+1}

            \draw[color=lightgray] (v_\i_\j) -- (v_\ip_\j);

            \coordinate (c_dr_\i_\j)
                at ($(v_\i_\j)!0.5!(v_\ip_\j)$);

        }
    }

    \foreach \i in {0,...,\numexpr\Ny-1} {
        \foreach \j in {1,...,\Nx} {

            \pgfmathtruncatemacro{\ip}{\i+1}
            \pgfmathtruncatemacro{\jm}{\j-1}

            \draw[color=lightgray] (v_\i_\j) -- (v_\ip_\jm);

            \coordinate (c_dl_\i_\j)
                at ($(v_\i_\j)!0.5!(v_\ip_\jm)$);

        }
    }
    \draw[orange, thick, decorate, opacity=0.5, decoration={snake, amplitude=.3mm, segment length=2mm}] ($(c_h_0_2)+0.5*(-0.5, -\b)$) node[left] {$\eta^\vee$} -- ($(c_h_1_2)+0.5*(0.5,\b)$);

    \filldraw[mygreen!60] (c_dr_0_0) node[below] {$X^{(\la)}$} circle (1pt);
    \filldraw[mygreen!60] (c_dr_0_1) node[below] {$X^{(\la)}$}  circle (1pt);
    \filldraw[mygreen!60] (c_dr_0_2) node[below] {$X^{(\la)}$}  circle (1pt);
    \filldraw[mygreen] (c_dr_0_3) node[below] {$X^{(\rho_h(\la))}$}  circle (1pt);
    \filldraw[mygreen] (c_dr_0_4) node[below] {$X^{(\rho_h(\la))}$}  circle (1pt);
    \filldraw[mygreen] (c_dr_0_5) node[below] {$X^{(\rho_h(\la))}$}  circle (1pt);

    \filldraw[mygreen!60] (c_dl_0_1) circle (1pt);
    \filldraw[mygreen!60] (c_dl_0_2) circle (1pt);
    \filldraw[mygreen!60] (c_dl_0_3) node[above] {$X^{(\la)}$} circle (1pt);
    \filldraw[mygreen] (c_dl_0_4) circle (1pt);
    \filldraw[mygreen] (c_dl_0_5) circle (1pt);
\end{tikzpicture}
\end{equation}
Therefore, moving a $\la$ 1-form symmetry defect through the $h$ 0-form symmetry defect transforms it into a $\rho_h(\la)$ 1-form symmetry defect.
This reproduces the interplay between split $\G$ symmetry defects well known in field theory~\cite{BH180309336}.

\subsection{Gauging and dual symmetries}\label{sec:split-gauging}

Thus far, we have introduced the lattice split 2-group $\G$ symmetry operators~\eqref{eq:split-2group-G-sym-op} and~\eqref{eq:split-2group-A-sym-op} and discussed their symmetry defects. We now gauge this $\G$ symmetry. We do so in two steps, first gauging the lattice $A$ 1-form symmetry and then gauging the $G$ 0-form symmetry.
Furthermore, we contextualize this gauging by tracking what the split $\G$-Potts model Hamiltonian~\eqref{eq:split-2group-pottsmodel} becomes after each step.

\subsubsection{\texorpdfstring{Gauging $A^{(1)}$: $\G\to G\ltimes \h{A}$}{Gauging \textit{A} 1-form symmetry}}\label{sec:split-gauge-a-1form}

We first gauge the lattice $A$ 1-form symmetry ${\{T^{(\ga)}\}_{\ga\in Z_1(\La^\vee; A)}}$ following the standard procedure. Namely, we introduce additional $A$-qudits onto the plaquettes of $\La$ and enforce a Gauss law constraint. This leads to a dual ${\h{A} = \text{Hom}(A,\Uone)}$ 0-form symmetry, and the split $\G$ symmetry becomes a ${G\ltimes \h{A}}$ 0-form symmetry. The action of $G$ on $\h{A}$ is given by~\eqref{rhoActonRep}.
This matches the field theory result~\cite{T171209542}.

\noindent\textbf{Enlarge Hilbert space.}
We introduce an $A$-qudit on each plaquette. This enlarges the Hilbert space to
\begin{equation}
    \scrH_{\mathrm{enlarged}~1} = \bigotimes_{i\in\La_0} \C[G]
    \otimes
    \bigotimes_{ij\in\La_1} \C[A]
    \otimes 
    \bigotimes_{ijk\in\La_2} \C[A].
\end{equation}
\noindent\textbf{Gauss law.}
Next, we define the Gauss operator that implements the gauging. We use the mutually commuting Gauss operators
\begin{equation}\label{eq:gauss-law-gauge-a-1form}
    G^{(\la)}_{jk} = 
    \begin{cases}
        X^{(\la)}_{ijk} 
        X^{(\la)}_{jk}
        X^{(\la)}_{jkl}
        \qquad &i<j<k<l,\\
        X^{(-\la)}_{jik} 
        X^{(\la)}_{jk}
        X^{(-\la)}_{jlk}
        \qquad &j<i<l<k.
    \end{cases}
\end{equation}
Graphically, depending on the link $jk$, the Gauss operator is
\begin{equation}\label{eq:gauss-law-gauge-a-1form-graphical}
    \begin{tikzpicture}[decoration={markings, mark=at position 0.55 with {\arrow{>}}}, scale=2]
        \draw[postaction=decorate,color=lightgray] (0, 0) -- (0.5, 0.866025);
        \draw[postaction=decorate,color=lightgray] (0, 0) node[below] {\footnotesize $i$} -- (1, 0); 
        \draw[postaction=decorate,color=lightgray] (1, 0) node[below,color=lightgray] {\footnotesize $j$} --node[color=black, anchor=mid] {\footnotesize $X^{(\la)}_{jk}$} (0.5, 0.866025) node[above] {\footnotesize $k$};
        \draw[postaction=decorate,color=lightgray] (0.5, 0.866025) -- (1.5, 0.866025) ;
        \draw[postaction=decorate,color=lightgray] (1, 0) -- (1.5, 0.866025) node[above] {\footnotesize $l$};
        \node () at (0.5, 0.188675) {\footnotesize $X^{(\la)}_{ijk}$};
        \node () at (1,  0.67735) {\footnotesize $X^{(\la)}_{jkl}$};
    \end{tikzpicture}
    ,\qquad
    \begin{tikzpicture}[decoration={markings, mark=at position 0.55 with {\arrow{>}}}, scale=2]
        \draw[postaction=decorate,color=lightgray] (0, 0) -- (0.5, 0.866025);
        \draw[postaction=decorate,color=lightgray] (0, 0) --node[anchor=mid,color=black] {\footnotesize $X^{(\la)}_{jk}$} (1, 0); 
        \draw[postaction=decorate,color=lightgray] (1, 0) -- (0.5, 0.866025) node[above] {\footnotesize $l$};
        \draw[postaction=decorate,color=lightgray] (0.5, -0.866025) -- (0,0) node[left,color=lightgray] {\footnotesize $j$};
        \draw[postaction=decorate,color=lightgray] (0.5, -0.866025) node[below] {\footnotesize $i$} -- (1,0) node[right] {\footnotesize $k$};
        \node () at (0.5, 0.288675) {\footnotesize $X^{(\la)}_{jkl}$};
        \node () at (0.5, -0.288675) {\footnotesize $X^{(\la)}_{ijk}$};
    \end{tikzpicture}
    ,\qquad
    \begin{tikzpicture}[decoration={markings, mark=at position 0.55 with {\arrow{>}}}, scale=2]
        \draw[postaction=decorate,color=lightgray] (0, 0) -- node[anchor=mid,color=black] {\footnotesize $X^{(\la)}_{jk}$} (0.5, 0.866025);
        \draw[postaction=decorate,color=lightgray] (0, 0) node[below,color=lightgray] {\footnotesize $j$} -- (1, 0); 
        \draw[postaction=decorate,color=lightgray] (1, 0) node[below] {\footnotesize $i$} -- (0.5, 0.866025) node[above] {\footnotesize $k$};
        \draw[postaction=decorate,color=lightgray] (-0.5, 0.866025) -- (0.5, 0.866025) ;
        \draw[postaction=decorate,color=lightgray] (0, 0) -- (-0.5, 0.866025) node[above] {\footnotesize $l$};
        \node () at (0.5, 0.188675) {\footnotesize $X^{(-\la)}_{jik}$};
        \node () at (0,  0.67735) {\footnotesize $X^{(-\la)}_{jlk}$};
    \end{tikzpicture}.
\end{equation}
The Hilbert space after gauging is the gauge-invariant subspace
\begin{equation}\label{2split2GrpGaugingAHilb}
    \{\ket{\psi}\in\scrH_{\mathrm{enlarged}~1}\mid G_{jk}^{(\la)}\ket{\psi} = \ket{\psi}~\forall~\la\in A, jk\in\La_1\}.
\end{equation}
Each lattice 1-form symmetry operator ${T^{(\ga)}}$ can be written as ${T^{(\ga)} = \prod_{ij\in\La_1} G^{(\ga_{ij})}_{ij}}$. Therefore, they act as the identity operator on the gauge-invariant subspace~\eqref{2split2GrpGaugingAHilb}. The dual $\h{A}$ 0-form symmetry arising from gauging is $\{V^{(\chi)}\}_{\chi\in\h{A}}$ with
\begin{equation}\label{eq:a-hat-0-form-sym-op}
    V^{(\chi)} = \prod_{ijk\in\La_2} (Z_{ijk}^{(\chi)})^{\eps_{ijk}},
\end{equation}
where ${\eps_{ijk} = +1}$ if the plaquette ${ijk}$ is a right side up triangle (i.e., $\tricorner{}$) and ${\eps_{ijk} = -1}$ if it is an upside down triangle (i.e., $\tricorner[down]{}$).
Note that every $V^{(\chi)}$ commutes with every $G_{jk}^{(\la)}$.

\noindent\textbf{Minimal coupling and dual symmetry.} Operators that commute with each $T^{(\ga)}$ before gauging get mapped to physical operators in the gauged theory by minimal coupling. 

For example, the $\G$-Potts model Hamiltonian~\eqref{eq:split-2group-pottsmodel} commutes with each $T^{(\ga)}$. Because it commutes with each $T^{(\ga)}$, minimal coupling modifies this Hamiltonian to make it commute with each $G_{jk}^{(\la)}$. This causes it to become
\begin{equation}\label{eq:split-2group-isingmodel-mc}
\begin{aligned}
    H_{\G\text{-Potts m.c.}} = 
    &- \frac{1}{|G|} \bigg(\sum_{ij\in\La_1} \sum_{\Ga\in\Irr G} d_\Ga\tr[Z_i^{(\Ga)} Z_{j}^{(\bar{\Ga})}]
    + h_G \sum_{i\in \La_0} \sum_{g\in G}  \overleftarrow{X}_{\!i}^{(g)}\bigg)\\
    &- \frac{1}{|A|} \bigg(\sum_{ijk\in\La_2} \sum_{\chi\in \h{A}} B_{ijk}^{(\chi)}\,Z_{ijk}^{(\bar{\chi})}
    + h_A \sum_{ij\in \La_1} \sum_{\la\in A} X^{(\la)}_{ij}\bigg).
\end{aligned}
\end{equation}
(We have dropped the ${\sum_{i\in\La_0} \sum_{\la\in A} A_i^{(\la)}}$ term from~\eqref{eq:split-2group-pottsmodel} since each $A_i^{(\la)}$ becomes the identity operator when acting on~\eqref{2split2GrpGaugingAHilb}.) Note that the $\h{A}$ 0-form symmetry operators~\eqref{eq:a-hat-0-form-sym-op} commute with this Hamiltonian.

More generally, an operator survives gauging if, after minimal coupling, it acts in a closed manner on the gauge-invariant subspace. A sufficient condition for an operator to act in a closed manner is for it to commute with the projector
\begin{equation}\label{gaugingAsplit2GrpGaugeInvProj}
    \mathsf{P}_G = \prod_{jk\in\La_1} \frac1{|A|}\sum_{\la\in A}G_{jk}^{(\la)}
\end{equation}
onto the gauge-invariant subspace~\eqref{2split2GrpGaugingAHilb}.\footnote{More generally, an operator $O$ acts in a closed manner on the gauge-invariant subspace if and only if it satisfies ${O\ket{\psi} = \mathsf{P}_G\,O\ket{\psi}}$ for all gauge-invariant states $\ket{\psi} = \mathsf{P}_G\ket{\psi}$.
This implies that $O$ acts in a closed manner if and only if it satisfies ${(1-\mathsf{P}_G)O \mathsf{P}_\mathsf{G} = 0}$ as an operator on the enlarged Hilbert space. The condition ${[\mathsf{P}_G,O] = 0}$ is a sufficient but not necessary condition.}
The criterion that an operator must commute with each Gauss operator is a special case of this more general criterion that it must commute with the projector~\eqref{gaugingAsplit2GrpGaugeInvProj}.\footnote{For example, the natural lattice translation operator will never commute with the Gauss law operators individually since they transform the lattice. However, it can be made to commute with the projector onto the gauge-invariant subspace.}
For example, after minimal coupling, the $G$ 0-form symmetry operator~\eqref{eq:split-2group-G-sym-op} becomes
\begin{equation}\label{eq:minimally-coupled-g-0-form}
    U^{(h)}_{\text{m.c.}} 
    =  \prod_{i\in\La_0} \overrightarrow{X}_i^{(h)}
    \prod_{ij\in\La_1}
    P_{ij}^{(\rho_h)}
    \prod_{ijk\in\La_2}
    P_{ijk}^{(\rho_h)}.
\end{equation}
It satisfies ${U^{(h)}_{\text{m.c.}} G^{(\la)}_{ij}  = G_{ij}^{(\rho_h(\la))}U^{(h)}_{\text{m.c.}}}$ and, therefore, commutes with the projector~\eqref{gaugingAsplit2GrpGaugeInvProj}. Furthermore, it satisfies
\begin{equation}
    U^{(h)}_{\text{m.c.}}\,V^{(\chi)}\, U^{(h)\dag}_{\text{m.c.}} = V^{\rho_h\triangleright \chi}.
\end{equation}
Therefore, after gauging, the split 2-group $\G$ symmetry has become the dual ${G\ltimes \h{A}}$ 0-form symmetry ${\{U^{(h)}_{\text{m.c.}}, V^{(\chi)}\}_{h\in G, \chi\in\h{A}}}$. One can check that gauging the $\h{A}$ 0-form symmetry maps the theory back to the original one with the split 2-group $\G$ symmetry.

\noindent\textbf{Change of basis.} After gauging, it is convenient to choose a unitary frame that disentangles the gauge-invariant Hilbert space (i.e., makes it into tensor-product Hilbert space).
For the Gauss operator~\eqref{eq:gauss-law-gauge-a-1form}, this can be achieved using the unitary operator
\begin{equation}\label{eq:disentangling-unitary-1}
    U_1 = \sum_{\substack{\left\{g_i\right\},\left\{a_{ij}\right\}\\ \{a_{ijk}\}}}
        \ket{\left\{g_i\right\},\left\{a_{ij}\right\},\{a_{ijk}-a_{ij} + a_{ik} - a_{jk}\}}\!\!\bra{ \left\{g_i\right\},\left\{a_{ij}\right\},\{a_{ijk}\}}.
\end{equation}
It satisfies
\begin{equation}
    U_1G_{jk}^{(\la)}U_1^\dag = X_{jk}^{(\la)}.
\end{equation}
In this unitary frame, the Gauss constraint polarizes each link $A$-qudit into the state ${\sum_{\la\in A}\ket{a_{ij} =\la}}$.
Thus, in this unitary frame, the link $A$-qudits decouple and the gauge-invariant Hilbert space~\eqref{2split2GrpGaugingAHilb} becomes
\begin{equation}\label{GsitesA2simpsHilb}
    \bigotimes_{i\in\La_0} \C[G]
    \otimes 
    \bigotimes_{ijk\in\La_2} \C[A].
\end{equation}
Furthermore, in this frame of the gauge-invariant subspace, the Hamiltonian~\eqref{eq:split-2group-isingmodel-mc} becomes
\begin{equation}\label{eq:split-2group-isingmodel-mc-unitary}
\begin{aligned}
    H_{G\ltimes\h{A}\text{-Potts}} =&\, U_1 H_{\G\text{-Potts m.c.}} U_1^\dag\\
    = 
    &- \frac{1}{|G|} \bigg(\sum_{ij\in\La_1} \sum_{\Ga\in\Irr G} d_\Ga\tr[Z_i^{(\Ga)} Z_{j}^{(\bar{\Ga})}]
    + h_G \sum_{i\in \La_0} \sum_{g\in G}  \overleftarrow{X}_{\!i}^{(g)}\bigg)\\
    &- \frac{1}{|A|} \bigg(\sum_{ijk\in\La_2} \sum_{\chi\in \h{A}} Z_{ijk}^{(\chi)} 
    + h_A \sum_{\<ijk, nml\>} \sum_{\la\in A} X_{ijk}^{(\la)} X_{nml}^{(\la)}\bigg),
\end{aligned}
\end{equation}
where $ \sum_{\<ijk, nml\>}$ is a sum over neighboring plaquettes. 
Similarly, the $G$ symmetry operator~\eqref{eq:minimally-coupled-g-0-form} becomes
\begin{equation}\label{eq:split-2group-G-sym-op-mc}
    U_1 U^{(h)}_{\text{m.c.}}U_1^\dag
    =  \prod_{i\in\La_0} \overrightarrow{X}_i^{(h)}
    \prod_{ijk\in\La_2}
    P_{ijk}^{(\rho_h)}.
\end{equation}
The $\h{A}$ symmetry operator~\eqref{eq:a-hat-0-form-sym-op} commutes with $U_1 $ and, therefore, is unchanged.

\subsubsection{\texorpdfstring{Gauging $G^{(0)}$: $G\ltimes \h{A} \to 2\Rep(\G)$}{Gauging \textit{G} 0-form symmetry}}\label{sec:split-gauge-g-0form}

We next gauge the $G$ 0-form symmetry~\eqref{eq:split-2group-G-sym-op-mc}, again following the standard procedure that involves first introducing a $G$-qudit on each link and then enforcing a Gauss law. As we'll see, this leads to a dual lattice $\Rep(G)$ 1-form symmetry, and the ${G\ltimes \h{A}}$ 0-form symmetry becomes a $2\Rep(\G)$ fusion 2-category symmetry.

\noindent\textbf{Enlarge Hilbert space.}
We introduce a $G$-qudit on each link. This enlarges the Hilbert space~\eqref{GsitesA2simpsHilb} to
\begin{equation}\label{gaugingGenlargedHilb}
    \scrH_{\mathrm{enlarged}~2} = \bigotimes_{i\in\La_0} \C[G]
    \otimes
    \bigotimes_{ij\in\La_1} \C[G]
    \otimes 
    \bigotimes_{ijk\in\La_2} \C[A].
\end{equation}
\noindent\textbf{Gauss law.}
Next, we define the Gauss operators
\begin{equation}\label{gaugingGsplit2grpsecGaussop}
    G_l^{(h)}
    =
    \begin{tikzpicture}[decoration={markings, mark=at position 0.55 with {\arrow{>}}}, scale=2.5]
    \coordinate (c) at (0,0);
    \coordinate (e) at (1,0);
    \coordinate (w) at (-1,0);
    \coordinate (nw) at (-0.5,{sqrt(3)/2});
    \coordinate (ne) at (0.5,{sqrt(3)/2});
    \coordinate (sw) at (-0.5,-{sqrt(3)/2});
    \coordinate (se) at (0.5,-{sqrt(3)/2});
    
    \filldraw[lightgray!20!white] (c) -- (nw) -- (ne);
    \filldraw[lightgray!20!white] (c) -- (ne) -- (e);
    
    \node () at (0, 0.57735+.14) {\footnotesize$P_{lno}^{(\rho_h)}$};
    \node () at (0.5+.06, 0.288675+.06) {\footnotesize$P_{l m o}^{(\rho_h)}$};
    
    \draw[postaction=decorate, color=lightgray] (c) -- node[anchor=mid, color=black] {\footnotesize $\overrightarrow{X}^{(h)}_{l m}$} (e) node[color=gray, right] {\footnotesize $m$};
    \draw[postaction=decorate, color=lightgray] (c) -- node[below, color=black, anchor=mid] {\footnotesize$\overrightarrow{X}^{(h)}_{l n}$} (nw) node[color=gray, above] {\footnotesize $n$};
    \draw[postaction=decorate, color=lightgray] (c) -- node[below, color=black, anchor=mid] {\footnotesize$\overrightarrow{X}^{(h)}_{l o}$} (ne) node[color=gray, above] {\footnotesize $o$};
    \draw[postaction=decorate, color=lightgray] (sw) node[color=gray, below] {\footnotesize $i$} -- node[below, color=black, anchor=mid] {\footnotesize$\overleftarrow{X}^{(h)}_{il}$} (c);
    \draw[postaction=decorate, color=lightgray] (se) node[color=gray, below] {\footnotesize $j$} -- node[below, color=black, anchor=mid] {\footnotesize$\overleftarrow{X}^{(h)}_{jl}$} (c);
    \draw[postaction=decorate, color=lightgray] (w) node[color=gray, left] {\footnotesize $k$} -- node[below, color=black, anchor=mid] {\footnotesize$\overleftarrow{X}^{(h)}_{kl}$} (c) node[anchor=mid, color=black] {\footnotesize$\overrightarrow{X}^{(h)}_l$};
\end{tikzpicture}.
\end{equation}
The gauge-invariant subspace of~\eqref{gaugingGenlargedHilb} is
\begin{equation}\label{gaugingGsplit2GrpCaseGaugeInvSubSpace}
    \{\ket{\psi}\in\scrH_{\mathrm{enlarged}~2}\mid G_{l}^{(h)}\ket{\psi} = \ket{\psi}~\forall~h\in G, l\in\La_0\}.
\end{equation}

The dual lattice $\Rep(G)$ 1-form symmetry arising from gauging is $\{W^{(\Ga)}_\ga\}_{\Ga\in\mathrm{irr}(G), \ga\in \cL(\La)}$, where $\cL(\La)$ denotes the set of oriented lattice loops in $\La$ and
\begin{equation}\label{eq:wilson-loop-operator}
    W^{(\Ga)}_\ga = 
    \Tr \left(\prod_{ij\in\ga} (Z_{ij}^{(\Ga)})^{\eps_{ij}(\ga)}\right).
\end{equation}
Here, ${\prod_{ij\in\ga}}$ is the path-ordered product and ${\eps_{ij}(\ga) = +1}$ if the orientation of $\ga$ is aligned with that of ${ij}$ while ${\eps_{ij}(\ga) = -1}$ if it is anti-aligned.
Note that every $W^{(\Ga)}_\ga$ commutes with every $G_{l}^{(h)}$. 
They satisfy the $\Rep(G)$ operator algebra
\begin{equation}\label{eq:repg-operator-algebra}
    W^{(\Ga_1)}_\ga\times W^{(\Ga_2)}_\ga = \sum_{\Ga\in\Irr G}N_{\Ga_1,\Ga_2}^\Ga
    W^{(\Ga)}_\ga,
\end{equation}
where ${N_{\Ga_1,\Ga_2}^\Ga\in\Z_{\geq 0}}$ are the multiplicities in the tensor product ${\Ga_1\otimes\Ga_2 = \bigoplus_{\Ga\in\Irr G}N_{\Ga_1,\Ga_2}^\Ga
    \Ga}$.
Furthermore, every $W^{(\Ga)}_\ga$ is a topological operator generating a 1-form symmetry in the subspace of~\eqref{gaugingGsplit2GrpCaseGaugeInvSubSpace} whose vectors $\ket{\psi}$ satisfy ${B_{lmn}\ket{\psi} = \ket{\psi}}$, where
\begin{equation}\label{Gflatnesssplit2grpgaugingGsec}
    B_{lmn}
    = \begin{cases}
        \frac1{|G|}\sum_{\Ga\in \Irr(G)} d_{\Ga}   \Tr\! \left[ 
        Z_{\triedge{s}}^{(\Ga)} 
        Z_{\triedge{nw}}^{(\Ga)} Z_{\triedge{ne}}^{(\bar{\Ga})}
        \right]\qquad
        & 
        \qquad lmn = \tricorner{}\,,
        \\[1em]
                 \frac1{|G|}\sum_{\Ga\in \Irr(G)} d_{\Ga}   \Tr\! \left[ 
        Z_{\triedge[down]{ne}}^{(\Ga)}
        Z_{\triedge[down]{s}}^{(\Ga)}
        Z_{\triedge[down]{nw}}^{(\bar{\Ga})}
        \right]\qquad
        & 
        \qquad lmn = \tricorner[down]{}\,.
    \end{cases}
\end{equation}
Indeed, the projector $B_{lmn}$ can be written as
\begin{equation}
        B_{lmn}        = 
        \sum_{\{g_i\},\{g_{ij}\},\{a_{ijk}\}}\, \del_{g_{l m} g_{mn}, g_{l n}} 
        \,\ketbra
        {\{g_{i}\},\{g_{ij}\}, \{a_{ijk}\}}
        {\{g_{i}\},\{g_{ij}\}, \{a_{ijk}\}}.
\end{equation}

\noindent\textbf{Minimal coupling and dual symmetry.} What operators become after gauging is found by minimal coupling.

For instance, let us first minimally couple the Hamiltonian~\eqref{eq:split-2group-isingmodel-mc-unitary}. The second and third terms commute with each $G_l^{(h)}$ and, therefore, need not be modified. Modifying the first and fourth terms such that they commute produces the minimally-coupled Hamiltonian
\begin{equation}\label{minimalCoupledsplit2GrpPottsaftergaugingG}
\begin{aligned}
    H^{(\rho)}_{G\ltimes\h{A}\text{-Potts m.c.}} = &- \frac{1}{|G|} \bigg(\sum_{ij\in\La_1} \sum_{\Ga\in\Irr G} d_\Ga\tr[Z_i^{(\Ga)} Z_{ij}^{(\bar{\Ga})} Z_{j}^{(\bar{\Ga})}]
    + h_G \sum_{i\in \La_0} \sum_{g\in G}  \overleftarrow{X}_{\!i}^{(g)}\bigg) - \sum_{ijk\in\La_2} B_{ijk}\\
    &- \frac{1}{|A|} \sum_{ijk\in\La_2} \sum_{\chi\in \h{A}} Z_{ijk}^{(\chi)} 
    - h_A  \sum_{ij\in\La_1} C^{(\rho)}_{ij}.
\end{aligned}
\end{equation}
where, in the final term,
\begin{equation}\label{eq:dual-a-ferromagnet-term-2rep-model}
    C^{(\rho)}_{ij} 
    =  
    \begin{cases}
        
    \frac{1}{|A|} \sum_{\la\in A} X_{\tricorner{}}^{(\,\rho_{g_{\triedge{s}}} (\la)\,)} 
    X_{\tricorner[down]{}}^{(\la)}
    & \qquad
    ij = \trilink{b},
    \\
   
    \frac{1}{|A|} \sum_{\la\in A} X_{\tricorner{}}^{(\la)}
     X_{\tricorner[down]{}}^{(\,\rho_{g_{\triedge[down]{ne}}} (\la)\,)} 
     &\qquad
     ij = \trilink{c},
     \\
    
    \frac{1}{|A|} \sum_{\la\in A} X_{\tricorner[up]{}}^{(\la)} 
    X_{\tricorner[down]{}}^{(\la)}
    \vphantom{X_{\tricorner[down]{}}^{\big(\rho_{g_{\tricorner[down]{left}}^{-1} g_{\tricorner[down]{apex}}}}}
    &\qquad
    ij = \trilink{a}.
    \end{cases}
\end{equation}
The Gauss operators $G_l^{(h)}$ and lattice $\Rep(G)$ 1-form symmetry operators both commute with~\eqref{minimalCoupledsplit2GrpPottsaftergaugingG}.
Furthermore, we have added the third term in~\eqref{minimalCoupledsplit2GrpPottsaftergaugingG} by hand. 
Since $B_{ijk}$ commutes with each term in the Hamiltonian, the third term energetically enforces the $\Rep(G)$ 1-form symmetry operators to be topological at low energies.

The $\h{A}$ 0-form symmetry operators $\{V^{(\chi)}\}_{\chi\in\h{A}}$ also get modified after gauging.
Since $\rho$ is nontrivial, a generic $V^{(\chi)}$ is not $G$-symmetric.
After gauging, its $G$-symmetric orbit sums give coset-type
non-invertible symmetry operators~\cite{HKZ240520401}.
App.~\ref{app:split-non-invertible-minimal-coupling} shows that minimally coupling the
$G$-symmetric sums of $V^{(\chi)}$ gives rise to symmetry operators $\mathsf{V}^{(K,\chi)}$ labeled by
a subgroup ${K\leq G}$ and $K$-invariant character ${\chi\in\h{A}^K = \{\chi\in\h{A}\mid \rho_k\triangleright \chi = \chi~\forall~k\in K\}}$.
The operator $\mathsf{V}^{(K,\chi)}$ is defined as
\begin{equation}\label{eq:non-invertible-2rep-op-before-unitary}
    \mathsf{V}^{(K,\chi)} \!=\!\! 
    \sum_{Kh\in K\backslash G}
    \mathsf{P}^{(K)}_{o, h}
    \left[\sum_{\substack{\left\{g_i\right\},\left\{g_{ij}\right\}\\ \{a_{ijk}\}}}
    \prod_{lmn\in\La_2}\!
    \chi(\rho_{h g_{\ga_{o\to l}}}(a_{lmn}))^{\eps_{lmn}}
    \ketbra
    {\{g_i\},\{g_{ij}\},\{a_{ijk}\}}
    {\{g_i\},\{g_{ij}\},\{a_{ijk}\}}\right]\!
    ,
\end{equation}
where $\mathsf{P}^{(K)}_{o, h}$ is the projector that satisfies
\begin{equation}\label{PKohmaintext}
    \mathsf{P}^{(K)}_{o, h} \ket{\{g_i\}, \{g_{ij}\},\{a_{ijk}\}} \!= 
    \begin{cases}
        \ket{\{g_i\}, \{g_{ij}\},\{a_{ijk}\}}
        \qquad &
    \prod_{ij\in\ga_{o}} g_{ij}^{\eps_{ij}(\ga_o)} \!\in h^{-1} K h~~\forall\text{ loops }\ga_o\text{ based at $o$}, \\
        0 \qquad &\text{otherwise}.
    \end{cases}
\end{equation}
In the subscript of $\rho$ in $\mathsf{V}^{(K,\chi)}$, we have defined the group element
\begin{equation}\label{eq:holonomy-group-element-g}
    g_{\ga_{o\to l}}  = \prod_{ij\in\ga_{o\to l}} g_{ij}^{\eps_{ij}(\ga_{o\to l})}\in G,
\end{equation}
where $\ga_{o\to l}$ is any lattice path from an arbitrary reference site $o$ to the site $l$. 
As shown in App.~\ref{app:split-non-invertible-minimal-coupling},
$\mathsf{V}^{(K,\chi)}$ does not depend on the choice of $o$, the path
$\ga_{o\to l}$, or the representative $h$ of the coset $Kh$, and it
commutes with each Gauss operator.
The operators~\eqref{eq:non-invertible-2rep-op-before-unitary} commute with the Hamiltonian~\eqref{minimalCoupledsplit2GrpPottsaftergaugingG}.

These 0-form symmetry operators satisfy the operator algebra
\begin{equation}\label{eq:non-inv-2rep-0-form-fusion-rule}
    \mathsf{V}^{(K_1,\chi_1)}
    \times
    \mathsf{V}^{(K_2,\chi_2)}
    =
    \sum_{K_1 r K_2 \,\in\, K_1 \backslash G / K_2} \mathsf{V}^{(K_1\cap \,r K_2 r^{-1},\, \chi_1\cdot (\rho_r\triangleright \chi_2))}.
\end{equation}
Note that ${\mathsf{V}^{(G,1)} = \mathsf{P}^{(G)}_{o,h} = 1}$.
On the other hand, ${\mathsf{V}^{(1,1)} = |G|\mathsf{P}^{(1)}_{o,1}}$, where $\mathsf{P}^{(1)}_{o,1}$ projects onto the subspace with trivial $G$ holonomy around every loop, equivalently the $\Rep(G)$ 1-form symmetric subspace. A generic operator $\mathsf{V}^{(K,\chi)}$ is non-invertible because of the projector~\eqref{PKohmaintext}. The operators $\mathsf{V}^{(G,\chi)}$, however, are invertible and satisfy the $\h{A}^G$ group law ${\mathsf{V}^{(G,\chi_1)}\times\mathsf{V}^{(G,\chi_2)}
=
\mathsf{V}^{(G,\,\chi_1\cdot\chi_2)}}$.
The algebra~\eqref{eq:non-inv-2rep-0-form-fusion-rule} is a subalgebra of the total fusion algebra formed by the simple objects of $2\Rep(\G)$~\cite{BBFP220805993}. 
More generally, a simple 0-form symmetry operator of $2\Rep(\G)$ is $\mathsf V^{(K,\chi,[\om])}$, with an additional class
${[\om]\in\cH^2(K,\Uone)}$ corresponding to a ${1+1}$d $K$-SPT
decoration of the symmetry operator.
We write down the explicit form of the general symmetry operator $\mathsf{V}^{(K,\chi,[\om])}$ in App.~\ref{app:split-non-invertible-minimal-coupling}.

The lattice 0-form symmetry operators $\mathsf{V}^{(K,\chi)}$ commute with the lattice 1-form symmetry operators $W^{(\Ga)}_\ga$:
\begin{equation}
    [\mathsf{V}^{(K,\chi)}, W^{(\Ga)}_\ga] = 0.
\end{equation}
The interplay between $\mathsf{V}^{(1,\chi)}$ and $W^{(\Ga)}_{\ga}$ is noteworthy. These operators satisfy
\begin{equation}\label{VchiactingonW}
    W^{(\Ga)}_{\ga} \times \mathsf{V}^{(1,\chi)}
    =
    \mathsf{V}^{(1, \chi)} \times W^{(\Ga)}_{\ga} 
    = d_\Ga\,\mathsf{V}^{(1,\chi)}.
\end{equation}
Thus, states carrying nontrivial $\Rep(G)$ 1-form symmetry charge---eigenstates of $W^{(\Ga)}_{\ga}$ with eigenvalue ${\neq d_\Ga}$---are annihilated by the $\mathsf{V}^{(1,\chi)}$ symmetry operators.

\noindent\textbf{Change of basis.} Finally, we choose a new unitary frame in which the site $ G $-qudits are decoupled and the Hilbert space admits an onsite tensor product factorization. We do so using the unitary operator
\begin{equation}\label{eq:gauge-g-0-form-unitary}
    U_2 = \sum_{\substack{\{g_i\},\{g_{ij}\}\\ \{a_{ijk}\}}}
        \ketbra{\{g_i\},\{g_i^{-1} g_{ij} g_{j}\},\{\rho_{g_i^{-1}}(a_{ijk})\}}{\{g_i\},\{g_{ij}\},\{a_{ijk}\}} 
        .
\end{equation}
It satisfies ${U_2 G_{i}^{(h)}U_2^\dag =
\overrightarrow{X}^{(h)}_i}$; thus the only physical site $G$-qudit configuration is fully the symmetrized state, and the site $G$-qudits are decoupled. The Hilbert space in this frame becomes
\begin{equation}
    \bigotimes_{ij\in\La_1} \C[G]
    \otimes 
    \bigotimes_{ijk\in\La_2} \C[A].
\end{equation}

In this new unitary frame, the Hamiltonian~\eqref{minimalCoupledsplit2GrpPottsaftergaugingG} becomes the $2\Rep(\G)$ Potts model Hamiltonian:
\begin{equation}
\begin{aligned}\label{eq:split-2rep-potts-model}
    H^{(\rho)}_{2\Rep(\G)\text{-Potts}} =& \,U_2 H^{(\rho)}_{G\ltimes\h{A}\text{-Potts m.c.}}U_2^\dag\\
    =&- \frac{1}{|G|} \sum_{ij\in\La_1} \sum_{\Ga\in\Irr G} d_\Ga\tr[Z_{ij}^{(\Ga)}] - h_G \sum_{i\in\La_0} A_i^{(\rho)} - \sum_{ijk\in\La_2} B_{ijk}\\
    &
    -
    \frac{1}{|A|} \sum_{ijk\in\La_2} \sum_{\chi\in \h{A}} Z_{ijk}^{(\chi)} 
    - 
    h_A \sum_{ij\in \La_1} C_{ij}^{(\rho)}.
\end{aligned}
\end{equation}
In the second term, we have defined the split $\G$ star operator
\begin{equation}\label{eq:split-star-term-def}
    A^{(\rho)}_{l} = \frac{1}{|G|} \sum_{h\in G} A^{(h,\rho)}_{l},
    \qquad\qquad
    A^{(h,\rho)}_{l}=
    \begin{tikzpicture}[decoration={markings, mark=at position 0.55 with {\arrow{>}}}, scale=2.5]
    \coordinate (c) at (0,0);
    \coordinate (e) at (1,0);
    \coordinate (w) at (-1,0);
    \coordinate (nw) at (-0.5,{sqrt(3)/2});
    \coordinate (ne) at (0.5,{sqrt(3)/2});
    \coordinate (sw) at (-0.5,-{sqrt(3)/2});
    \coordinate (se) at (0.5,-{sqrt(3)/2});
    
    \filldraw[lightgray!20!white] (c) -- (nw) -- (ne);
    \filldraw[lightgray!20!white] (c) -- (ne) -- (e);
    
    \node () at (0, 0.57735+.14) {\footnotesize$P_{lno}^{(\rho_h)}$};
    \node () at (0.5+.06, 0.288675+.06) {\footnotesize$P_{l m o}^{(\rho_h)}$};
    
    \draw[postaction=decorate, color=lightgray] (c) node[color=gray, below] {\footnotesize $l$} -- node[anchor=mid, color=black] {\footnotesize $\overrightarrow{X}^{(h)}_{l m}$} (e) node[color=gray, right] {\footnotesize $m$};
    \draw[postaction=decorate, color=lightgray] (c) -- node[below, color=black, anchor=mid] {\footnotesize$\overrightarrow{X}^{(h)}_{l n}$} (nw) node[color=gray, above] {\footnotesize $n$};
    \draw[postaction=decorate, color=lightgray] (c) -- node[below, color=black, anchor=mid] {\footnotesize$\overrightarrow{X}^{(h)}_{l o}$} (ne) node[color=gray, above] {\footnotesize $o$};
    \draw[postaction=decorate, color=lightgray] (sw) node[color=gray, below] {\footnotesize $i$} -- node[below, color=black, anchor=mid] {\footnotesize$\overleftarrow{X}^{(h)}_{il}$} (c);
    \draw[postaction=decorate, color=lightgray] (se) node[color=gray, below] {\footnotesize $j$} -- node[below, color=black, anchor=mid] {\footnotesize$\overleftarrow{X}^{(h)}_{jl}$} (c);
    \draw[postaction=decorate, color=lightgray] (w) node[color=gray, left] {\footnotesize $k$} -- node[below, color=black, anchor=mid] {\footnotesize$\overleftarrow{X}^{(h)}_{kl}$} (c);
\end{tikzpicture}.
\end{equation}
The $2\Rep(\G)$ Potts model is a lattice 2-group gauge theory. In the limit of large $h_G,h_A$, it is in a topological phase whose ground-state subspace is described by split $\G$ gauge theory. (We will discuss this limit more in the next subsection.) In the limit of small $h_G,h_A$, it is in a Higgs phase.

Furthermore, in this frame, the $2\Rep(\G)$ symmetry operators are left unchanged.
The $2\Rep(\G)$ Potts model Hamiltonian commutes with the $2\Rep(\G)$ symmetry operators for all parameters $h_A,h_G$.

\subsection{Split 2-group quantum double model}\label{sec:split-quantum-double}

The Hamiltonian~\eqref{eq:split-2rep-potts-model} in its deconfined 2-group gauge theory phase, where ${h_G,h_A\gg 1}$, is represented at a convenient exactly-solvable point by
\begin{equation}\label{eq:split-quantum-double}
    H_{\G}^{(\rho)} = -\sum_{i\in \La_0} A^{(\rho)}_i 
    -\sum_{ijk\in\La_2} B_{ijk}
    -\sum_{ij\in \La_1} C^{(\rho)}_{ij}.
\end{equation}
This Hamiltonian is a commuting-projector model and is a generalization of Kitaev's quantum double model~\cite{K9707021} for split 2-groups. 

The Hamiltonian $H_{\G}^{(\rho)}$ is a gapped Hamiltonian.
As we show in App.~\ref{app:double-from-path-integral}, its ground-state space realizes untwisted split $\G$ gauge theory. 
We also show that its ground-state degeneracy on a spatial sphere is 
\begin{equation}\label{maintextGSDsplitQDMsphere}
    \mathrm{GSD}_{S^2}
    =
    |\h{A}/G|,
\end{equation}
where $\h{A}/G$ is the set of $G$-orbits ${[\chi] = \{\rho_g\triangleright \chi\mid g\in G\}}$. 
When
$\rho$ is trivial, GSD$_{S^2}$ simplifies to ${|A|}$.
On a spatial torus, the ground-state degeneracy is
\begin{equation}\label{splitQDGSDmaintext}
    \mathrm{GSD}_{T^2}
    =
    \sum_{[\chi]\in\h{A}/G}
    \left|
        \Hom(\Z^2,G_\chi)/G_\chi
    \right|,
\end{equation}
where the action of ${G_\chi = \{g\in G \mid \rho_g\triangleright \chi = \chi\}}$ on ${\Hom(\Z^2,G_\chi)\cong \{(g_x,g_y)\in G^2\mid g_xg_y = g_y g_x\}}$ is ${(g_x,g_y) \mapsto (h g_xh^{-1},\,hg_yh^{-1})}$.
When
$\rho$ is trivial, GSD$_{T^2}$ simplifies to ${|A|\,|\Hom(\Z^2,G)/G|}$. For ${A=1}$, this reduces to the torus ground-state degeneracy of the ordinary $G$ quantum
double model: ${|\Hom(\Z^2,G)/G|}$.
Generally, each $G$-orbit $[\chi]$ contributes to~\eqref{splitQDGSDmaintext} the torus ground-state degeneracy of the ordinary $G_\chi$ quantum double model.

\subsection{Example: Charge conjugation}\label{sec:split-example}

We end this section on split 2-groups by specializing our general results to the simple case where ${G = \Z_2}$ and ${A = \Z_n}$. (We use multiplicative notation for ${G = \Z_2}$ and additive notation for ${A = \Z_n}$ in this example.) Here, we focus on the nontrivial $G$-action where ${\rho_{1}(\la) = \la}$ and ${\rho_{-1}(\la) = -\la}$ for all ${\la\in A}$. For this split 2-group symmetry, the $\Z_2$ 0-form symmetry acts as charge conjugation for the $\Z_n$ 1-form symmetry.

For these groups, the 2-group symmetry operators from Sec.~\ref{sec:split-2group-sym-ops} act on a Hilbert space with a qubit on each site and a $\Z_n$-qudit on each link. We denote the Pauli operators for the qubit at site $i$ by ${Z_i, X_i}$ and the clock and shift operators for the $\Z_n$-qudit at link $ij$ by ${\cZ_{ij}, \cX_{ij}}$.\footnote{The clock and shift operators of a $\Z_n$-qudit satisfy ${\cX^n = \cZ^n = 1}$ and ${\cZ\cX = \ee^{2\pi\ii/n}\cX\cZ}$. They are related to the $\Z_n$-qudit generalized Pauli operators as follows:
\begin{equation}
    X^{(\la)} \equiv \cX^\la,
    \qquad
    Z^{(\chi)} = \cZ^{k_{\chi}},
\end{equation}
where $\la\in \Z_n \cong\{0,1,\cdots, n-1\}$ and $k_\chi$ is the integer appearing in ${\chi(\la) = \ee^{\frac{2\pi \ii }n k_\chi \la}}$.
}
The $\Z_2$ 0-form symmetry~\eqref{eq:split-2group-G-sym-op} is generated by
\begin{equation}
    U = \prod_{i\in \La_0} X_i \prod_{ij\in\La_1}\cC_{ij},
\end{equation}
where $\cC_{ij}$ transforms only the $\Z_n$-qudit at link $ij$ and satisfies 
\begin{equation}
    \cC_{ij}^2 = 1,
    \qquad
    \cC_{ij}\cX_{ij}\cC_{ij}^\dag = \cX_{ij}^\dag,
    \qquad
    \cC_{ij}\cZ_{ij}\cC_{ij}^\dag = \cZ_{ij}^\dag.
\end{equation}
It is equal to $P^{(\rho_{-1})}_{ij}$ in the generalized qudit notation.
The lattice $\Z_n$ 1-form symmetry operators~\eqref{eq:split-2group-A-sym-op} can be written 
\begin{equation}
    T^{(\ga)} = \prod_{ij\in\La_1} \cX_{ij}^{\ga_{ij}},   
\end{equation}
where ${\ga\equiv \sum_{ij\in\La_1} \ga_{ij} [ij]^\vee}$ is a $\Z_n$-valued dual 1-cycle. They satisfy
\begin{equation}
    UT^{(\ga)}U^\dag = (T^{(\ga)})^\dag.
\end{equation}

Gauging this split 2-group $\G$ symmetry leads to a new model with a qubit on each link and a $\Z_n$-qudit on each plaquette, and a dual $2\Rep(\G)$ symmetry. The 1-form part of this $2\Rep(\G)$ symmetry is a $\Z_2$ 1-form symmetry, and its symmetry operators~\eqref{eq:wilson-loop-operator} can be written as
\begin{equation}\label{eq:z2-1-form-symmetry}
    W^{(\Upsilon)} = \prod_{ij\in\La_1} Z_{ij}^{\Upsilon{ij}},
\end{equation}
where ${\Upsilon\equiv \sum_{ij\in\La_1} \Upsilon_{ij} [ij]}$ is a $\Z_2$-valued 1-cycle. 
The 0-form operators~\eqref{eq:non-invertible-2rep-op-before-unitary} are labeled by a subgroup ${K\leq \Z_2}$ and a $K$-invariant character ${\chi\in\Hom(\Z_n,\Uone)^K}$ subject to the equivalence relation ${\chi \sim \bar{\chi}}$ where ${\bar{\chi}(a) = \chi(a)^{-1}}$.\footnote{As discussed in App.~\ref{app:split-non-invertible-minimal-coupling}, the simple objects of $2\Rep(\G)$ for split $\G$ are generally labeled by ${(K,\chi,[\om])}$, with ${[\om]\in\cH^2(K,\Uone)}$, modulo ${(K,\chi,[\om]) \sim (gKg^{-1},\rho_g\triangleright\chi,\mathrm{Conj}_g[\om])}$. In this example, because $\cH^2(K,\Uone) = 1$ for all ${K\leq \Z_2}$, the simples are labeled by $(K,\chi)$. Furthermore, because $G$ is an abelian group and ${\rho_{-1}\triangleright\chi = \bar{\chi}}$, the equivalence relation simplifies to ${(K,\chi) \sim (K,\bar{\chi})}$.}
The operators with ${K=1}$ can be written as
\begin{equation}
    \mathsf{V}^{(1,\chi)} = 
    \mathsf{P}
    \!\!\!\sum_{ \{g_{ij}\}, \{a_{ijk}\} }
    \!\!\!\left(\,
    \prod_{lmn\in\La_2}
    \!\!
    \chi(a_{lmn})^{\eps_{lmn} g_{\ga_{o\to l}}}
    +
    \prod_{lmn\in\La_2}
    \!\!
    \bar{\chi}(a_{lmn})^{\eps_{lmn} g_{\ga_{o\to l}}}
    \!\right)
    \ketbra
    {\{g_{ij}\},\{a_{ijk}\}}
    {\{g_{ij}\},\{a_{ijk}\}}
    ,
\end{equation}
where the projector ${\mathsf{P} = |Z_1(\La;\Z_2)|^{-1}\sum_{\Upsilon\in Z_1(\La;\Z_2)} W^{(\Upsilon)}}$.\footnote{\label{condOpFootnote}In the topological subspace for the $\Z_2$ 1-form symmetry, ${W^{(\Upsilon)} = 1}$ if ${\Upsilon\in B_1(\La;\Z_2)}$ and the projector $\mathsf{P}$ simplifies to ${\mathsf{P}_\mathrm{top} = \frac14\sum_{[\Upsilon]\in H_1(\La;\Z_2)} W^{(\Upsilon)}}$, where we used that ${|Z_1(\La;\Z_2)|/|B_1(\La;\Z_2)| = |H_1(\La;\Z_2)| = 4}$. In this topological subspace, $\mathsf{V}^{(1,1)}$ becomes ${\mathsf{V}^{(1,1)}_\mathrm{top} = \frac12\sum_{[\Upsilon]\in H_1(\La;\Z_2)} W^{(\Upsilon)}}$, which is the condensation operator for the $\Z_2$ 1-form symmetry on a torus~\cite{RSS220402407}.} The operators with ${K=\Z_2}$ are
\begin{equation}
    \mathsf{V}^{(\Z_2,\chi)} =
    \sum_{\{g_{ij}\}, \{a_{ijk}\}}
    ~\prod_{lmn\in\La_2}
    \chi(a_{lmn})^{\eps_{lmn} g_{\ga_{o\to l}}}
    \ketbra
    {\{g_{ij}\},\{a_{ijk}\}}
    {\{g_{ij}\},\{a_{ijk}\}}
    ,
\end{equation}
where ${\chi\in \Hom(\Z_n,\Uone)^{\Z_2} = \{\chi\in \Hom(\Z_n,\Uone)\mid \bar{\chi} = \chi\}}$. These 0-form symmetry operators satisfy the operator algebra
\begin{gather}
    \mathsf{V}^{(1,\chi_1)}
    \times
    \mathsf{V}^{(1,\chi_2)}
    =
    \mathsf{V}^{(1,\chi_1\cdot \chi_2)}
    +
    \mathsf{V}^{(1,\chi_1\cdot \bar{\chi}_2)},\\
    \mathsf{V}^{(\Z_2,\chi_1)}
    \times
    \mathsf{V}^{(\Z_2,\chi_2)}
    =
    \mathsf{V}^{(\Z_2,\chi_1\chi_2)},\\
    \mathsf{V}^{(1,\chi_1)}
    \times
    \mathsf{V}^{(\Z_2,\chi_2)}
    =
    \mathsf{V}^{(\Z_2,\chi_2)}
    \times
    \mathsf{V}^{(1,\chi_1)}
    =
    \mathsf{V}^{(1,\chi_1\cdot\chi_2)}.
\end{gather}

For this example, the split 2-group quantum double model Hamiltonian~\eqref{eq:split-quantum-double} simplifies to
\begin{equation}
    H^{(\rho)}_\G = -\frac12 \sum_{i\in\La_0} (1+A^{(\cC)}_i) 
        - \frac12\sum_{ijk\in\La_2} (1+Z_{ij} Z_{jk} Z_{ik})
        - \sum_{ij\in \La_1} C_{ij}^{(\rho)},
\end{equation}
where\footnote{We use the shorthand ${\cX^{Z} \equiv \frac12(1+Z)
    \cX
    + \frac12(1-Z) \cX^\dag}$ in this expression for $C^{(\rho)}_{ij}$.}
\begin{equation}
    A^{(\cC)}_{l} = 
    \begin{tikzpicture}[decoration={markings, mark=at position 0.55 with {\arrow{>}}}, scale=2.5]
    \coordinate (c) at (0,0);
    \coordinate (e) at (1,0);
    \coordinate (w) at (-1,0);
    \coordinate (nw) at (-0.5,{sqrt(3)/2});
    \coordinate (ne) at (0.5,{sqrt(3)/2});
    \coordinate (sw) at (-0.5,-{sqrt(3)/2});
    \coordinate (se) at (0.5,-{sqrt(3)/2});
    
    \filldraw[lightgray!20!white] (c) -- (nw) -- (ne);
    \filldraw[lightgray!20!white] (c) -- (ne) -- (e);
    
    \node () at (0, 0.57735+.14) {\footnotesize$\cC_{l n o}$};
    \node () at (0.5+.06, 0.288675+.06) {\footnotesize$\cC_{l m o}$};
    
    \draw[postaction=decorate, color=lightgray] (c) node[color=gray, below] {\footnotesize $l$} -- node[anchor=mid, color=black] {\footnotesize $X_{l m}$} (e) node[color=gray, right] {\footnotesize $m$};
    \draw[postaction=decorate, color=lightgray] (c) -- node[below, color=black, anchor=mid] {\footnotesize$X_{l n}$} (nw) node[color=gray, above] {\footnotesize $n$};
    \draw[postaction=decorate, color=lightgray] (c) -- node[below, color=black, anchor=mid] {\footnotesize$X^{(h)}_{l o}$} (ne) node[color=gray, above] {\footnotesize $o$};
    \draw[postaction=decorate, color=lightgray] (sw) node[color=gray, below] {\footnotesize $i$} -- node[below, color=black, anchor=mid] {\footnotesize$X_{il}$} (c);
    \draw[postaction=decorate, color=lightgray] (se) node[color=gray, below] {\footnotesize $j$} -- node[below, color=black, anchor=mid] {\footnotesize$X_{jl}$} (c);
    \draw[postaction=decorate, color=lightgray] (w) node[color=gray, left] {\footnotesize $k$} -- node[below, color=black, anchor=mid] {\footnotesize$X_{kl}$} (c);
\end{tikzpicture},
\qquad
C^{(\rho)}_{ij} 
    =  
    \begin{cases}
    \frac{1}{n} \sum_{k=0}^{n-1} (\cX_{\tricorner{}}^{Z_{\triedge{s}}} 
    \cX_{\tricorner[down]{}})^{k}
    & \qquad
    ij = \trilink{b},
    \vspace{3pt}\\
   
    \frac{1}{n} \sum_{k=0}^{n-1} (\cX_{\tricorner{}}
     \cX_{\tricorner[down]{}}^{Z_{\triedge[down]{ne}}})^{k} 
     &\qquad
     ij = \trilink{c},
     \vspace{3pt}\\
    
    \frac{1}{n} \sum_{k=0}^{n-1} (\cX_{\tricorner[up]{}} 
    \cX_{\tricorner[down]{}})^{k}
    &\qquad
    ij = \trilink{a}.
    \end{cases}
\end{equation}
The operator $A_i^{(\cC)}$ is the toric code star term decorated with $\Z_n$-qudit charge conjugation operators $\cC$.
The $C_{ij}^{(\rho)}$ term is the $n$-state Potts model term controlled by the link qubits. The ground-state degeneracy on a spatial sphere~\eqref{maintextGSDsplitQDMsphere} and torus~\eqref{splitQDGSDmaintext} for this example simplifies to
\begin{equation}
    \mathrm{GSD}_{S^2} = 
    \begin{cases}
        \dfrac{n+1}{2}\qquad &\text{odd }n,\vspace{5pt}\\
        1+\dfrac{n}{2}\qquad &\text{even }n,
    \end{cases}
    \qquad
    \mathrm{GSD}_{T^2} = 
    \begin{cases}
        4+\dfrac{n-1}{2}\qquad &\text{odd }n,\vspace{5pt}\\
        7+\dfrac{n}{2}\qquad &\text{even }n.
    \end{cases}
\end{equation}
If we instead had chosen trivial $\rho$, this Hamiltonian would be the sum of two decoupled Hamiltonians: the triangular lattice toric code and a Hamiltonian unitarily equivalent to the $n$-states Potts model. In this case, GSD$_{S^2}$ would be $n$ and GSD$_{T^2}$ would be $4n$.

\section{Central 2-group symmetry}\label{sec:central-2group-symmetry}

This section presents and explores symmetry operators for a central 2-group symmetry. As in Sec.~\ref{sec:split-2group-symmetry}, we assume that the 0-form symmetry group $G$ is finite and that the 1-form symmetry group $A$ is finite abelian. We use additive notation for $A$.

By central 2-group, we mean a 2-group whose monodromy action of $G$ on $A$ is trivial.
Such a 2-group $\G$ may be viewed as an extension
\begin{equation}\label{central2grpext}
    1\longrightarrow BA \longrightarrow \G \longrightarrow G \longrightarrow 1
\end{equation}
that is classified by the Postnikov class 
\begin{equation}
    [\bt] \in \cH^3(G,A).
\end{equation}
Hence, the data specifying a central 2-group $\G$ is ${(G,A,[\bt])}$. 

For each Postnikov class, we choose a representative 3-cocycle ${\bt\in\cZ^3(G,A)}$. Thus, $\bt$ is a map ${\bt\colon G\times G\times G \to A}$ satisfying the 3-cocycle condition 
\begin{equation}
    \bt(h,k,l) + \bt(g,hk,l) + \bt(g,h,k) = \bt(gh,k,l) + \bt(g,h,kl),
\end{equation}
for all ${g,h,k,l\in G}$.
The Postnikov class $[\bt]$ is the class of $\bt$ subject to the equivalence relation
\begin{equation}
    \bt(g,h,k) \sim \bt(g,h,k) + \eta(h,k) - \eta(gh,k) + \eta(g,hk) - \eta(g,h),
\end{equation}
for all maps ${\eta\colon G\times G \to A}$.
We take the representative $\bt$ to be a normalized 3-cocycle: ${\bt(g,h,k) = 0}$ if any of ${g,h,k}$ is equal to the identity.

\subsection{Symmetry operators}\label{sec:central-2group-sym-ops}

We consider the same class of quantum lattice systems as in Sec.~\ref{sec:split-2group-sym-ops}. Namely, the two-dimensional spatial lattice is a triangular lattice, which we model as a simplicial complex $\La$ of a torus (see Fig.~\ref{fig:lattice-branching-structure}). We place a single $G$-qudit on each site and a single $A$-qudit on each link.

\noindent\textbf{Aside: central extension symmetry.} To motivate the central 2-group symmetry operators, let us first recall
the analogous construction for an ordinary central extension. Let $\cG$ be a finite group fitting into an extension
\begin{equation}\label{centralext}
    1
    \longrightarrow A
    \longrightarrow \cG
    \longrightarrow G
    \longrightarrow 1.
\end{equation}
Because the extension is central, the action of $G$ on $A$ is trivial. Equivalence classes of such extensions, with $G$ and $A$ fixed, are classified by an extension class
\begin{equation}
    [\om]\in\cH^2(G,A).
\end{equation}
We choose a normalized representative
${\om\in\cZ^2(G,A)}$ of $[\om]$. The underlying set of $\cG$ may be identified with
${G\times A}$ whose group multiplication is
\begin{equation}
    (h_1,\la_1)(h_2,\la_2)
    =
    \bigl(
        h_1h_2,\,
        \la_1+\la_2+\om(h_1,h_2)
    \bigr).
\end{equation}

A set of operators acting on a single $G$- and $A$-qudit realizing this ordinary central extension is~\cite{PLA240918113}
\begin{equation}\label{centralExtOps}
    \big\{
    \overrightarrow{X}^{(h,\la)} = \sum_{g\in G,a\in A}\ketbra{hg,a+\la+\om(h,g)}{g,a}
    \big\}_{h\in G, \la\in A},
\end{equation}
Indeed, using the 2-cocycle condition on $\om$, they satisfy ${\overrightarrow{X}^{(h_1,\la_1)}\overrightarrow{X}^{(h_2,\la_2)} = \overrightarrow{X}^{(h_1h_2,\la_1+\la_2+\om(h_1,h_2))}}$. Consequently, in a lattice system with one $G$-qudit and one $A$-qudit on each site, the following onsite unitaries are $\cG$ symmetry operators:
\begin{align}
    \prod_{i\in\La_0} \overrightarrow{X}^{(h,0)}_i &= \sum_{\{g_i\},\{a_i\}}\ketbra{\{hg_i\},\{a_i+\om(h,g_i)\}}{\{g_i\},\{a_i\}},\label{central1grpsymopG}\\
    \prod_{i\in\La_0} \overrightarrow{X}^{(1,\la)}_i &= \sum_{\{g_i\},\{a_i\}}\ketbra{\{g_i\},\{a_i+\la\}}{\{g_i\},\{a_i\}}.\label{central1grpsymopA}
\end{align}

\noindent\textbf{Back to central 2-group symmetry.} 
The central 2-group extension~\eqref{central2grpext} is a categorified analog of the ordinary central extension~\eqref{centralext}, with the Postnikov class $[\bt]$ taking the role played by the extension class $[\om]$. Motivated by~\eqref{central1grpsymopG} and~\eqref{central1grpsymopA}, we introduce the lattice operators
\begin{align}
    U^{(h)} 
    &=
    \sum_{\{g_i\},\{a_{ij}\}}
    \ketbra
    {\{hg_i\}, \{a_{ij} + \bt(h,g_i, g_i^{-1}g_j)\}}
    {\{g_i\}, \{a_{ij}\}}
    ,\label{eq:central-2group-G-sym-op}\\
    T^{(\ga)} &= \prod_{ij\in\La_1} X^{(\ga_{ij})}_{ij},\label{eq:central-2group-A-1sym-op}
\end{align}
where ${\ga}$ is an $A$-valued $1$-cycle of the dual lattice. 
The operator~\eqref{eq:central-2group-G-sym-op} is the generalization of~\eqref{central1grpsymopG} to the $A$-qudits residing on links rather than sites and the cocycle being a 3-cocycle $\bt$ rather than a 2-cocycle $\om$.\footnote{
The analogy between~\eqref{central1grpsymopG} and~\eqref{eq:central-2group-G-sym-op} is especially transparent in homogeneous-cochain notation. For trivial $G$-action on $A$, an
inhomogeneous $n$-cochain
${\nu^{(n)}\colon G^n\to A}$ corresponds to the homogeneous cochain
${\widetilde{\nu}^{(n)}\colon G^{n+1}\to A}$ defined by
\begin{equation*}
    \widetilde{\nu}^{(n)}(g_0,g_1,\ldots,g_n)
    =
    \nu^{(n)}
    (
        g_0^{-1}g_1,\,
        g_1^{-1}g_2,\,
        \ldots,\,
        g_{n-1}^{-1}g_n
    ).
\end{equation*}
In terms of their corresponding homogeneous cochains,
\begin{equation*}
    \om(h,g_i)
    =
    \widetilde{\om}(1,h,hg_i),
    \qquad
    \bt\bigl(h,g_i,g_i^{-1}g_j\bigr)
    =
    \widetilde{\bt}(1,h,hg_i,hg_j).
\end{equation*}
} The operator~\eqref{eq:central-2group-A-1sym-op} is the generalization of~\eqref{central1grpsymopA} to a 1-form symmetry.
Both $U^{(h)}$ and $T^{(\ga)}$ are unitary operators.
(While $U^{(h)}$ is unitary, ${(U^{(h)})^{-1}}$ does not generally equal ${U^{(h^{-1})}}$; their multiplication law will be derived below.)

For the central 2-group construction, we also introduce the
operators
\begin{equation}\label{eq:central-Sf}
    S_f
    =
    \sum_{\{g_i\},\{a_{ij}\}}
    \ketbra{
        \{g_i\},
        \{a_{ij}-f(g_i)+f(g_j)\}
    }{
        \{g_i\},
        \{a_{ij}\}
    },
\end{equation}
where ${f\in\mathrm{Map}(G,A)\cong A^{|G|}}$. 
The operators~\eqref{eq:central-2group-G-sym-op} are generically not $G$ 0-form symmetry operators. Instead, they satisfy
\begin{equation}\label{eq:tilde-u-algebra}
    U^{(h)}U^{(k)}
        = U^{(hk)} S_{b_{h,k}},
\end{equation}
where ${b_{h,k}\in \mathrm{Map}(G,A)}$ satisfies
\begin{equation}
    b_{h,k}(g) = \bt(h,k,g).
\end{equation}
On the other hand, the operators $\{S_f\}_{f\in\mathrm{Map}(G,A)}$ satisfy
\begin{equation}
    S_{f_1}\times S_{f_2} = S_{f_1+f_2},\qquad S_f^{-1} = S_{-f},\qquad S_{f} = 1
    \ \Longleftrightarrow\
    f\text{ is constant}. 
\end{equation}
Therefore, they furnish a faithful representation of the finite abelian group 
\begin{equation}
    \text{Map}(G,A)/A_\text{const} \cong A^{|G|-1},
\end{equation}
where $A_\text{const}$ is the subgroup of constant maps in ${\text{Map}(G,A)}$.\footnote{A constant map ${f\colon G\to A}$ is one satisfying ${f(g) = a}$, with fixed ${a\in A}$, for all ${g\in G}$.}

For the lattice central 2-group symmetry we consider, we take the 0-form symmetry part to be formed by all $U^{(h)}$ and $S_f$. We denote a generic 0-form operator by
\begin{equation}\label{GbetaSymOps}
    U^{(h, [f])} \equiv U^{(h)}S_{f}, \qquad h\in G,~[f]\in \mathrm{Map}(G,A)/A_\mathrm{const}.
\end{equation}
It satisfies
\begin{equation}\label{UgSfmult}
    U^{(g_1, [f_1])}
    U^{(g_2, [f_2])}
    =
    U^{(g_1 g_2, [f_1\triangleleft g_2+f_2 + b_{g_1,g_2}])},
\end{equation}
where ${(f\triangleleft g)(h) = f(gh)}$.
The operators $\{ U^{(h, [f])}\}$ are $\t{G}$ 0-form symmetry operators, where the finite group $\t{G}$ is described by the group extension
\begin{equation}\label{extensionofGNbt}
    1 \longrightarrow \mathrm{Map}(G,A)/A_{\mathrm{const}} \longrightarrow \t{G} \xlongrightarrow{p} G \longrightarrow 1.
\end{equation}
This is generally not a central extension since ${U^{(g)}S_{f}(U^{(g)})^{-1} = S_{f\triangleleft g^{-1}}}$. As shown in App.~\ref{app:central-beta-dependence}, changing the representative $\bt$ of the Postnikov class $[\bt]$ in~\eqref{eq:central-2group-G-sym-op} changes the representative of the extension class $[b]$ of~\eqref{extensionofGNbt}.
Note that this extension splits if and only if $[\beta]$ is trivial. Therefore this group extension is a lattice manifestation of the nontriviality of the Postnikov class $[\beta]$.

The operators $T^{(\ga)}$ are the same lattice $A$ 1-form symmetry
operators introduced in the split 2-group case in
Eq.~\eqref{eq:split-2group-A-sym-op}. They satisfy the group law
\begin{equation}
    T^{(\ga_1)}T^{(\ga_2)}
    =
    T^{(\ga_1+\ga_2)}
\end{equation}
of $Z_1(\La^\vee;A)$, as in~\eqref{latticeA1FSgrouplaw}. 
Since
$U^{(h,[f])}$ and $T^{(\ga)}$ act on each $A$-qudit by commuting
translations, they obey
\begin{equation}\label{central2Grp0fsand1fscommute}
    U^{(h, [f])} T^{(\ga)} = T^{(\ga)} U^{(h, [f])}.
\end{equation}
As discussed in Sec.~\ref{sec:split-2group-sym-ops}, $T^{(\ga)}$ are lattice 1-form symmetry operators on the full Hilbert space, but become topological and implement a 1-form symmetry in the subspace $\scrH_\mathrm{top}$ given by~\eqref{HtopDefSplit2GrpSection}.

The symmetry formed by $\{U^{(g,[f])}, T^{(\ga)}\}_{(g,[f])\in \t{G},\, \ga\in Z_1(\La^\vee,A)}$ is the lattice avatar of the central 2-group $\G$ symmetry.\footnote{One could treat the symmetry $\{U^{(g,[f])}, T^{(\ga)}\}_{(g,[f])\in \t{G},\, \ga\in Z_1(\La^\vee,A)}$ as its own central 2-group symmetry formed by a $\t{G}$ 0-form symmetry and an $A$ 1-form symmetry. This is a bit misguided, however. The fact that a central 2-group depends on the cohomology class of $\bt$ and not the 3-cocycle relies on the 1-form symmetry being topological. That said, we do note that if it were treated as a central 2-group symmetry, its Postnikov class would be given by $p^*[\bt]$ where $p^*$ is the pullback of the quotient homomorphism $p$ in~\eqref{extensionofGNbt}~\cite{BH180309336}. In App.~\ref{app:pullback-beta-trivialization}, we show that ${p^*[\bt] = [0]}$.} This is in the same sense in which $\{T^{(\ga)}\}_{\ga\in Z_1(\La^\vee,A)}$ is the lattice analog of an $A$ 1-form symmetry: In the topological subspace, $\{U^{(g,[f])}, T^{(\ga)}\}_{(g,[f])\in \t{G},\, \ga\in Z_1(\La^\vee,A)}$ are central 2-group symmetry operators. 
Denote by\footnote{In the topological subspace, changing the choice of representative normalized 3-cocycle $\bt$ of the Postnikov class conjugates $U^{(h)}_\mathrm{top}$ by a unitary. The symmetry group represented by $U^{(h)}_\mathrm{top}$ is $G$, independent of $\bt$. See App.~\ref{app:central-beta-dependence} for more discussion.}
\begin{equation}\label{Utopdef}
    U^{(h)}_\mathrm{top} \equiv U^{(h,[f])}\big|_{\scrH_\mathrm{top}},
    \qquad
    T^{([\ga])}_\mathrm{top} \equiv T^{(\ga)}\big|_{\scrH_\mathrm{top}}.
\end{equation}
When restricted to the topological subspace $\scrH_\mathrm{top}$, these symmetry operators satisfy
\begin{equation}\label{central2GrpSymAlg}
    U^{(h)}_\mathrm{top}U^{(k)}_\mathrm{top} = U^{(hk)}_\mathrm{top},\qquad
    T^{([\ga_1])}_\mathrm{top}T^{([\ga_2])}_\mathrm{top} = T^{([\ga_1+\ga_2])}_\mathrm{top},
    \qquad
    U^{(h)}_\mathrm{top} T^{([\ga])}_\mathrm{top} = T^{([\ga])}_\mathrm{top} U^{(h)}_\mathrm{top},
\end{equation}
Indeed, every $S_f$ acts trivially on the basis vectors~\eqref{Htopbasisvectors} of $\scrH_\mathrm{top}$ and, therefore, every $S_f$ acts trivially in $\scrH_\mathrm{top}$:
\begin{equation}\label{SfonTopSub}
    S_f \big|_{\scrH_\mathrm{top}} = 1.
\end{equation}
This causes $\{U^{(h)}\}_{h\in G}$ to become $G$ symmetry operators on $\scrH_\mathrm{top}$ and, hence, restricting to $\scrH_\mathrm{top}$ implements the quotient homomorphism ${p\colon \t{G}\to G}$ in~\eqref{extensionofGNbt}. 
The symmetry algebra~\eqref{central2GrpSymAlg} is the algebra of a central $\G$ symmetry. It is not affected by the Postnikov class $[\bt]$. We will see the role of the Postnikov class throughout this section, through these lattice operators' charged operators in Sec.~\ref{sec:obstructionToSubSym}, symmetry defects in Sec.~\ref{sec:central-2group-sym-defects}, and through their gauging in Secs.~\ref{sec:central-gauging} and~\ref{sec:central-quantum-double}.

\subsubsection{Onsiteability}\label{sec:onsiteability}

The $\t{G}$ 0-form symmetry operators~\eqref{GbetaSymOps} are not onsite. This is different from the split 2-group symmetry operators in Sec.~\ref{sec:split-2group-sym-ops}, which were all onsite. Here we discuss whether, and how, the $\t{G}$ 0-form symmetry operators can be made onsite.

In the simplest scenario, a non-onsite symmetry operator can be made onsite using a locality-preserving unitary transformation. If the operators $\{U^{(h,[f])}\}_{(h,[f])\in \t{G}}$ can be made onsite, then ${U^{(1,[f]) }= S_f}$ is necessarily onsiteable too. However, we show in App.~\ref{app:Sf-non-onsiteability} that there does not exist a quantum cellular automaton (QCA)\footnote{Recall that, from a Hilbert space perspective, a QCA is a unitary operator $U$ that maps every operator with finite support to another operator with finite support. That is, if $O_X$ has finite support $X$, then $UO_XU^\dagger$ has finite support $\cN_R(X)$, where $R$ is the range of the QCA $U$ and $\cN_R(X)$ denotes the $R$-neighborhood of $X$. From an operator algebra perspective, a QCA is a $*$-automorphism satisfying this strictly locality-preserving condition. See, for example, the beginning of~\cite{FHH191007998} for an introduction.} $W$ such that ${W S_f W^{\dagger}}$ is onsite. Therefore, there does not exist a QCA on the Hilbert space~\eqref{GsitesAlinksHilb1} that can make each $\{U^{(h,[f])}\}_{(h,[f])\in \t{G}}$ onsite.

\noindent\textbf{Adding ancilla qudits.} The next simplest possibility is that, while $\{U^{(h,[f])}\}_{(h,[f])\in \t{G}}$ cannot be made onsite using a QCA, it can be made onsite using finite-dimensional ancillae and a QCA.
A prototypical example of such a symmetry operator that requires ancilla qudits to be onsiteable is the ${1+1}$d $\Z_2$ symmetry operator $\prod_{j} \mathsf{CZ}_{j,j+1}$~\cite{ZLL241105004}.
We now show that the symmetry $\{U^{(h,[f])}\}_{(h,[f])\in \t{G}}$ can be made onsite using ancilla qudits and a QCA. In fact, this QCA will turn out to be a finite-depth quantum circuit.

The ancillae we consider are $\t{G}$-qudits on the sites of the lattice. Thus, adding the ancilla qudits modifies the Hilbert space~\eqref{GsitesAlinksHilb1} as
\begin{equation}
    \scrH \longmapsto 
    \bigotimes_{i\in\La_0} (\C[G]\otimes \C[\t{G}])
    \otimes 
    \bigotimes_{ij\in\La_1} \C[A] \equiv \t{\scrH}.
\end{equation}
We note that $|\t{G}|$ is always finite and satisfies ${|\t{G}| = |G||A|^{|G|-1}}$.
We assume the ancillae $\t{G}$ symmetry operators are ${\prod_{i\in\La_0}\overrightarrow{\mathsf{X}}_i^{(g,[f])}}$ where ${\overrightarrow{\mathsf{X}}_i^{(g,[f])}}$ acts on the $\t{G}$-qudit at site $i$ by left-multiplication by ${(g,[f])\in \t{G}}$. After adding the ancillae, the $\t{G}$ symmetry is generated by the diagonal subgroup of the original and ancilla $\t{G}$ symmetry. Therefore, adding these ancillae modifies the $\t{G}$ symmetry operators~\eqref{GbetaSymOps} as
\begin{equation}
    U^{(g, [f])} \longmapsto U^{(g)}S_{f} \prod_{i\in\La_0}\overrightarrow{\mathsf{X}}_i^{(g,[f])} \equiv \t{U}^{(g, [f])}.
\end{equation}
The new symmetry operator $\t{U}^{(h, [f])}$ acts on $\ket{\{g_i\}, \{k_i,[s_i]\}, \{a_{ij}\}}$ by
\begin{equation}\label{diagSymAction}
    \begin{gathered}
        g_i \longmapsto h g_i,\qquad k_i \longmapsto hk_i, \qquad [s_i] \longmapsto [s_i+f \triangleleft k_i + b_{h,k_i}],\\
        a_{ij} \longmapsto a_{ij} + \bt(h,g_i,g_i^{-1}g_j) - f(g_i) + f(g_j).
    \end{gathered}
\end{equation}

We now present a QCA $W$ that simultaneously makes each $\t{U}^{(h, [f])}$ onsite. For each basis vector $\ket{\{g_i\}, \{k_i,[s_i]\}, \{a_{ij}\}}$, we define the $A$ group elements
\begin{equation}
    \xi_{ij} = s_i(k_i^{-1}g_i) - s_i(k_i^{-1} g_j ) - \bt(k_i, k_i^{-1}g_i, g_i^{-1}g_j).
\end{equation}
Using the 3-cocycle condition on $\bt$, each element $\xi_{ij}$ transforms under~\eqref{diagSymAction} as
\begin{equation}\label{xiFtransformation}
    \xi_{ij} \longmapsto \xi_{ij} - \left( \bt(h,g_i,g_i^{-1}g_j) - f(g_i) + f(g_j)\right).
\end{equation}
The element of $A$ that $\xi_{ij}$ transforms by is the inverse of the element of $A$ that each $a_{ij}$ transforms by. Therefore, we define the QCA
\begin{equation}\label{eq:Gbeta-onsiteizing-circuit}
    W
    =
    \sum_{\{g_i\},\{k_i,[s_i]\},\{a_{ij}\}}
    \ketbra{
        \{g_i\},
        \{k_i,[s_i]\},
        \{a_{ij}+\xi_{ij}\}
    }{
        \{g_i\},
        \{k_i,[s_i]\},
        \{a_{ij}\}
    }.
\end{equation}
This is a finite-depth quantum circuit.
From~\eqref{diagSymAction} and~\eqref{xiFtransformation}, it satisfies
\begin{equation}
    W \t{U}^{(g, [f])} W^\dag = \prod_{i\in\La_0} \overrightarrow{X}_i^{(g)}\,\overrightarrow{\mathsf{X}}_i^{(g,[f])}.
\end{equation}
Therefore, $W$ makes $\t{U}^{(g, [f])}$ onsite, and $\t{U}^{(g, [f])}$ is onsiteable using ancillae and a finite-depth quantum circuit.

\subsubsection{Postnikov obstruction to 0-form subsymmetry}\label{sec:obstructionToSubSym}

Although the Postnikov class enters the multiplication law of the lattice $\t{G}$ 0-form symmetry operators, it is not visible in their mutual commutation relation with the $A$ 1-form symmetry operators; see~\eqref{central2Grp0fsand1fscommute}.
This does not mean, however, that the 0-form and 1-form symmetries act independently. The Postnikov class becomes visible in the action of the 0-form symmetry on operators carrying 1-form symmetry charge. This is an operator-level manifestation of the obstruction to treating the 0-form symmetry as a subsymmetry of the central 2-group~\cite{CI180204790, BH180309336}.

On the full tensor-product Hilbert space, operators carrying $A$ 1-form symmetry charge are the Wilson operators
\begin{equation}\label{genericWilsonOpAqudits}
    W^{(\Xi)}
    =
    \prod_{ij\in\La_1}
    Z_{ij}^{(\Xi_{ij})\,\dag},
    \qquad
    \Xi\in C_1(\La;\h{A}),
\end{equation}
where ${\h{A} = \Hom(A,\Uone)}$.
Indeed, they satisfy
\begin{equation}\label{eq:Wilson-one-form-charge}
    T^{(\ga)}
    W^{(\Xi)}
    T^{(\ga)\,\dagger}
    =
    \<\Xi,\ga\>\,
    W^{(\Xi)},
    \qquad
    \<\Xi,\ga\>
    =
    \prod_{ij\in\La_1}
    \Xi_{ij}(\ga_{ij})
    \in\Uone.
\end{equation}
A Wilson operator restricts to the topological subspace~\eqref{HtopDefSplit2GrpSection} if and only if it commutes with every star operator $A_i^{(\la)}$, which requires ${\Xi\in Z_1(\La;\h{A})}$.

It is sufficient to consider Wilson operators carrying a fixed charge ${\chi\in\widehat A}$. Let
\begin{equation}
    \eps
    =
    \sum_{ij\in\La_1}
    \eps_{ij}\,[ij],
    \qquad
    \eps_{ij}=0,\pm1,
\end{equation}
be a 1-chain describing an oriented lattice path, and define
\begin{equation}\label{eq:elementary-Wilson-eps}
    W_{\eps}^{(\chi)}
    :=
    \prod_{ij\in\La_1}
    \left(
        Z_{ij}^{(\chi)\,\dagger}
    \right)^{\eps_{ij}}.
\end{equation}
Products of such operators generate the general Wilson
operators~\eqref{genericWilsonOpAqudits}. The operator
$W_{\eps}^{(\chi)}$ preserves $\scrH_{\mathrm{top}}$ precisely when
$\eps$ is closed: ${\pp\eps=0}$.

The $\t{G}$ 0-form symmetry operators act nontrivially on these Wilson operators.
Indeed, define the operator
\begin{equation}\label{Dopdef}
    D^{(h,[f],\chi)}_{\eps} = \sum_{\{g_i\},\{a_{ij}\}}
    \prod_{ij\in\La_1} 
    \chi
    \big(
    \bt(h,h^{-1}g_i, g_i^{-1}g_j)
    -f(h^{-1}g_i)
    +f(h^{-1}g_j)
    \big)^{\eps_{ij}}
    \ketbra{\{g_i\},\{a_{ij}\}}{\{g_i\},\{a_{ij}\}}.
\end{equation}
The $\t{G}$ 0-form symmetry operator~\eqref{GbetaSymOps} acts on $W^{(\chi)}_{\eps}$ as
\begin{equation}\label{GbetaactiononW}
    U^{(h,[f])}
    \,
    W^{(\chi)}_{\eps}
    U^{(h,[f])\,\dag} = 
    D^{(h,[f],\chi)}_{\eps}
    \,
    W^{(\chi)}_{\eps}.
\end{equation}
Thus, the $0$-form symmetry does not generally map an $A$-qudit Wilson line to another operator acting only on the $A$-qudits. Instead, it attaches the $G$-configuration-dependent dressing ${D_{\eps}^{(h,[f],\chi)}}$.

For a closed lattice path, $\eps$ satisfies ${\pp\eps=0}$ and the $f$-dependent factor in~\eqref{Dopdef} telescopes:
${\prod_{ij\in\La_1}\chi(
-f(h^{-1}g_i)+f(h^{-1}g_j)
)^{\eps_{ij}} = 1}$.
Consequently, when ${\pp\eps = 0}$, 
\begin{equation}\label{DopTopSubSpace}
    D^{(h,[f],\chi)}_{\eps} = \sum_{\{g_i\},\{a_{ij}\}}
    \prod_{ij\in\La_1} 
    \chi
    \big(
    \bt(h,h^{-1}g_i, g_i^{-1}g_j))
    \big)^{\eps_{ij}}
    \ketbra{\{g_i\},\{a_{ij}\}}{\{g_i\},\{a_{ij}\}}
    \equiv
    D^{(h,\chi)}_{\eps},
\end{equation}
and, therefore,
\begin{equation}\label{WtransunderGTopSub}
    U^{(h,[f])}
    \,
    W^{(\chi)}_{\eps}
    U^{(h,[f])\,\dag} = 
    D^{(h,\chi)}_{\eps}
    \,
    W^{(\chi)}_{\eps}.
\end{equation}
This implies that
\begin{equation}
    U^{(h)}_\mathrm{top}
    \,
    \left(W^{(\chi)}_{\eps}\mid_{\scrH_{\mathrm{top}}}\right)
    U^{(h)\,\dag}_\mathrm{top} = 
    (D^{(h,\chi)}_{\eps}
    \,
    W^{(\chi)}_{\eps})
    \mid_{\scrH_{\mathrm{top}}}.
\end{equation}
The operators ${\{D_{\eps}^{(h,\chi)}\}_{h\in G}}$ furnish the non-onsite cocycle dressing of a ${1+1}$d $G$ symmetry action on the Wilson loop operator. The anomaly of the resulting ${1+1}$d $G$-symmetry action is ${[\chi\circ\bt] \in \cH^3(G,\Uone)}$.\footnote{\label{footnote:D-ops}
More precisely, consider a periodic one-dimensional lattice with a $G$-qudit on each site and a normalized cocycle ${\om\in\cZ^3(G,\Uone)}$.
Define 
${D^{(h,\om)} = \sum_{\{g_i\}}
\prod_{ij\in\La_1} 
\om
(h,h^{-1}g_i, g_i^{-1}g_j)
\ketbra{\{g_i\}}{\{g_i\}}
}$. Multiplying the anomaly-free $G$ symmetry operator ${U^{(h)}_{[1]} = \prod_{i\in\La_0} \overrightarrow{X}^{(h)}_i}$ by $D^{(h,\om)}$ yields the $G$ symmetry operator ${U^{(h)}_{[\om]} = D^{(h,\om)}U^{(h)}_{[1]}}$. This has an anomaly $[\om]$. One way to see this is as follows. Let ${V^{(\om)} = \sum_{\{g_i\}}
\prod_{ij\in\La_1} 
\om
(g_i, g_i^{-1} g_j, g_j^{-1})^{-1}
\ketbra{\{g_i\}}{\{g_i\}}}$. The operator $U^{(h)}_{[\om]}$ satisfies ${V^{(\om)} U^{(h)}_{[\om]} V^{(\om)\,\dag} = 
\sum_{\{g_i\}}
\prod_{ij\in\La_1} 
\om(g_i^{-1} g_{j}, g_{j}^{-1} h^{-1}, h)
\ketbra{\{hg_i\}}{\{g_i\}}}$, which is a fixed-point boundary symmetry action of a ${2+1}$d $G$ SPT with class ${[\om]}$~\cite{CGL11064772}.
See~\cite{WWW170506728} for the generalization of $D^{(h,\om)}$ to arbitrary spatial dimension.
}

The invariant content of~\eqref{WtransunderGTopSub} depends only on the cohomology class ${[\chi\circ\bt]}$:
\begin{itemize}
    \item Suppose $\chi$ satisfies ${\chi\circ\bt = \del c}$ for some ${c\in \cC^2(G,\Uone)}$. 
Defining
\begin{equation}
    \cR_{\eps}^{(c)}
    =
    \sum_{\{g_i\},\{a_{ij}\}}
    \prod_{ij\in\La_1}
    c(g_i,g_i^{-1}g_j)^{\eps_{ij}}
    \ketbra{\{g_i\},\{a_{ij}\}}{\{g_i\},\{a_{ij}\}},
\end{equation}
the operator $D^{(h,\chi)}_{\eps}$ for such $\chi$ can be written as
\begin{equation}
    D_{\eps}^{(h,\chi)}
    =
    U^{(h,[f])}
    \cR_{\eps}^{(c)}
    U^{(h,[f])\,\dag}
    \cR_{\eps}^{(c)\,\dag}
\end{equation}
for all ${[f]\in \mathrm{Map}(G,A)/A_\mathrm{const}}$.
Consequently, the locally dressed Wilson loop
\begin{equation}\label{localDressingofW}
    \cW_{\eps}^{(\chi,c)}
    =
    \cR_{\eps}^{(c)\,\dag}\,
    W_{\eps}^{(\chi)}
\end{equation}
is invariant under the lattice $\t{G}$ 0-form symmetry:
\begin{equation}
    U^{(h,[f])}
    \cW_{\eps}^{(\chi,c)}
    U^{(h,[f])\,\dag}
    =
    \cW_{\eps}^{(\chi,c)}.
\end{equation}
Upon restriction to $\scrH_{\mathrm{top}}$, this becomes invariance under the quotient $G$ symmetry. Because $\cR_{\eps}^{(c)}$ acts only on the $G$-qudits, the dressed operator $\cW_{\eps}^{(\chi,c)}$ carries the same $A$ 1-form symmetry charge as $W_{\eps}^{(\chi)}$.
\item If ${[\chi\circ\bt]\neq[1]}$ in $\cH^3(G,\Uone)$, however, then no such local dressing~\eqref{localDressingofW} exists. Then, the image of a Wilson loop of charge $\chi$ under the 0-form symmetry carries an intrinsically ${1+1}$d $G$ symmetry action whose anomaly is the pushforward ${[\chi\circ\bt]}$.
Thus, a $\chi$ Wilson loop detects the Postnikov class $[\bt]$ through its pushforward $[\chi\circ\bt]$. This is a charged operator manifestation of the Postnikov class's obstruction to treating the $G$ 0-form symmetry as an independent subsymmetry of the central 2-group, and agrees with the QFT analysis of~\cite{HJS260804248}.
\end{itemize}

\subsubsection{Central 2-group Ising model}\label{sec:central-2group-ising-model}

As in the split case, there are infinitely many local Hamiltonians that commute with the central 2-group symmetry operators~\eqref{GbetaSymOps} and~\eqref{eq:central-2group-A-1sym-op}. In Sec.~\ref{sec:split-2group-ising-model}, we introduced one such Hamiltonian, the split $\G$-Ising model. 
We now construct its analog for a central 2-group. 
We denote by ${A_i = \frac1{|A|}\sum_{\la\in A} A_i^{(\la)}}$, where $A_i^{(\la)}$ is defined in Eq.~\eqref{HtopDefSplit2GrpSection}, and also define the $\bt$-twisted $A$-qudit plaquette operator
\begin{equation}
    B^{(\chi, \bt)}_{ijk} = \sum_{\{g_l\},\{a_{lm}\}} \chi\big(a_{ij} - a_{ik} + a_{jk} - \bt(g_i,\, g_i^{-1}g_j,\, g_j^{-1}g_k)\big)\, \ketbra{\{g_l\},\{a_{lm}\}}.
\end{equation}
The $\bt$-dependent shift in $B^{(\chi, \bt)}_{ijk}$ is required for ${[B^{(\chi, \bt)}_{ijk},U^{(h)}] = 0}$.

The central 2-group Ising model is described by the local Hamiltonian
\begin{align}\label{eq:central-2group-isingmodel}
    H_{\G\text{-Ising}}^{(\bt)} = 
    -\frac12\bigg(& \sum_{ij\in\La_1} \sum_{\Ga\in\Irr G} J^{(\Ga)}_G \tr[Z_i^{(\Ga)} Z_{j}^{(\bar{\Ga})}]
    + \sum_{i\in \La_0} \sum_{h\in G}  h^{\!(h)}_G\, \overleftarrow{X}_{i}^{(h)} \, A_i
    M_i^{(h,\bt)}
    \\
    &
    +  \sum_{ijk\in\La_2} \sum_{\chi\in \h{A}}  J_A^{(\chi)} B_{ijk}^{(\chi,\bt)} 
    + \sum_{ij\in \La_1} \sum_{\la\in A} h^{\!(\la)}_A X^{(\la)}_{ij}
    + \sum_{i\in\La_0} A_i\bigg)
    +
    \mathrm{H.c.}
    ,\nonumber
\end{align}
where $J_G^{(\Gamma)}$, $h_G^{(h)}$, $J_A^{(\chi)}$, $h_A^{(\la)}\in \R$. 
The ordinary $G$-paramagnetic term ${\sum_{i\in\La_0}\sum_{h\in G} h^{\!(h)}_G \overleftarrow{X}^{(h)}_i}$, which appears in the $G$ Ising model and in the split 2-group Ising model~\eqref{eq:split-2group-isingmodel}, is dressed in $H_{\G\text{-Ising}}^{(\bt)}$ by ${A_iM_i^{(h,\bt)}}$ where
\begin{equation}
    M_{l}^{(h,\bt)}
    \equiv
   \begin{tikzpicture}[decoration={markings, mark=at position 0.55 with {\arrow{>}}}, scale=3.2]
    \coordinate (c) at (0,0);
    \coordinate (e) at (1,0);
    \coordinate (w) at (-1,0);
    \coordinate (nw) at (-0.5,{sqrt(3)/2});
    \coordinate (ne) at (0.5,{sqrt(3)/2});
    \coordinate (sw) at (-0.5,-{sqrt(3)/2});
    \coordinate (se) at (0.5,-{sqrt(3)/2});
    
    \draw[postaction=decorate, color=lightgray] (c) -- node[color=black,align=center, yshift=0.1em, xshift=0.55em] 
    {\footnotesize $X^{(\bt(g_l, h^{-1}, h g_l^{-1} g_m))}$} 
    (e) node[color=gray, right] {\footnotesize $m$};
    \draw[postaction=decorate, color=lightgray] (c) -- 
    node[color=black, xshift=1.5em, yshift=1em, align=center] 
    {\footnotesize$ X^{(\bt(g_l, h^{-1}, h g_l^{-1} g_n))}$} 
    (nw) node[color=gray, above] {\footnotesize $n$};
    \draw[postaction=decorate, color=lightgray] (c) -- 
    node[color=black, yshift=-0.5em, xshift=2em, align=center] 
    {\footnotesize$ X^{(\bt(g_l, h^{-1}, h g^{-1}_l g_o))}$} 
    (ne) node[color=gray, above] {\footnotesize $o$};
    \draw[postaction=decorate, color=lightgray] (sw) node[color=gray, below] {\footnotesize $i$} -- 
    node[color=black,xshift=1.5em, yshift=0.5em, align=center] 
    {\footnotesize$ X^{(- \bt(g_i, g_i^{-1} g_l, h^{-1}))}$} 
    (c);
    \draw[postaction=decorate, color=lightgray] (se) node[color=gray, below] {\footnotesize $j$} -- 
    node[color=black,xshift=1.9em, yshift=-1em, align=center] 
    {\footnotesize$ X^{(- \bt(g_j, g_j^{-1} g_l, h^{-1}))}$} 
    (c);
    \draw[postaction=decorate, color=lightgray] (w) node[color=gray, left] {\footnotesize $k$} -- 
    node[color=black, align=center, yshift=0.1em,] 
    {\footnotesize$ X^{( - \bt(g_k, g_k^{-1} g_l, h^{-1}))}$} 
    (c) node[color=gray, below] {\footnotesize $l$};
\end{tikzpicture}.
\end{equation}
This dressing is required to make the $G$-paramagnetic term $\t{G}$ symmetric.
The Hamiltonian commutes with the lattice central 2-group symmetry operators~\eqref{GbetaSymOps} and \eqref{eq:central-2group-A-1sym-op}. The final term in~\eqref{eq:central-2group-isingmodel} energetically favors the topological subspace $\scrH_{\mathrm{top}}$ of the lattice $A$ 1-form symmetry.

As with the split $\G$-Ising model, the Hamiltonian $H_{\G\text{-Ising}}^{(\bt)}$ has a rich phase diagram. Consider the choice of couplings ${J_G^{(\Ga)} = d_\Ga / |G|}$, where $d_\Ga$ is the dimension of the irrep $\Ga$, ${h_G^{(h)} = h_G/|G|}$, ${J_{A}^{(\chi)} = 1/|A|}$, and ${h_A^{(\la)} = h_A/|A|}$. In this limit, the central $\G$-Ising model becomes 
\begin{equation}\label{eq:central-2group-pottsmodel}
\begin{aligned}
    H_{\G\text{-Potts}}^{(\bt)} = 
    & -\frac{1}{|G|}\bigg(\sum_{ij\in\La_1} \sum_{\Ga\in\Irr G} 
    d_\Ga \tr[Z_i^{(\Ga)} Z_{j}^{(\bar{\Ga})}]
    + 
    \frac{h_G}2 \sum_{i\in \La_0} \sum_{h\in G} (\overleftarrow{X}_{i}^{(h)}  M_i^{(h,\bt)}
    +
     M_i^{(h^{-1},-\bt)}   \overleftarrow{X}_{i}^{(h)}
     )A_i
    \bigg)
    \\
    &
    -
    \frac{1}{|A|}
    \bigg(  \sum_{ijk\in\La_2} \sum_{\chi\in \h{A}} B_{ijk}^{(\chi,\bt)} 
    + h_A \sum_{ij\in \La_1} \sum_{\la\in A} X^{(\la)}_{ij}
    \bigg)
    - \sum_{i\in\La_0} A_i
    ,
\end{aligned}
\end{equation}
which we refer to as the central $\G$-Potts model. 
While the split 2-group Potts model~\eqref{eq:split-2group-pottsmodel} had four exactly-solvable fixed points, the central 2-group Potts model~\eqref{eq:central-2group-pottsmodel} has three. 
Namely, its ground states are exactly-solvable in the limits: ${h_G=h_A=0}$, ${h_G=0,h_A\to\infty}$, ${h_G,h_A\to\infty}$. In these limits, it is straightforward to show that the SSB pattern and ground-state degeneracies are the same as the split case:
\setlength{\tabcolsep}{8pt} \renewcommand{\arraystretch}{1.5} 
\begin{center}
\begin{tabular}{c|c|c} 
  \quad Parameters \quad   &  \quad  $\G$-SSB pattern \quad  & \quad  ground-state degeneracy\quad  \\ 
\hhline{=|=|=}
$h_G = h_A = 0$  & $\G\ssb 1$ & $|H^0(\La,G)| \, |H^1(\La,A)|$  \\
\hline 
$h_G = 0, h_A \to\infty$  & $\G\ssb A^{(1)}$ &  $|H^0(\La,G)|$ \\
\hline 
$h_G, h_A \to\infty $  & $\G\ssb \G$ &  1 \\
\end{tabular}
\end{center}
\renewcommand{\arraystretch}{1}
However, the Hamiltonian is not exactly-solvable in the limit ${h_G\to\infty}$, ${h_A = 0}$.
Indeed, the second term---the dressed $G$-paramagnetic term---fails to commute with the third term---the dressed $A$-plaquette operator term. This is true even after projecting into the topological subspace where every ${A_i = 1}$.

\subsection{Symmetry defects}\label{sec:central-2group-sym-defects}

A characteristic manifestation of a nontrivial Postnikov class is
that 0-form symmetry defects can source 1-form symmetry defects.
Namely, a 1-form symmetry defect $\bt(g,h,k)$ can end on a junction between $g$, $h$, and $k$ 0-form symmetry defects~\cite{BH180309336}.
As we show in App.~\ref{sec:defectSectorFusionApp}, in the presence of a 0-form symmetry defect, this modifies the 0-form symmetry fusion rule by a 1-form symmetry defect. (See Fig.~\ref{fig:beta-junction-resolution}.)
In particular, after inserting a ${k\in G}$ 0-form symmetry defect, multiplying the ${g,h\in C_G(k)}$ 0-form symmetry operators yields a $gh$ 0-form symmetry operator and a ${\iota_k\bt(g,h)\in A}$ 1-form symmetry operator, where the slant product of $\bt$ with $k$ is
\begin{equation}
     \iota_k\bt(g,h) \equiv \bt(k,g,h) - \bt(g,k,h) + \bt(g,h,k).
\end{equation}
In this subsection, we show how this appears when the lattice central 2-group symmetry operators introduced above are restricted to the topological subspace.

\subsubsection{Full tensor-product Hilbert space}

We begin with the full tensor-product Hilbert space~\eqref{GsitesAlinksHilb1}.
We will insert a general $\t{G}$ 0-form symmetry defect, which will be labeled by the group element ${(k,[s])\in \t{G}}$.
For every ${[f]\in \mathrm{Map}(G,A)/A_\mathrm{const}}$, we choose a normalized representative satisfying ${f(1) = 0}$.
To insert the ${(k,[s])}$ symmetry defect along a loop $\eta^\vee$, we consider a truncation of the $U^{(k,[s])}$ symmetry operator that acts only on qudits belonging to the sites and links within a region whose boundary is $\eta^\vee$. In the thermodynamic limit, we consider the following truncation:
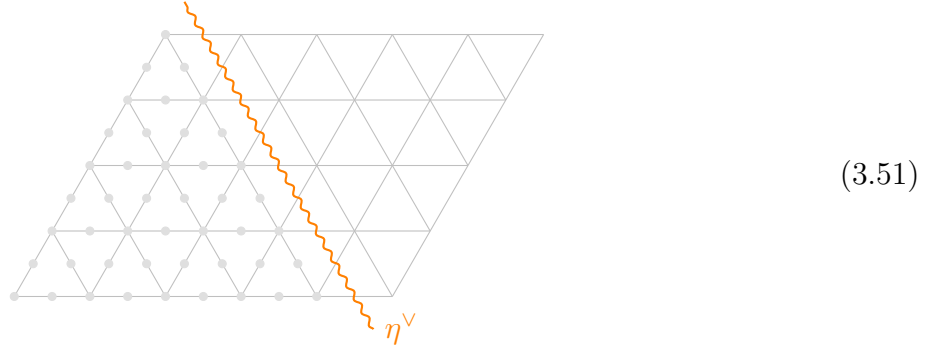
\begin{equation}
\begin{tikzpicture}[scale=1,
    decoration={markings, mark=at position 0.55 with {\arrow{>}}},
    d/.style={postaction=decorate},
    dot/.style={lightgray!50!white, opacity=1}
]

    \def\Nx{5}
    \def\Ny{4}

    \pgfmathsetmacro{\a}{1}
    \pgfmathsetmacro{\b}{0.8660254}  

    \foreach \i in {0,...,\Ny} {
        \foreach \j in {0,...,\Nx} {

            \coordinate (v_\i_\j)
                at (\j*\a + 0.5*\i*\a, \i*\b);

        }
    }

    \foreach \i in {0,...,\Ny} {
        \foreach \j in {0,...,\numexpr\Nx-1} {

            \pgfmathtruncatemacro{\jp}{\j+1}

            \draw[color=lightgray] (v_\i_\j) -- (v_\i_\jp);

            \coordinate (c_h_\i_\j)
                at ($(v_\i_\j)!0.5!(v_\i_\jp)$);

        }
    }

    \foreach \i in {0,...,\numexpr\Ny-1} {
        \foreach \j in {0,...,\Nx} {

            \pgfmathtruncatemacro{\ip}{\i+1}

            \draw[color=lightgray] (v_\i_\j) -- (v_\ip_\j);

            \coordinate (c_dr_\i_\j)
                at ($(v_\i_\j)!0.5!(v_\ip_\j)$);

        }
    }

    \foreach \i in {0,...,\numexpr\Ny-1} {
        \foreach \j in {1,...,\Nx} {

            \pgfmathtruncatemacro{\ip}{\i+1}
            \pgfmathtruncatemacro{\jm}{\j-1}

            \draw[color=lightgray] (v_\i_\j) -- (v_\ip_\jm);

            \coordinate (c_dl_\i_\j)
                at ($(v_\i_\j)!0.5!(v_\ip_\jm)$);

        }
    }

    \filldraw[dot] (c_dr_0_0) circle (1.5pt);
    \filldraw[dot] (c_dr_1_0) circle (1.5pt);
    \filldraw[dot] (c_dr_2_0) circle (1.5pt);
    \filldraw[dot] (c_dr_3_0) circle (1.5pt); 
    \filldraw[dot] (c_dr_0_1) circle (1.5pt);
    \filldraw[dot] (c_dr_1_1) circle (1.5pt);
    \filldraw[dot] (c_dr_2_1) circle (1.5pt);
    \filldraw[dot] (c_dr_0_2) circle (1.5pt);
    \filldraw[dot] (c_dr_1_2) circle (1.5pt);
    \filldraw[dot] (c_dr_0_3) circle (1.5pt);
    \filldraw[dot] (v_0_0) circle (1.5pt);
    \filldraw[dot] (v_1_0) circle (1.5pt); 
    \filldraw[dot] (v_2_0) circle (1.5pt);
    \filldraw[dot] (v_3_0) circle (1.5pt);
    \filldraw[dot] (v_4_0) circle (1.5pt);
    \filldraw[dot] (v_0_1) circle (1.5pt);
    \filldraw[dot] (v_1_1) circle (1.5pt);
    \filldraw[dot] (v_2_1) circle (1.5pt); 
    \filldraw[dot] (v_3_1) circle (1.5pt);
    \filldraw[dot] (v_0_2) circle (1.5pt);
    \filldraw[dot] (v_1_2) circle (1.5pt);
    \filldraw[dot] (v_2_2) circle (1.5pt);
    \filldraw[dot] (v_0_3) circle (1.5pt); 
    \filldraw[dot] (v_1_3) circle (1.5pt);
    \filldraw[dot] (v_0_4) circle (1.5pt);


    \filldraw[dot] (c_dl_0_1) circle (1.5pt);
    \filldraw[dot] (c_dl_1_1) circle (1.5pt);
    \filldraw[dot] (c_dl_2_1) circle (1.5pt);
    \filldraw[dot] (c_dl_3_1) circle (1.5pt);
    \filldraw[dot] (c_dl_0_2) circle (1.5pt);
    \filldraw[dot] (c_dl_1_2) circle (1.5pt);
    \filldraw[dot] (c_dl_2_2) circle (1.5pt);
    \filldraw[dot] (c_dl_0_3) circle (1.5pt);
    \filldraw[dot] (c_dl_1_3) circle (1.5pt);
    \filldraw[dot] (c_dl_0_4) circle (1.5pt);

    
    \filldraw[dot] (c_h_0_0) circle (1.5pt);
    \filldraw[dot] (c_h_1_0) circle (1.5pt);
    \filldraw[dot] (c_h_2_0) circle (1.5pt);
    \filldraw[dot] (c_h_3_0) circle (1.5pt);
    \filldraw[dot] (c_h_0_1) circle (1.5pt);
    \filldraw[dot] (c_h_1_1) circle (1.5pt);
    \filldraw[dot] (c_h_2_1) circle (1.5pt);
    \filldraw[dot] (c_h_0_2) circle (1.5pt);
    \filldraw[dot] (c_h_1_2) circle (1.5pt);
    \filldraw[dot] (c_h_0_3) circle (1.5pt);


    \draw[orange, thick, decorate, decoration={snake, amplitude=.3mm, segment length=2mm}] ($(c_h_0_4) + 0.5*(0.5,-\b)$) node[right] {$\eta^\vee $} -- ($(c_h_4_0) + (-0.5*0.5, 0.5*\b)$);
\end{tikzpicture}
\end{equation}
The gray dots indicate the qudits on which the truncated symmetry operator acts.

A symmetry operator $U^{(g,[f])}$ remains a symmetry in the presence of the ${(k,[s])}$ defect if its group element $(g,[f])$ is in the centralizer $C_{\t{G}}\big( (k,[s]) \big)$. 
Then, ${(h,[f])}$ commutes with $(k,[s])$ and is in ${C_{\t{G}}\big( (k,[s]) \big)}$ if and only if the element $h$ and representative $f$ satisfy
\begin{equation}\label{eq:Gbeta-defect-centralizer}
    kh = hk,
    \qquad
    [b_{k,h}+s\triangleleft h+f] = [b_{h,k}+f\triangleleft k+s].
\end{equation}
For normalized representatives $f$ and $s$, the second condition is equivalent to
\begin{equation}\label{eq:Gbeta-defect-centralizer2}
    \bt(h, k, g)-\bt(k, h, g) =  f(k) + f(g) - f(k g) - s(h) - s(g) + s(h g)\qquad\forall~g\in G.
\end{equation}

Conjugating $U^{(h,[f])}$, where ${(h,[f])\in C_{\t{G}}\big( (k,[s]) \big)}$, by this truncated $U^{(k,[s])}$ operator yields
\begin{equation}\label{eq:central-defect-symmetry-operator}
    U_{\eta^\vee; (k,[s])}^{(h,[f])}
    =
    \sum_{\left\{g_i\right\},\left\{a_{ij}\right\}}
    \ketbra{
    \{ h g_i\}, 
    \{
    a_{ij} 
    + \bt(h,\t{g}_{i;ij},\t{g}_{i;ij}^{\,-1}g_j)
    - f(\t{g}_{i;ij})+f(g_j)
    \}
    }
    {
    \{ g_i\}, \{a_{ij}\}
    },
\end{equation}
where
\begin{equation}
    \t{g}_{i;ij} = \begin{cases}
        g_i\quad &ij\not\in \eta^\vee,\\
        k^{-1} g_i\quad &ij\in \eta^\vee.
    \end{cases}
\end{equation}
This can also be written as
\begin{equation}
    U_{\eta^\vee; (k,[s])}^{(h,[f])} = 
    U^{(h,[f])}
    \prod_{ij\in\eta^\vee}X_{ij}^{(\bt(h,k^{-1}g_i,g_i^{-1} k g_j) - \bt(h,g_i,g_i^{-1}g_j)
    -
    f(k^{-1}g_i) + f(g_i))}.
\end{equation}
This operator is the ${(h,[f]) \in C_{\t{G}}\big( (k,[s]) \big)}$ symmetry operator in the presence of a ${{(k,[s])\in \t{G}}}$ symmetry defect at $\eta^\vee$.
Away from the defect line, it reduces locally to the defect-free operator $U^{(h,[f])}$. On links $ij$ crossed by the defect, the transformation is modified by the appearance of $k^{-1}$. Although $s$ does not appear explicitly in~\eqref{eq:central-defect-symmetry-operator}, it does determine which pairs ${(h,[f])}$ are allowed through the centralizer condition~\eqref{eq:Gbeta-defect-centralizer}.
The defect symmetry operators form a representation of the centralizer $C_{\t{G}}\big( (k,[s]) \big)$:\footnote{For two generic normalized maps ${f_1,f_2\in\mathrm{Map}(G,A)}$, the map ${b_{h_1, h_2}+f_1 \triangleleft h_2+f_2}$ is not normalized. Indeed, ${b_{h_1, h_2}+f_1 \triangleleft h_2+f_2}$ maps ${1\in G}$ to $f_1(h_2)$. Therefore, the normalized representative of ${[b_{h_1, h_2}+f_1 \triangleleft h_2+f_2]}$ is ${b_{h_1, h_2}+f_1 \triangleleft h_2+f_2 - c_{f_1(h_2)}}$ where the constant map ${c_{f_1(h_2)}\colon g \mapsto f_1(h_2)}$ for all ${g\in G}$.}
\begin{equation}
    U_{\eta^\vee ;(k,[s])}^{(h_1,[f_1])} U_{\eta^\vee ;(k,[s])}^{(h_2,[f_2])}=U_{\eta^\vee ;(k,[s])}^{(h_1 h_2,[b_{h_1, h_2}+f_1 \triangleleft h_2+f_2])}.
\end{equation}

\subsubsection{Topological subspace}

We now restrict the defect-sector symmetry operators to the
topological subspace $\scrH_{\mathrm{top}}$.
Recall that the topological subspace~\eqref{HtopDefSplit2GrpSection} of the $A$ 1-form symmetry is the subspace of the full tensor-product Hilbert space on which every star operator ${A^{(\la_i)}_i = 1}$.
This local constraint gives rise to the redundancy ${a_{ij}\sim a_{ij} + \la_j - \la_i}$ on the $A$-qudits.
The truncated $\t{G}$ symmetry operators commute with every $A^{(\la_i)}_i$. Therefore, inserting a $(k,[s])$ 0-form symmetry defect does not modify the topological subspace. 

In the defect-free sector, every operator ${U^{(1,[f])} = S_f}$ acts trivially on $\scrH_{\mathrm{top}}$ (see Eq.~\eqref{SfonTopSub}). In the presence of a ${(k,[s])}$ symmetry defect, however,
\begin{equation}
    U_{\eta^\vee; (k,[s])}^{(1,[f])}\big|_{\scrH_\mathrm{top}} =     \prod_{ij\in\eta^\vee}X_{ij}^{(f(k))}.
\end{equation}
where we used that ${(1,[f]) \in C_{\t{G}}\big( (k,[s]) \big)}$ requires ${f(k) + f(g) = f(k g)}$ for all $g$. Therefore, in the ${(k,[s])}$-defect sector, $S_f\big|_{\scrH_\mathrm{top}}$ acts as a ${f(k)\in A}$ 1-form symmetry operator along the 0-form symmetry defect.
Informally, this means that the extension~\eqref{extensionofGNbt} defining $\t{G}$ in the topological subspace of the ${(k,[s])}$-defect sectors becomes an extension of 0-form symmetry operators by 1-form symmetry operators. This is the mechanism on the lattice that gives rise to the defect network deformation shown in the right-hand side of Fig.~\ref{fig:beta-junction-resolution}.

We now work within the topological subspace and demonstrate this. For each $U_{\eta^\vee; (k,[s])}^{(h,[f])}$, we denote by
\begin{equation}\label{mathcalUdef}
    \mathcal{U}_{\eta^\vee; (k,[s])}^{(h)} = 
    U_{\eta^\vee; (k,[s])}^{(h,[f])}\big|_{\scrH_\mathrm{top}}
    \,
    \prod_{ij\in\eta^\vee}X_{ij}^{(s(h) - f(k))}.
\end{equation}
This is the operator $U_{\eta^\vee; (k,[s])}^{(h,[f])}$ restricted to the topological subspace and dressed by an $A$ 1-form symmetry operator along the symmetry defect. In the absence of the $(k,[s])$ symmetry defect, this operator is ${U_\mathrm{top}^{(h)} \equiv U^{(h,[f])}\big|_{\scrH_\mathrm{top}}}$.
It is $f$-independent even in the presence of the symmetry defect due to the dressing by the 1-form symmetry operator. 
Indeed, decomposing the group element ${(h,[f])}$ as ${(h,[f]) = (h,[\t{f}])(1,[f-\t{f}])}$ and using ${U_{\eta^\vee; (k,[s])}^{(h,[f])}\big|_{\scrH_\mathrm{top}} = 
U_{\eta^\vee; (k,[s])}^{(h,[\t{f}])}\big|_{\scrH_\mathrm{top}}
\prod_{ij\in\eta^\vee}X_{ij}^{(f(k)-\t{f}(k))}}$, then 
\begin{equation}
    U_{\eta^\vee; (k,[s])}^{(h,[f])}\big|_{\scrH_\mathrm{top}}
    \,
    \prod_{ij\in\eta^\vee}X_{ij}^{(s(h) - f(k))}
    =
    U_{\eta^\vee; (k,[s])}^{(h,[\t{f}])}\big|_{\scrH_\mathrm{top}}
    \,
    \prod_{ij\in\eta^\vee}X_{ij}^{(s(h) - \t{f}(k))}.
\end{equation}
The allowed elements $h$ for $\mathcal{U}_{\eta^\vee; (k,[s])}^{(h)}$ form the group
\begin{equation}\label{CGksubgroup}
     \big\{h \in C_G(k) \mid \exists~ f\colon G\to A\text{~~satisfying~~}[b_{k,h}+s\triangleleft h+f] = [b_{h,k}+f\triangleleft k+s]\big\}.
\end{equation}
In general, this forms a proper subgroup of $C_G(k)$.

Using the 3-cocycle condition on $\bt$, one can show that these operators satisfy
\begin{equation}\label{ghFusionRuleTopSubspace}
    \mathcal{U}_{\eta^\vee; (k,[s])}^{(g)}
    \mathcal{U}_{\eta^\vee; (k,[s])}^{(h)}
    =
    \mathcal{U}_{\eta^\vee; (k,[s])}^{(gh)}
    \,
    \prod_{ij\in\eta^\vee}X_{ij}^{(\iota_k\bt(g,h))}.
\end{equation}
Therefore, ${\{U_\mathrm{top}^{(h)}\}}$ are 0-form symmetry operators of a central 2-group symmetry, which arise from the lattice 2-group operators from Sec.~\ref{sec:central-2group-sym-ops}. Redefining ${\mathcal{U}_{\eta^\vee; (k,[s])}^{(g)}\!\mapsto \mathcal{U}_{\eta^\vee; (k,[s])}^{(g)}\prod_{ij\in\eta^\vee} X^{(\mu(g))}_{ij}}$ modifies the slant product in~\eqref{ghFusionRuleTopSubspace} by ${\iota_k\bt(g,h)\mapsto \iota_k\bt(g,h) + \del\mu(g,h)}$. 
The ${s(h)\in A}$ 1-form symmetry operator in~\eqref{mathcalUdef} was added to have ${\iota_k\bt(g,h)}$ appear in~\eqref{ghFusionRuleTopSubspace} without any added 2-coboundary.
Therefore, whenever $\iota_k\bt$ restricted to the subgroup~\eqref{CGksubgroup} of $C_G(k)$ is cohomologically nontrivial, there is a 1-form symmetry operator appearing in~\eqref{ghFusionRuleTopSubspace} that cannot be redefined away.

\subsection{Gauging and dual symmetries}\label{sec:central-gauging}

Thus far in this section, we have introduced lattice central 2-group symmetry operators~\eqref{GbetaSymOps} and \eqref{eq:central-2group-A-1sym-op}, and have demonstrated manifestations of the Postnikov class. We now gauge this lattice 2-group symmetry, using the same strategy as Sec.~\ref{sec:split-gauging}. Namely, we first gauge the lattice $A$ 1-form symmetry, after which the $\t{G}$-symmetry operators become $G$-symmetry operators. We then gauge the resulting 0-form $G$-symmetry. Likewise, we also show how the $\G$-Potts model~\eqref{eq:central-2group-pottsmodel} changes under this gauging procedure.

\subsubsection{\texorpdfstring{Gauging $A^{(1)}$: $\G\to\text{anomalous }G\times \h{A}$}{Gauging \textit{A} 1-form symmetry}}\label{sec:central-gauge-a-1form}

We begin, as in Sec.~\ref{sec:split-gauge-a-1form}, by gauging the lattice $A$ 1-form symmetry $\left\{ T^{(\ga)} \right\}_{\ga\in Z_1(\La^\vee;A)}$. 
As we'll see, this yields a dual 0-form $\h{A}$ symmetry, and causes the central $\G$ symmetry to become a ${G\times \h{A}}$ 0-form symmetry with anomaly class $[\bt]$. This agrees with the results from QFT~\cite{T171209542, BH180309336}.

\noindent\textbf{Enlarge Hilbert space.} First, we introduce $A$-qudits on plaquettes; the enlarged Hilbert space is 
\begin{equation}\label{eq:central-gauge-a-invariant-hilbert-space}
    \scrH_{\mathrm{enlarged}~1} = \bigotimes_{i\in\La_0} \C[G]
    \otimes
    \bigotimes_{ij\in\La_1} \C[A]
    \otimes 
    \bigotimes_{ijk\in\La_2} \C[A].
\end{equation}
This Hilbert space is spanned by the computational basis $\ket{\{g_i\},\{a_{ij}\},\{a_{ijk}\}}$. 

\noindent\textbf{Gauss law.} We gauge the lattice 1-form symmetry using the Gauss operators
\begin{equation}\label{eq:central-gauss-law-gauge-a-1form}
    G^{(\la)}_{jk} = 
    \begin{cases}
        X^{(\la)}_{ijk} 
        X^{(\la)}_{jk}
        X^{(\la)}_{jkl}
        \qquad &i<j<k<l,\\
        X^{(-\la)}_{jik} 
        X^{(\la)}_{jk}
        X^{(-\la)}_{jlk}
        \qquad &j<i<l<k,
    \end{cases}
\end{equation}
which are depicted graphically in~\eqref{eq:gauss-law-gauge-a-1form-graphical}.
The gauge-invariant subspace of~\eqref{eq:central-gauge-a-invariant-hilbert-space} is 
\begin{equation}\label{eq:central-gauge-a-invariant-hilbert-space2}
    \{\ket{\psi} \in \scrH_{\text{enlarged 1}} 
    \mid G_{jk}^{(\la)}\ket{\psi} = \ket{\psi} 
    \quad 
    \forall \la\in A, \; jk\in \La_1\}.
\end{equation}
By writing ${T^{(\ga)} = \prod_{ij\in\La_1} G^{(\ga_{ij})}_{ij}}$, one can see the lattice 1-form symmetry operators $T^{(\ga)}$ act as the identity in the gauge-invariant Hilbert space. The dual $\h{A}$ 0-form symmetry is implemented by the operators 
\begin{equation}\label{eq:central-dual-ahat-symmetry}
    V^{(\chi)} = \prod_{ijk\in \La_2} (Z_{ijk}^{(\chi)})^{\eps_{ijk}},
\end{equation}
where ${\eps_{ijk} = 1}$ if $ijk$ is a right-side-up triangle ($\tricorner{}$) and ${\eps_{ijk} = -1}$ if $ijk$ is an upside-down triangle ($\tricorner[down]{}$). 

\noindent\textbf{Minimal coupling and dual symmetry.} After gauging, operators on~\eqref{GsitesAlinksHilb1} that are invariant under the $A$ 1-form symmetry become operators on~\eqref{eq:central-gauge-a-invariant-hilbert-space2} via minimal coupling.
For example, minimally coupling the central $\G$-Potts model Hamiltonian \eqref{eq:central-2group-pottsmodel} yields the Hamiltonian
\begin{equation}\label{eq:central-2group-pottsmodel-mc}
\begin{aligned}
    H_{\G\text{-Potts m.c.}}^{(\bt)} = 
    &- \frac{1}{|G|} \bigg(\sum_{ij\in\La_1} \sum_{\Ga\in\Irr G} d_\Ga\tr[Z_i^{(\Ga)} Z_{j}^{(\bar{\Ga})}]
    + \frac{h_G}2 \sum_{i\in \La_0} \sum_{h\in G}  (\overleftarrow{X}_{\!i}^{(h)} M_i^{(h,\bt)}
    +
    M_i^{(h^{-1},-\bt)} \overleftarrow{X}_{\!i}^{(h)}
    )
    \bigg) \\
    &- \frac{1}{|A|} \bigg(\sum_{ijk\in\La_2} \sum_{\chi\in \h{A}} B_{ijk}^{(\chi,\bt)}\,Z_{ijk}^{(\bar{\chi})}
    + h_A \sum_{ij\in \La_1} \sum_{\la\in A} X^{(\la)}_{ij}\bigg).
    \end{aligned}
\end{equation}
Here, in addition to minimal coupling, we have also used that the Gauss law ${G_{jk}^{(\la)} = 1}$ sets every ${A_i^{(\la)} =1}$.

The $\t{G}$ 0-form symmetry operators~\eqref{GbetaSymOps} are $A$ 1-form symmetric and commute with every Gauss operator $G_{jk}^{(\la)}$. Thus, they are operators on the gauge-invariant subspace without minimal coupling.
However, because the Gauss operators set every ${A_i^{(\la)} =1}$, they also set every ${S_f = 1}$. Therefore, gauging the lattice $A$ 1-form symmetry operator causes the $\t{G}$ 0-form symmetry to become a $G$ 0-form symmetry ${\{U^{(h)}\}_{h\in G}}$. 
Thus, this gauging causes the lattice central 2-group symmetry to become a symmetry with operators ${\{U^{(h)}, V^{(\chi)}\}_{h\in G,\chi\in\h{A}}}$. These operators satisfy
\begin{equation}
    U^{(g)}U^{(h)} = U^{(gh)}, 
    \qquad
    V^{(\chi_1)}V^{(\chi_2)} = V^{(\chi_1\chi_2)},
    \qquad
    U^{(g)}V^{(\chi)}
    =
    V^{(\chi)}U^{(g)}.
\end{equation}
Therefore, the dual symmetry is a ${G\times\h{A}}$ 0-form symmetry. 
These dual ${G\times\h{A}}$ symmetry operators commute with ${H_{\G\text{-Potts m.c.}}^{(\bt)}}$.
Furthermore, as we will next show, this ${G\times\h{A}}$ 0-form symmetry has an anomaly described by ${[\bt]\in \cH^3(G,A)\leq\cH^4(G\times\h{A},\Uone)}$.\footnote{\label{footnote:anomaly}By the K{\"u}nneth formula, 
\begin{equation}
    \cH^4(G\times\h{A},\Uone) \cong \prod_{p+q=4}\cH^p(G,\cH^q(\h{A},\Uone)).
\end{equation}
Consider the ${\cH^3(G,\cH^1(\h{A},\Uone))\cong \cH^3(G,A)}$ subgroup. 
A representative 4-cocycle ${\Om_\bt\in \cZ^4(G\times\h{A},\Uone)}$ of the cohomology class ${[\bt]\in\cH^3(G,A)\leq \cH^4(G\times\h{A},\Uone)}$ is ${\Om_\bt\left((g_1,\chi_1),(g_2,\chi_2),(g_3,\chi_3),(g_4,\chi_4)\right) = \chi_4(\bt(g_1,g_2,g_3))}$.
In terms of invertible field theory, denoting by $\cA_{\h{A}}$ and $\cA_{G}$ background gauge fields for the $\h{A}$ and $G$ symmetries, respectively, the ${3+1}$d ${G\times\h{A}}$ SPT corresponding to $[\bt]$ is ${\exp[2\pi \ii\int \<\bt(\cA_G),\cup \cA_{\h{A}}\>]}$ with the evaluation pairing ${\<-,-\>\colon A\times\h{A}\to\R/\Z}$.
Physically, it describes how $\h{A}$ symmetry defects are decorated by ${2+1}$d $G$-SPTs, as determined by $\bt$. Namely, a ${\chi\in\h{A}}$ symmetry defect is dressed by a ${[\chi\circ\bt]}$ $G$-SPT. This implies that the corresponding anomaly in ${2+1}$d manifests through the $G$ symmetry operators carrying a ${[\chi\circ\bt]}$ ${1+1}$d $G$-anomaly along a ${\chi\in\h{A}}$ symmetry defect.
}

\noindent\textbf{Change of basis and anomaly.} We may now transform to a unitary frame disentangling the gauge-invariant Hilbert space into a tensor-product Hilbert space. 
Consider the unitary
\begin{equation}\label{eq:central-disentangling-unitary-1}
    U_1 = \sum_{\substack{\left\{g_i\right\},\left\{a_{ij}\right\}\\ \{a_{ijk}\}}}
        \ket{\left\{g_i\right\},\left\{a_{ij}\right\},\{a_{ijk}-a_{ij} + a_{ik} - a_{jk}+ \bt(g_i, g_{i}^{-1}g_j, g_j^{-1}g_k)\}}\!\!\bra{ \left\{g_i\right\},\left\{a_{ij}\right\},\{a_{ijk}\}}.
\end{equation}
Note that this simplifies to~\eqref{eq:disentangling-unitary-1} when ${\bt=0}$. 
It satisfies ${U_1 G^{(\la)}_{jk} U_1^\dag = X_{jk}^{(\la)}}$ and, thus, the Gauss law in this frame causes the link $A$-qudits to decouple. The gauge-invariant Hilbert space is then~\eqref{GsitesA2simpsHilb}. (This shift by $\bt$ in~\eqref{eq:central-disentangling-unitary-1} is not necessary for disentangling the Hilbert space, but simplifies the dual symmetry operators and minimally-coupled Hamiltonian~\eqref{eq:central-2group-pottsmodel-mc}.) 

In this new frame, the ${G\times\h{A}}$ 0-form symmetry operators ${\{U^{(h)}, V^{(\chi)}\}_{h\in G,\chi\in\h{A}}}$ on the gauge-invariant subspace~\eqref{GsitesA2simpsHilb} are
\begin{align}
    \t{U}^{(h)} &= U_1 U^{(h)} U_1^\dag = \prod_{i\in\La_0} \overrightarrow{X}_i^{(h)},\label{eq:disentangled-G-anomalous-sym-ops0}\\
    \t{V}^{(\chi)} &= U_1 V^{(\chi)}U_1^\dag 
    =
    (U_{G-\text{SPT}}^{(\chi\circ\bt)})^\dag
    \prod_{ijk\in \La_2} (Z_{ijk}^{(\chi)})^{\eps_{ijk}},\label{eq:disentangled-Ahat-anomalous-sym-ops0}
\end{align}
where, for ${\om\in\cZ^3(G,\Uone)}$, 
\begin{equation}
    U_{G-\text{SPT}}^{(\om)}
    = 
    \sum_{\{g_i\},\{a_{ijk}\}}
    \prod_{lmn\in\La_2}
    \om(g_l,g_l^{-1} g_m, g_m^{-1} g_n)^{\eps_{lmn}}
    \ketbra{\{g_i\},\{a_{ijk}\}}.
\end{equation}
The unitary operator $U_{G-\text{SPT}}^{(\om)}$ is a $G$-SPT entangler for the $G$ symmetry ${\{\prod_{i\in\La_0} \overrightarrow{X}_i^{(h)}\}_{h\in G}}$ that prepares the $G$-SPT classified by ${[\om]\in \cH^3(G,\Uone)}$~\cite{CGL11064772}.

With the $G\times\h{A}$ symmetry operators~\eqref{eq:disentangled-G-anomalous-sym-ops0} and~\eqref{eq:disentangled-Ahat-anomalous-sym-ops0} in this frame, it is easy to show that the dual ${G\times\h{A}}$ 0-form symmetry has an anomaly described by ${[\bt]}$.
Informally, the anomaly arises because the ${\t{V}^{(\chi)}}$ symmetry operators are decorated by a $G$-SPT entangler for the $U^{(h)}$ symmetry operators, which commonly gives rise to mixed anomalies~\cite{B190505790, TTV211007599, SS240401369, PV250920431}.
More precisely, the mixed anomaly can be detected by inserting an $\h{A}$ symmetry defect. Consider the lattice path $\eta$:
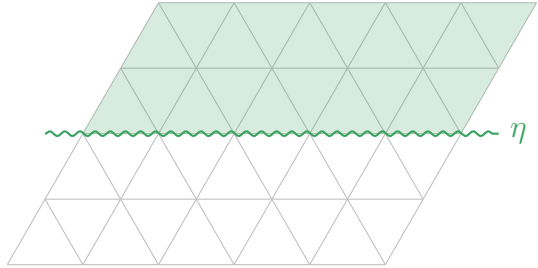
\begin{equation*}
  \begin{tikzpicture}[scale=1,
    decoration={markings, mark=at position 0.55 with {\arrow{>}}},
    d/.style={postaction=decorate}
]

    \def\Nx{5}
    \def\Ny{4}

    \pgfmathsetmacro{\a}{1}
    \pgfmathsetmacro{\b}{0.8660254}  

    \foreach \i in {0,...,\Ny} {
        \foreach \j in {0,...,\Nx} {

            \coordinate (v_\i_\j)
                at (\j*\a + 0.5*\i*\a, \i*\b);

        }
    }

    \foreach \i in {0,...,\Ny} {
        \foreach \j in {0,...,\numexpr\Nx-1} {

            \pgfmathtruncatemacro{\jp}{\j+1}

            \draw[color=lightgray] (v_\i_\j) -- (v_\i_\jp);

            \coordinate (c_h_\i_\j)
                at ($(v_\i_\j)!0.5!(v_\i_\jp)$);

        }
    }

    \foreach \i in {0,...,\numexpr\Ny-1} {
        \foreach \j in {0,...,\Nx} {

            \pgfmathtruncatemacro{\ip}{\i+1}

            \draw[color=lightgray] (v_\i_\j) -- (v_\ip_\j);

            \coordinate (c_dr_\i_\j)
                at ($(v_\i_\j)!0.5!(v_\ip_\j)$);

        }
    }

    \foreach \i in {0,...,\numexpr\Ny-1} {
        \foreach \j in {1,...,\Nx} {

            \pgfmathtruncatemacro{\ip}{\i+1}
            \pgfmathtruncatemacro{\jm}{\j-1}

            \draw[color=lightgray] (v_\i_\j) -- (v_\ip_\jm);

            \coordinate (c_dl_\i_\j)
                at ($(v_\i_\j)!0.5!(v_\ip_\jm)$);

        }
    }

    \draw[color=mygreen, decorate, thick, decoration={snake, amplitude=.3mm, segment length=2mm}]
        ($(v_2_0) + (-0.5, 0)$) -- ($(v_2_5) + (0.5, 0)$) node[right] {$\eta$};

    \fill[color=mygreen, opacity=0.2] (v_2_0) -- (v_2_5) -- (v_4_5) -- (v_4_0) -- cycle;
\end{tikzpicture}
\end{equation*}
We insert a $\chi\in\h{A}$ symmetry defect by truncating $\t{V}^{(\chi)}$ to act nontrivially only above $\eta$---only on plaquettes shaded in green above.
Conjugating the $G$ symmetry operators $\t{U}^{(h)}$ by this truncated $V^{(\chi)}$ operator yields 
\begin{equation}
    \t{U}_{\eta, \chi}^{(h)}
    =
    D_{\eta}^{(h,\chi)}
    \widetilde{U}^{(h)}, 
\end{equation}
where 
\begin{equation}
    D_{\eta}^{(h,\chi)}
    \equiv 
    \sum_{\{g_i\},\{a_{ijk}\}}
    \prod_{lm\in\eta} 
    \chi\left(\bt(h,h^{-1}g_l,g_l^{-1} g_m)\right)
    \ketbra{\{g_i\},\{a_{ijk}\}}.
\end{equation}
The operators $\t{U}_{\eta, \chi}^{(h)}$ are the $G$ symmetry operators in the presence of the ${\chi\in\h{A}}$ symmetry defect at $\eta$.
As explained in Footnote~\ref{footnote:D-ops}, the operator $D_{\eta}^{(h,\chi)}$ causes the $G$ symmetry operator $\t{U}_{\eta, \chi}^{(h)}$ to act along the locus of the ${\chi\in\h{A}}$ symmetry defect as an anomalous $G$ symmetry with ${1+1}$d anomaly ${[\chi\circ\bt]\in\cH^3(G,\Uone)}$. As explained in Footnote~\ref{footnote:anomaly}, this is a manifestation of the mixed anomaly between the $G$ and $\h{A}$ symmetry operators corresponding to ${[\bt]\in\cH^3(G,A)\leq \cH^4(G\times\h{A},\Uone)}$.

In this new frame, the minimally-coupled $\G$-Potts Hamiltonian given in \eqref{eq:central-2group-pottsmodel-mc} becomes
\begin{align}\label{eq:central-gxa-pottsmodel}
    H^{(\bt)}_{G\times \h{A}\text{-Potts}}
    =
    & \; \; U_1 H_{\G\text{-Potts m.c.}}^{(\bt)} U_1^{\dag} \\
    =
    &- \frac{1}{|G|} \bigg(\sum_{ij\in\La_1} \sum_{\Ga\in\Irr G} d_\Ga\tr[Z_i^{(\Ga)} Z_{j}^{(\bar{\Ga})}]
    + 
    \frac{h_G}2 \sum_{i\in \La_0} \sum_{h\in G} (\overleftarrow{X}_{\!i}^{(h)} N_i^{(h,\bt)}
    +
    N_i^{(h^{-1},-\bt)}\overleftarrow{X}_{\!i}^{(h)} 
    )\bigg)
    \nonumber
    \\
    &- \frac{1}{|A|} \bigg(\sum_{ijk\in\La_2} \sum_{\chi\in \h{A}} Z_{ijk}^{(\chi)}
    + 
    h_A \sum_{\langle ijk, lmn \rangle} \sum_{\la\in A} X_{ijk}^{(\la)} X_{lmn}^{(\la)}\bigg).\nonumber
\end{align}
The $G$-paramagnetic term is now dressed by the operator $N_l^{(h,\bt)}$ that acts as
\begin{equation}
\begin{tikzpicture}[decoration={markings, mark=at position 0.55 with {\arrow{>}}}, scale=3.2]
    \coordinate (c) at (0,0);
    \coordinate (e) at (1,0);
    \coordinate (w) at (-1,0);
    \coordinate (nw) at (-0.5,{sqrt(3)/2});
    \coordinate (ne) at (0.5,{sqrt(3)/2});
    \coordinate (sw) at (-0.5,-{sqrt(3)/2});
    \coordinate (se) at (0.5,-{sqrt(3)/2});

    \coordinate (t1) at ($(c)!0.333!(e) + (c)!0.333!(ne) - (c)$);   
    \coordinate (t2) at ($(c)!0.333!(ne) + (c)!0.333!(nw) - (c)$);  
    \coordinate (t3) at ($(c)!0.333!(nw) + (c)!0.333!(w) - (c)$);   
    \coordinate (t4) at ($(c)!0.333!(w) + (c)!0.333!(sw) - (c)$);   
    \coordinate (t5) at ($(c)!0.333!(sw) + (c)!0.333!(se) - (c)$);  
    \coordinate (t6) at ($(c)!0.333!(se) + (c)!0.333!(e) - (c)$);   
    
    \node at (t1) {\tiny $a_{l m o}$};
    \node at (t2) {\tiny $a_{l n o}$};
    \node at (t3) {\tiny $a_{kl n}$};
    \node at (t4) {\tiny $a_{ikl}$};
    \node at (t5) {\tiny $a_{ijl}$};
    \node at (t6) {\tiny $a_{jl m}$};

    \node at (sw) {\footnotesize $g_i$}; 
    \node at (se) {\footnotesize $g_j$}; 
    \node at (w)  {\footnotesize $g_k$};
    \node at (e)  {\footnotesize $g_m$}; 
    \node at (nw) {\footnotesize $g_n$};
    \node at (ne) {\footnotesize $g_o$};

    \draw[postaction=decorate, color=lightgray] (c) -- (e);
    \draw[postaction=decorate, color=lightgray] (c) -- (nw);
    \draw[postaction=decorate, color=lightgray] (c) -- (ne);
    \draw[postaction=decorate, color=lightgray] (sw) -- (c);
    \draw[postaction=decorate, color=lightgray] (se) -- (c);
    \draw[postaction=decorate, color=lightgray] (w) -- (c) node[anchor=mid, color=black] {\footnotesize $g_l$};
\end{tikzpicture}
\quad
\xrightarrow{N_l^{(h,\bt)}}
\quad
\begin{tikzpicture}[decoration={markings, mark=at position 0.55 with {\arrow{>}}}, scale=3.2]
    \coordinate (c) at (0,0);
    \coordinate (e) at (1,0);
    \coordinate (w) at (-1,0);
    \coordinate (nw) at (-0.5,{sqrt(3)/2});
    \coordinate (ne) at (0.5,{sqrt(3)/2});
    \coordinate (sw) at (-0.5,-{sqrt(3)/2});
    \coordinate (se) at (0.5,-{sqrt(3)/2});

    \coordinate (t1) at ($(c)!0.333!(e) + (c)!0.333!(ne) - (c)$);   
    \coordinate (t2) at ($(c)!0.333!(ne) + (c)!0.333!(nw) - (c)$);  
    \coordinate (t3) at ($(c)!0.333!(nw) + (c)!0.333!(w) - (c)$);   
    \coordinate (t4) at ($(c)!0.333!(w) + (c)!0.333!(sw) - (c)$);   
    \coordinate (t5) at ($(c)!0.333!(sw) + (c)!0.333!(se) - (c)$);  
    \coordinate (t6) at ($(c)!0.333!(se) + (c)!0.333!(e) - (c)$);   

    \node at (sw) {\footnotesize $g_i$}; 
    \node at (se) {\footnotesize $g_j$}; 
    \node at (w)  {\footnotesize $g_k$};
    \node at (e)  {\footnotesize $g_m$}; 
    \node at (nw) {\footnotesize $g_n$};
    \node at (ne) {\footnotesize $g_o$};

    \node[align=center, xshift=1em, yshift=-0.5em] at (t1) {\tiny $a_{l m o}$ \\[-1.2em] \tiny ${\color{myblue} \vphantom{} + \bt(h^{-1},h g_l^{-1} g_m, g_m^{-1} g_o)}$};
    \node[align=center, yshift=1em] at (t2) {\tiny $a_{l n o}$ \\[-1.2em] \tiny $ {\color{myblue} \vphantom{} + \bt(h^{-1},h g_l^{-1} g_n, g_n^{-1} g_o)}$};
    \node[align=center, xshift=-1em, yshift=-0.5em] at (t3) {\tiny $a_{kl n}$ \\[-1.2em] \tiny ${\color{myblue} \vphantom{} -\bt(g_k^{-1} g_l, h^{-1}, h g_l^{-1} g_n)}$};
    \node[align=center, xshift=-1em, yshift=0.5em] at (t4) {\tiny $a_{ikl}$ \\[-1.2em] \tiny ${\color{myblue} \vphantom{} + \bt(g_i^{-1} g_k, g_k^{-1} g_l, h^{-1})}$};
    \node[align=center, yshift=-1em] at (t5) {\tiny $a_{ijl}$ \\[-1.2em] \tiny ${\color{myblue} \vphantom{} + \bt(g_i^{-1} g_j, g_j^{-1} g_l, h^{-1})}$};
    \node[align=center, xshift=1em, yshift=0.5em] at (t6) {\tiny $a_{jl m}$\\[-1.2em] \tiny ${\color{myblue} \vphantom{} -\bt(g_j^{-1} g_l, h^{-1}, h g_l^{-1} g_m)}$};
    
    \draw[postaction=decorate, color=lightgray] (c) -- (e);
    \draw[postaction=decorate, color=lightgray] (c) -- (nw);
    \draw[postaction=decorate, color=lightgray] (c) -- (ne);
    \draw[postaction=decorate, color=lightgray] (sw) -- (c);
    \draw[postaction=decorate, color=lightgray] (se) -- (c);
    \draw[postaction=decorate, color=lightgray] (w) -- (c) node[anchor=mid, color=black] {\footnotesize $g_l {\color{myblue}}$};
\end{tikzpicture}.
\end{equation}
The Hamiltonian~\eqref{eq:central-gxa-pottsmodel} is a ${G\times \h{A}}$ Potts model for an anomalous ${G\times \h{A}}$ symmetry. The nontrivial decoration $N_l^{(h,\bt)}$ arises due to the anomaly. When ${\bt=0}$, the Hamiltonian simplifies to the ${G\times \h{A}}$ Potts model for an anomaly-free ${G\times \h{A}}$ symmetry.

\subsubsection{\texorpdfstring{Gauging $G^{(0)}$: $\text{anomalous }G\times \h{A} \to 2\Rep(\G)$}{Gauging \textit{G} 0-form symmetry}}\label{sec:central-gauge-g-0form}

We now gauge the $G$ 0-form subsymmetry of the anomalous ${G\times\h{A}}$ symmetry obtained in the preceding subsection.
Although the full ${G\times\h{A}}$ symmetry has a mixed anomaly, its $G$ subgroup is anomaly-free and is implemented by onsite operators~\eqref{eq:disentangled-G-anomalous-sym-ops0}.
Gauging this symmetry produces a dual lattice $\Rep(G)$ 1-form symmetry together with non-invertible 0-form symmetry operators. 
In the subspace in which the dual lattice $\Rep(G)$ 1-form symmetry is topological, these operators realize the $2\Rep(\G)$ fusion 2-category symmetry.

\noindent\textbf{Enlarge Hilbert space.} We first enlarge the Hilbert space by introducing a $G$-qudit on each link. The Hilbert space becomes 
\begin{equation}\label{eq:central-enlarged-hilbert-space-2}
    \scrH_{\text{enlarged 2}} 
        = 
        \bigotimes_{i\in \La_0} \C[G]
        \otimes
        \bigotimes_{ij \in \La_1} \C[G]
        \otimes
        \bigotimes_{ijk \in \La_2} \C[A].
\end{equation}

\noindent\textbf{Gauss law.} We now define the following Gauss operators: 
\begin{equation}\label{eq:gauss-law-central-gauge-G}
    G_l^{(h)}
    =
    \begin{tikzpicture}[decoration={markings, mark=at position 0.55 with {\arrow{>}}}, scale=2.5]
    \coordinate (c) at (0,0);
    \coordinate (e) at (1,0);
    \coordinate (w) at (-1,0);
    \coordinate (nw) at (-0.5,{sqrt(3)/2});
    \coordinate (ne) at (0.5,{sqrt(3)/2});
    \coordinate (sw) at (-0.5,-{sqrt(3)/2});
    \coordinate (se) at (0.5,-{sqrt(3)/2});
    
    \draw[postaction=decorate, color=lightgray] (c) -- node[anchor=mid, color=black] {\footnotesize $\overrightarrow{X}^{(h)}_{l m}$} (e) node[color=gray, right] {\footnotesize $m$};
    \draw[postaction=decorate, color=lightgray] (c) -- node[below, color=black, anchor=mid] {\footnotesize$\overrightarrow{X}^{(h)}_{l n}$} (nw) node[color=gray, above] {\footnotesize $n$};
    \draw[postaction=decorate, color=lightgray] (c) -- node[below, color=black, anchor=mid] {\footnotesize$\overrightarrow{X}^{(h)}_{l o}$} (ne) node[color=gray, above] {\footnotesize $o$};
    \draw[postaction=decorate, color=lightgray] (sw) node[color=gray, below] {\footnotesize $i$} -- node[below, color=black, anchor=mid] {\footnotesize$\overleftarrow{X}^{(h)}_{il}$} (c);
    \draw[postaction=decorate, color=lightgray] (se) node[color=gray, below] {\footnotesize $j$} -- node[below, color=black, anchor=mid] {\footnotesize$\overleftarrow{X}^{(h)}_{jl}$} (c);
    \draw[postaction=decorate, color=lightgray] (w) node[color=gray, left] {\footnotesize $k$} -- node[below, color=black, anchor=mid] {\footnotesize$\overleftarrow{X}^{(h)}_{kl}$} (c) node[anchor=mid, color=black] {\footnotesize$\overrightarrow{X}^{(h)}_l$};
\end{tikzpicture}.
\end{equation}
This is the same Gauss operator as in~\eqref{gaugingGsplit2grpsecGaussop} but with trivial $\rho$.
The physical Hilbert space is the gauge-invariant subspace
\begin{equation}\label{eq:central-gauge-g-invariant-hilbert-space}
    \{\ket{\psi}\in \scrH_{\text{enlarged 2}} \mid G_{l}^{(h)}\ket{\psi} = \ket{\psi} \quad \forall l\in \La_0, \; h\in G\}.
\end{equation}

After gauging, there is a dual lattice $\Rep(G)$ 1-form symmetry formed by the Wilson loop operators $W_{\ga}^{(\Ga)}$ defined in~\eqref{eq:wilson-loop-operator}.
The Wilson loop operators are topological in the subspace whose vectors $\ket{\psi}$ satisfy ${B_{ijk}\ket{\psi} = \ket{\psi}}$ with $B_{ijk}$ defined in Eq.~\eqref{Gflatnesssplit2grpgaugingGsec}.

\noindent\textbf{Minimal coupling and dual symmetry.} We next perform minimal coupling. For instance, minimally coupling the Potts model Hamiltonian~\eqref{eq:central-gxa-pottsmodel} such that it commutes with the Gauss operators yields
\begin{equation}\label{anomalousAGpotts}
\begin{aligned}
    H^{(\bt)}_{G\times \h{A}\text{-Potts, m.c.}} 
    =
    &- \frac{1}{|G|} \bigg(\sum_{ij\in\La_1} \sum_{\Ga\in\Irr G} d_\Ga\tr[Z_i^{(\Ga)} Z_{ij}^{(\bar{\Ga})} Z_{j}^{(\bar{\Ga})}]
    \\
    &
    \qquad\qquad + 
    \frac{h_G}2 \sum_{i\in \La_0} \sum_{h\in G} 
    (\overleftarrow{X}_{\!i}^{(h)} N_{i,\text{m.c.}}^{(h,\bt)} 
    +
    N_{i,\text{m.c.}}^{(h^{-1},-\bt)} 
    \overleftarrow{X}_{\!i}^{(h)} 
    )
    \bigg)
     - \sum_{ijk\in\La_2} B_{ijk}
    \\
    &
    - \frac{1}{|A|} \bigg(\sum_{ijk\in\La_2} \sum_{\chi\in \h{A}} Z_{ijk}^{(\chi)}
    + 
    h_A \sum_{\langle ijk, lmn \rangle} \sum_{\la\in A} X_{ijk}^{(\la)} X_{lmn}^{(\la)}\bigg)
\end{aligned}.
\end{equation}
The operator $N_{i,\text{m.c.}}^{(h,\bt)} $ is the minimally-coupled $N_{i}^{(h,\bt)}$ and it acts as
\begin{equation}
\begin{tikzpicture}[decoration={markings, mark=at position 0.55 with {\arrow{>}}}, scale=3.2]
    \coordinate (c) at (0,0);
    \coordinate (e) at (1,0);
    \coordinate (w) at (-1,0);
    \coordinate (nw) at (-0.5,{sqrt(3)/2});
    \coordinate (ne) at (0.5,{sqrt(3)/2});
    \coordinate (sw) at (-0.5,-{sqrt(3)/2});
    \coordinate (se) at (0.5,-{sqrt(3)/2});

    \coordinate (t1) at ($(c)!0.333!(e) + (c)!0.333!(ne) - (c)$);   
    \coordinate (t2) at ($(c)!0.333!(ne) + (c)!0.333!(nw) - (c)$);  
    \coordinate (t3) at ($(c)!0.333!(nw) + (c)!0.333!(w) - (c)$);   
    \coordinate (t4) at ($(c)!0.333!(w) + (c)!0.333!(sw) - (c)$);   
    \coordinate (t5) at ($(c)!0.333!(sw) + (c)!0.333!(se) - (c)$);  
    \coordinate (t6) at ($(c)!0.333!(se) + (c)!0.333!(e) - (c)$);   
    
    \node at (t1) {\tiny $a_{l m o}$};
    \node at (t2) {\tiny $a_{l n o}$};
    \node at (t3) {\tiny $a_{kl n}$};
    \node at (t4) {\tiny $a_{ikl}$};
    \node at (t5) {\tiny $a_{ijl}$};
    \node at (t6) {\tiny $a_{jl m}$};

    \node at (sw) {\footnotesize $g_i$}; 
    \node at (se) {\footnotesize $g_j$}; 
    \node at (w)  {\footnotesize $g_k$};
    \node at (e)  {\footnotesize $g_m$}; 
    \node at (nw) {\footnotesize $g_n$};
    \node at (ne) {\footnotesize $g_o$};

    \draw[postaction=decorate, color=lightgray] (c) -- (e);
    \draw[postaction=decorate, color=lightgray] (c) -- (nw);
    \draw[postaction=decorate, color=lightgray] (c) -- (ne);
    \draw[postaction=decorate, color=lightgray] (sw) -- (c);
    \draw[postaction=decorate, color=lightgray] (se) -- (c);
    \draw[postaction=decorate, color=lightgray] (w) -- (c) node[anchor=mid, color=black] {\footnotesize $g_l$};
\end{tikzpicture}
\quad
\xrightarrow{N_{l,\text{m.c.}}^{(h,\bt)}}
\quad
\begin{tikzpicture}[decoration={markings, mark=at position 0.55 with {\arrow{>}}}, scale=3.2]
    \coordinate (c) at (0,0);
    \coordinate (e) at (1,0);
    \coordinate (w) at (-1,0);
    \coordinate (nw) at (-0.5,{sqrt(3)/2});
    \coordinate (ne) at (0.5,{sqrt(3)/2});
    \coordinate (sw) at (-0.5,-{sqrt(3)/2});
    \coordinate (se) at (0.5,-{sqrt(3)/2});

    \node at (sw) {\footnotesize $g_i$}; 
    \node at (se) {\footnotesize $g_j$}; 
    \node at (w)  {\footnotesize $g_k$};
    \node at (e)  {\footnotesize $g_m$}; 
    \node at (nw) {\footnotesize $g_n$};
    \node at (ne) {\footnotesize $g_o$};

    \coordinate (t1) at ($(c)!0.333!(e) + (c)!0.333!(ne) - (c)$);   
    \coordinate (t2) at ($(c)!0.333!(ne) + (c)!0.333!(nw) - (c)$);  
    \coordinate (t3) at ($(c)!0.333!(nw) + (c)!0.333!(w) - (c)$);   
    \coordinate (t4) at ($(c)!0.333!(w) + (c)!0.333!(sw) - (c)$);   
    \coordinate (t5) at ($(c)!0.333!(sw) + (c)!0.333!(se) - (c)$);  
    \coordinate (t6) at ($(c)!0.333!(se) + (c)!0.333!(e) - (c)$);   
    
    \node[align=center, xshift=1em, yshift=-0.5em] at (t1) {\tiny $a_{l m o}$ \\[-1.2em] \tiny ${\color{myblue} \vphantom{} + \bt(h^{-1},h g_l^{-1} g_{lm} g_m, g_m^{-1} g_{mo} g_o)}$};
    \node[align=center, yshift=1em] at (t2) {\tiny $a_{l n o}$ \\[-1.2em] \tiny $ {\color{myblue} \vphantom{} +\bt(h^{-1},h g_l^{-1} g_{ln} g_n, g_n^{-1} g_{no} g_o)}$};
    \node[align=center, xshift=-1em, yshift=-0.5em] at (t3) {\tiny $a_{kl n}$ \\[-1.2em] \tiny ${\color{myblue} \vphantom{} -\bt(g_k^{-1} g_{kl} g_l, h^{-1}, h g_l^{-1} g_{ln} g_n)}$};
    \node[align=center, xshift=-1em, yshift=0.5em] at (t4) {\tiny $a_{ikl}$ \\[-1.2em] \tiny ${\color{myblue} \vphantom{} + \bt(g_i^{-1} g_{ik} g_k, g_k^{-1} g_{kl} g_l, h^{-1})}$};
    \node[align=center, yshift=-1em] at (t5) {\tiny $a_{ijl}$ \\[-1.2em] \tiny ${\color{myblue} \vphantom{} + \bt(g_i^{-1} g_{ij} g_j, g_j^{-1} g_{jl} g_l, h^{-1})}$};
    \node[align=center, xshift=1em, yshift=0.5em] at (t6) {\tiny $a_{jl m}$\\[-1.2em] \tiny ${\color{myblue} \vphantom{} -\bt(g_j^{-1} g_{jl} g_l, h^{-1}, h g_l^{-1} g_{lm} g_m)}$};
    
    \draw[postaction=decorate, color=lightgray] (c) -- (e);
    \draw[postaction=decorate, color=lightgray] (c) -- (nw);
    \draw[postaction=decorate, color=lightgray] (c) -- (ne);
    \draw[postaction=decorate, color=lightgray] (sw) -- (c);
    \draw[postaction=decorate, color=lightgray] (se) -- (c);
    \draw[postaction=decorate, color=lightgray] (w) -- (c) node[anchor=mid, color=black] {\footnotesize $g_l {\color{myblue}}$};
\end{tikzpicture}.
\end{equation}
We have added the term ${-\sum_{ijk}B_{ijk}}$ by hand. It commutes with every other
term and energetically favors the subspace in which the
$\Rep(G)$ Wilson loops are topological.

The $\h{A}$ 0-form symmetry operators $\t{V}^{(\chi)}$ in~\eqref{eq:disentangled-Ahat-anomalous-sym-ops0} are $G$ symmetric but do not commute with the Gauss operator~\eqref{eq:gauss-law-central-gauge-G}. Therefore, they must be minimally-coupled.
While every $\t{V}^{(\chi)}$ is invertible, after minimal coupling, the resulting operators include both invertible and non-invertible operators. 
There are multiple, inequivalent ways to minimally couple a given operator $\t{V}^{(\chi)}$.
Here, we highlight two special cases and refer the reader to App.~\ref{app:central-non-invertible-minimal-coupling} for the more general case.

We first highlight a minimal coupling that applies to every $\t{V}^{(\chi)}$. Let $\mathsf{P}_{o,h}^{(\{1\})}$ be the projector defined in Eq.~\eqref{PKohmaintext} with ${K=\{1\}}$, and $g_{\ga_{o\to l}}$ be the element of $G$ defined in Eq.~\eqref{eq:holonomy-group-element-g}. 
As shown in App.~\ref{app:central-non-invertible-minimal-coupling}, each $\t{V}^{(\chi)}$ can be minimally-coupled to become the operator
\begin{equation}\label{eq:central-m.c.-ahat-operator-K=G}
\begin{aligned}
    \t{V}^{(\chi)}_{\mathrm{m.c.}} 
    = 
    |G|
    \sum_{\substack{\{g_i\}, \{g_{ij}\}\\ \{a_{ijk}\} }}\prod_{l mn}
    &
    \,\chi\left(a_{l mn} - \bt(g_{\ga_{0\to l}}g_l,\, g_l^{-1}g_{l m}g_m,\, g_m^{-1}g_{mn}g_n)\right)^{\eps_{l mn}}\\[-1em]
    &
    \hspace{5em}\times \mathsf{P}^{(\{1\})}_{o,1}\, 
    \ketbra{\{g_i\}, \{g_{ij}\}, \{a_{ijk}\}}.
\end{aligned}
\end{equation}
These symmetry operators~\eqref{eq:central-m.c.-ahat-operator-K=G} are non-invertible because of the projector $\mathsf{P}^{(\{1\})}_{o,1}$. They satisfy the symmetry algebra
\begin{equation}
    \t{V}_{\text{m.c.}}^{(\chi_1)} 
    \times 
    \t{V}_{\text{m.c.}}^{(\chi_2)} 
    = 
    |G|\; 
    \t{V}_{\text{m.c.}}^{(\chi_1\chi_2)} .
\end{equation}

The operator~\eqref{eq:central-m.c.-ahat-operator-K=G} is not the only possible outcome of minimal coupling. 
Consider symmetry operators $\t{V}^{(\chi)}$ with ${\chi}$ satisfying ${[\chi\circ\bt] = [1]}$. For such $\chi$, there exists a ${c\in \cC^2(G,\Uone)}$ such that ${\chi\circ\bt = \delta c}$. As we show in App.~\ref{app:central-non-invertible-minimal-coupling}, these operators can be minimally-coupled to become
\begin{equation}\label{eq:central-m.c.-ahat-operator-K=1}
    \t{V}^{(\chi,c_\chi)}_{\mathrm{m.c.}} =\!\! 
    \sum_{
    \substack{\{g_i\},\{g_{ij}\} \\ \{a_{ijk}\}}}
    \,\prod_{l m n\in\La_2}\!
    \left(
    \frac{\chi(a_{l mn})}
    {c_\chi(g_l^{-1}g_{l m}g_m,\, g_m^{-1}g_{mn}g_n)}
    \right)^{\eps_{lmn}}
    \ketbra{\{g_i\},\{g_{ij}\},\{a_{ijk}\}}.
\end{equation}
These are diagonal unitary operators and they, therefore, form an invertible symmetry. 
They satisfy
\begin{equation}
    \t{V}_{\text{m.c.}}^{(\chi_1, c_1)} 
    \times 
    \t{V}_{\text{m.c.}}^{(\chi_2, c_2)} 
    = 
    \t{V}_{\text{m.c.}}^{(\chi_1\chi_2, c_1 c_2)}.
\end{equation}
Note that ${\t{V}_{\text{m.c.}}^{(1, 1)} = 1}$.
For fixed $\chi$, the set of
solutions to ${\delta c=\chi\circ\bt}$ is a
$\cZ^2(G,\Uone)$-torsor. Indeed, if ${\del c = \chi\circ \bt}$, then ${c' = c\,\om}$ with ${\om\in\cZ^2(G,\Uone)}$ also satisfies ${\del c' = \chi\circ \bt}$. 
On the full physical Hilbert space, each operator $\t{V}^{(\chi,c)}_{\mathrm{m.c.}}$ depends on the chosen 2-cochain $c$.
In the lattice $\Rep(G)$ 1-form symmetry's topological subspace, however, each operator $\t{V}^{(\chi,c)}_{\mathrm{m.c.}}$ is invariant under ${c\mapsto c\,\del\mu}$ for any ${\mu\in\cC^1(G,\Uone)}$. Thus, fixing $\chi$, inequivalent choices of $c$ for ${\t{V}^{(\chi,c)}_{\mathrm{m.c.}}}$ in the topological subspace form an ${\cH^2(G,\Uone)}$-torsor.

We present the most general minimal coupling procedure within the ${B_{ijk} =1}$ subspace in App.~\ref{app:central-non-invertible-minimal-coupling}.
The resulting 0-form symmetry operators make up the 0-form symmetry part of the $2\Rep(\G)$ fusion 2-category symmetry.
They are labeled by triples
\begin{equation}\label{eq:central-2Rep-labels-preview}
    (K,\chi,[c]),
    \qquad
    K\leq G,
    \qquad
    \delta c
    =
    (\chi\circ\bt)\big|_K.
\end{equation}
Here $[c]$ denotes an equivalence class of 2-cochains, which forms an $\cH^2(K,\Uone)$-torsor. 
These are, indeed, the labels for the simples of $2\Rep(\G)$ for central $\G$~\cite{E0408120}.
The two cases highlighted above correspond to ${K=\{1\}}$ and ${K = G}$, respectively.

\noindent\textbf{Change of basis.} We now transform to a unitary frame where the site $G$-qudits are decoupled and the gauge-invariant Hilbert space admits an onsite tensor product factorization. We use the unitary
\begin{equation}\label{eq:central-gauge-g-0-form-unitary}
    U_2 = \sum_{\substack{\{g_i\},\{g_{ij}\}\\ \{a_{ijk}\}}}
        \ketbra{\{g_i\},\{g_i^{-1} g_{ij} g_{j}\},\{a_{ijk}\}}{\{g_i\},\{g_{ij}\},\{a_{ijk}\}} 
        .
\end{equation}
It satisfies ${U_2 G_{i}^{(h)} U_2^\dag = \overrightarrow{X}_{i}^{(h)}}$. Thus, in this unitary frame, the only gauge-invariant site $G$-qudit configuration consists of fully $G$-symmetrized states on each site and the gauge-invariant Hilbert space \eqref{eq:central-gauge-g-invariant-hilbert-space} is 
\begin{equation}
    \bigotimes_{ij\in\La_1} \C[G]
    \otimes
    \bigotimes_{ijk\in\La_2} \C[A].
\end{equation}
In this frame, the Hamiltonian~\eqref{anomalousAGpotts} becomes
\begin{equation}\label{eq:central-2rep-potts-model}
\begin{aligned}
    H_{2\Rep \text{-Potts}} 
    =
    & 
    \;\; U_2 H_{G\times \h{A}\text{-Potts, m.c.}} U_2^\dag \\
    =
    &- \frac{1}{|G|} \bigg(\sum_{ij\in\La_1} \sum_{\Ga\in\Irr G} d_\Ga\tr[ Z_{ij}^{(\Ga)}]
    + 
    \frac{h_G}2 \sum_{i\in \La_0} \sum_{h\in G} (
    A_i^{(h,\bt)} + A_i^{(h,\bt)\dag}
    ) 
    \bigg)
    -
    \sum_{ijk\in\La_2} B_{ijk}
    \\
    &- \frac{1}{|A|} \bigg(\sum_{ijk\in\La_2} \sum_{\chi\in \h{A}} Z_{ijk}^{(\chi)}
    + 
    h_A \sum_{\langle ijk, lmn \rangle} \sum_{\la\in A} X_{ijk}^{(\la)} X_{lmn}^{(\la)}\bigg).
\end{aligned}
\end{equation}
We have defined the star term $A_i^{(h,\bt)}$ that acts as
\begin{equation}
\begin{tikzpicture}[decoration={markings, mark=at position 0.55 with {\arrow{>}}}, scale=3]
    \coordinate (c) at (0,0);
    \coordinate (e) at (1,0);
    \coordinate (w) at (-1,0);
    \coordinate (nw) at (-0.5,{sqrt(3)/2});
    \coordinate (ne) at (0.5,{sqrt(3)/2});
    \coordinate (sw) at (-0.5,-{sqrt(3)/2});
    \coordinate (se) at (0.5,-{sqrt(3)/2});

    \coordinate (t1) at ($(c)!0.333!(e) + (c)!0.333!(ne) - (c)$);   
    \coordinate (t2) at ($(c)!0.333!(ne) + (c)!0.333!(nw) - (c)$);  
    \coordinate (t3) at ($(c)!0.333!(nw) + (c)!0.333!(w) - (c)$);   
    \coordinate (t4) at ($(c)!0.333!(w) + (c)!0.333!(sw) - (c)$);   
    \coordinate (t5) at ($(c)!0.333!(sw) + (c)!0.333!(se) - (c)$);  
    \coordinate (t6) at ($(c)!0.333!(se) + (c)!0.333!(e) - (c)$);   
    
    \node at (t1) {\tiny $a_{l m o}$};
    \node at (t2) {\tiny $a_{l n o}$};
    \node at (t3) {\tiny $a_{kl n}$};
    \node at (t4) {\tiny $a_{ikl}$};
    \node at (t5) {\tiny $a_{ijl}$};
    \node at (t6) {\tiny $a_{jl m}$};
    
    \draw[postaction=decorate, color=lightgray] (c) node[below, color=gray] {\footnotesize $l$} -- node[anchor=mid, color=black] {\footnotesize $g_{l m}$} (e) node[color=gray, right] {\footnotesize $m$};
    \draw[postaction=decorate, color=lightgray] (c) -- node[anchor=mid, color=black] {\footnotesize $g_{l n}$} (nw) node[color=gray, above] {\footnotesize $n$};
    \draw[postaction=decorate, color=lightgray] (c) -- node[anchor=mid, color=black] {\footnotesize $g_{l o}$} (ne) node[color=gray, above] {\footnotesize $o$};
    \draw[postaction=decorate, color=lightgray] (sw) node[color=gray, below] {\footnotesize $i$} -- node[anchor=mid, color=black] {\footnotesize $g_{il}$} (c);
    \draw[postaction=decorate, color=lightgray] (se) node[color=gray, below] {\footnotesize $j$} -- node[anchor=mid, color=black] {\footnotesize $g_{jl}$} (c);
    \draw[postaction=decorate, color=lightgray] (w) node[color=gray, left] {\footnotesize $i$} -- node[anchor=mid, color=black] {\footnotesize $g_{k l}$} (c);
\end{tikzpicture}
\quad
\xrightarrow{A_l^{(h,\bt)}}
\quad
\begin{tikzpicture}[decoration={markings, mark=at position 0.55 with {\arrow{>}}}, scale=3]
    \coordinate (c) at (0,0);
    \coordinate (e) at (1,0);
    \coordinate (w) at (-1,0);
    \coordinate (nw) at (-0.5,{sqrt(3)/2});
    \coordinate (ne) at (0.5,{sqrt(3)/2});
    \coordinate (sw) at (-0.5,-{sqrt(3)/2});
    \coordinate (se) at (0.5,-{sqrt(3)/2});

    \coordinate (t1) at ($(c)!0.333!(e) + (c)!0.333!(ne) - (c)$);   
    \coordinate (t2) at ($(c)!0.333!(ne) + (c)!0.333!(nw) - (c)$);  
    \coordinate (t3) at ($(c)!0.333!(nw) + (c)!0.333!(w) - (c)$);   
    \coordinate (t4) at ($(c)!0.333!(w) + (c)!0.333!(sw) - (c)$);   
    \coordinate (t5) at ($(c)!0.333!(sw) + (c)!0.333!(se) - (c)$);  
    \coordinate (t6) at ($(c)!0.333!(se) + (c)!0.333!(e) - (c)$);   
    
    \node[align=center, xshift=1em, yshift=-0.5em] at (t1) {\tiny $a_{l m o}$ \\[-1.2em] \tiny ${\color{myblue} \vphantom{} + \zeta_{\t{h}}(g_{lm}, g_{mo})}$};
    \node[align=center, yshift=1em] at (t2) {\tiny $a_{l n o}$ \\[-1.2em] \tiny $ {\color{myblue} \vphantom{} + \zeta_{\t{h}}(g_{ln}, g_{no})}$};
    \node[align=center, xshift=-1em, yshift=-0.5em] at (t3) {\tiny $a_{kl n}$ \\[-1.2em] \tiny ${\color{myblue} \vphantom{} + \zeta_{\t{h}}(g_{kl}, g_{ln}) }$};
    \node[align=center, xshift=-1em, yshift=0.5em] at (t4) {\tiny $a_{ikl}$ \\[-1.2em] \tiny ${\color{myblue} \vphantom{} + \zeta_{\t{h}}(g_{ik}, g_{kl}) }$};
    \node[align=center, yshift=-1em] at (t5) {\tiny $a_{ijl}$ \\[-1.2em] \tiny ${\color{myblue} \vphantom{} + \zeta_{\t{h}}(g_{ij}, g_{jl})}$};
    \node[align=center, xshift=1em, yshift=0.5em] at (t6) {\tiny $a_{jl m}$\\[-1.2em] \tiny ${\color{myblue} \vphantom{} + \zeta_{\t{h}}(g_{jl}, g_{lm})}$};
    
    \draw[postaction=decorate, color=lightgray] (c) node[below, color=gray] {\footnotesize $l$} -- node[anchor=mid, color=black] {\footnotesize ${\color{myblue}h} g_{l m}$} (e) node[color=gray, right] {\footnotesize $m$};
    \draw[postaction=decorate, color=lightgray] (c) -- node[anchor=mid, color=black] {\footnotesize ${\color{myblue}h}g_{l n}$} (nw) node[color=gray, above] {\footnotesize $n$};
    \draw[postaction=decorate, color=lightgray] (c) -- node[anchor=mid, color=black] {\footnotesize ${\color{myblue}h}g_{l o}$} (ne) node[color=gray, above] {\footnotesize $o$};
    \draw[postaction=decorate, color=lightgray] (sw) node[color=gray, below] {\footnotesize $i$} -- node[anchor=mid, color=black] {\footnotesize $g_{il}{\color{myblue}h^{-1}}$} (c);
    \draw[postaction=decorate, color=lightgray] (se) node[color=gray, below] {\footnotesize $j$} -- node[anchor=mid, color=black] {\footnotesize $g_{jl}{\color{myblue}h^{-1}}$} (c);
    \draw[postaction=decorate, color=lightgray] (w) node[color=gray, left] {\footnotesize $i$} -- node[anchor=mid, color=black] {\footnotesize $g_{k l}{\color{myblue}h^{-1}}$} (c);
\end{tikzpicture},
\end{equation}
where the descendant
\begin{equation}
   \zeta_{\t h}(g_{ij}, g_{jk}) = \bt(g_{ij}, g_{jk}, \t{h}_k^{-1}) 
        - \bt(g_{ij},\t{h}_j^{-1},\t{h}_j g_{jk} \t{h}_k^{-1}) 
        + \bt(\t{h}_i^{-1}, \t{h}_i g_{ij} \t{h}_j^{-1}, \t{h}_j g_{jk} \t{h}_k^{-1}).
\end{equation}
and
\begin{equation}
    \t{h}_i = \begin{cases}
        h\quad &i=l,\\
        1\quad&\text{otherwise}.
    \end{cases}
\end{equation}
See App.~\ref{app:double-from-path-integral} for discussion of the properties of the descendant $\zeta_h$.

Furthermore, the 0-form $2\Rep(\G)$ symmetry operators given in Eq.~\eqref{eq:central-m.c.-ahat-operator-K=G} and~\eqref{eq:central-m.c.-ahat-operator-K=1} are also independent of site $G$-qudits in this frame. They now take the form
\begin{equation}\label{eq:central-m.c.-ahat-operator-K=G2}
    \begin{aligned}
        \mathsf{V}^{(\chi)}
            & = U_2\t{V}^{(\chi)}_{\text{m.c.}} U_2^\dag
            \\
            &= 
            |G| \sum_{\{g_{ij}\},\{a_{ijk}\}}\prod_{ijk\in\La_2} 
                \chi(a_{ijk} - \bt(g_{o\to i}, g_{ij}, g_{jk}))^{\eps_{ijk}}
                    \;
                    \mathsf{P}_{o,1}^{(\{1\})}
                    \ket{\{g_{ij}\}, \{a_{ijk}\}}\!\!\bra{\{g_{ij}\}, \{a_{ijk}\}},
    \end{aligned}
\end{equation}
and
\begin{equation}\label{eq:central-m.c.-ahat-operator-K=12}
    \begin{aligned}
        \mathsf{V}^{(\chi,c_\chi)} &= 
        U_2 \t{V}^{(\chi,c_\chi)}_\text{m.c.} U_2^\dagger\\
        &=
        \sum_{\{g_{ij}\},\{a_{ijk}\}}
        \prod_{l m n\in \La_2}
        \;
        \chi(a_{l mn})^{\eps_{l mn}}\;
        c_\chi(g_{l m},\,g_{mn})^{-\eps_{l mn}}
        \ketbra{\{g_{ij}\},\{a_{ijk}\}}.
    \end{aligned}
\end{equation}

\subsection{Central 2-group quantum double model}\label{sec:central-quantum-double}

The Hamiltonian~\eqref{eq:central-2rep-potts-model} lies in the deconfined 2-group gauge theory phase when ${h_G,h_A\gg 1}$. In this limit, it is represented at a convenient exactly-solvable point by
\begin{equation}\label{eq:central-quantum-double}
    H_{\G}^{(\bt)} 
    =
    - \frac{1}{2|G|}\sum_{i\in \La_0} \sum_{h\in G} (A_i^{(h,\bt)} + A_i^{(h,\bt)\dag} )
    -
    \sum_{ijk\in\La_2} B_{ijk}
    - \frac{1}{|A|} \sum_{\langle ijk, lmn \rangle} \sum_{\la\in A} X_{ijk}^{(\la)} X_{lmn}^{(\la)}.
\end{equation}
This Hamiltonian is a generalization of Kitaev's quantum double model~\cite{K9707021} for central 2-groups. 
While the second and third terms form a commuting projector Hamiltonian and both commute with the first term, the star operators ${\sum_{h\in G} A_j^{(h,\bt)}}$ fail to commute with each other. 
However, in the  subspace in which ${B_{ijk} = 1}$ and ${X_{ijk}^{(\la)} X_{lmn}^{(\la)} = 1}$, each ${\sum_{h\in G} A_j^{(h,\bt)}}$ becomes Hermitian and the first term forms a commuting projector Hamiltonian. Therefore, the ground-state space of the Hamiltonian $H_{\G}^{(\bt)}$ is exactly-solvable.

$H_{\G}^{(\bt)}$ is a gapped Hamiltonian and, as shown in App.~\ref{app:double-from-path-integral}, its ground-state space realizes untwisted central 2-group $\G$ gauge theory. App.~\ref{app:double-from-path-integral} also includes the calculation of the ground-state degeneracy. For instance, the ground-state degeneracy on the spatial sphere is 
\begin{equation}
    \text{GSD}_{S^2} = |A|.
\end{equation}
Notice that this is independent of the Postnikov class $[\bt]$.
In contrast, the ground-state degeneracy of $H_{\G}^{(\bt)}$ on a spatial torus is 
\begin{equation}\label{eq:central-quantum-double-gsd-torus}
    \text{GSD}_{T^2}
    =
    \sum_{\chi\in\h{A}}
    \sum_{[g]\in \operatorname{Conj}(G)} 
    |\operatorname{Irr}_{\iota_g(\chi\circ\bt)}(C_G(g))|,
\end{equation}
where $\operatorname{Conj}(G)$ denotes the conjugacy classes of $G$, $C_G(g)$ is the centralizer of $g$, and $\operatorname{Irr}_{\iota_g \om}(C_G(g))$ is the set of inequivalent irreducible $\iota_g\om$-projective representations of $C_G(g)$ with $\iota_g\om$ defined in Eq.~\eqref{eq:slant-product-app}.
Importantly, ${\sum_{[g]\in \operatorname{Conj}(G)}|\operatorname{Irr}_{\iota_g(\chi\circ\bt)}(C_G(g))|}$ is equal to the ground-state degeneracy of the ${\chi\circ\bt}$-twisted $G$-quantum double model~\cite{HWW12113695}. Therefore, each character $\chi$ contributes to $\text{GSD}_{T^2}$ the ground-state degeneracy of the ${\chi\circ\bt}$-twisted $G$-quantum double. 
This is the torus Hilbert-space
realization of the decomposition conjecture of~\cite{PRS220413708,PS230316220}.
When $[\bt]$ is trivial, ${\iota_g[\chi\circ\bt] = [1]}$ and $\text{GSD}_{T^2}$ becomes ${|A|\sum_{[g]\in \operatorname{Conj}(G)} |\operatorname{Irr}(C_G(g))| = |A||\Hom(\Z^2, G)/G|}$. Therefore, when ${A=1}$, $\text{GSD}_{T^2}$ reduces to the torus ground-state degeneracy of the ordinary $G$-quantum double model.

\subsection{Example: Type-I Postnikov cocycle}\label{sec:central-example}

We end this section on central 2-group symmetry by illustrating our general results in the simplest case. 
Let ${G = \Z_2 = \{1,-1\}}$ and ${A = \Z_2 = \{0,1\}}$ with the action of $G$ on $A$ trivial. 
There is one qubit per site and one qubit per link.
For these groups, ${\cH^3(\Z_2,\Z_2) \cong \Z_2}$. We choose the following normalized representative of the nontrivial cohomology class
\begin{equation}\label{eq:type-i-cocycle-beta}
    \bt_\mathrm{I}(g,h,k) = 
    \begin{cases}
        1\quad g=h=k=-1,\\
        0\quad \text{otherwise}.
    \end{cases}
\end{equation}
This is the $\Z_2$-valued type-I cocycle. Together, this data defines a central 2-group ${\G = (\Z_2, \Z_2, \bt_\mathrm{I})}$.

The 0-form symmetry part~\eqref{GbetaSymOps} involves the group $\Map(G,A)/A_\mathrm{const}$. For ${G=A=\Z_2}$, $\Map(G,A)/A_\mathrm{const}\cong \Z_2$. Each $\t{G}$ symmetry operator is formed by products of
\begin{align}
    U 
    &=
    \sum_{\{g_i\},\{a_{ij}\}}
    \ketbra
    {\{-g_i\}, \{a_{ij} + 
    \del_{g_i = -1}
    \del_{g_j = 1}\}}
    {\{g_i\}, \{a_{ij}\}},\\
    S 
    &=
    \sum_{\{g_i\},\{a_{ij}\}}
    \ketbra
    {\{g_i\}, \{a_{ij} - 
    \del_{g_i = -1}
    +\del_{g_j = -1}\}}
    {\{g_i\}, \{a_{ij}\}}.
\end{align}
These operators can be written in terms of conventional quantum gates. Namely, introducing the Toffoli and CNOT gates
\begin{align}
    \mathsf{CCX}_{i,j, ij} &= 
1 + \frac14(X_{ij}-1)(1-Z_i)(1-Z_j),\\
\mathsf{CX}_{i, ij} &= \frac12\left[\left(1+Z_i\right)
+
X_{ij}\left(1-Z_i\right)\right],
\end{align}
they satisfy
\begin{align}
    U &= 
    \prod_{ij\in\La_1}
    ( X_i \,\mathsf{CCX}_{i,j, ij}\, X_i )\, 
    \prod_{i\in\La_0}X_i,\\
    S&= \prod_{ij\in\La_1}
    \mathsf{CX}_{i,ij}
    \mathsf{CX}_{j,ij}.
\end{align}
Furthermore, $U$ satisfies
\begin{equation}
    U^2 
    =
    S,
    \qquad U^4 = 1.
\end{equation}
Therefore, $U$ generates a $\Z_4$ 0-form symmetry on the full tensor-product Hilbert space. Namely, the extension~\eqref{extensionofGNbt} for this example has ${G=\Z_2}$ extended by ${\Z_2}$ to yield ${\t{G} = \Z_4}$.

The lattice $\Z_2$ 1-form symmetry operator is
\begin{equation}
    T^{(\ga)} = \prod_{ij\in\La_1} X^{\ga_{ij}}_{ij},
\end{equation}
where ${\ga\equiv\sum_{ij\in\La_1}\ga_{ij}\,[ij]^\vee}$ with ${\ga_{ij}\in\{0,1\}}$ satisfying ${\ga_{ij} +\ga_{jk} + \ga_{ik} = 0\bmod 2}$ for all ${ijk\in\La_2}$. The topological subspace $\scrH_\mathrm{top}$ is formed by all vectors $\ket{\psi}$ satisfying $A_i^{\text{TC}} \ket{\psi} = \ket{\psi}$ where the toric code star operator
\begin{equation}
    A_i^{\text{TC}} 
    =
    X_{\trivertex{w}}
    X_{\trivertex{sw}}
    X_{\trivertex{se}}
    X_{\trivertex{e}}
    X_{\trivertex{ne}}
    X_{\trivertex{nw}}.
\end{equation}
Note that the operator $S$ on the full Hilbert space can be written as
\begin{equation}
    S = \prod_{i\in\La_0} \left(\frac{1+Z_i + (1-Z_i)A_i^\mathrm{TC}}2\right).
\end{equation}
Therefore, ${S\mid_{\scrH_\mathrm{top}} = 1}$ and the $\Z_4$ 0-form symmetry ${\{U^n\}_{n=0}^3}$ acts as a $\Z_2$ 0-form symmetry on $\scrH_\mathrm{top}$.

The central 2-group $\G$-Potts model Hamiltonian~\eqref{eq:central-2group-pottsmodel} for this example can be written in terms of conventional qubit operators. We define the dressed toric code plaquette operator 
\begin{equation}
    B^{(\bt_\mathrm{I})}_{ijk} = 
    Z_{ij}Z_{ik}Z_{jk}
    \,
    \mathsf{CCZ}_{i,j,k},
\end{equation}
where the controlled-controlled-Z gate ${\mathsf{CCZ}_{i,j,k} = 1-\frac14(1-Z_i)(1-Z_j)(1-Z_k)}$, and further define
\begin{equation}
    M_{l}^{(\bt_\mathrm{I})}
    \equiv
   \begin{tikzpicture}[decoration={markings, mark=at position 0.55 with {\arrow{>}}}, scale=2.5]
    \coordinate (c) at (0,0);
    \coordinate (e) at (1,0);
    \coordinate (w) at (-1,0);
    \coordinate (nw) at (-0.5,{sqrt(3)/2});
    \coordinate (ne) at (0.5,{sqrt(3)/2});
    \coordinate (sw) at (-0.5,-{sqrt(3)/2});
    \coordinate (se) at (0.5,-{sqrt(3)/2});
    
    \draw[postaction=decorate, color=lightgray] (c) 
    node[color=gray, below] {\footnotesize $l$}
    -- node[color=black,align=center] 
    {\footnotesize $\mathsf{CCX}_{l,m,lm}$} 
    (e) node[color=gray, right] {\footnotesize $m$};
    \draw[postaction=decorate, color=lightgray] (c) -- 
    node[color=black, align=center, xshift=-0.5em, yshift=0.5em,] 
    {\footnotesize$ 
    \mathsf{CCX}_{l,n,ln}$} 
    (nw) node[color=gray, above] {\footnotesize $n$} ;
    \draw[postaction=decorate, color=lightgray] (c) -- 
    node[color=black, align=center, xshift=0.5em, yshift=0.5em] 
    {\footnotesize$ 
    \mathsf{CCX}_{l,o,lo}$} 
    (ne) node[color=gray, above] {\footnotesize $o$};
    \draw[postaction=decorate, color=lightgray] (sw) node[color=gray, below] {\footnotesize $i$} -- 
    node[color=black,xshift=-1.5em, yshift=-0.5em, align=center] 
    {\footnotesize$ 
    X_l\mathsf{CCX}_{i,l,il}X_l$} 
    (c);
    \draw[postaction=decorate, color=lightgray] (se) node[color=gray, below] {\footnotesize $j$} -- 
    node[color=black,xshift=1.5em, yshift=-0.5em, align=center] 
    {\footnotesize$ 
    X_l\mathsf{CCX}_{j,l,jl}X_l
    $} 
    (c);
    \draw[postaction=decorate, color=lightgray] (w) node[color=gray, left] {\footnotesize $k$} -- 
    node[color=black, align=center] 
    {\footnotesize$
    X_l\mathsf{CCX}_{k,l,kl}X_l
    $} 
    (c);
\end{tikzpicture}.
\end{equation}
The corresponding 2-group Potts model Hamiltonian can be written as
\begin{equation}\label{eq:central-potts-typei-example}
\begin{aligned}
    H_{\G\text{-Potts}}^{(\bt_I)}
    = 
    &
    -
    \frac{1}{2}
    \left(
    \sum_{ij\in\La_1}
    (1+Z_i Z_j)
    + 
    \frac{h_G}{4}
    \sum_{i\in\La_0}
    (1+A_i^{\text{TC}})
    (2+ X_iM_i^{(\bt_I)}
    +
    M_i^{(\bt_I)}X_i)
    \right)
    \\ 
    &
    -
    \frac{1}{2}
    \left(
    \sum_{ijk\in\La_2}
    (1+ B^{(\bt_I)}_{ijk})
    +
    \sum_{i\in\La_0}
    (1+A_i^{\text{TC}})
    + 
    h_A
    \sum_{ij\in\La_1}
    (1+X_{ij})
    \right).
\end{aligned}
\end{equation}
The first term in~\eqref{eq:central-potts-typei-example} is the Ising ferromagnet term and the second is a dressed Ising paramagnet term. The last three terms make up a $\Z_2$ abelian Higgs model whose plaquette term is dressed.

Gauging this central 2-group $\G$ symmetry leads to a model with one qubit on each link and plaquette.
The dual lattice $2\Rep(\G)$ is made up of the following. There is a lattice $\Z_2$ 1-form symmetry formed by the Wilson loop operators $W^{(\Upsilon)}$ defined in~\eqref{eq:z2-1-form-symmetry}.
The non-invertible 0-form operators~\eqref{eq:central-m.c.-ahat-operator-K=G2} and~\eqref{eq:central-m.c.-ahat-operator-K=12} also simplify. 
For ${A=\Z_2}$, the character group ${\h{A} = \{1,\chi_\text{sgn}\}}$ with $\chi_\text{sgn}$ the sign irrep. The operator~\eqref{eq:central-m.c.-ahat-operator-K=G2} for these two irreps can be simplified to
\begin{equation}
    \mathsf{V}^{(1)}
    = 
    2 \mathsf{P},
\qquad
    \mathsf{V}^{(\chi_\mathrm{sgn})}
    = 
        2\mathsf{P}
        \,
        \prod_{ijk\in\La_2} 
        \left(
        Z_{ijk}
        \mathsf{CCZ}_{\gamma_{o\to i}, ij,jk}
        \right)^{\epsilon_{ijk}},
\end{equation}
where ${\mathsf{P} = |Z_1(\La;\Z_2)|^{-1}\sum_{\Upsilon\in Z_1(\La;\Z_2)} W^{(\Upsilon)}}$ and
\begin{equation}
     \mathsf{CCZ}_{ij,jk, \gamma_{o\to i}}
     = 
     1 
     - 
     \frac{1}{4}
     (1 - Z_{ij})
     (1 - Z_{jk})\left(1- \prod_{lm\in\gamma_{o\to i}} Z_{lm}^{\epsilon_{lm}(\gamma_{o\to i})}\right).
\end{equation}
(See footnote~\ref{condOpFootnote}.)
They satisfy
\begin{equation}
    \mathsf{V}^{(\chi_\mathrm{sgn})}\times \mathsf{V}^{(\chi_\mathrm{sgn})} = 2\mathsf{V}^{(1)}.
\end{equation}
The operator~\eqref{eq:central-m.c.-ahat-operator-K=12} is defined only for $\chi$ such that ${[\chi\circ\bt] = [1]}$.
However, ${[\chi_\mathrm{sgn}\circ\bt_\mathrm{I}] \neq [1]}$. Thus,~\eqref{eq:central-m.c.-ahat-operator-K=12} holds only for the trivial irrep, where it simplifies to
\begin{equation}
    \mathsf{V}^{(1,c)} =
        \sum_{\{g_{ij}\},\{a_{ijk}\}}
        \prod_{l m n\in\La_2}
        c(g_{l m},\,g_{mn})^{-\eps_{l mn}}
        \ketbra{\{g_{ij}\},\{a_{ijk}\}},
\end{equation}
where ${c\in\cZ^2(\Z_2,\Uone)}$.

Lastly, we discuss the central 2-group quantum double model~\eqref{eq:central-quantum-double}. For this example, it takes the form
\begin{equation}\label{eq:central-quantum-double-typei-example}
    H^{(\bt_I)}_\G = -\frac14 \sum_{i\in\La_0} (2+\t{A}_i+\t{A}_i^{\,\dag}) 
        - \frac12\sum_{ijk\in\La_2} (1+Z_{ij} Z_{jk} Z_{ik})
        - \frac12\sum_{\<ijk,lmn\>} (1+X_{ijk}X_{lmn}),
\end{equation}
where $\t{A}_i = \t{A}^{(1)}_i\t{A}^{(2)}_i\t{A}^{(3)}_i$ with
\begin{equation}
\begin{gathered}
    \t{A}^{(1)}_l
    =
    \begin{tikzpicture}[decoration={markings, mark=at position 0.55 with {\arrow{>}}}, scale=1.9]
    \coordinate (c) at (0,0);
    \coordinate (e) at (1,0);
    \coordinate (w) at (-1,0);
    \coordinate (nw) at (-0.5,{sqrt(3)/2});
    \coordinate (ne) at (0.5,{sqrt(3)/2});
    \coordinate (sw) at (-0.5,-{sqrt(3)/2});
    \coordinate (se) at (0.5,-{sqrt(3)/2});
    %
    \coordinate (t1) at ($(c)!0.333!(e) + (c)!0.333!(ne) - (c)$);   
    \coordinate (t2) at ($(c)!0.333!(ne) + (c)!0.333!(nw) - (c)$);  
    \coordinate (t3) at ($(c)!0.333!(nw) + (c)!0.333!(w) - (c)$);   
    \coordinate (t4) at ($(c)!0.333!(w) + (c)!0.333!(sw) - (c)$);   
    \coordinate (t5) at ($(c)!0.333!(sw) + (c)!0.333!(se) - (c)$);  
    \coordinate (t6) at ($(c)!0.333!(se) + (c)!0.333!(e) - (c)$);   
    %
    \draw[postaction=decorate, color=lightgray] (c) node[color=gray, below] {\footnotesize $l$} -- (e) node[color=gray, right] {\footnotesize $m$};
    \draw[postaction=decorate, color=lightgray] (c) -- (nw) node[color=gray, above] {\footnotesize $n$};
    \draw[postaction=decorate, color=lightgray] (c) -- node[anchor=mid, color=black] {$X_{l o}$} (ne) node[color=gray, above] {\footnotesize $o$};
    \draw[postaction=decorate, color=lightgray] (sw) node[color=gray, below] {\footnotesize $i$} -- node[anchor=mid, color=black] {$X_{il}$} (c);
    \draw[postaction=decorate, color=lightgray] (se) node[color=gray, below] {\footnotesize $j$} -- node[anchor=mid, color=black] {$X_{jl}$} (c);
    \draw[postaction=decorate, color=lightgray] (w) node[color=gray, left] {\footnotesize $k$} -- node[anchor=mid, color=black] {$X_{k l}$} (c);
\end{tikzpicture},
\qquad \t{A}^{(2)}_l
    =
\begin{tikzpicture}[decoration={markings, mark=at position 0.55 with {\arrow{>}}}, scale=1.9]
    \coordinate (c) at (0,0);
    \coordinate (e) at (1,0);
    \coordinate (w) at (-1,0);
    \coordinate (nw) at (-0.5,{sqrt(3)/2});
    \coordinate (ne) at (0.5,{sqrt(3)/2});
    \coordinate (sw) at (-0.5,-{sqrt(3)/2});
    \coordinate (se) at (0.5,-{sqrt(3)/2});
    %
    \coordinate (t1) at ($(c)!0.333!(e) + (c)!0.333!(ne) - (c)+(.1,0)$);   
    \coordinate (t2) at ($(c)!0.333!(ne) + (c)!0.333!(nw) - (c)+(0,.1)$);  
    \coordinate (t3) at ($(c)!0.333!(nw) + (c)!0.333!(w) - (c)-(.1,0)$);   
    \coordinate (t4) at ($(c)!0.333!(w) + (c)!0.333!(sw) - (c)-(.1,0)$);   
    \coordinate (t5) at ($(c)!0.333!(sw) + (c)!0.333!(se) - (c)-(0,.1)$);  
    \coordinate (t6) at ($(c)!0.333!(se) + (c)!0.333!(e) - (c)+(.1,0)$);   
    %
    \node at (t1) { \footnotesize $\mathsf{CCX}_{lmo}$};
    \node at (t2) { \footnotesize $\mathsf{CCX}_{lno}$};
    \node at (t3) { \footnotesize $\mathsf{CCX}_{kln}$};
    \node at (t4) { \footnotesize $\mathsf{CCX}_{ikl}$};
    \node at (t5) { \footnotesize $\mathsf{CCX}_{ijl}$};
    \node at (t6) { \footnotesize $\mathsf{CCX}_{jlm}$};
    %
    \draw[postaction=decorate, color=lightgray] (c) node[color=gray, below] {\footnotesize $l$} -- (e) node[color=gray, right] {\footnotesize $m$};
    \draw[postaction=decorate, color=lightgray] (c) -- (nw) node[color=gray, above] {\footnotesize $n$};
    \draw[postaction=decorate, color=lightgray] (c) -- (ne) node[color=gray, above] {\footnotesize $o$};
    \draw[postaction=decorate, color=lightgray] (sw) node[color=gray, below] {\footnotesize $i$} -- (c);
    \draw[postaction=decorate, color=lightgray] (se) node[color=gray, below] {\footnotesize $j$} -- (c);
    \draw[postaction=decorate, color=lightgray] (w) node[color=gray, left] {\footnotesize $k$} -- (c);
\end{tikzpicture},\\
\t{A}^{(3)}_l
    =
\begin{tikzpicture}[decoration={markings, mark=at position 0.55 with {\arrow{>}}}, scale=1.9]
    \coordinate (c) at (0,0);
    \coordinate (e) at (1,0);
    \coordinate (w) at (-1,0);
    \coordinate (nw) at (-0.5,{sqrt(3)/2});
    \coordinate (ne) at (0.5,{sqrt(3)/2});
    \coordinate (sw) at (-0.5,-{sqrt(3)/2});
    \coordinate (se) at (0.5,-{sqrt(3)/2});
    %
    \coordinate (t1) at ($(c)!0.333!(e) + (c)!0.333!(ne) - (c)$);   
    \coordinate (t2) at ($(c)!0.333!(ne) + (c)!0.333!(nw) - (c)$);  
    \coordinate (t3) at ($(c)!0.333!(nw) + (c)!0.333!(w) - (c)$);   
    \coordinate (t4) at ($(c)!0.333!(w) + (c)!0.333!(sw) - (c)$);   
    \coordinate (t5) at ($(c)!0.333!(sw) + (c)!0.333!(se) - (c)$);  
    \coordinate (t6) at ($(c)!0.333!(se) + (c)!0.333!(e) - (c)$);   
    %
    \draw[postaction=decorate, color=lightgray] (c) node[color=gray, below] {\footnotesize $l$} -- node[anchor=mid, color=black] {$X_{l m}$} (e) node[color=gray, right] {\footnotesize $m$};
    \draw[postaction=decorate, color=lightgray] (c) -- node[anchor=mid, color=black] {$X_{l n}$} (nw) node[color=gray, above] {\footnotesize $n$};
    \draw[postaction=decorate, color=lightgray] (c) -- (ne) node[color=gray, above] {\footnotesize $o$};
    \draw[postaction=decorate, color=lightgray] (sw) node[color=gray, below] {\footnotesize $i$} -- (c);
    \draw[postaction=decorate, color=lightgray] (se) node[color=gray, below] {\footnotesize $j$} -- (c);
    \draw[postaction=decorate, color=lightgray] (w) node[color=gray, left] {\footnotesize $k$} -- (c);
\end{tikzpicture}.
\end{gathered}
\end{equation}
Here, we use the shorthand ${\mathsf{CCX}_{ijk}\equiv \mathsf{CCX}_{ij,jk,ijk}}$. The first two terms of \eqref{eq:central-quantum-double-typei-example} reduces to the toric code if the decoration $\tilde{A}_i^{(2)}$ is removed from $\t{A}_i$. The third term is precisely an Ising ferromagnetic term.
The ground-state degeneracy of ${H^{(\bt_I)}_\G}$ in~\eqref{eq:central-quantum-double-typei-example} is a special case of~\eqref{eq:central-quantum-double-gsd-torus}. 
Applying the general formula, we find
\begin{equation}
    \text{GSD}_{S^2} = 2, \qquad \text{GSD}_{T^2} = 8.
\end{equation}
We note that, in this case, these ground-state degeneracies happen to be equal to the degeneracy for trivial $[\bt]$.

\section{Outlook}\label{sec:conclusion}

In this work, we developed and systematically analyzed lattice realizations of two broad classes of finite 2-group symmetries in ${2+1}$d quantum lattice systems with tensor-product Hilbert spaces. We studied split 2-groups in Sec.~\ref{sec:split-2group-symmetry} and central 2-groups with nontrivial Postnikov class in Sec.~\ref{sec:central-2group-symmetry}. In both cases, we constructed explicit symmetry operators and symmetric Hamiltonians, investigated the associated symmetry defects, and tracked the symmetries and Hamiltonians under gauging. A summary of these results is given in Sec.~\ref{Sec:summary}.

This work suggests several directions for further study. We highlight
three particularly natural ones.
\begin{enumerate}
    \item \textit{2-group Potts models:} 
    For both split and central 2-groups, we introduced a $\G$-Potts Hamiltonian and its gauged $2\Rep(\G)$-Potts counterpart. These are given by Eqs.~\eqref{eq:split-2group-pottsmodel} and~\eqref{eq:split-2rep-potts-model} in the split case, and by Eqs.~\eqref{eq:central-2group-pottsmodel} and ~\eqref{eq:central-2rep-potts-model} in the central case.
    The models possess lattice $\G$ and $2\Rep(\G)$ symmetries, respectively, and are expected to exhibit rich phase diagrams involving generalized symmetry breaking and topological order. 
    For instance, SymTFT analyses of $2$-group and $2\Rep(\G)$ symmetries have identified a wide range of gapped phases, including phases in which distinct symmetry-breaking vacua carry different topological orders~\cite{BSTW250220440}. It would be interesting to determine which of these phases are realized in the lattice Hamiltonians introduced here.
    
    Related generalized Ising and Potts models in ${1+1}$d have been studied using invertible and non-invertible symmetries~\cite{AFM200808598, BBS240505302, CAW240505331,BBS240505964,CBN250811003,LCT260210183}.
    We identified several exactly-solvable limits of our ${2+1}$d models, but a systematic characterization of their phase diagrams remains open. 
    It would be particularly interesting to study the split and central examples of Secs.~\ref{sec:split-example} and~\ref{sec:central-example}, determine their critical behavior and universality classes, and investigate whether they possess any generalized Kramers--Wannier self-dualities.
    \item \textit{Anomalies:} 
    Anomalies of symmetries in quantum lattice systems have recently become an active area of research. In ${2+1}$d lattice systems, anomalies have recently been investigated for both 0-form symmetries~\cite{PKC250504684, KX250504719, KS250707430,GM250716475, TLE250721209,SZJ250721267,CGT251202105,PB260211266, OE260402856} and lattice 1-form symmetries~\cite{KS250716966, FKCR250912304, FCH251023701}.
    The lattice 2-group symmetries constructed in this work were anomaly-free: they could be gauged and admitted a trivially gapped symmetric phase. 
    It would be interesting to construct and classify anomalous lattice 2-group symmetries and explore their relationship to onsiteability. Of particular interest are genuinely 2-group anomalies that are not detected by restricting to the 0-form or lattice 1-form symmetries separately. 
    \item \textit{Generalized gauging:} There are generally several inequivalent ways to gauge a finite symmetry. For an invertible symmetry in quantum field theory, one may first stack with an SPT and then gauge, or equivalently include a discrete-torsion counterterm in the gauging procedure~\cite{Vafa:1986wx,GW14125148}. 
    This is often referred to as twisted gauging. 
    Related generalized gauging procedures have been explored in ${1+1}$d quantum lattice systems for invertible 0-form symmetry~\cite{LSY240514939} and non-invertible 0-form symmetry~\cite{SSY250302925}, as well as for invertible 0-form symmetry in ${2+1}$d~\cite{VD250116301}.

    The gauging procedures for lattice 2-group symmetry in Secs.~\ref{sec:split-gauging} and~\ref{sec:central-gauging} correspond to untwisted gauging, and it would be interesting to implement twisted gauging procedures of these lattice 2-group symmetries.
    More generally, it would be interesting to explore generalized gauging of generalized symmetries in greater than ${2+1}$ dimensional quantum lattice systems.
\end{enumerate}

\section*{Acknowledgements}

We thank
Arkya Chatterjee,
Clay C{\' o}rdova,
Theodore Jacobson,
Elias Riedel G\aa rding,
Shu-Heng Shao,
Wilbur Shirley,
Nathanan Tantivasadakarn,
Bram Vancraeynest--De Cuiper,
Xiao-Gang Wen,
and
Xinping Yang
for discussion.
S.D.P. is particularly grateful for multiple discussions with Wilbur Shirley and Xinping Yang on higher-group symmetries in lattice systems.
We also thank 
{\" O}mer Aksoy,
Arkya Chatterjee,
Ryohei Kobayashi,
Sakura Sch{\"a}fer-Nameki,
and
Wilbur Shirley
for comments on the draft.
S.D.P. acknowledges support from the Simons Collaboration on Ultra-Quantum Matter, which is a grant from the Simons Foundation (651446, XGW), the Marvin L. Goldberger Membership, the William Loughlin Membership, and the IBM Einstein Fellowship Fund.

\appendix

\section{Review of group qudits}\label{app:group-qudit-review}

In this appendix, we review the group-qudit Hilbert spaces and some of their commonly used operators. 

Let $G$ be a finite group. A $G$-qudit is a quantum mechanical system with Hilbert space
\begin{equation}
    \scrH_G = \mathbb C[G]
    \cong
    \mathbb C^{|G|}.
\end{equation}
Its orthonormal computational basis is
\begin{equation}
    \left\{
        \ket{g}
    \right\}_{g\in G},
    \qquad
    \braket{g}{h}
    =
    \delta_{g,h}.
\end{equation}

A $G$-qudit has two types of generalized Pauli operators.

For general $G$, there are distinct left and right shift operators defined by
\begin{equation}\label{eq:group-qudit-left-right-X}
    \overrightarrow X^{(h)}
    =
    \sum_{g\in G}
    \ketbra{hg}{g},
    \qquad
    \overleftarrow X^{(h)}
    =
    \sum_{g\in G}
    \ketbra{gh^{-1}}{g}.
\end{equation}
They are generalizations of the Pauli $X$ matrix, and are the left and right regular representations of $G$~\cite{AA211112096}.

The analog of the ordinary Pauli-$Z$ operator should be diagonal in the basis $\{\ket g\}$. For a non-abelian group, however, an irrep can have dimension greater than one. Accordingly, the generalized $Z$-operator carries auxiliary representation indices. Let $\Ga$ be a unitary irrep of dimension $d_\Ga$. Its matrices obey ${\Ga(g)\Ga(h) = \Ga(gh)}$.
We define the corresponding matrix-valued $Z$-operator by
\begin{equation}\label{eq:group-qudit-Z-matrix}
    Z^{(\Gamma)}
    =
    \sum_{g\in G}
    \Gamma(g)\otimes\ketbra{g}{g}
    \in
    \operatorname{End}
    \left(
        \C^{d_\Ga}\otimes\scrH_G
    \right).
\end{equation}
Equivalently, its matrix entries are operators on $\scrH_G$:
\begin{equation}\label{eq:group-qudit-Z-components}
    [Z^{(\Gamma)}]_{\al\bt}
    =
    \sum_{g\in G}
    [\Gamma(g)]_{\al\bt}
    \ketbra{g}{g}.
\end{equation}
The $X$-type operators act on
$[Z^{(\Gamma)}]_{\alpha\beta}$ as
\begin{align}
    \overrightarrow X^{(g)}
    [Z^{(\Gamma)}]_{\alpha\beta}
    \overrightarrow X^{(g)\,\dagger}
    &=
    \sum_{\ga = 1}^{d_\Ga}
    [\Gamma(g^{-1})]_{\alpha\gamma}
    [Z^{(\Gamma)}]_{\gamma\beta},
    \label{eq:group-qudit-left-XZ-algebra}
    \\
    \overleftarrow X^{(g)}
    [Z^{(\Gamma)}]_{\alpha\beta}
    \overleftarrow X^{(g)\,\dagger}
    &=
    \sum_{\ga = 1}^{d_\Ga}
    [Z^{(\Gamma)}]_{\alpha\gamma}
    [\Gamma(g)]_{\gamma\beta}.
    \label{eq:group-qudit-right-XZ-algebra}
\end{align}

When $G$ is abelian, the left and right shift operators are related. Namely,
\begin{equation}
    \overrightarrow{X}^{(g)} = \overleftarrow{X}^{(g^{-1})} \equiv X^{(g)}.
\end{equation}
Also, for abelian $G$, every irrep is one-dimensional and is labeled by a character
\begin{equation}
    \chi:G\to \Uone.
\end{equation}
Eq.~\eqref{eq:group-qudit-Z-components} then reduces to
\begin{equation}\label{eq:abelian-group-qudit-Z}
    Z^{(\chi)}
    =
    \sum_{g\in G}
    \chi(g)\ketbra{g}{g}.
\end{equation}
The characters form the Pontryagin-dual group $\h{G}$, and the
corresponding operators satisfy
\begin{equation}
    Z^{(\chi_1)}
    Z^{(\chi_2)}
    =
    Z^{(\chi_1\chi_2)},
    \qquad
    Z^{(\chi)\,\dagger}
    =
    Z^{(\bar\chi)}
    =
    Z^{(\chi^{-1})}.
\end{equation}

An important class of group-qudit operators we consider in the main text are the automorphism operators. Let $A$ be a finite abelian group, written additively, and let
\begin{equation}
    \rho:G\to\Aut(A)
\end{equation}
be an action of $G$ on $A$. Consider an $A$-qudit. The automorphism operator associated with $\rho_h$ for this $A$-qudit  is
\begin{equation}\label{eq:group-qudit-automorphism-operator}
    P^{(\rho_h)}
    =
    \sum_{a\in A}
    \ketbra{\rho_h(a)}{a}.
\end{equation}
It is a unitary operator satisfying
\begin{equation}
    P^{(\rho_g)}
    P^{(\rho_h)}
    =
    P^{(\rho_{gh})},
\end{equation}
and
\begin{equation}\label{eq:automorphism-conjugates-X}
    P^{(\rho_h)}
    X^{(\lambda)}
    P^{(\rho_h)\,\dagger}
    =
    X^{(\rho_h(\lambda))}.
\end{equation}
For a character ${\chi\in\h{A}}$, it similarly satisfies
\begin{equation}
    P^{(\rho_h)}
    Z^{(\chi)}
    P^{(\rho_h)\,\dagger}
    =
    Z^{(\chi\circ\rho_h^{-1})}.
\end{equation}

Another set of important qudit operators are controlled operators. 
Suppose a $G$-qudit is located on a simplex $I_m$ and an $A$-qudit is located on a simplex $J_n$. We define the automorphism of the $A$-qudit controlled by the $G$-qudit as
\begin{equation}\label{eq:controlled-automorphism-operator}
    P_{J_n}^{(\rho_{g_{I_m}})}
    =
    \sum_{g\in G}
    \ketbra{g}{g}_{I_m}
    \otimes
    P_{J_n}^{(\rho_g)}.
\end{equation}
Its action is
\begin{equation}
    P_{J_n}^{(\rho_{g_{I_m}})}
    \ket{g}_{I_m}\otimes\ket{a}_{J_n}
    =
    \ket{g}_{I_m}
    \otimes \ket{\rho_g(a)}_{J_n}.
\end{equation}
Similarly, for a fixed $\lambda\in A$, we define the controlled translation
\begin{equation}\label{eq:controlled-A-translation}
    X_{J_n}^{(\rho_{g_{I_m}}(\lambda))}
    =
    \sum_{g\in G}
    \ketbra{g}{g}_{I_m}
    \otimes
    X_{J_n}^{(\rho_g(\lambda))}.
\end{equation}
It acts as
\begin{equation}
    X_{J_n}^{(\rho_{g_{I_m}}(\lambda))}
    \ket{g}_{I_m}\otimes\ket{a}_{J_n}
    =
    \ket{g}_{I_m}\otimes
    \ket{a+\rho_g(\lambda)}_{J_n}.
\end{equation}
Thus, $X_{J_n}^{(\rho_h(\lambda))}$ is an ordinary translation whose action is fixed independently of the lattice configuration, whereas $X_{J_n}^{(\rho_{g_{I_m}}(\lambda))}$ is a controlled translation whose action depends on the state of the $G$-qudit at $I_m$.

\section{2-group quantum double model from Euclidean path integral}\label{app:double-from-path-integral}

In this appendix, we explain how the quantum double Hamiltonians
\eqref{eq:split-quantum-double} and
\eqref{eq:central-quantum-double} arise from the Euclidean path integral of
untwisted ${2+1}$d 2-group gauge theory, and we compute their ground-state degeneracies.

Let ${\G=(G,A,\rho,[\bt])}$ be a finite 2-group, where the groups $G$ and $A$ are finite,
$A$ is abelian, ${\rho\colon G\to \Aut(A)}$, and
${[\bt]\in \cH^3_\rho(G,A)}$. We choose a normalized representative
${\bt\in \cZ_\rho^3(G,A)}$. Let $K$ be a simplicial triangulation of three-dimensional Euclidean spacetime $X$. 
A $\G$ gauge field on $X$ is a map ${X\to B\G}$, where $B\G$ is the classifying space of $\G$.
It is equivalent to the pair of simplicial cochains~\cite{BL0307200, BS0412325, BS0511710}
\begin{equation}
    g\in C^1(K,G), \qquad a\in C^2(K,A),
\end{equation}
satisfying the $\G$-flatness constraints
\begin{align}\label{General 2Grp flatness Cond}
     g_{ij} g_{jk} = g_{ik},\qquad 
     (\dd_g a)_{ijkl}\equiv \rho_{g_{ij}} (a_{jkl}) - a_{ikl} + a_{ijl} - a_{ijk} = 
    \bt(g_{ij}, g_{jk}, g_{kl}).
\end{align}

A $\G$ gauge transformation is specified by the simplicial cochains ${f\in C^0(K,G)}$ and ${\la\in C^1(K,A)}$ subject to the equivalence relation ${\la_{ij} \sim \la_{ij} + \rho_{g_{ij}}(\al_j) - \al_i}$ for all ${\al\in C^0(K,A)}$. 
It corresponds to a homotopy of the map ${X\to B\G}$.
There are two types of $\G$ gauge transformations:
\begin{enumerate}
    \item A $\G$ gauge transformation with nontrivial $f$ and trivial $\la$ (i.e., ${\la_{ij} = \rho_{g_{ij}}(\al_j) - \al_i}$) is
\begin{equation}\label{0FormGaugeTrans}
    g_{ij} \mapsto f_i g_{ij} f^{-1}_j,
    \qquad 
    a_{ijk} \mapsto \rho_{f_{i}}(a_{ijk}) 
    + \zeta_f(g_{ij},g_{jk}),
\end{equation}
where ${\zeta_f\colon G\times G \to A}$ satisfies ${\zeta_{f=1}(g_{ij},g_{jk}) = 0}$.
For~\eqref{General 2Grp flatness Cond} to be invariant under this transformation, $\zeta_f$ must also satisfy
\begin{equation}\label{CondOnZeta}
\begin{aligned}
    &\rho_{f_i g_{ij} f_j^{-1}}(\zeta_f(g_{jk},g_{kl})) - \zeta_f(g_{ik}, g_{kl}) + \zeta_f(g_{ij}, g_{jl}) - \zeta_f(g_{ij}, g_{jk}) \\
    &\hspace{100pt}= 
    \bt(f_ig_{ij} f^{-1}_j,f_jg_{jk} f^{-1}_k,f_kg_{kl} f^{-1}_l) 
    - \rho_{f_i}(\bt(g_{ij},g_{jk},g_{kl})).
\end{aligned}
\end{equation}
One explicit expression for $\zeta_f$ that satisfies~\eqref{CondOnZeta} is
\begin{equation}\label{eq:zeta-f-app-defn}
   \zeta_f(g_{ij}, g_{jk}) = \rho_{f_i} \left[ \bt(g_{ij}, g_{jk}, f_k^{-1}) 
        - \bt(g_{ij},f_j^{-1},f_j g_{jk} f_k^{-1}) 
        + \bt(f_i^{-1}, f_i g_{ij} f_j^{-1}, f_j g_{jk} f_k^{-1})
        \right].
\end{equation}
Furthermore, $\zeta_f$ composes up to coboundaries under successive gauge transformations: 
\begin{equation}\label{eq:zeta-composition}
    \rho_{h_i}[\zeta_f(g_{ij}, g_{jk})]
    + 
    \zeta_h(f_i g_{ij} f_j^{-1}, f_j g_{jk} f_k^{-1})
    - 
    \zeta_{hf}(g_{ij}, g_{jk}) 
    = 
    \rho_{h_i f_i}[\del_\rho \theta_{f,h}(g_{ij}, g_{jk})],
\end{equation}
where
\begin{align}
    \theta_{f,h}(g_{ij})
    &= 
    - 
    \bt(g_{ij}, f_j^{-1}, h_j^{-1})
    +
    \bt(f_i^{-1}, f_i g_{ij} f_j^{-1}, h_j^{-1})\\
    &\qquad 
    -
    \bt(f_i^{-1}, h_i^{-1}, (h_i f_i) \, g_{ij}\, (h_j f_j)^{-1}),
    \\ 
    \del_\rho\theta_{f,h} (g_{ij}, g_{jk})
    &= 
    \rho_{g_{ij}} \theta_{f,h}\theta(g_{jk})
    - 
    \theta_{f,h}(g_{ij}g_{jk})
    +
    \theta_{f,h}(g_{ij}).
\end{align}
\item The $\G$ gauge transformation with trivial $f$ and nontrivial $\la$ is
\begin{equation}\label{1FormGaugeTrans}
    g_{ij} \mapsto g_{ij},
    \qquad 
    a_{ijk} \mapsto a_{ijk} + \rho_{g_{ij}}(\la_{jk})
    -
    \la_{ik}
    +
    \la_{ij}
    .
\end{equation}
The combination ${\rho_{g_{ij}}(\la_{jk})
    -
    \la_{ik}
    +
    \la_{ij}}$ is invariant under ${\la_{ij} \mapsto \la_{ij} + \rho_{g_{ij}}(\al_j) - \al_i}$ when~\eqref{General 2Grp flatness Cond} is satisfied.
\end{enumerate}
A general $\G$-gauge transformation ${(f,\la) \equiv (f,0)\circ (1,\la)}$ is
\begin{equation}
    g_{ij} \mapsto f_i g_{ij} f^{-1}_j,
    \qquad 
    a_{ijk} \mapsto \rho_{f_{i}}(a_{ijk} + \rho_{g_{ij}}(\la_{jk})
    -
    \la_{ik}
    +
    \la_{ij}) 
    + \zeta_f(g_{ij},g_{jk}).
\end{equation}
Homotopy classes ${[X,B\G]}$ of maps ${X\to B\G}$ correspond to gauge equivalence classes of ${(g,a)}$.

The Euclidean path integral of general $\G$ gauge theory is
\begin{equation}\label{EuclidPathIntGen2GrpGaugeThy}
    \cZ[X] \propto \sum_{(g,a)\in[X,B\G]}\, \prod_{ijkl\in\La_3} \,\om(g_{ij}, g_{jk}, g_{kl}; a_{ijk}, a_{ijl}, a_{ikl})^{\si_{ijkl}},
\end{equation}
where ${\om\in Z^3(B\G,\Uone)}$ is the allowed twist and ${\si_{ijkl} = \pm1}$ is the orientation of the 3-simplex ${ijkl}$. We suppress the overall normalization, which will not be relevant below.
In what follows, we set ${\om = 1}$.

We now specialize to a particular spacetime triangulation $K$ formed by triangular prisms that are each subdivided into three 3-simplices. For example:
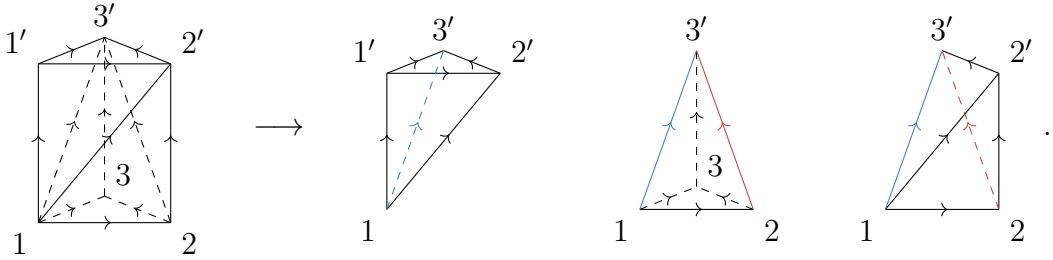
\begin{equation*}
    \begin{tikzpicture}[decoration={markings, mark=at position 0.55 with {\arrow{>}}}, scale=1.75]
        \draw[postaction=decorate] (0, 0) -- (1, 0) node[below right] {$2$};
        \draw[dashed, postaction=decorate] (0, 0) -- (0.5, 0.2) node[above right] {$3$};
        \draw[dashed, postaction=decorate] (1, 0) -- (0.5, 0.2);
        \draw[postaction=decorate] (0, 0) node[below left] {$1$} -- (0, 1.2) node[above left]{$1'$};
        \draw[postaction=decorate] (1, 0) -- (1, 1.2) node[above right, color=black] {$2'$};
        \draw[dashed, postaction=decorate] (0.5, 0.2) -- (0.5, 1.4) node[above, color=black] {$3'$};
        \draw[postaction=decorate] (0, 1.2) -- (1, 1.2);
        \draw[postaction=decorate] (0, 1.2) -- (0.5, 1.4);
        \draw[postaction=decorate] (1, 1.2) -- (0.5, 1.4);
        \draw[postaction=decorate] (0, 0) -- (1, 1.2);
        
        \draw[dashed, postaction=decorate] (0, 0) -- (0.5, 1.4);
        \draw[dashed, postaction=decorate] (1, 0) -- (0.5, 1.4);
    \end{tikzpicture}
    \quad 
    \longrightarrow
    \quad
    \begin{tikzpicture}[decoration={markings, mark=at position 0.55 with {\arrow{>}}}, scale=1.5]
        \draw[postaction=decorate] (0, 0) node[below left] {$1$} -- (0, 1.2) node[above left]{$1'$};
        \draw[postaction=decorate] (0, 1.2) -- (1, 1.2) node[above right] {$2'$};
        \draw[postaction=decorate] (0, 1.2) -- (0.5, 1.4) node[above] {$3'$};
        \draw[postaction=decorate] (1, 1.2) -- (0.5, 1.4);
        \draw[postaction=decorate] (0, 0) -- (1, 1.2);
        
        \draw[color=myblue, dashed, postaction=decorate] (0, 0) -- (0.5, 1.4);
    \end{tikzpicture}\hspace{2em}%
    \begin{tikzpicture}[decoration={markings, mark=at position 0.55 with {\arrow{>}}}, scale=1.5]
        \draw[postaction=decorate] (0, 0) node[below left] {$1$} -- (1, 0) node[below right] {$2$};
        \draw[postaction=decorate, dashed] (0, 0) -- (0.5, 0.2) node[above right] {$3$};
        \draw[dashed, postaction=decorate] (1, 0) -- (0.5, 0.2);
        \draw[dashed, postaction=decorate] (0.5, 0.2) -- (0.5, 1.4) node[above, color=black] {$3'$};
        \draw[color=myblue, postaction=decorate] (0, 0) -- (0.5, 1.4);
        \draw[color=myred, postaction=decorate] (1, 0) -- (0.5, 1.4);
    \end{tikzpicture}\hspace{2em}
    \begin{tikzpicture}[decoration={markings, mark=at position 0.55 with {\arrow{>}}}, scale=1.5]
        \draw[postaction=decorate] (0, 0) -- (1, 0) node[below right] {$2$};
        \draw[postaction=decorate] (1, 0) -- (1, 1.2) node[above right, color=black] {$2'$};
        \draw[postaction=decorate] (1, 1.2) -- (0.5, 1.4);
        \draw[postaction=decorate] (0, 0) node[below left] {$1$} -- (1, 1.2);
        
        \draw[postaction=decorate, color=myblue] (0, 0) -- (0.5, 1.4) node[color=black, above] {$3'$};
        \draw[color=myred, dashed, postaction=decorate] (1, 0) -- (0.5, 1.4);
    \end{tikzpicture}.
\end{equation*}
We interpret the vertical direction as the Euclidean time direction, which makes each time slice a triangular lattice $\La$. The flatness conditions in the time direction constrain gauge field configurations at different time slices to differ by $\G$ gauge transformations. For example, the timelike flatness conditions associated to the above triangular prism are
\begin{equation}
\begin{gathered}
    g_{11'}g_{1'2'} = g_{12'},
    \qquad
    g_{12}g_{22'} = g_{12'},
    \qquad
    g_{11'}g_{1'3'} = g_{13'},
    \\
    g_{13}g_{33'} = g_{13'},
    \qquad
    g_{22'}g_{2'3'} = g_{23'},
    \qquad
    g_{23}g_{33'} = g_{23'},
    \\
    (\dd_g a)_{11'2'3'} = 
    \bt(g_{11'}, g_{1'2'}, g_{2'3'}),
    \qquad
    (\dd_g a)_{1233'} = 
    \bt(g_{12}, g_{23}, g_{33'})\\
    (\dd_g a)_{122'3'} = 
    \bt(g_{12}, g_{22'}, g_{2'3'}).
\end{gathered}
\end{equation}
These constrain each spatial component $g_{i'j'}$ and $a_{i'j'k'}$ to equal $g_{ij}$ and $a_{ijk}$, respectively, up to the $\G$ gauge transformation with $\{f_i=g_{ii'}^{-1}\}$ and $\{\la_{ij}=a_{ii'j'}-a_{ijj'}\}$.

Now take ${X = Y\times S^1}$, where $Y$ is a closed spatial surface triangulated by $\La$.
The Hilbert space on $Y$ of untwisted $\G$ gauge theory is ${\scrH_\G^{(\rho,\bt)} = \C^{|[Y,B\G]|}}$, where ${|[Y,B\G]|}$ is the number of homotopy classes ${[Y,B\G]}$.
This Hilbert space is equivalent to a subspace of the larger Hilbert space
\begin{equation}\label{2GrpQDHilb}
    \scrH = \bigotimes_{ij\in\La_1} \C[G]
    \otimes
    \bigotimes_{ijk\in\La_2} \C[A].
\end{equation}
Indeed, this larger space admits a natural basis formed by ${\bigotimes_{ij\in\La_1} \ket{g_{ij}}\bigotimes_{ijk\in\La_2} \ket{a_{ijk}}\equiv \ket{\{g_{ij}\}, \{a_{ijk}\}}}$. (Here and below, ${ij}$ and ${ijk}$ refer only to links and plaquettes of $\La$.) Consider the subspace
\begin{equation}\label{FlatHilb}
    \scrH_{\text{flat}} = \mathrm{span}_\C\{ \ket{\{g_{ij}\}, \{a_{ijk}\}} \mid g_{ij} g_{jk} = g_{ik}\text{ for all }ijk\in\La_2\}\subset \scrH.
\end{equation}
States in $\scrH_\G^{(\rho,\bt)}$ correspond to states in $\scrH_{\text{flat}}$ invariant under generic $\G$ gauge transformations
\begin{equation}
    \ket{\{g_{ij}\}, \{a_{ijk}\}}
    \mapsto
    \ket{\{f_i g_{ij} f^{-1}_j\}, \{\rho_{f_{i}}(a_{ijk} + \rho_{g_{ij}}(\la_{jk})
    -
    \la_{ik}
    +
    \la_{ij}) 
    + \zeta_f(g_{ij},g_{jk})\}}.
\end{equation}
We now construct Hamiltonians acting on $\scrH$ whose ground-state space is equivalent to $\scrH_\G^{(\rho,\bt)}$.
We specialize to the 2-groups considered in the main text; constructing the Hamiltonian for the most general $\G$ is a straightforward generalization.

\subsection{\texorpdfstring{$G$ quantum double model}{\textit{G}-quantum double model}}\label{app:group-double-path-integral}

As a warm-up, we first set ${A = 1}$, in which case $\G$ gauge theory becomes ordinary $G$ gauge theory.
The full Hilbert space~\eqref{2GrpQDHilb} reduces to ${\scrH = \bigotimes_{ij\in\La_1} \C[G]}$.
Let us define the operators
\begin{align}
     A_{i} &\equiv \frac{1}{|G|} \sum_{h\in G} \overleftarrow{X}_{\trivertex{w}_i}^{(h)} 
    \overleftarrow{X}_{\trivertex{sw}_i}^{(h)} 
    \overleftarrow{X}_{\trivertex{se}_i}^{(h)} 
    \overrightarrow{X}_{\trivertex{e}_i}^{(h)} 
    \overrightarrow{X}_{\trivertex{ne}_i}^{(h)} 
    \overrightarrow{X}_{\trivertex{nw}_i}^{(h)} ,\label{OrdinaryQDmodelStarTerm}\\
    B_{ijk}
    &= \begin{cases}
        \frac1{|G|}\sum_{\Ga\in \Irr(G)} d_{\Ga}   \Tr\! \left[ 
        Z_{\triedge{s}}^{(\Ga)} 
        Z_{\triedge{nw}}^{(\Ga)} Z_{\triedge{ne}}^{(\bar{\Ga})}
        \right]\qquad
        & 
        \qquad ijk = \tricorner{}\,,
        \\[1em]
                 \frac1{|G|}\sum_{\Ga\in \Irr(G)} d_{\Ga}   \Tr\! \left[ 
        Z_{\triedge[down]{ne}}^{(\Ga)}
        Z_{\triedge[down]{s}}^{(\Ga)}
        Z_{\triedge[down]{nw}}^{(\bar{\Ga})}
        \right]\qquad
        & 
        \qquad ijk = \tricorner[down]{}\,.\label{OrdinaryQDmodelPlaqTerm}
    \end{cases}
\end{align}
These are mutually commuting projectors. The projector $\prod_{ijk\in\La_2}B_{ijk}$ projects $\scrH$ to $\scrH_\text{flat}$, and the projector $\prod_{i\in\La_0}A_{i}$ further projects $\scrH_\text{flat}$ to $\scrH_G$.
Therefore, the commuting projector Hamiltonian
\begin{equation}
    H_G = -\sum_{i\in \La_0} A_i 
    -\sum_{ijk\in\La_2} B_{ijk}
\end{equation}
has a ground-state subspace equivalent to ${\scrH_G\cong \C^{|[Y,BG]|}}$. This Hamiltonian is Kitaev's quantum double model~\cite{K9707021}.

\subsection{Split 2-group quantum double model}\label{app:split-2group-double-path-integral}

We now take $\G$ to be the split 2-group determined by $(G,A,\rho)$. 
The full Hilbert space is~\eqref{2GrpQDHilb}.
The projectors~\eqref{OrdinaryQDmodelPlaqTerm} from the ordinary quantum double model still project into the $\scrH_\mathrm{flat}$ subspace~\eqref{FlatHilb}. The gauge transformations, however, are now
\begin{align}
   \ket{\{g_{ij}\}, \{a_{ijk}\}}
    &\mapsto
    \ket{\{f_i g_{ij} f^{-1}_j\}, \{\rho_{f_{i}}(a_{ijk}) \}},\label{split2Grp0FormGaugeTransfState}\\
    \ket{\{g_{ij}\}, \{a_{ijk}\}}
    &\mapsto
    \ket{\{g_{ij}\}, \{a_{ijk} + \rho_{g_{ij}}(\la_{jk})
    -
    \la_{ik}
    +
    \la_{ij}
    \}}.\label{split2Grp1FormGaugeTransfState}
\end{align}

We introduce two types of projectors to project into the $\G$ gauge invariant subspace: one for~\eqref{split2Grp0FormGaugeTransfState} and another for~\eqref{split2Grp1FormGaugeTransfState}.
The first projector is
\begin{equation}
     A^{(\rho)}_{i} \equiv \frac{1}{|G|} \sum_{h\in G} \overleftarrow{X}_{\trivertex{w}_i}^{(h)} 
    \overleftarrow{X}_{\trivertex{sw}_i}^{(h)} 
    \overleftarrow{X}_{\trivertex{se}_i}^{(h)} 
    \overrightarrow{X}_{\trivertex{e}_i}^{(h)} 
    \overrightarrow{X}_{\trivertex{ne}_i}^{(h)} 
    \overrightarrow{X}_{\trivertex{nw}_i}^{(h)} 
    P_{ijk}^{(\rho_h)}
    P_{ijl}^{(\rho_h)}.
\end{equation}
Its action on a state is depicted as
\begin{equation}
\begin{tikzpicture}[decoration={markings, mark=at position 0.55 with {\arrow{>}}}, scale=2.5]
    \coordinate (c) at (0,0);
    \coordinate (e) at (1,0);
    \coordinate (w) at (-1,0);
    \coordinate (nw) at (-0.5,{sqrt(3)/2});
    \coordinate (ne) at (0.5,{sqrt(3)/2});
    \coordinate (sw) at (-0.5,-{sqrt(3)/2});
    \coordinate (se) at (0.5,-{sqrt(3)/2});
    
    \filldraw[lightgray!20!white] (c) -- (nw) -- (ne);
    \filldraw[lightgray!20!white] (c) -- (ne) -- (e);
    
    \node () at (0, 0.57735) {\tiny $a_{l n o}$};
    \node () at (0.5, 0.288675) {\tiny $a_{l m o}$};
    
    \draw[postaction=decorate, color=lightgray] (c) -- node[anchor=mid, color=black] {\footnotesize $g_{l m}$} (e);
    \draw[postaction=decorate, color=lightgray] (c) -- node[below, color=black, anchor=mid] {\footnotesize$g_{l n}$} (nw);
    \draw[postaction=decorate, color=lightgray] (c) -- node[below, color=black, anchor=mid] {\footnotesize$g_{l o}$} (ne);
    \draw[postaction=decorate, color=lightgray] (sw) -- node[below, color=black, anchor=mid] {\footnotesize$g_{il}$} (c);
    \draw[postaction=decorate, color=lightgray] (se) -- node[below, color=black, anchor=mid] {\footnotesize$g_{jl}$} (c);
    \draw[postaction=decorate, color=lightgray] (w) -- node[below, color=black, anchor=mid] {\footnotesize$g_{kl}$} (c) node[anchor=mid, color=black] {$l$};
\end{tikzpicture}
\quad
\xrightarrow{A^{(\rho)}_l}
\quad
    \frac{1}{|G|}
    \sum_{h\in G} 
\begin{tikzpicture}[decoration={markings, mark=at position 0.55 with {\arrow{>}}}, scale=2.5]
    \coordinate (c) at (0,0);
    \coordinate (e) at (1,0);
    \coordinate (w) at (-1,0);
    \coordinate (nw) at (-0.5,{sqrt(3)/2});
    \coordinate (ne) at (0.5,{sqrt(3)/2});
    \coordinate (sw) at (-0.5,-{sqrt(3)/2});
    \coordinate (se) at (0.5,-{sqrt(3)/2});
    
    \filldraw[lightgray!20!white] (c) -- (nw) -- (ne);
    \filldraw[lightgray!20!white] (c) -- (ne) -- (e);
    
    \node[color=myblue] () at (0, 0.57735) {\tiny $\rho_h({\color{black} a_{l n o}})$};
    \node[color=myblue] () at (0.5, 0.288675) {\tiny $\rho_{h}({\color{black}a_{l m o}})$};
    
    \draw[postaction=decorate, color=lightgray] (c) -- node[below, color=black] {\footnotesize ${\color{myblue}h}g_{l m}$} (e);
    \draw[postaction=decorate, color=lightgray] (c) -- node[below, color=black, anchor=mid] {\footnotesize${\color{myblue}h}g_{l n}$} (nw);
    \draw[postaction=decorate, color=lightgray] (c) -- node[below, color=black, anchor=mid] {\footnotesize${\color{myblue}h}g_{l o}$} (ne);
    \draw[postaction=decorate, color=lightgray] (sw) -- node[below, color=black, anchor=mid] {\footnotesize$g_{il}{\color{myblue}h^{-1}}$} (c);
    \draw[postaction=decorate, color=lightgray] (se) -- node[below, color=black, anchor=mid] {\footnotesize$g_{jl}{\color{myblue}h^{-1}}$} (c);
    \draw[postaction=decorate, color=lightgray] (w) -- node[below, color=black, anchor=mid] {\footnotesize$g_{kl}{\color{myblue}h^{-1}}$} (c) node[anchor=mid, color=black] {$l$};
\end{tikzpicture}.
\end{equation}
The automorphism $\rho_h$ in $A_l^{(\rho)}$ acts only on $A$-qudits whose plaquette starts with $l$.
Note that $A^{(\rho)}_i$ simplifies to~\eqref{OrdinaryQDmodelStarTerm} when every $\rho_h$ is the identity $A$-automorphism.
The operators $\{A^{(\rho)}_i\}_{i\in \La_0}$ are mutually commuting projectors and generalizations of the star operators~\eqref{OrdinaryQDmodelStarTerm}.
The projector $\prod_{i\in\La_0} A^{(\rho)}_i$ projects onto the subspace invariant under~\eqref{split2Grp0FormGaugeTransfState}.
The second projector is
\begin{equation}\label{Split2grpCops}
    C^{(\rho)}_{ij} 
    =  
    \begin{cases}
        
    \frac{1}{|A|} \sum_{\la\in A} X_{\tricorner{}}^{(\,\rho_{g_{\triedge{s}}} (\la)\,)} 
    X_{\tricorner[down]{}}^{(\la)}
     & \qquad
    ij = \trilink{b} 
    \\
   
    \frac{1}{|A|} \sum_{\la\in A} X_{\tricorner{}}^{(\la)}
     X_{\tricorner[down]{}}^{(\,\rho_{g_{\triedge[down]{ne}}} (\la)\,)} 
     &\qquad
     ij = \trilink{c} 
     \\
    
    \frac{1}{|A|} \sum_{\la\in A} X_{\tricorner[up]{}}^{(\la)} 
    X_{\tricorner[down]{}}^{(\la)}
    \vphantom{X_{\tricorner[down]{}}^{\big(\rho_{g_{\tricorner[down]{left}}^{-1} g_{\tricorner[down]{apex}}}}}
    &\qquad
    ij = \trilink{a} 
    \end{cases}
\end{equation}
We depict the action of these projectors on states as follows:
\begin{gather}
    \begin{tikzpicture}[decoration={markings, mark=at position 0.55 with {\arrow{>}}}, scale=2.5]
        \draw[postaction=decorate,color=lightgray] (0, 0) -- (0.5, 0.866025) node[midway, left, color=lightgray] {\footnotesize $g_{ik}$};
        \draw[postaction=decorate,color=lightgray] (0, 0) -- (1, 0) node[midway, below, color=black] {\footnotesize $g_{ij}$}; 
        \draw[postaction=decorate,color=lightgray] (1, 0) -- (0.5, 0.866025);
        \draw[postaction=decorate,color=lightgray] (0.5, 0.866025) -- (1.5, 0.866025) node[midway, above, color=lightgray] {\footnotesize $g_{kl}$};
        \draw[postaction=decorate,color=lightgray] (1, 0) -- (1.5, 0.866025) node[midway, right, color=lightgray] {\footnotesize $g_{jl}$};
        \node () at (0.5, 0.288675) {\footnotesize $a_{ijk}$};
        \node () at (1,  0.57735) {\footnotesize $a_{jkl}$};
    \end{tikzpicture}   
    \xrightarrow{C^{(\rho)}_{jk}}\frac{1}{|A|} \sum_{\la\in A}\,
    \begin{tikzpicture}[decoration={markings, mark=at position 0.55 with {\arrow{>}}}, scale=2.5]
        \draw[postaction=decorate,color=lightgray] (0, 0) -- (0.5, 0.866025) node[midway, left, color=lightgray] {\footnotesize $g_{ik}$};
        \draw[postaction=decorate,color=lightgray] (0, 0) -- (1, 0) node[midway, below, color=black] {\footnotesize $g_{ij}$}; 
        \draw[postaction=decorate,color=lightgray] (1, 0) -- (0.5, 0.866025);
        \draw[postaction=decorate,color=lightgray] (0.5, 0.866025) -- (1.5, 0.866025) node[midway, above, color=lightgray] {\footnotesize $g_{kl}$};
        \draw[postaction=decorate,color=lightgray] (1, 0) -- (1.5, 0.866025) node[midway, right, color=lightgray] {\footnotesize $g_{jl}$};
        \node[align=center] () at (0.5, 0.288675) {\footnotesize $a_{ijk} $\\[-1em] \footnotesize ${\color{mygreen}\vphantom{}+\rho_{g_{ij}}(\la)}$};
        \node () at (1,  0.57735) {\footnotesize $a_{jkl} {\color{mygreen}\vphantom{}+\la}$};
    \end{tikzpicture} \\ 
    \begin{tikzpicture}[decoration={markings, mark=at position 0.55 with {\arrow{>}}}, scale=2.5]
        \draw[postaction=decorate,color=lightgray] (0, 0) -- (0.5, 0.866025) node[midway, left, color=lightgray] {\footnotesize $g_{jl}$};
        \draw[postaction=decorate,color=lightgray] (0, 0) -- (1, 0); 
        \draw[postaction=decorate,color=lightgray] (1, 0) -- (0.5, 0.866025) node[midway, right, color=lightgray] {\footnotesize $g_{kl}$};
        \draw[postaction=decorate,color=lightgray] (0.5, -0.866025) -- (0,0) node[midway, left, color=black] {\footnotesize $g_{ij}$};
        \draw[postaction=decorate,color=lightgray] (0.5, -0.866025) -- (1,0) node[midway, right, color=lightgray] {\footnotesize $g_{ik}$};
        \node () at (0.5, 0.288675) {\footnotesize $a_{jkl}$};
        \node () at (0.5, -0.288675) {\footnotesize $a_{ijk}$};
    \end{tikzpicture} 
    \xrightarrow{C^{(\rho)}_{jk}}\frac{1}{|A|} \sum_{\la\in A}\,
    \begin{tikzpicture}[decoration={markings, mark=at position 0.55 with {\arrow{>}}}, scale=2.5]
        \draw[postaction=decorate,color=lightgray] (0, 0) -- (0.5, 0.866025) node[midway, left, color=lightgray] {\footnotesize $g_{jl}$};
        \draw[postaction=decorate,color=lightgray] (0, 0) -- (1, 0); 
        \draw[postaction=decorate,color=lightgray] (1, 0) -- (0.5, 0.866025) node[midway, right, color=lightgray] {\footnotesize $g_{kl}$};
        \draw[postaction=decorate,color=lightgray] (0.5, -0.866025) -- (0,0) node[midway, left, color=black] {\footnotesize $g_{ij}$};
        \draw[postaction=decorate,color=lightgray] (0.5, -0.866025) -- (1,0) node[midway, right, color=lightgray] {\footnotesize $g_{ik}$};
        \node () at (0.5, 0.288675) {\footnotesize $a_{jkl} {\color{mygreen}\vphantom{}+\la}$};
        \node[align=center] () at (0.5, -0.288675) {\footnotesize $a_{ijk}$\\[-1em] \footnotesize ${\color{mygreen}\vphantom{}+\rho_{g_{ij}}(\la)}$};
    \end{tikzpicture} 
    \\ 
    \begin{tikzpicture}[decoration={markings, mark=at position 0.55 with {\arrow{>}}}, scale=2.5]
        \draw[postaction=decorate,color=lightgray] (0, 0) -- (0.5, 0.866025);
        \draw[postaction=decorate,color=lightgray] (0, 0) -- (1, 0) node[midway, below, color=lightgray] {\footnotesize $g_{ij}$}; 
        \draw[postaction=decorate,color=lightgray] (1, 0) -- (0.5, 0.866025) node[midway, right, color=lightgray] {\footnotesize $g_{ik}$};
        \draw[postaction=decorate,color=lightgray] (-0.5, 0.866025) -- (0.5, 0.866025) node[midway, above, color=lightgray] {\footnotesize $g_{kl}$};
        \draw[postaction=decorate,color=lightgray] (0, 0) -- (-0.5, 0.866025) node[midway, left, color=lightgray] {\footnotesize $g_{jl}$};
        \node () at (0.5, 0.288675) {\footnotesize $a_{ijk}$};
        \node () at (0,  0.57735) {\footnotesize $a_{jkl}$};
    \end{tikzpicture}   
    \xrightarrow{C^{(\rho)}_{jk}}\frac{1}{|A|} \sum_{\la\in A}\,
    \begin{tikzpicture}[decoration={markings, mark=at position 0.55 with {\arrow{>}}}, scale=2.5]
        \draw[postaction=decorate,color=lightgray] (0, 0) -- (0.5, 0.866025);
        \draw[postaction=decorate,color=lightgray] (0, 0) -- (1, 0) node[midway, below, color=lightgray] {\footnotesize $g_{ij}$}; 
        \draw[postaction=decorate,color=lightgray] (1, 0) -- (0.5, 0.866025) node[midway, right, color=lightgray] {\footnotesize $g_{ik}$};
        \draw[postaction=decorate,color=lightgray] (-0.5, 0.866025) -- (0.5, 0.866025) node[midway, above, color=lightgray] {\footnotesize $g_{kl}$};
        \draw[postaction=decorate,color=lightgray] (0, 0) -- (-0.5, 0.866025) node[midway, left, color=lightgray] {\footnotesize $g_{jl}$};
        \node () at (0.5, 0.288675) {\footnotesize $a_{ijk} {\color{mygreen}\vphantom{}+ \la}$};
        \node () at (0,  0.57735) {\footnotesize $a_{jkl} {\color{mygreen}\vphantom{}+ \la}$};
    \end{tikzpicture}   
\end{gather}
The operators $\{C^{(\rho)}_{ij}\}_{ij\in \La_1}$ are mutually commuting projectors, and the projector $\prod_{ij\in\La_1} C^{(\rho)}_{ij}$ projects onto the subspace invariant under~\eqref{split2Grp1FormGaugeTransfState}.

Using these three projectors, we define the Hamiltonian
\begin{equation}\label{appendixsplit2grpQD}
    H_{\G}^{(\rho)} = -\sum_{i\in \La_0} A^{(\rho)}_i 
    -\sum_{ijk\in\La_2} B_{ijk}
    -\sum_{ij\in \La_1} C^{(\rho)}_{ij} .
\end{equation}
Since ${\{A^{(\rho)}_i\}_{i\in\La_0}\cup \{B_{ijk}\}_{ijk\in\La_2} \cup \{C^{(\rho)}_{ij}\}_{ij\in\La_1}}$ are mutually commuting projectors, this is a commuting projector Hamiltonian.
The ground-state subspace of this Hamiltonian is equivalent to $\scrH_\G^{(\rho)}$ for split 2-group gauge theory. Hence, this Hamiltonian is the generalization of Kitaev's quantum double model for split 2-groups.

\subsubsection{Torus ground-state degeneracy}

The ground-state degeneracy can be evaluated explicitly when ${Y=T^2}$. 
It suffices to do so using the triangulation of $T^2$:\footnote{Strictly speaking, this is not a triangulation (i.e., a simplicial complex). Instead, it is a $\Del$-complex. This distinction does not matter for our purposes, and we will abuse terminology throughout this appendix by referring to $\Del$-complexes as triangulations.}
\begin{equation}\label{onesimplextorustrianglulation}
\begin{tikzpicture}[decoration={markings, mark=at position 0.55 with {\arrow{>}}}, scale=2.5]
    \draw[postaction=decorate,color=black] (0, 0) node[left, color=black] {$i$} -- (0.5, 0.866025);
    \draw[postaction=decorate,color=black] (0, 0) -- (1, 0) node[right, color=black] {$j$}; 
    \draw[postaction=decorate,color=black] (1, 0) -- (0.5, 0.866025);
    \draw[postaction=decorate,color=black] (-0.5, 0.866025) -- (0.5, 0.866025) node[right, color=black] {$l$};
    \draw[postaction=decorate,color=black] (0, 0) -- (-0.5, 0.866025) node[left, color=black] {$k$};
    ;
\end{tikzpicture}
\end{equation}
where the sites ${i=j=k=l}$ and the links ${ij=kl}$, ${ik=jl}$. The full Hilbert space is spanned by the vectors $\ket{\{g_{x}, g_{y}, g_{il}\}, \{a_{\tricorner{}}, a_{\tricorner[down]{}}\}}$ where ${g_x = g_{ij} \equiv g_{kl}}$, ${g_y = g_{jl} \equiv g_{ik}}$, ${a_{\tricorner{}} = a_{ijl}}$ and ${a_{\tricorner[down]{}} = a_{ikl}}$. The projector $B_{\tricorner{}}B_{\tricorner[down]{}}$ enforces ${g_{il} = g_{x}g_{y} = g_{y}g_{x}}$ in the ground-state space. Thus, for this triangulation of $T^2$,
\begin{equation}\label{Hflatonesimptrosutriang}
    \scrH_\mathrm{flat} = \mathrm{span}_\C\{\ket{\{g_x,g_y\},\{a_{\tricorner{}}, a_{\tricorner[down]{}}\}}\mid (g_x,g_y)\in \Hom(\Z^2,G),~ a_{\tricorner{}}, a_{\tricorner[down]{}}\in A\},
\end{equation}
where ${\Hom(\Z^2,G) \cong \{(g_x,g_y)\in G^2 \mid g_xg_y = g_yg_x\}}$.

The ground-state subspace is the gauge-invariant subspace of $\scrH_\mathrm{flat}$.
The gauge transformations~\eqref{split2Grp1FormGaugeTransfState} with ${\la_x\equiv\la_{ij}=\la_{kl}}$, ${\la_y\equiv\la_{ik}=\la_{jl}}$ are
\begin{equation}
    a_{\tricorner{}}
    \mapsto
    a_{\tricorner{}}
    +\rho_{g_x}(\la_y)-\la_{il}+\la_x,
    \qquad
    a_{\tricorner[down]{}}
    \mapsto
    a_{\tricorner[down]{}}
    +\rho_{g_y}(\la_x)-\la_{il}+\la_y.
\end{equation}
The projector $C^{(\rho)}_{ij}C^{(\rho)}_{jl}C^{(\rho)}_{il}$ averages over these.
Gauge fixing using ${\la_x=\la_y = 0}$ and ${\la_{il} = a_{\tricorner[down]{}}}$ causes ${\{a_{\tricorner{}}, a_{\tricorner[down]{}}\} \mapsto \{a,0\}}$ with ${a \equiv a_{\tricorner{}} - a_{\tricorner[down]{}}}$. Under the remaining gauge transformation, 
\begin{equation}
    a
    \mapsto
    a
    +(\mathrm{id}_A-\rho_{g_y})(\la_x)
    +(\rho_{g_x}-\mathrm{id}_A)(\la_y).
\end{equation}
Therefore, for a fixed commuting pair ${\{g_x,g_y\}}$, the gauge-inequivalent $A$-qudit
configurations are labeled by
\begin{equation}\label{split2groupcaAgxgy}
    \cA_{g_x,g_y}
    \equiv
    \frac{A}{
        \Im(\rho_{g_x}-\mathrm{id}_A)
        +\Im(\rho_{g_y}-\mathrm{id}_A)
    },
\end{equation}
where ${\Im(\rho_{g_x}-\mathrm{id}_A) + \Im(\rho_{g_y}-\mathrm{id}_A) 
=
\{a_1-\rho_{g_y}(a_1) + \rho_{g_x}(a_2)-a_2 
\mid a_1,a_2\in A\}\leq A}$. Thus, the subspace of $\scrH_\mathrm{flat}$ invariant under~\eqref{split2Grp1FormGaugeTransfState} is spanned by vectors $\ket{g_x,g_y,[a]_{g_x,g_y}}$ with commuting elements ${g_x,g_y\in G}$ and ${[a]_{g_x,g_y}\in \cA_{g_x,g_y}}$. The triples ${(g_x,g_y,[a]_{g_x,g_y})}$ form the set 
\begin{equation}
    \cA \equiv \bigsqcup_{(g_x,g_y)\in \Hom(\Z^2,G)}\cA_{g_x,g_y}.
\end{equation}
It remains to impose the $G$ gauge invariance~\eqref{split2Grp0FormGaugeTransfState}. Since the triangulation
has a single site, the gauge transformation is 
\begin{equation}\label{Ggaugetransformationsplit2grp4}
    (g_x,g_y,[a]_{g_x,g_y})
    \mapsto
    \big(hg_xh^{-1},hg_yh^{-1},[\rho_h(a)]_{hg_x h^{-1},h g_yh^{-1}}\big).
\end{equation}
This transformation defines an action of $G$ on $\cA$. Thus, the ground-state degeneracy of~\eqref{appendixsplit2grpQD} on a spatial torus is
\begin{equation}\label{split2grpGSDAmodG}
    \mathrm{GSD}_{T^2} = \left| \cA/G \right|,
\end{equation}
where ${\cA/G}$ is the set of orbits under this $G$ action on $\cA$. The ground-state subspace is spanned by the vectors ${\sum_{h\in G}\ket{hg_xh^{-1},hg_yh^{-1},[\rho_h(a)]_{hg_x h^{-1},h g_yh^{-1}}}}$. 

The expression~\eqref{split2grpGSDAmodG} admits a more transparent
form. To find it, we first perform the Fourier transformation
\begin{equation}\label{split2grpAFT}
    \ket{g_x,g_y,\eta_{g_x,g_y}}
    \equiv
    \frac{1}{\sqrt{|\cA_{g_x,g_y}|}}
    \sum_{[a]_{g_x,g_y}\in\cA_{g_x,g_y}}
    \eta_{g_x,g_y}\big([a]_{g_x,g_y}\big)
    \ket{g_x,g_y,[a]_{g_x,g_y}},
\end{equation}
where ${\eta_{g_x,g_y} \in \h{\cA}_{g_x,g_y} \equiv \Hom(\cA_{g_x,g_y},\Uone)}$.
The $G$ gauge transformation~\eqref{Ggaugetransformationsplit2grp4} acts on the vector~\eqref{split2grpAFT} as
\begin{equation}\label{Gactioncharacterbasissplit}
    \ket{g_x,g_y,\eta_{g_x,g_y}}
    \mapsto
    \ket{
        hg_xh^{-1},
        hg_yh^{-1},
        h\triangleright\eta_{g_x,g_y}
    },
\end{equation}
where ${(h\triangleright\eta_{g_x,g_y})([a]_{hg_xh^{-1},hg_yh^{-1}})\equiv\eta_{g_x,g_y}([\rho_{h^{-1}}(a)]_{g_x,g_y})}$.
Defining ${\cA^\vee \equiv \bigsqcup_{(g_x,g_y)\in\Hom(\Z^2,G)} \h{\cA}_{g_x,g_y}}$,
the ground-state degeneracy can be written as ${\mathrm{GSD}_{T^2} = |\cA^\vee/G|}$.

We next use the fact that a character of $\cA_{g_x,g_y}$ is equivalently a character of $A$ that is trivial on the subgroup ${\Im(\rho_{g_x}-\mathrm{id}_A)+\Im(\rho_{g_y}-\mathrm{id}_A)}$. 
Therefore, there is a bijection
\begin{equation}
    \h{\cA}_{g_x,g_y}
    \cong
    \left\{
        \chi\in\hat A
        \mid
        \chi\circ\rho_{g_x}=\chi,\quad
        \chi\circ\rho_{g_y}=\chi
    \right\}
    =
    \left\{
        \chi\in\hat A
        \mid
        g_x,g_y\in G_\chi
    \right\},
\end{equation}
where ${G_\chi = \{g\in G \mid \rho_g\triangleright\chi=\chi\}}$.
This implies that
\begin{equation}
    \cA^\vee
    \cong
    \left\{
        (g_x,g_y,\chi)
        \mid
        \chi\in\h{A},~
        (g_x,g_y)\in\Hom(\Z^2,G_\chi)
    \right\}.
\end{equation}
The $G$-action~\eqref{Gactioncharacterbasissplit} induces the $G$-action ${(g_x,g_y,\chi)
\mapsto
(hg_xh^{-1},hg_yh^{-1},\rho_h \triangleright \chi)}$. Choose one representative $\chi$ from each orbit ${[\chi]\in\h A/G}$. The
transformations preserving this representative form its stabilizer $G_\chi$ and act
on $\Hom(\Z^2,G_\chi)$ by simultaneous conjugation. Hence,
\begin{equation}
    \cA^\vee/G
    \cong
    \bigsqcup_{[\chi]\in\h A/G}
    \Hom(\Z^2,G_\chi)/G_\chi,
\end{equation}
and therefore
\begin{equation}
    \mathrm{GSD}_{T^2}
    =
    \sum_{[\chi]\in\h A/G}
    \left|
        \Hom(\Z^2,G_\chi)/G_\chi
    \right|.
\end{equation}
Thus, each $G$-orbit of characters $\chi\in\h A$ contributes the torus
ground-state degeneracy of the ordinary $G_\chi$ quantum double model.

\subsubsection{Sphere ground-state degeneracy}

To illustrate the dependence of the ground-state degeneracy on the spatial topology, we now
take ${Y=S^2}$. We consider the triangulation of $S^2$ formed by two plaquettes
\begin{equation}\label{twosimplexspheretriangulation}
\begin{tikzpicture}[
    decoration={markings, mark=at position 0.55 with {\arrow{>}}},
    scale=1.8
]
    \coordinate (i) at (0,0);
    \coordinate (j) at (1.2,0);
    \coordinate (k) at (0.6,1.039);

    \draw[postaction=decorate] (i) node[left] {$i$} -- (j) node[right] {$j$};
    \draw[postaction=decorate] (i) -- (k) node[above] {$k$};
    \draw[postaction=decorate] (j) -- (k);
    \node at (0.6,0.38) {$\tricorner{}_+$};

    \coordinate (l) at (2.4,0);
    \coordinate (m) at (3.6,0);
    \coordinate (n) at (3.0,1.039);

    \draw[postaction=decorate] (l) node[left] {$l$} -- (m) node[right] {$m$};
    \draw[postaction=decorate] (l) -- (n) node[above] {$n$};
    \draw[postaction=decorate] (m) -- (n);
    \node at (3.0,0.38) {$\tricorner{}_-$};
\end{tikzpicture}
\end{equation}
where the $0$-simplices and $1$-simplices are identified as
\begin{equation}
    i=l,\qquad j=m,\qquad k=n,
    \qquad
    ij=lm,\qquad ik=ln,\qquad jk=mn,
\end{equation}
and the two $2$-simplices have opposite orientations. The full Hilbert space is spanned by vectors $\ket{g_{ij}, g_{jk}, g_{ik}, a_{\tricorner{}_+}, a_{\tricorner{}_-}}$. The projector $B_{ijk}$ imposes
${g_{ij}g_{jk}=g_{ik}}$, and $\scrH_\mathrm{flat}$ is spanned by the vectors $\ket{g_{ij}, g_{jk}, a_{\tricorner{}_+}, a_{\tricorner{}_-}}$.

The gauge transformation~\eqref{split2Grp1FormGaugeTransfState} acts identically on both $A$-qudits as
\begin{equation}
    a_{\tricorner{}_\pm}
    \mapsto
    a_{\tricorner{}_\pm}
    +\rho_{g_{ij}}(\la_{jk})-\la_{ik}+\la_{ij}.
\end{equation}
Gauge-inequivalent $A$-qudit configurations are labeled by ${a \equiv a_{\tricorner{}_+} - a_{\tricorner{}_-}}$, and the subspace of $\scrH_\mathrm{flat}$ invariant under~\eqref{split2Grp1FormGaugeTransfState} is spanned by vectors $\ket{g_{ij}, g_{jk}, a}$.
The $G$ gauge transformation~\eqref{split2Grp0FormGaugeTransfState} acts on these vectors as
\begin{equation}
    \ket{g_{ij}, g_{jk}, a} \mapsto \ket{f_i g_{ij} f_j^{-1}, f_j g_{jk} f_k^{-1}, \rho_{f_i}(a)}.
\end{equation}
Every $G$-qudit configuration is gauge-equivalent to the trivial one in $\scrH_\mathrm{flat}$. Indeed, the $G$ gauge transformation with ${f_i=1}$, ${f_j=g_{ij}}$, and ${f_k=g_{ik}}$ maps each $G$-qudit state to $1$ while leaving $a$ unchanged.
After this gauge
choice, the remaining basis states are $\ket{a}$ with ${a\in A}$.
The remaining $G$ gauge transformation, corresponding to ${f_i=f_j=f_k\equiv h}$, acts as
${\ket{a}\mapsto \ket{\rho_h(a)}}$. 
Hence the
ground-state subspace is spanned by vectors labeled by the $G$-orbits ${[a] = \{\rho_h(a)\mid h\in G\}\in A/G}$, and
\begin{equation}
    \mathrm{GSD}_{S^2}
    =
    |A/G|.
\end{equation}
Equivalently, after Fourier
transforming, the vectors spanning the ground-state space can be labeled by the $G$-orbits ${[\chi] = \{\rho_h\triangleright\chi\mid h\in G\}\in \h{A}/G}$, and
\begin{equation}
    \mathrm{GSD}_{S^2}
    =
    |\h{A}/G|.
\end{equation}

\subsection{Central 2-group quantum double model}\label{app:central-2group-double-path-integral}

We next take $\G$ to be the central 2-group determined by ${(G,A,\bt)}$. 
The full Hilbert space is again~\eqref{2GrpQDHilb}.
The projectors~\eqref{OrdinaryQDmodelPlaqTerm} from the ordinary quantum double model still project into the $\scrH_\mathrm{flat}$ subspace~\eqref{FlatHilb}. The gauge transformations, however, are now
\begin{align}
   \ket{\{g_{ij}\}, \{a_{ijk}\}}
    &\mapsto
    \ket{\{f_i g_{ij} f^{-1}_j\}, \{a_{ijk} 
    + \zeta_f(g_{ij},g_{jk})\}},\label{central2Grp0FormGaugeTransfState}\\
    \ket{\{g_{ij}\}, \{a_{ijk}\}}
    &\mapsto
    \ket{\{g_{ij} \}, \{a_{ijk} + \la_{jk}
    -
    \la_{ik}
    +
    \la_{ij}
    \}},\label{central2Grp1FormGaugeTransfState}
\end{align}
where $\zeta_f$ satisfies
\begin{equation}
   \zeta_f(g_{ij}, g_{jk}) = \bt(g_{ij}, g_{jk}, f_k^{-1}) 
        - \bt(g_{ij},f_j^{-1},f_j g_{jk} f_k^{-1}) 
        + \bt(f_i^{-1}, f_i g_{ij} f_j^{-1}, f_j g_{jk} f_k^{-1}).
\end{equation}

The operator that projects onto the subspace invariant under~\eqref{central2Grp1FormGaugeTransfState} is given by~\eqref{Split2grpCops} with each $\rho_h$ the trivial $A$-automorphism.
We denote it by ${C_{ij}^{(\mathrm{id})}\equiv C_{ij}}$.
The operator that projects onto the subspace invariant under~\eqref{central2Grp0FormGaugeTransfState} is a generalization of the star operator of the ordinary quantum double model. Let us define the operator $A_i^{(\bt)}$ which acts on a state as
\begin{equation}\label{eq:a_i-central-2group}
\begin{tikzpicture}[decoration={markings, mark=at position 0.55 with {\arrow{>}}}, scale=3.2]
    \coordinate (c) at (0,0);
    \coordinate (e) at (1,0);
    \coordinate (w) at (-1,0);
    \coordinate (nw) at (-0.5,{sqrt(3)/2});
    \coordinate (ne) at (0.5,{sqrt(3)/2});
    \coordinate (sw) at (-0.5,-{sqrt(3)/2});
    \coordinate (se) at (0.5,-{sqrt(3)/2});

    \coordinate (t1) at ($(c)!0.333!(e) + (c)!0.333!(ne) - (c)$);   
    \coordinate (t2) at ($(c)!0.333!(ne) + (c)!0.333!(nw) - (c)$);  
    \coordinate (t3) at ($(c)!0.333!(nw) + (c)!0.333!(w) - (c)$);   
    \coordinate (t4) at ($(c)!0.333!(w) + (c)!0.333!(sw) - (c)$);   
    \coordinate (t5) at ($(c)!0.333!(sw) + (c)!0.333!(se) - (c)$);  
    \coordinate (t6) at ($(c)!0.333!(se) + (c)!0.333!(e) - (c)$);   
    
    \node at (t1) {\tiny $a_{l m o}$};
    \node at (t2) {\tiny $a_{l n o}$};
    \node at (t3) {\tiny $a_{kl n}$};
    \node at (t4) {\tiny $a_{ikl}$};
    \node at (t5) {\tiny $a_{ijl}$};
    \node at (t6) {\tiny $a_{jl m}$};
    
    \draw[postaction=decorate, color=lightgray] (c) -- node[anchor=mid, color=black] {\footnotesize $g_{l m}$} (e);
    \draw[postaction=decorate, color=lightgray] (c) -- node[anchor=mid, color=black] {\footnotesize $g_{l n}$} (nw);
    \draw[postaction=decorate, color=lightgray] (c) -- node[anchor=mid, color=black] {\footnotesize $g_{l o}$} (ne);
    \draw[postaction=decorate, color=lightgray] (sw) -- node[anchor=mid, color=black] {\footnotesize $g_{il}$} (c);
    \draw[postaction=decorate, color=lightgray] (se) -- node[anchor=mid, color=black] {\footnotesize $g_{jl}$} (c);
    \draw[postaction=decorate, color=lightgray] (w) -- node[anchor=mid, color=black] {\footnotesize $g_{k l}$} (c);
\end{tikzpicture}
\quad
\xrightarrow{A^{(\bt)}_l}
\quad
\frac{1}{|G|}
\sum_{h\in G}
\begin{tikzpicture}[decoration={markings, mark=at position 0.55 with {\arrow{>}}}, scale=3.2]
    \coordinate (c) at (0,0);
    \coordinate (e) at (1,0);
    \coordinate (w) at (-1,0);
    \coordinate (nw) at (-0.5,{sqrt(3)/2});
    \coordinate (ne) at (0.5,{sqrt(3)/2});
    \coordinate (sw) at (-0.5,-{sqrt(3)/2});
    \coordinate (se) at (0.5,-{sqrt(3)/2});

    \coordinate (t1) at ($(c)!0.333!(e) + (c)!0.333!(ne) - (c)$);   
    \coordinate (t2) at ($(c)!0.333!(ne) + (c)!0.333!(nw) - (c)$);  
    \coordinate (t3) at ($(c)!0.333!(nw) + (c)!0.333!(w) - (c)$);   
    \coordinate (t4) at ($(c)!0.333!(w) + (c)!0.333!(sw) - (c)$);   
    \coordinate (t5) at ($(c)!0.333!(sw) + (c)!0.333!(se) - (c)$);  
    \coordinate (t6) at ($(c)!0.333!(se) + (c)!0.333!(e) - (c)$);   
    
    \node[align=center, xshift=1em, yshift=-0.5em] at (t1) {\tiny $a_{l m o}$ \\[-1.2em] \tiny ${\color{myblue} \vphantom{} + \bt(h^{-1},h g_{l m}, g_{m o})}$};
    \node[align=center, yshift=1em] at (t2) {\tiny $a_{l n o}$ \\[-1.2em] \tiny $ {\color{myblue} \vphantom{} + \bt(h^{-1},h g_{l n}, g_{n o})}$};
    \node[align=center, xshift=-1em, yshift=-0.5em] at (t3) {\tiny $a_{kl n}$ \\[-1.2em] \tiny ${\color{myblue} \vphantom{} -\bt(g_{k l}, h^{-1}, h g_{l n})}$};
    \node[align=center, xshift=-1em, yshift=0.5em] at (t4) {\tiny $a_{ikl}$ \\[-1.2em] \tiny ${\color{myblue} \vphantom{} + \bt(g_{i k}, g_{k l}, h^{-1})}$};
    \node[align=center, yshift=-1em] at (t5) {\tiny $a_{ijl}$ \\[-1.2em] \tiny ${\color{myblue} \vphantom{} + \bt(g_{ij}, g_{jl}, h^{-1})}$};
    \node[align=center, xshift=1em, yshift=0.5em] at (t6) {\tiny $a_{jl m}$\\[-1.2em] \tiny ${\color{myblue} \vphantom{} -\bt(g_{jl}, h^{-1}, h g_{l m})}$};
    
    \draw[postaction=decorate, color=lightgray] (c) -- node[anchor=mid, color=black] {\footnotesize ${\color{myblue}h} g_{l m}$} (e);
    \draw[postaction=decorate, color=lightgray] (c) -- node[anchor=mid, color=black] {\footnotesize ${\color{myblue}h}g_{l n}$} (nw);
    \draw[postaction=decorate, color=lightgray] (c) -- node[anchor=mid, color=black] {\footnotesize ${\color{myblue}h}g_{l o}$} (ne);
    \draw[postaction=decorate, color=lightgray] (sw) -- node[anchor=mid, color=black] {\footnotesize $g_{il}{\color{myblue}h^{-1}}$} (c);
    \draw[postaction=decorate, color=lightgray] (se) -- node[anchor=mid, color=black] {\footnotesize $g_{jl}{\color{myblue}h^{-1}}$} (c);
    \draw[postaction=decorate, color=lightgray] (w) -- node[anchor=mid, color=black] {\footnotesize $g_{k l}{\color{myblue}h^{-1}}$} (c);
\end{tikzpicture}
\end{equation}
The operators $\{A_i^{(\bt)}\}_{i\in\La_0}$ are generically not mutually commuting projectors. They become mutually commuting projectors, however, when acting on the subspace spanned by vectors $\ket{\psi}$ satisfying ${C_{ij}\ket{\psi} = B_{ijk}\ket{\psi} = \ket{\psi}}$.
Then, in this subspace, $\prod_{i}A^{(\bt)}_i$ is a projector that projects onto the subspace invariant under~\eqref{central2Grp0FormGaugeTransfState}.

Using these operators, we define the Hamiltonian
\begin{equation}\label{appendixcentral2grpQD}
    H^{(\bt)}_\G = -\frac12\sum_{i\in \La_0} (A^{(\bt)}_i + A^{(\bt)\dag}_i )
    -\sum_{ijk\in\La_2} B_{ijk}
    -\sum_{ij\in \La_1} C_{ij} .
\end{equation}
The operators $\{A^{(\bt)}_i\}_{i\in\La_0}$ are not mutually commuting, so this is not a commuting projector Hamiltonian. However, its ground-state subspace is still exactly-solvable and is equivalent to the Hilbert space $\scrH_\G^{(\bt)}$ of central 2-group $\G$ gauge theory. Indeed, the operators $\{B_{ijk}\}_{ijk\in\La_2}\cup \{C_{ij}\}_{ij\in\La_1}$ are mutually commuting projectors and commute with every $A^{(\bt)}_i$. Since $\{A^{(\bt)}_i\}_{i\in\La_0}$ are mutually commuting projectors in the ${C_{ij} = B_{ijk} = 1}$ subspace, $H^{(\bt)}_\G$ is a commuting projector Hamiltonian in the ${C_{ij} = B_{ijk} = 1}$ subspace, and its spectrum can be exactly computed in that subspace. 

\subsubsection{Torus ground-state degeneracy}

We now compute the ground-state degeneracy for ${Y=T^2}$. Following the split 2-group case, we use the triangulation~\eqref{onesimplextorustrianglulation} of a torus. The subspace $\scrH_\text{flat}$ is still given by~\eqref{Hflatonesimptrosutriang}. Furthermore, it follows from the split 2-group case that the subspace of $\scrH_\text{flat}$ invariant under~\eqref{central2Grp1FormGaugeTransfState} is spanned by vectors $\ket{g_x,g_y,a}$ (see Eq.~\eqref{split2groupcaAgxgy} with ${\rho_g = \text{id}_A}$). The triples ${(g_x,g_y,a)}$ form the set ${\Hom(\Z^2,G)\times A}$ .

It remains to impose the $G$ gauge invariance~\eqref{central2Grp0FormGaugeTransfState}. Since the triangulation
has a single site, a gauge transformation is specified by
one element ${h\in G}$ and acts as
\begin{equation}\label{central2grpGactiontorus}
    (g_x,g_y,a)
    \mapsto
    \left(
        hg_xh^{-1},
        hg_yh^{-1},
        a+\tau_\bt(g_x,g_y,h)
    \right),
\end{equation}
where
\begin{equation}
    \begin{aligned}
        \tau_\bt(g_{x},g_{y},h)
        =&~
        \bt(g_{x}, g_{y}, h^{-1}) 
    - \bt(g_{x},h^{-1},h g_{y} h^{-1}) 
    + \bt(h^{-1}, h g_{x} h^{-1}, h g_{y} h^{-1})\\
    &-\bt(g_{y}, g_{x}, h^{-1}) 
    + \bt(g_{y},h^{-1},h g_{x} h^{-1}) 
    - \bt(h^{-1}, h g_{y} h^{-1}, h g_{x} h^{-1}).
    \end{aligned}
\end{equation}
This transformation defines an action of $G$ on $\Hom(\Z^2,G)\times A$. Thus, the ground-state degeneracy of~\eqref{appendixcentral2grpQD} on a spatial torus is
\begin{equation}\label{GSDquotientcentral}
    \mathrm{GSD}_{T^2} = \left| (\Hom(\Z^2,G)\times A)/G \right|,
\end{equation}
where the quotient denotes the orbit set under~\eqref{central2grpGactiontorus}.
The ground-state subspace is spanned by the vectors ${\sum_{h\in G}\ket{hg_xh^{-1},hg_yh^{-1},a + \tau_\bt(g_{x},g_{y},h)}}$.

To obtain a more transparent form of this result, let us consider Fourier transformation
\begin{equation}\label{central2grpAFT}
    \ket{g_x,g_y,\chi}
    \equiv
    \frac{1}{\sqrt{|A|}}
    \sum_{a\in A}
    \chi(a)\ket{g_x,g_y,a},
    \qquad
    \chi\in\hat A.
\end{equation}
The $G$ gauge transformation~\eqref{central2grpGactiontorus} acts in this basis as
\begin{equation}\label{central2grpGgaugeTransGSDder}
    \ket{g_x,g_y,\chi} \mapsto \chi(\tau_\bt(g_x,g_y,h))^{-1}\ket{h g_xh^{-1},\, hg_yh^{-1},\,\chi}.
\end{equation}
For each ${\chi\in\hat A}$, let us define the normalized $3$-cocycle
\begin{equation}
    \om_\chi
    \equiv
    \chi\circ\bt
    \in
    \cZ^3(G,\Uone).
\end{equation}
The phase in~\eqref{central2grpGgaugeTransGSDder} can then be written as
\begin{equation}
    \chi(\tau_\bt(g_x,g_y,h))^{-1} = \frac{
    \om_\chi(g_{y}, g_{x}, h^{-1})\,
    \om_\chi(g_{x},h^{-1},h g_{y} h^{-1})\,
    \om_\chi(h^{-1}, h g_{y} h^{-1}, h g_{x} h^{-1})
    }{
    \om_\chi(g_{x}, g_{y}, h^{-1})\,
    \om_\chi(g_{y},h^{-1},h g_{x} h^{-1}) \,
    \om_\chi(h^{-1}, h g_{x} h^{-1}, h g_{y} h^{-1})
    }.
\end{equation}
This is precisely the cocycle phase multiplying states ${\{\ket{g_x,g_y}\mid (g_x,g_y)\in \Hom(\Z^2,G)\}}$ in a $G$ gauge transformation of the ${2+1}$d
$\om_\chi$-twisted $G$ quantum double model on the torus triangulation~\eqref{onesimplextorustrianglulation}~\cite[Eq.~(31)]{HWW12113695}.

Let us now define the Fourier block
\begin{equation}
    \mathscr{K}_\chi
    \equiv
    \mathrm{span}_{\C}
    \left\{
        \ket{g_x,g_y,\chi}
        \mid
        (g_x,g_y)\in\Hom(\Z^2,G)
    \right\}.
\end{equation}
Since the gauge transformation~\eqref{central2grpGgaugeTransGSDder} does not modify the $\chi$ label of the state $\ket{g_x,g_y,\chi}$, every $\mathscr{K}_\chi$ is preserved by the $G$ gauge projector $A_i^{(\bt)}$.
Furthermore, the restriction of $A_i^{(\bt)}$ to $\mathscr{K}_\chi$ is exactly the gauge projector
of the $\om_\chi$-twisted quantum double model.
Therefore, the torus ground-state space of~\eqref{appendixcentral2grpQD} decomposes as
\begin{equation}\label{centralHilbdecomposition}
    \scrH_\G^{(\bt)}
    \cong
    \bigoplus_{\chi\in\hat A}
    \scrH_{D^{\om_\chi}(G)},
\end{equation}
where $\scrH_{D^{\om_\chi}(G)}$ denotes the torus ground-state space of the $\om_\chi$-twisted $G$ quantum double model $D^{\om_\chi}(G)$. Eq.~\eqref{centralHilbdecomposition} is the torus Hilbert-space
realization of the decomposition conjecture of~\cite{PRS220413708,PS230316220}.

The decomposition~\eqref{centralHilbdecomposition} implies that the ground-state degeneracy~\eqref{GSDquotientcentral} can be written as
\begin{equation}
    \mathrm{GSD}_{T^2}
    =
    \sum_{\chi\in\hat A}
    \mathrm{GSD}_{T^2}(D^{\om_\chi}(G)).
\end{equation}
The torus ground-state degeneracy of the $\om$-twisted $G$ quantum double is~\cite{HWW12113695}
\begin{equation}
    \mathrm{GSD}_{T^2}(D^\om(G))
    =
    \sum_{[g]\in\mathrm{Conj}(G)}
    \left|
        \Irr_{\iota_g \om}\!\big(C_G(g)\big)
    \right|,
\end{equation}
where the slant product of $\om$ with $g$ satisfies
\begin{equation}\label{eq:slant-product-app}
    \iota_g \om(x,y)
    \equiv
    \om(g,x,y)\,
    \om(x,g,y)^{-1}\,
    \om(x,y,g),
    \qquad
    x,y\in C_G(g)
\end{equation}
and is a U(1)-valued $2$-cocycle, and
${\Irr_{\iota_g \om}(C_G(g))}$ is the set of inequivalent irreducible
$\iota_g \om$-projective representations of $C_G(g)$.
Hence,
\begin{equation}
    \mathrm{GSD}_{T^2}
    =
    \sum_{\chi\in\hat A}\,
    \sum_{\,[g]\in\mathrm{Conj}(G)}
    \left|
        \Irr_{\iota_g (\chi\circ\bt)}\!\big(C_G(g)\big)
    \right|.
\end{equation}

\subsubsection{Sphere ground-state degeneracy}

We now compute the ground-state degeneracy for ${Y=S^2}$ using the triangulation~\eqref{twosimplexspheretriangulation}. As in the split case, enforcing the $G$-flatness condition and gauge invariance~\eqref{central2Grp1FormGaugeTransfState} gives rise to a subspace spanned by vectors ${\ket{g_{ij},g_{jk},a}}$ where ${a \equiv a_{\tricorner{}_+} - a_{\tricorner{}_-}}$.

It remains to impose the $G$ gauge transformation~\eqref{central2Grp0FormGaugeTransfState}. Under the identifications defining~\eqref{twosimplexspheretriangulation}, ${\zeta_f(g_{ij},g_{jk}) = \zeta_f(g_{lm},g_{mn})}$. Therefore, the $G$ gauge transformation acts as
\begin{equation}
    \ket{g_{ij},g_{jk},a}
    \mapsto
    \ket{
        f_i g_{ij}f_j^{-1},
        f_j g_{jk}f_k^{-1},
        a
    }.
\end{equation}
Since flatness implies ${g_{ik}=g_{ij}g_{jk}}$, the gauge transformation with ${f_i=1}$, ${f_j=g_{ij}}$, and ${f_k=g_{ik}}$ maps each $\ket{g_{ij},g_{jk},a}$ to $\ket{1,1,a}$. Therefore, each
element ${a\in A}$ labels one gauge-equivalence class and hence one ground-state basis vector. It
follows that
\begin{equation}
    \mathrm{GSD}_{S^2}
    =
    |A|.
\end{equation}
Equivalently, since $A$ is finite abelian,
\begin{equation}
    \mathrm{GSD}_{S^2}
    =
    |\hat A|.
\end{equation}
Note that the sphere ground-state degeneracy is independent of the Postnikov class $[\bt]$. Furthermore, this is also consistent with the decomposition conjecture:
\begin{equation}
    \mathrm{GSD}_{S^2}
    =
    \sum_{\chi\in\hat A}
    \mathrm{GSD}_{S^2}(D^{\om_\chi}(G)),
\end{equation}
since ${\mathrm{GSD}_{S^2}(D^{\om_\chi}(G)) = 1}$.

\section{\texorpdfstring{$2\Rep(\G)$ 0-form symmetry operators}{2 rep 2 group 0-form symmetry operators}}\label{app:non-invertible-minimal-coupling}

\subsection{\texorpdfstring{Split 2-group $\G$}{Split 2-groups}}\label{app:split-non-invertible-minimal-coupling}

In this appendix, we show how the $\h{A}$ 0-form symmetry operator~\eqref{eq:a-hat-0-form-sym-op} is affected by gauging the $G$ 0-form symmetry operator~\eqref{eq:split-2group-G-sym-op-mc}. 
Because the $G$ symmetry acts on $\hat A$, an individual operator
$V^{(\chi)}$ is generally not $G$-symmetric, and cannot be minimally-coupled. However, the sum
\begin{equation}\label{symmetrizedVchi0}
    \sum_{h\in G} V^{(\rho_{h^{-1}}\triangleright\chi)}
\end{equation}
is symmetric, and can be minimally-coupled. 
More generally, for each subgroup ${K\leq G}$ and $K$-invariant character ${\chi\in\h{A}^K = \{\chi\in\h{A}\mid \rho_k\triangleright\chi = \chi~~\forall~~k\in K\}}$, the following is $G$-symmetric and can be minimally-coupled: 
\begin{equation}\label{symmetrizedVchi}
    \sum_{Kh\in K\backslash G} V^{(\rho_{h^{-1}}\triangleright\chi)}.
\end{equation}
Note that under ${h\mapsto kh}$, the operator ${V^{(\rho_{h^{-1}}\triangleright\chi)} \mapsto V^{(\rho_{h^{-1}}\triangleright\rho_{k^{-1}}\triangleright\chi)} = V^{(\rho_{h^{-1}}\triangleright\chi)}}$ for ${\chi\in\h{A}^K}$.
The operator~\eqref{symmetrizedVchi} reduces to~\eqref{symmetrizedVchi0} when ${K=1}$.

The operator~\eqref{symmetrizedVchi} fails to commute with the Gauss operator~\eqref{gaugingGsplit2grpsecGaussop}. Indeed, conjugating~\eqref{symmetrizedVchi} by $\prod_{i\in\La_0}G_{i}^{(h_i)}$ maps it to
\begin{equation}
    \sum_{Kh\in K\backslash G} \,
    \prod_{lmn\in\La_2}
    \sum_{\substack{\left\{g_i\right\},\left\{g_{ij}\right\}\\ \{a_{ijk}\}}}\!\!\!\!
    \chi(\rho_{h h_l^{-1}}(a_{lmn}))^{\eps_{lmn}}
    \ketbra
    {\{g_i\},\{g_{ij}\},\{a_{ijk}\}}
    {\{g_i\},\{g_{ij}\},\{a_{ijk}\}}.
\end{equation}
It is tempting to redefine the summation variable $h$ to absorb the problematic group element $h_l^{-1}$ appearing in the subscript of $\rho$, but this cannot be done since $h_l^{-1}$ depends on the site $l$.

Motivated by this, however, we minimally couple~\eqref{symmetrizedVchi} such that it becomes
\begin{equation}\label{mathsfKchi}
    \mathsf{V}^{(K,\chi)} \!=\!\! 
    \sum_{Kh\in K\backslash G}
    \mathsf{P}^{(K)}_{o, h}
    \left[\sum_{\substack{\left\{g_i\right\},\left\{g_{ij}\right\}\\ \{a_{ijk}\}}}
    \prod_{lmn\in\La_2}\!
    \chi(\rho_{h g_{\ga_{o\to l}}}(a_{lmn}))^{\eps_{lmn}}
    \ketbra
    {\{g_i\},\{g_{ij}\},\{a_{ijk}\}}
    {\{g_i\},\{g_{ij}\},\{a_{ijk}\}}\right]\!
    ,
\end{equation}
where $\mathsf{P}^{(K)}_{o, h}$ is the projector that satisfies
\begin{equation}\label{Projectorsplit2grouprepop}
    \mathsf{P}^{(K)}_{o, h} \ket{\{g_i\}, \{g_{ij}\},\{a_{ijk}\}} \!= 
    \begin{cases}
        \ket{\{g_i\}, \{g_{ij}\},\{a_{ijk}\}}
        \qquad &
    \prod_{ij\in\ga_{o}} g_{ij}^{\eps_{ij}(\ga_o)} \!\in h^{-1} K h~~\forall\text{ loops }\ga_o\text{ based at $o$}, \\
        0 \qquad &\text{otherwise}.
    \end{cases}
\end{equation}
In the subscript of $\rho$ in $\mathsf{V}^{(K,\chi)}$, we have defined the group element
\begin{equation}
    g_{\ga_{o\to l}}  = \prod_{ij\in\ga_{o\to l}} g_{ij}^{\eps_{ij}(\ga_{o\to l})}\in G,
\end{equation}
where $\ga_{o\to l}$ is any lattice path from an arbitrary reference site $o$ to the site $l$. 
The operator $\mathsf{V}^{(K,\chi)}$ does not depend on the choice of lattice path because ${\chi\in\h{A}^K}$ and the projector~\eqref{Projectorsplit2grouprepop} enforces ${g_{\ga_{o\to l}}g^{-1}_{\t{\ga}_{o\to l}}\in h^{-1} K h}$ for two different paths $\ga_{o\to l}$ and $\t{\ga}_{o\to l}$.
The operator $\mathsf{V}^{(K,\chi)}$ is also independent of the choice of reference site $o$ and the representative $h$ of the coset $Kh$.
Each Gauss operator $G_i^{(h)}$ commutes with $\mathsf{V}^{(K, \chi)}$.
Note that ${\mathsf{V}^{(1,1)} = |G|\mathsf{P}^{(1)}_{o,1}}$, where $\mathsf{P}^{(1)}_{o,1}$ projects onto the $\Rep(G)$ 1-form symmetric subspace, and ${\mathsf{V}^{(G,1)} = \mathsf{P}^{(G)}_{o,h} = 1}$.

The symmetry operators $\mathsf{V}^{(K,\chi)}$ do not capture all of the 0-form symmetry operators of the $2\Rep(\G)$ fusion 2-category symmetry.
To obtain the most general symmetry operator, we note that the
operators $\mathsf V^{(K,1)}$ can be interpreted as condensation operators
for the $\Rep(G)$ 1-form symmetry.  
(See~\cite{BCH220807367, DT230101259, TC230703180, CSS240513105, GST240612978, VD250116301} for further discussion of condensation operators and condensation defects in quantum lattice systems.)
Namely, the projectors
$\mathsf P^{(K)}_{o,h}$ restrict the symmetry operator to configurations
whose $G$ holonomies based at $o$ lie in the conjugate subgroup
$h^{-1}Kh$. After conjugating by $h$, these become $K$ holonomies.  
From the field-theory perspective, this operator is supported on a defect surface in spacetime.
Once the $G$ gauge field on this surface is reduced to $K$,
the surface may be dressed by a ${1+1}$d $K$-SPT, i.e., a gauged SPT defect.
Such decorations
are classified by ${\cH^2(K,\Uone)}$.
Thus the general 0-form symmetry operators are labeled by triples ${(K,\chi,[\om])}$, with
${\chi\in\h{A}^K}$ and ${[\om]\in \cH^2(K,\Uone)}$.\footnote{Strictly speaking, the triple ${(K,\chi,[\om])}$ does not uniquely label a symmetry operator. In particular, the symmetry operators are labeled by ${(K,\chi,[\om])}$ modulo the equivalence relation ${(K,\chi,[\om]) \sim (gKg^{-1},\rho_g\triangleright\chi,\mathrm{Conj}_g[\om])}$. Indeed, the explicit expression~\eqref{mathsfKchiomega} we find satisfies ${\mathsf{V}^{(K,\chi,[\om])} = \mathsf{V}^{(gKg^{-1},\rho_g\triangleright\chi,\mathrm{Conj}_g[\om])}}$.
}
The operators $\mathsf V^{(K,\chi)}$ above correspond to the trivial class
${[\om]=[1]}$.

We now write down the decorated operator explicitly.  We restrict to the
subspace in which the $\Rep(G)$ 1-form symmetry operators are topological,
namely the image of the flatness projector
${\prod_{ijk\in\La_2}B_{ijk}}$.

To write down the general symmetric operator, we must find the lattice expression for the aforementioned reduced $K$ gauge field. 
Consider the coset ${Kh g_{\ga_{o\to i}}\in K\backslash G}$.
The projector $\mathsf P^{(K)}_{o,h}$ makes this coset independent of the
path $\ga_{o\to i}$ from $o$ to $i$.  Equivalently, for every edge
$ij$, it imposes ${Kh g_{\ga_{o\to i}}g_{ij}=Kh g_{\ga_{o\to j}}}$.
To extract a $K$ lattice gauge field from the coset $Kh g_{\ga_{o\to i}}$, we choose a section ${s:K\backslash G\to G}$ satisfying ${s(Kh)\in Kh}$ and define 
\begin{equation}\label{eq:kappa-K-gauge-field-app}
    \ka_{ij}
    =
    s(Kh g_{\ga_{o\to i}})
    \,g_{ij}\,
    s(Kh g_{\ga_{o\to j}})^{-1}.
\end{equation}
(We leave the dependence of $\ka_{ij}$ on $Kh$, $o$, and $\{g_{ij}\}$ implicit to avoid overloading notation.)
$\ka_{ij}$ is generally a generic group element of $G$. However, on the support of $\mathsf P^{(K)}_{o,h}$:\footnote{By the support of $\mathsf P^{(K)}_{o,h}$, we mean configurations $\{\{g_i\}, \{g_{ij}\}, \{a_{ijk}\}\}$ such that the vector $\ket{\{g_i\}, \{g_{ij}\}, \{a_{ijk}\}}$ lies in the image of $\mathsf P^{(K)}_{o,h}$.}
\begin{enumerate}
    \item Each ${\ka_{ij}\in K}$ because ${Kh g_{\ga_{o\to j}} = Kh g_{\ga_{o\to i}}g_{ij}}$. (In general, ${s(Kh_1)h_2\,s(Kh_1h_2)^{-1}\in K}$ for all ${h_1,h_2\in G}$ since the group elements $s(Kh_1)h_2$ and $s(Kh_1h_2)$ both lie in the coset $Kh_1h_2$. Here, we use this formula with ${h_1 = h g_{\ga_{o\to i}}}$ and ${h_2 = g_{ij}}$.)
    \item Under the gauge transformation ${g_{ij}\mapsto h_i^{-1} g_{ij} h_j}$, $\ka_{ij}$ transforms as ${\ka_{ij} \mapsto r_i^{-1} \ka_{ij} \,r_j}$ where each ${r_i = s(Khh_o^{-1}g_{\ga_{o\to i}})h_{i}\, s(Khh_o^{-1}g_{\ga_{o\to i}}h_i)^{-1} \in K}$. The $\ka_{ij}$ appearing in ${r_i^{-1} \ka_{ij} \,r_j}$ is now defined with respect to the coset $Khh_{o}^{-1}$.
    \item Each $\ka_{ij}$ satisfies the flatness condition ${\ka_{ij}\ka_{jk} = \ka_{ik}}$ for all ${ijk\in\La_2}$. This follows from $\{g_{ij}\}$ satisfying the flatness condition in the ${B_{ijk} = 1}$ subspace in which we are working.
\end{enumerate}
Therefore, the group elements $\ka_{ij}$ on the support of $\mathsf P^{(K)}_{o,h}$ are the $K$-valued edge variables of the reduced $K$ gauge field on the symmetry operator.

We now choose a normalized
2-cocycle ${\om\in \cZ^2(K,\Uone)}$,
and define the operator\footnote{
We note that the U(1) phase $\prod_{lmn\in\La_2} \om(\ka_{lm},\ka_{mn})^{\eps_{lmn}}$ appearing in the operator $\mathsf{C}^{(K,[\om])}_{o,h}$ is the Euclidean path-integral weight of ${1+1}$d $K$ Dijkgraaf--Witten theory. More broadly, unitary operators of the form $\sum_{\{g_{ij\}}} \left(\prod_{lmn\in\La_2} \nu(g_{lm},g_{mn})^{\eps_{lmn}} \right) \ketbra{\{g_{ij}\}}{\{g_{ij}\}}$, with ${\nu\in\cZ^2(G,\Uone)}$, are the lattice symmetry operators corresponding to gauged $G$ SPT defect surfaces studied in ${2+1}$d Euclidean field theories. 
}
\begin{equation}
    \mathsf{C}^{(K,[\om])}_{o,h}
    =
    \bigg(\sum_{\left\{g_i\right\},\left\{g_{ij}\right\}, \{a_{ijk}\}}
    \,
    \prod_{lmn\in\La_2} \om(\ka_{lm},\ka_{mn})^{\eps_{lmn}}
     \ketbra
     {\{g_i\},\{g_{ij}\},\{a_{ijk}\}}
     {\{g_i\},\{g_{ij}\},\{a_{ijk}\}}\bigg)
     \mathsf{P}^{(K)}_{o, h}.
\end{equation}
This operator projects onto the ${\mathsf{P}^{(K)}_{o, h}=1}$ subspace and then decorates states by the ${1+1}$d $K$-SPT phase $\om$. Due to the flatness condition satisfied by $\ka_{ij}$ on the support of $\mathsf{P}^{(K)}_{o, h}$, this operator is invariant under the transformation ${\om(k_1,k_2)\mapsto \om(k_1,k_2) \mu(k_1)\mu(k_2)/\mu(k_1k_2)}$ for arbitrary ${\mu\colon K\to\Uone}$ and, therefore, depends only on the cohomology class ${[\om]}$.
Using it, we define the general symmetry operator
\begin{equation}\label{mathsfKchiomega}
    \mathsf{V}^{(K,\chi,[\om])} \!=\!\! 
    \!\!\!\sum_{Kh\in K\backslash G}
    \!\!\!\!\mathsf{C}^{(K,[\om])}_{o, h}
    \!\!\left[\!\sum_{\substack{\left\{g_i\right\},\left\{g_{ij}\right\}\\ \{a_{ijk}\}}}
    \prod_{lmn\in\La_2}\!\!
    \chi(\rho_{h g_{\ga_{o\to l}}}(a_{lmn}))^{\eps_{lmn}}
    \ketbra
    {\{g_i\},\{g_{ij}\},\{a_{ijk}\}}
    {\{g_i\},\{g_{ij}\},\{a_{ijk}\}}\right]\!\!
    .
\end{equation}
When ${[\om]=[1]}$, this becomes the original operator
$\mathsf V^{(K,\chi)}$.
Furthermore, one can check that it commutes with each Gauss operator $G_{i}^{(h)}$ using the usual relabeling ${Kh \mapsto Kh h_0}$ of the summation variable $Kh$.

It is straightforward, albeit tedious, to check that the symmetry operators $\mathsf{V}^{(K,\chi,[\om])}$ satisfy 
\begin{equation}\label{2RepG0FSalgSpit}
    \mathsf{V}^{(K_1,\chi_1,[\om_1])}
    \times
    \mathsf{V}^{(K_2,\chi_2,[\om_2])}
    =
    \sum_{K_1rK_2 \in K_1 \backslash G / K_2} \mathsf{V}^{(K_r,\, \chi_r, \, [\om_r] )},
\end{equation}
where ${K_r = K_1\cap \,r K_2 r^{-1}}$, ${\chi_r = \chi_1\cdot (\rho_r\triangleright \chi_2)}$, and ${[\om_r] = \mathrm{Res}^{K_1}_{K_r}([\om_1])\cdot \mathrm{Res}^{r K_2 r^{-1}}_{K_r}(\mathrm{Conj}_r([\om_2]))}$. The cohomology class $[\om_r]$ is defined using the cohomological operations $\mathrm{Res}$ and $\mathrm{Conj}$. 
At the level of 2-cochains, the former acts as
\begin{equation}\label{eq:res-cohom-op-defn}
    \mathrm{Res}^{K_1}_{K_r}\om_1(k_r, k_r') = \om_1(\iota(k_r), \iota(k_r')),
\end{equation}
where ${k_r,k_r'\in K_r}$ and ${\iota\colon K_r\hookrightarrow K_1}$ is the subgroup inclusion homomorphism.
The latter acts as
\begin{equation}\label{eq:conj-cohom-op-defn}
    \mathrm{Conj}_r\om_2(k,k') = \om_2(r^{-1} k r, r^{-1}k' r).
\end{equation}
The algebra~\eqref{2RepG0FSalgSpit} is the general 0-form symmetry algebra of a $2\Rep(\G)$ fusion 2-category symmetry~\cite{BBFP220805993}.
Thus, the symmetry operators $\mathsf{V}^{(K,\chi,[\om])}$ make up the 0-form symmetry part of the $2\Rep(\G)$ fusion 2-category symmetry.

\subsection{\texorpdfstring{Central 2-group $\G$}{Central 2-groups}}\label{app:central-non-invertible-minimal-coupling}

We now investigate how the central 0-form $\h{A}$ symmetry operators $\t{V}^{(\chi)}$ given in Eq.~\eqref{eq:disentangled-Ahat-anomalous-sym-ops0} are affected by gauging the $G$ 0-form symmetry~\eqref{eq:disentangled-G-anomalous-sym-ops0}. 
We restrict ourselves to the image of ${\prod_{ijk\in\La_2}B_{ijk}}$, which is the lattice $\Rep(G)$ 1-form symmetry topological subspace.
The form of the resulting operator depends on the subgroup $K$ for which the restriction ${\chi\circ\bt\vert_K}$ is a coboundary. 

We begin by considering two edge cases. Suppose ${K=G}$, i.e., ${[\chi\circ\bt]=[1]}$. Then ${\chi\circ\bt = \del c_\chi}$ for some ${c_{\chi}\in \cC^2(G,U(1))}$, and we have 
\begin{equation}
    \chi(\bt(g_l,\, g_l^{-1}g_m,\, g_m^{-1}g_n))
    = 
    \frac{c_{\chi}(g_l^{-1}g_m,\, g_m^{-1}g_n)\, c_{\chi}(g_l,\, g_l^{-1}g_n)}
       {c_{\chi}(g_m,\, g_m^{-1}g_n)\, c_{\chi}(g_l,\, g_l^{-1}g_m)}.
\end{equation}
Correspondingly, the 0-form $\h{A}$ symmetry operator is
\begin{equation}
    \t{V}^{(\chi)} 
    = 
    \sum_{\{g_{i}\}, \{a_{ijk}\}} 
    \prod_{l mn\in\La_2}
    \chi(a_{l mn})^{\eps_{l mn}}
    \;
    c_{\chi}(g_l^{-1}g_m,\, g_m^{-1}g_n)^{-\eps_{l mn}}
    \ketbra{\{g_{i}\}, \{a_{ijk}\}}.
\end{equation}
This operator may be minimally-coupled to commute with the Gauss law operators~\eqref{eq:gauss-law-central-gauge-G} by modifying it as follows:
\begin{equation}\label{eq:app-central-m.c.-ahat-operator-K=G}
    \begin{aligned}
    \t{V}^{(\chi,[c_{\chi}])}_{\mathrm{m.c.}} = 
    \sum_{\{g_i\},\{g_{ij}\},\{a_{ijk}\}}
    \prod_{l m n\in\La_2}
     &
    \;
    \chi(a_{l mn})^{\eps_{l mn}}\;
    c_{\chi}(g_l^{-1}g_{l m}g_m,\, g_m^{-1}g_{mn}g_n)^{-\eps_{l mn}}
    \\[-1em]
    &
    \qquad 
    \times 
    \ketbra{\{g_i\},\{g_{ij}\},\{a_{ijk}\}}.
    \end{aligned}
\end{equation}
In the image of ${\prod_{ijk\in\La_2}B_{ijk}}$, this operator is labeled by an element $[c_{\chi}]$ of an ${\cH^2(G,\Uone)}$-torsor since it is invariant under multiplying $c_{\chi}$ by a 2-coboundary. 

On the other hand, for every $\chi$, the operator ${\t{V}^{(\chi)}}$ may be minimally-coupled to commute with~\eqref{eq:gauss-law-central-gauge-G} by modifying it as follows:
\begin{align}\label{eq:app-central-m.c.-ahat-operator-K=1}
    \t{V}^{(\chi)}_{\mathrm{m.c.}} 
    = 
    |G|\sum_{ \{g_i\}, \{g_{ij}\}, \{a_{ijk}\} }\prod_{l mn\in\La_2}
    &\chi\!\left(a_{l mn} - \bt(g_{o\to l}g_l,\, g_l^{-1}g_{l m}g_m,\, g_m^{-1}g_{mn}g_n)\right)^{\eps_{l mn}}\\[-1em]
    &\hspace{4em}\times \mathsf{P}^{(\{1\})}_{o,1}\, 
    \ketbra{\{g_i\}, \{g_{ij}\}, \{a_{ijk}\}}
    .
    \nonumber
\end{align}
This operator commutes with the Gauss law operator $G_o^{(h)}$ and is independent of the choice of reference site $o$. This is because the projector $\mathsf{P}_{o,1}^{(\{1\})}$ ensures that, using cocycle conditions,
\begin{equation}
    \begin{aligned}
        \frac{\chi(\bt(h\, g_{o\to l}g_l,\, g_l^{-1}g_{l m}g_m,\,             g_m^{-1}g_{mn}g_n)}{
        \chi(\bt(g_{o\to l}g_l,\, g_l^{-1}g_{l m}g_m,\, g_m^{-1}g_{mn}g_n)) }
        =
        \frac{
        \chi(\beta(h,g_{o\to l} g_l, g_l^{-1} g_{lm} g_m)) 
        \chi(\beta(h,g_{o\to m} g_m, g_m^{-1} g_{mn} g_n))
        }{
        \chi(\beta(h,g_{o\to l} g_l, g_l^{-1} g_{ln} g_n))
        }
    \end{aligned}
\end{equation}
on the support of $\mathsf{P}_{o,1}^{(\{1\})}$. Then, because
\begin{equation}    
    \prod_{lmn\in\La_2} 
            \left(\frac{
        \chi(\beta(h,g_{o\to l} g_l, g_l^{-1} g_{lm} g_m)) 
        \chi(\beta(h,g_{o\to m} g_m, g_m^{-1} g_{mn} g_n))
        }{
        \chi(\beta(h,g_{o\to l} g_l, g_l^{-1} g_{ln} g_n))
        }\right)^{\eps_{lmn}}
    = 
    1.
\end{equation}
the commutator ${[\t{V}^{(\chi)}_{\mathrm{m.c.}} ,\, G_o^{(h)}]=0}$. 

Most generally, $\chi$ may satisfy ${[\chi\circ\bt]\big|_K=[1]}$ for some subgroup ${K\leq G}$. For a section $s: K\backslash G \to G$, let us define $v_i$ and $\mu_i$ satisfying
\begin{equation}
    v_i = s(Kg_i)\in G,
    \qquad 
    \mu_i = g_i v_i^{-1}\in K
\end{equation}
where again we leave dependence on $g_i$ and $K$ implicit for ease of notation. Using Eq.~\eqref{eq:zeta-composition}, we find that $\bt$ can be decomposed as the sum of a restriction $\bt\vert_K$ and the descendant $\zeta_v$: 
\begin{equation}\label{eq:app-mc-central-2rep-factorization-beta}
    \begin{aligned}  
    \bt(g_i,\, g_i^{-1}g_j,\, g_j^{-1}g_k)
    &= 
    -\zeta_v(g_i^{-1}g_j,\, g_j^{-1}g_k)
    + \bt(\mu_i,\, \mu_i^{-1}\mu_j,\, \mu_j^{-1}\mu_k)
    \\
    &\qquad 
    + \xi(g_i, g_j, v_i, v_j)
    - \xi(g_i, g_k, v_i, v_k)
    + \xi(g_j, g_k, v_j, v_k)
    \end{aligned}
\end{equation}
for
\begin{equation}
    \xi_{ij}
    =
    \beta(g_i, g_i^{-1} g_j, v_j^{-1})
    - 
    \beta(g_i, v_i^{-1}, \mu_i^{-1} \mu_j)
    .
\end{equation}

Because ${\chi\circ\bt\big|_K = \del c_{\chi}}$ for some ${c_{\chi}\in \cC^2(K,\Uone)}$, we can use~\eqref{eq:app-mc-central-2rep-factorization-beta} to write
\begin{align}
\t{V}^{(\chi)} 
    &= \sum_{\{g_i\},\{a_{ijk}\}} \prod_{l mn\in\La_2}
    \chi\!\left(a_{l mn} - \bt(g_l,\, g_l^{-1}g_m,\, g_m^{-1}g_n)\right)^{\eps_{l mn}}
    \ketbra{\{g_i\},\{a_{ijk}\}}
    \\
    &= \sum_{\{g_i\},\{a_{ijk}\}} \prod_{l mn\in\La_2}
    \chi\!\left(a_{l mn} + \zeta_v(g_l^{-1}g_m,\, g_m^{-1}g_n)\right)^{\eps_{l mn}}
    c_{\chi}(\mu_l^{-1}\mu_m,\, \mu_m^{-1}\mu_n)^{-\eps_{l mn}}
    \\[-1em]
    &\hspace{10em} \times
    \ketbra{\{g_i\},\{a_{ijk}\}}\nonumber
\end{align}
where we've used $\prod_{lmn\in\La_2} \chi(\xi_{lm} - \xi_{ln} + \xi_{mn})^{\eps_{lmn}} = 1$.

We now minimally couple this operator to ensure it commutes with the Gauss law operator~\eqref{eq:gauss-law-central-gauge-G}. In what follows, we restrict to the subspace in which the $\Rep(G)$ 1-form symmetry operators are topological, that is, the image of the projector ${\prod_{ijk\in\La_2} B_{ijk}}$. 
This operator is minimally-coupled by modifying it as\footnote{As with the split $0$-form ${2\Rep(\G)}$ operators in~\eqref{mathsfKchiomega}, the central 0-form $2\Rep (\G)$ are, strictly speaking, labeled by equivalence classes $(K,\chi,[c_{\chi}]) \sim (rKr^{-1}, \chi, \operatorname{Conj}^{\beta}_{r,\chi}c)$ as the operators satisfy $\t{V}^{(K,\chi,[c_{\chi}])}=\t{V}^{(rKr^{-1}, \chi, \operatorname{Conj}^{\beta}_{r,\chi}c)}$ for $\operatorname{Conj}_{r,\chi}^{\beta}$ defined in Eq.~\eqref{eq:conj-beta-defn}.}
\begin{align}
    \t{V}^{(K,\chi,[c_{\chi}])}_{\mathrm{m.c.}}
    = 
    \sum_{Kh\in K\backslash G} 
    \mathsf{P}^{(K)}_{o,h} 
    \sum_{\{g_i\},\{g_{ij}\},\{a_{ijk}\}} \prod_{l mn}
    &
    \;
    \chi\!\left(a_{l mn} + \zeta_{v'}(g_l^{-1}g_{l m}g_m,\,                   g_m^{-1}g_{mn}g_n)\right)^{\eps_{l mn}} \nonumber
    \\[-1em]
    &\times\, 
    c_\chi (\mu_l'^{-1}\kappa_{l m}\mu_m',\, \mu_m'^{-1}\kappa_{mn}\mu_n')^{-       \eps_{l mn}}
    \\
    &\times\, 
    \ketbra{\{g_i\},\{g_{ij}\},\{a_{ijk}\}},\nonumber
\end{align}
where $\kappa_{ij}$ is defined in~\eqref{eq:kappa-K-gauge-field-app}, ${v_i' = s(Khg_{o\to i} g_i)}$, ${\mu_i' = u_i g_i v_i'^{-1}}$ and ${u_i = s(Khg_{o\to i})}$.
One can verify this operator commutes with each Gauss law operator $G_{i}^{(h)}$ by relabeling ${Kh \mapsto Khh_0}$ and using~\eqref{eq:zeta-composition}. Furthermore, it is independent of reference site $o$, representative $h$ of the coset $Kh$, and choice of section $s: K\backslash G \to G$.

The operator $\t{V}^{(K,\chi,[c_{\chi}])}_{\mathrm{m.c.}}$ is invariant under shifting the cochain $c_{\chi}(\kappa_{lm},\kappa_{mn})$ by a coboundary thanks to the product $\prod_{lmn}$. The set of such equivalence classes $[c_{\chi}]$ of ${c_{\chi}\in \cC^2(K,\Uone)}$ satisfying ${(\chi\circ \bt)\vert_K =  \del c_{\chi}}$ forms a $\cH^2(K,\Uone)$-torsor.
Note that ${\t{V}^{(\{1\},\chi,[c_{\chi}])}_{\mathrm{m.c.}} = \t{V}^{(\chi)}_{\mathrm{m.c.}}}$
where $\t{V}_{\mathrm{m.c.}}$ is given in~\eqref{eq:app-central-m.c.-ahat-operator-K=1} and ${\t{V}^{(G, \chi,[c_{\chi}])}_{\mathrm{m.c.}} = \t{V}^{(\chi,c_{\chi})}_{\mathrm{m.c.}}}$ as given in~\eqref{eq:app-central-m.c.-ahat-operator-K=G}.
Furthermore, in the disentangling frame given by~\eqref{eq:central-gauge-g-0-form-unitary}, we have the following central 0-form $2\Rep(\G)$ symmetry operator:
\begin{align}
    \mathsf{V}^{(K,\chi,[c_{\chi}])}
    &= U_2 \t{V}^{(K,\chi,[c_{\chi}])}_{\mathrm{m.c.}} U_2^\dagger \\ 
    &= 
    \sum_{Kh\in K\backslash G} \mathsf{P}^{(K)}_{o,h} 
    \sum_{\{g_{ij}\}, \{a_{ijk}\}}
    \prod_{l m n\in\La_2}
    \chi\!\left(a_{l mn} + \zeta_u(g_{l m},\, g_{mn})\right)^{\eps_{l mn}} c_{\chi}(\kappa_{l m},\, \kappa_{mn})^{-\eps_{l mn}}
    \\[-1em]
    &
    \hspace{15em}\times 
    \ketbra{{\{g_{ij}\}, \{a_{ijk}\}}}.\nonumber
\end{align}

One can verify that these operators satisfy an equation analogous to~\eqref{2RepG0FSalgSpit} for the split $2\Rep(\G)$ 0-form symmetry operators:
\begin{equation}
    \mathsf{V}^{(K_1,\chi_1,[c_{\chi_1}])}
    \times
    \mathsf{V}^{(K_2,\chi_2,[c_{\chi_2}])}
    =
    \sum_{K_1rK_2 \in K_1 \backslash G / K_2} \mathsf{V}^{(K_r,\, \chi_1 \chi_2, \, [c_{r,\chi_1\chi_2}] )},
\end{equation}
where ${K_r = K_1\cap \,r K_2 r^{-1}}$ and ${[c_{r,\chi_1\chi_2}] = \mathrm{Res}^{K_1}_{K_r}([c_{\chi_1}])\cdot \mathrm{Res}^{r K_2 r^{-1}}_{K_r}([\mathrm{Conj}^\bt_{r,\chi_2}(c_{\chi_2})])}$. $\mathrm{Res}$ is the operation defined in~\eqref{eq:res-cohom-op-defn} and $\mathrm{Conj}^\bt_{r,\chi} (c_{\chi_2})$ satisfies 
\begin{equation}\label{eq:conj-beta-defn}
    \mathrm{Conj}_{r,\chi_2}^\bt (c_{\chi_2})\,(k, k') = 
    \frac{\mathrm{Conj}_r(c_{\chi_2})\,(k, k')}
    {\chi_2(\varsigma_r(\iota (k), \iota(k')))},
\end{equation}
for $\mathrm{Conj}$ defined in~\eqref{eq:conj-cohom-op-defn}, ${k, k'\in rK_2r^{-1}}$ and ${\iota: rK_2r^{-1} \hookrightarrow G}$ is the group inclusion homomorphism. ${\varsigma_r\in C^2(G,A)}$ is a 2-cochain
\begin{equation}
    \varsigma_r(g,h) 
    = 
    \beta(r,r^{-1} g r, r^{-1} hr)
    - \beta(g, r, r^{-1} h r)
    + \beta(g,h,r)
\end{equation}
satisfying $\delta \varsigma_r (g_{ij}, g_{jk}, g_{kl})
        = \mathrm{Conj}_r  \bt(g_{ij}, g_{jk}, g_{kl})
            - \bt(g_{ij}, g_{jk}, g_{kl})$.
The operation $\mathrm{Conj}^\bt_{r,\chi_2}$ in the definition of $[c_{r,\chi_1\chi_2}]$ ensures that ${\delta c_{r,\chi_1\chi_2} = (\chi_1\chi_2)\circ\bt\mid_{K_r}}$.

\section{Dependence on the Postnikov cocycle representative}\label{app:central-beta-dependence}

The symmetry operators $\{U^{(h,[f])}\}_{(h,[f])\in \t{G}}$ defined by Eq.~\eqref{GbetaSymOps} depend on the chosen representative normalized cocycle
${\bt\in\cZ^3(G,A)}$ of the Postnikov class $[\bt]$.
In this appendix, we describe this dependence both on the full tensor-product Hilbert space and on
the topological subspace $\scrH_{\mathrm{top}}$. We make the $\bt$ dependence on $U^{(h,[f])}$ explicit by denoting it as $U^{(\bt; h,[f])}$ in this appendix. Similarly, we denote the operators $U^{(h)}$ given in \eqref{eq:central-2group-G-sym-op} by $U^{(\beta;h)}$. 

\subsection{Tensor-product Hilbert space}

Let $\bt$ and $\bt'$ be normalized representatives of the same
Postnikov class: ${[\bt] = [\bt']}$. There is a normalized 2-cochain ${\eta\colon G\times G\to A}$ such that
\begin{equation}
    \bt'(g,h,k) = \bt(g,h,k) + \eta(h,k) - \eta(gh,k) + \eta(g,hk) - \eta(g,h).
\end{equation}
Given this normalized 2-cochain $\eta$, we define the unitary
\begin{equation}\label{eq:Weta-def}
    V_\eta
    =
    \sum_{\{g_i\},\{a_{ij}\}}
    \ketbra{
        \{g_i\},
        \left\{
            a_{ij}
            +\eta(g_i,g_i^{-1}g_j)
        \right\}
    }{
        \{g_i\},
        \{a_{ij}\}
    }.
\end{equation}
A direct calculation shows
\begin{equation}\label{eq:U-beta-representative-change}
    U^{(\bt'; h, [f])}
    =
    V_\eta^{-1}
    U^{(\bt; h, [f])}
    V_\eta
    \,S_{\eta(h,-)},
\end{equation}
where $S_f$ is defined by Eq.~\eqref{eq:central-Sf} and ${\eta(h,-)\colon g \mapsto \eta(h,g)}$. 
Recall that the group generated by $U^{(\bt; h, [f])}$ is the extension $\t{G}$ given by \eqref{extensionofGNbt} with extension class $[b_{-,-}^{(\beta)}]$, where we've made the dependence of $b_{-,-}$ on $\beta$ explicit. The symmetry groups of both operators $U^{(\bt; h, [f])}$ and $U^{(\bt'; h, [f])}$ are isomorphic to $\t{G}$ because $b^{(\beta')}_{-,-} $ and $b^{(\beta)}_{-,-} $ differ by a 2-coboundary plus a constant map. In particular,
\begin{equation}
    b^{(\beta')}_{h,k} 
    =
    b_{h,k}^{(\beta)}
    + 
    c_k 
    -
    c_{hk}
    + 
    c_h \triangleleft k
    +
    a_{h,k}
\end{equation}
where $c_h(k) = \eta(h,k)$ and $a_{h,k}$ is a constant map. This constant map does not affect the symmetry group since the operators $S_f$ are invariant under shifting $f$ by a constant map.

\subsection{Topological subspace}

In the topological subspace~\eqref{HtopDefSplit2GrpSection}, each operator $S_f$ acts as the identity (see Eq.~\eqref{SfonTopSub} and surrounding discussion). Furthermore, the unitary $V_\eta$ commutes with the $A_i^{(\la)}$ defining the topological subspace, and, therefore, restricts to a unitary
\begin{equation}
    V_{\eta,\mathrm{top}}
    =
    \left.V_\eta\right|_{\scrH_{\mathrm{top}}}.
\end{equation}
Restricting Eq.~\eqref{eq:U-beta-representative-change} to
$\scrH_{\mathrm{top}}$ gives
\begin{equation}\label{eq:U-beta-top-representative-change}
    U_{\mathrm{top}}^{(\bt';h)}
    =
    V_{\eta,\mathrm{top}}^{-1}
    U_{\mathrm{top}}^{(\bt;h)}
    V_{\eta,\mathrm{top}},
\end{equation}
Therefore, although the operators $U^{(\bt; h)}$ depend on the representative
$\bt$ on the full Hilbert space, their induced action on
$\scrH_{\mathrm{top}}$ depends only on the Postnikov class $[\bt]$, up
to the local unitary equivalence
\eqref{eq:U-beta-top-representative-change}.

\section{Onsiteability of \texorpdfstring{$S_f$}{Sf}}
\label{app:Sf-non-onsiteability}

In this appendix, we show that a nontrivial $S_f$ cannot
be made onsite by conjugation with a QCA.
We start by considering the trace of $S_f$. 
For each ${a\in A}$, let
\begin{equation}
    n_f(a)=|\{g\in G \mid f(g) = a\}|.
\end{equation}
A straightforward calculation then shows that
\begin{equation}\label{eq:Sf-trace}
    \operatorname{Tr}(S_f)
    =
    |A|^{|\La_1|}
    \sum_{b\in A}n_f(b)^{|\La_0|}.
\end{equation}

For any unitary $V$ with nonzero trace, we define
\begin{equation}
    \om_V(O)
    =
    \frac{\operatorname{Tr}(VO)}
    {\operatorname{Tr}(V)},
\end{equation}
for all ${O\in \mathrm{End}(\scrH)}$.
From~\eqref{eq:Sf-trace}, ${\operatorname{Tr}(S_f)>0}$ and, therefore, $\om_{S_f}$ is well-defined.
Consider the local projector
\begin{equation}
    P_i^{(a)}
    =
    \sum_{g\in G\,\mid\, f(g)=a}
    \ketbra{g}{g}_i.
\end{equation}
It satisfies
\begin{equation}\label{eq:Sf-long-range-correlation}
    \om_{S_f}(P_i^{(a)})
    =
    \om_{S_f}(P_i^{(a)}P_j^{(a)})
    =
   \frac{n_f(a)^{|\La_0|}}
    {\sum_{b\in A}n_f(b)^{|\La_0|}}.
\end{equation}

We now prove that $S_f$ cannot be made onsite using a QCA. We argue by contradiction. Suppose that there exists a QCA $W$ such that
\begin{equation}\label{eq:Sf-onsite-assumption}
    W S_f W^\dagger
    =
    S_{\mathrm{on}},
    \qquad
    S_{\mathrm{on}}
    =
    \bigotimes_x s_x,
\end{equation}
where $x$ labels the local tensor factors of the Hilbert space $\scrH$. By the cyclic property of the trace, 
\begin{align}
    \operatorname{Tr}(S_{\mathrm{on}})
    &=
    \operatorname{Tr}(S_f)
    \neq0,\label{eq:nonzeroSharedTrace}\\
    \om_{S_f}(O)
    &=
    \om_{S_{\mathrm{on}}}
    \left(
        W O W^\dagger
    \right).\label{eq:twisted-trace-conjugation}
\end{align}
Since $S_{\mathrm{on}}$ is a tensor product, ${\operatorname{Tr}(S_\mathrm{on}) = \prod_x \operatorname{Tr}(s_x)}$ and~\eqref{eq:nonzeroSharedTrace} implies that every
${\operatorname{Tr}(s_x)\neq 0}$.
The tensor product structure of $S_\mathrm{on}$ also implies that for operators $O_X$ and $O_Y$
whose finite supports $X$ and $Y$, respectively, are disjoint collections of local tensor factors,
\begin{equation}\label{eq:onsite-twisted-factorization}
    \om_{S_{\mathrm{on}}}(O_XO_Y)
    =
    \om_{S_{\mathrm{on}}}(O_X)
    \om_{S_{\mathrm{on}}}(O_Y).
\end{equation}

Since $W$ is a QCA, we can always take $i$ and $j$ to be sufficiently far apart such that 
\begin{equation}
    W P_i^{(a)}W^\dagger
    \qquad\text{and}\qquad
    W P_j^{(a)}W^\dagger
\end{equation}
have disjoint supports. Eqs.
\eqref{eq:twisted-trace-conjugation} and
\eqref{eq:onsite-twisted-factorization} would, therefore, imply
\begin{equation}
    \om_{S_f}
    (
        P_i^{(a)}P_j^{(a)}
    )
    =
    \om_{S_f}
    (
        P_i^{(a)}
    )
    \om_{S_f}
    (
        P_j^{(a)}
    ).
\end{equation}
Using~\eqref{eq:Sf-long-range-correlation}, this requires
\begin{equation}\label{pa=pa^2}
    \frac{n_f(a)^{|\La_0|}}
    {\sum_{b\in A}n_f(b)^{|\La_0|}}=\left(\frac{n_f(a)^{|\La_0|}}
    {\sum_{b\in A}n_f(b)^{|\La_0|}}\right)^2
    \implies
    \frac{n_f(a)^{|\La_0|}}
    {\sum_{b\in A}n_f(b)^{|\La_0|}} = 0\text{ or }1.
\end{equation}
Because $f$ is non-constant, there are at least two distinct elements ${a_1,a_2\in A}$ such that ${n_f(a_1)\neq 0}$ and ${n_f(a_2)\neq 0}$. Therefore, there exists an ${a\in A}$ for which
\begin{equation}
    \frac{n_f(a)^{|\La_0|}}
    {\sum_{b\in A}n_f(b)^{|\La_0|}} \neq 0\text{ or }1,
\end{equation}
and~\eqref{pa=pa^2} is a contradiction.
Hence, no QCA can conjugate a nontrivial
$S_f$ to an onsite operator.

\section{Pullback of the Postnikov cocycle}\label{app:pullback-beta-trivialization}

In this appendix, we prove that the pullback of the Postnikov cocycle ${\bt\in\cZ^3(G,A)}$ by the quotient homomorphism ${p\colon \t{G}\to G}$ from~\eqref{extensionofGNbt} is cohomologically trivial. 
Recall that the pullback $p^*$ induces a group homomorphism
\begin{equation}
    p^*\colon \cH^3(G,A) \to \cH^3(\t{G},A).
\end{equation}
In what follows, we will construct a 2-cochain ${\al\in\cC^2(\t{G},A)}$ satisfying ${\del\al=p^*\bt}$, which implies ${p^*[\bt]=[0]\in \cH^3(\t{G},A)}$.

Recall from Sec.~\ref{sec:central-2group-sym-ops} that every group element of $\t{G}$ can be written as ${(g,[f])}$ with ${g\in G}$ and ${[f]\in \mathrm{Map}(G,A)/A_\mathrm{const}}$, and the quotient homomorphism satisfies ${p(g,[f])=g}$. The multiplication law of $\t{G}$ is
\begin{equation}\label{eq:app-Gbeta-multiplication}
    (g,[f])(h,[s])
    =
    \left(
        gh,\,
        [b_{g,h}+f\triangleleft h+s]
    \right),
\end{equation}
where ${b_{g,h}(k)=\bt(g,h,k)}$ and ${(f\triangleleft h)(k) = f(hk)}$.

Consider the 2-cochain ${\al\in \cC^2(\t{G},A)}$ satisfying
\begin{equation}\label{eq:app-alpha-def}
    \al\bigl((g,[f]),(h,[s])\bigr)
    =
    f(1)-f(h).
\end{equation}
This expression is independent of the representative $f$ of $[f]$ as it is invariant under ${f\mapsto f + c}$ with ${c\colon G\to A}$ a constant map. The coboundary of $\al$ satisfies
\begin{equation}
    \del\al(x,y,z)
    =
    \al(y,z)
    -\al(xy,z)
    +\al(x,yz)
    -\al(x,y),\qquad x,y,z\in \t{G}.
\end{equation}
Let
\begin{equation}
    x=(g,[f]),
    \qquad
    y=(h,[s]),
    \qquad
    z=(k,[r]).
\end{equation}
Then, defining ${q = b_{g,h}+f\triangleleft h+s}$ and using the definition of $\al$ and the group law~\eqref{eq:app-Gbeta-multiplication}, 
\begin{align}
    \del\al(x,y,z)
    &=
    s(1) - s(k)
    -
    (q(1) - q(k))
    +
    f(1) - f(hk)
    -(f(1) - f(h))\\
    &= \bt(g,h,k),
\end{align}
where we used that $\bt$ is a normalized 3-cocycle.
Since
\begin{equation}
    (p^*\bt)(x,y,z)
    =
    \bt\big(p(x),p(y),p(z)\big)
    =
    \bt(g,h,k),
\end{equation}
we conclude that ${\del\al = p^*\bt}$. Therefore, ${p^*[\bt]=[0]}$.

\section{Defect-sector symmetry algebra from \texorpdfstring{$F$}{F}-moves}\label{sec:defectSectorFusionApp}

In this Appendix, we consider a central 2-group symmetry ${\G=(G,A,[\bt])}$ and show how the Postnikov class modifies the $G$ 0-form symmetry fusion rule in the presence of a $G$ 0-form symmetry defect. In particular, we will show that in the presence of a ${k\in G}$ symmetry defect, the fusion of 0-form symmetry defects ${g,h\in C_G(k)}$ yields the 0-form symmetry defect $gh$ as well as a $1$-form symmetry defect ${\al_k(g,h)\in A}$.
We take Euclidean time to point upward and adopt the convention that a right-oriented horizontal defect labeled by $g$ corresponds to a ${g}$ symmetry operator upon quantization.
Then, graphically, in ${1+1}$d Euclidean spacetime:\footnote{The results of the Appendix apply to central 2-group symmetries in general spacetime dimension. We draw the defect networks in ${1+1}$d for simplicity.}
\begin{equation}\label{Postnikovfromfusion}
    \begin{tikzpicture}[decoration={markings, mark=at position 0.55 with {\arrow{>}}}, scale=1.5]
        \draw[postaction=decorate,color=black] (0, 0) -- (0, 0.866025) node[midway, above left, color=black] {$k$};
        \draw[postaction=decorate,color=black] (0, 0.866025) -- (0, 2*0.866025) node[midway, above left, color=black] {$k$};
        \draw[postaction=decorate,color=black] (0, 2*0.866025) -- (0, 3*0.866025) node[midway, above left, color=black] {$k$};
        \draw[postaction=decorate,color=black] (-1.5, 0.866025) -- (0, 0.866025) node[midway, above left, color=black] {$h$}; 
        \draw[postaction=decorate,color=black] (0, 0.866025) -- (1.5, 0.866025) node[midway, above right, color=black] {$h$}; 
        \draw[postaction=decorate,color=black] (-1.5, 2*
        0.866025) -- (0, 2*0.866025) node[midway, above left, color=black] {$g$};
        \draw[postaction=decorate,color=black] (0, 2*
        0.866025) -- (1.5, 2*0.866025) node[midway, above right, color=black] {$g$}; 
    \end{tikzpicture} 
    \quad = \quad
    \begin{tikzpicture}[decoration={markings, mark=at position 0.55 with {\arrow{>}}}, scale=1.5]
        \draw[postaction=decorate,color=black] (0, 0) -- (0, 1.5*0.866025) node[midway, above left, color=black] {$k$};
        \draw[postaction=decorate,color=black] (0, 1.5*0.866025) -- (0, 3*0.866025) node[midway, above left, color=black] {$k$};
        \draw[postaction=decorate,color=black] (-1.5, 1.5*0.866025) -- (0, 1.5*0.866025) node[midway, above left, color=black] {$gh$}; 
        \draw[postaction=decorate,color=black] (0, 1.5*0.866025) -- (1.5, 1.5*0.866025) node[midway, above right, color=black] {$gh$}; 
        \fill[black] (-0.75,2.5*0.866025) circle (1.75pt) node[above left, color=black] {$\al_k(g,h)$};
    \end{tikzpicture} \,.
\end{equation}
We show that ${\al_k\colon C_G(k)\times C_G(k)\to A}$ is a 2-cocycle with cohomology class ${[\al_k] = [\iota_k\bt]}$, where the slant product of $\bt$ with $k$ satisfies
\begin{equation}
    \iota_k\bt(g,h) = \bt(k,g,h) - \bt(g,k,h) + \bt(g,h,k).
\end{equation}
(The diagrammatic calculus of $\G$ we use applies straightforwardly to the fusion category $\Vec_G^{\om}$ with ${[\om]\in \cH^3(G,\Uone)}$. Thus, the result of this Appendix implies that the 't Hooft anomaly $[\om]$ of a $G$ 0-form symmetry in ${1+1}$d manifests as a projective representation $[\iota_k\om]$ of $C_G(k)$ in the $k$ symmetry defect Hilbert space. This is consistent with an $[\om]$ $G$-SPT in ${2+1}$d becoming an $[\iota_k\om]$ $C_G(k)$-SPT when compactified on a $k$ symmetry flux~\cite{T170609769}.)

Categorically, a central 2-group ${\G}$ is equivalent to a skeletal grouplike-monoidal
groupoid with objects labeled by ${g\in G}$ and morphism sets 
\begin{equation}
    \Hom_\G(g,h) = \begin{cases}
        \emptyset \quad &\text{if }g\neq h,\\
        A \quad &\text{if }g = h.
    \end{cases}
\end{equation}
Both composition
of morphisms and the monoidal product of morphisms are given by addition in $A$, while the monoidal product of objects is ${g\otimes h=gh}$. For each ${g,h\in G}$, we choose a fusion junction ${v_{g,h}\in \Hom_\G(g\otimes h, gh)\cong A}$. Within the graphical calculus of $\G$, $v_{g,h}$ is represented as  
\begin{equation}
    \begin{tikzpicture}[decoration={markings, mark=at position 0.55 with {\arrow{>}}}, scale=1.5]
        \draw[postaction=decorate,color=black] (0.5, 0.866025) -- (0.5, 2*0.866025) node[midway, left, color=black] {$gh$};
        \draw[postaction=decorate,color=black] (0, 0) -- (0.5, 0.866025) node[midway, left, color=black] {$g\,$};
        \draw[postaction=decorate,color=black] (2*0.5, 0) -- (0.5, 0.866025) node[midway, right, color=black] {$\,h$};
        \node[draw, fill=white, inner sep=1.5pt, text height=1.5ex] at (0.5, 0.866025){$v_{g,h}$};
    \end{tikzpicture}.
\end{equation}
Every other fusion junction in ${\Hom_\G(g\otimes h, gh)}$ is uniquely of the form ${a\otimes v_{g,h} = a + v_{g,h}}$ for some ${a\in A \cong \Hom_\G(1,1)}$. Within the graphical calculus of $\G$, we represent each ${a\in \Hom_\G(1,1)}$ by
\begin{equation}
    \begin{tikzpicture}[decoration={markings, mark=at position 0.55 with {\arrow{>}}}, scale=1.5]
        \draw[postaction=decorate,color=black] (0.5, 0.866025) -- (0.5, 2*0.866025) node[midway, left, color=black] {$1$};
        \draw[postaction=decorate,color=black] (0.5, 0) -- (0.5, 0.866025) node[midway, left, color=black] {$1$};
        \node[draw, fill=white, inner sep=1.5pt, text height=1.5ex] at (0.5, 0.866025){$\,a_{\,}$};
    \end{tikzpicture}
    \quad\equiv\quad
    \begin{tikzpicture}[decoration={markings, mark=at position 0.55 with {\arrow{>}}}, scale=1.5]
        \fill[black] (0.5, 0.866025) circle (1.75pt) node[above left, color=black] {$a$};
    \end{tikzpicture}\,.
\end{equation}
Relative to the choice of $\{v_{g,h}\}_{g,h\in G}$, the associator of $\G$ is represented by an
$F$-move:
\begin{equation}\label{AFmove}
    \begin{tikzpicture}[decoration={markings, mark=at position 0.55 with {\arrow{>}}}, scale=1.5]
        \draw[postaction=decorate,color=black] (0, 0) -- (2*0.5, 2*0.866025) node[midway, above left, color=black] {$g$}; 
        \draw[postaction=decorate,color=black] (2*0.5, 2*0.866025) -- (3*0.5, 3*0.866025) node[midway, above left, color=black] {$ghk$}; 
        \draw[postaction=decorate,color=black] (1, 0) -- (3*0.5, 0.866025) node[midway, left, color=black] {$h\,$};
        \draw[postaction=decorate,color=black] (2, 0) -- (3*0.5, 0.866025) node[midway, right, color=black] {$\,k$}; 
        \draw[postaction=decorate,color=black] (3*0.5, 0.866025) -- (2*0.5, 2*0.866025) node[midway, right, color=black] {$\,hk$}; 
        \node[draw, fill=white, inner sep=1.5pt, text height=1.5ex] at (3*0.5, 0.866025) {$v_{h,k}$};
        \node[draw, fill=white, inner sep=1.5pt, text height=1.5ex] at (2*0.5, 2*0.866025){$v_{g,hk}$};
    \end{tikzpicture}
    \quad = \quad 
    \begin{tikzpicture}[decoration={markings, mark=at position 0.55 with {\arrow{>}}}, scale=1.5]
        \draw[postaction=decorate,color=black] (0, 0) -- (0.5, 0.866025) node[midway, left, color=black] {$g\,$}; 
        \draw[postaction=decorate,color=black] (0.5, 0.866025) -- (2*0.5, 2*0.866025) node[midway, left, color=black] {$gh\,$}; 
        \draw[postaction=decorate,color=black] (2*0.5, 2*0.866025) -- (3*0.5, 3*0.866025) node[midway, above left, color=black] {$ghk$}; 
        \draw[postaction=decorate,color=black] (1, 0) -- (0.5, 0.866025) node[midway, right, color=black] {$\,h$};
        \draw[postaction=decorate,color=black] (2, 0) -- (2*0.5, 2*0.866025) node[midway, above right, color=black] {$k$}; 
        \node[draw, fill=white, inner sep=1.5pt, text height=1.5ex] at (0.5, 0.866025){$v_{g,h}$};
        \node[draw, fill=white, inner sep=1.5pt, text height=1.5ex] at (2*0.5, 2*0.866025){$v_{gh,k}$};
        \fill[black] (0,2*0.866025) circle (1.75pt) node[above left, color=black] {$\bt(g,h,k)$};
    \end{tikzpicture} 
     .
\end{equation}

To derive~\eqref{Postnikovfromfusion}, we must first place the left-hand side of the equation into the graphical calculus of $\G$. We follow the convention that
\begin{equation}\label{defectnetGraphCalc}
    \begin{tikzpicture}[decoration={markings, mark=at position 0.55 with {\arrow{>}}}, scale=1.5]
        \draw[postaction=decorate,color=black] (0, 0) -- (0, 0.866025) node[midway, above left, color=black] {$k$};
        \draw[postaction=decorate,color=black] (0, 0.866025) -- (0, 2*0.866025) node[midway, above left, color=black] {$k$};
        \draw[postaction=decorate,color=black] (0, 2*0.866025) -- (0, 3*0.866025) node[midway, above left, color=black] {$k$};
        \draw[postaction=decorate,color=black] (-1.5, 0.866025) -- (0, 0.866025) node[midway, above left, color=black] {$h$}; 
        \draw[postaction=decorate,color=black] (0, 0.866025) -- (1.5, 0.866025) node[midway, above right, color=black] {$h$}; 
        \draw[postaction=decorate,color=black] (-1.5, 2*
        0.866025) -- (0, 2*0.866025) node[midway, above left, color=black] {$g$};
        \draw[postaction=decorate,color=black] (0, 2*
        0.866025) -- (1.5, 2*0.866025) node[midway, above right, color=black] {$g$}; 
    \end{tikzpicture} 
    \quad \equiv \quad
    \begin{tikzpicture}[decoration={markings, mark=at position 0.55 with {\arrow{>}}}, scale=1.5]
        \draw[postaction=decorate,color=black] (0, 0) -- (3*0.5, 3*0.866025) node[midway, above left, color=black] {$g$}; 
        \draw[postaction=decorate,color=black] (1, 0) -- (3*0.5, 0.866025) node[midway, left, color=black] {$h\,$};
        \draw[postaction=decorate,color=black] (2, 0) -- (3*0.5, 0.866025) node[midway, right, color=black] {$\,k$}; 
        \draw[postaction=decorate,color=black] (3*0.5, 0.866025) -- (4*0.5, 2*0.866025) node[midway, left, color=black] {$hk\,$};
        \draw[postaction=decorate,color=black] (4*0.5, 2*0.866025) -- (3, 0) node[midway, above right, color=black] {$h$}; 
        \draw[postaction=decorate,color=black] (4*0.5, 2*0.866025) -- (3*0.5, 3*0.866025) node[midway, right, color=black] {$\,k$}; 
        \draw[postaction=decorate,color=black] (3*0.5, 3*0.866025) -- (4*0.5, 4*0.866025) node[midway, left, color=black] {$gk\,$}; 
        \draw[postaction=decorate,color=black] (4*0.5, 4*0.866025) -- (8*0.5, 0) node[midway, above right, color=black] {$g$}; 
        \draw[postaction=decorate,color=black] (4*0.5, 4*0.866025) -- (3*0.5, 5*0.866025) node[midway, above right, color=black] {$k$}; 
        \node[draw, fill=white, inner sep=1.5pt, text height=1.5ex] at (3*0.5, 0.866025){$v_{h,k}$};
        \node[draw, fill=white, inner sep=1.5pt, text height=1.5ex] at (4*0.5, 2*0.866025) {$v_{hk,h^{-1}}$};
        \node[draw, fill=white, inner sep=1.5pt, text height=1.5ex] at (3*0.5, 3*0.866025){$v_{g,k}$};
        \node[draw, fill=white, inner sep=1.5pt, text height=1.5ex] at (4*0.5, 4*0.866025){$v_{gk,g^{-1}}$};
    \end{tikzpicture}.
\end{equation}
We assume all defect networks are on a cylinder and the two ${g}$ lines are connected through the compact direction of the cylinder, and likewise for the two ${h}$ lines.
Reversing the orientations of the rightmost $g$ and $h$ lines, and then performing the $F$-move~\eqref{AFmove} three times, changes the right-hand side of~\eqref{defectnetGraphCalc} to
\begin{equation}\label{fusiontreepostFmove}
    \begin{tikzpicture}[decoration={markings, mark=at position 0.55 with {\arrow{>}}}, scale=1.5]
        \draw[postaction=decorate,color=black] (0, 0) -- (0.5, 0.866025) node[midway, left, color=black] {$g\,$}; 
        \draw[postaction=decorate,color=black] (1, 0) -- (0.5, 0.866025) node[midway, right, color=black] {$\,h$};
        \draw[postaction=decorate,color=black] (2, 0) -- (2*0.5, 2*0.866025) node[midway, above right, color=black] {$k$};
        \draw[postaction=decorate,color=black] (0.5, 0.866025) -- (2*0.5, 2*0.866025) node[midway, left, color=black] {$gh\,$};
        \draw[postaction=decorate,color=black] (6*0.5, 0) -- (7*0.5, 0.866025) node[midway, left, color=black] {$h^{-1}$}; 
        \draw[postaction=decorate,color=black] (2*0.5, 2*0.866025) -- (4*0.5, 4*0.866025) node[midway, above left, color=black] {$ghk$}; 
        \draw[postaction=decorate,color=black] (8*0.5, 0) -- (7*0.5, 0.866025) node[midway, right, color=black] {$\,g^{-1}$}; 
        \draw[postaction=decorate,color=black] (7*0.5, 0.866025) -- (4*0.5, 4*0.866025) node[midway, above right, color=black] {$h^{-1}g^{-1}$}; 
        \draw[postaction=decorate,color=black] (4*0.5, 4*0.866025) -- (3*0.5, 5*0.866025) node[midway, above right, color=black] {$k$}; 
        \fill[white] (0,3*0.866025) circle (1.75pt) node[above left, color=black] {$\bt(g,hk,h^{-1})-\bt(ghk,h^{-1},g^{-1})~$};
        \fill[black] (0,3*0.866025) circle (1.75pt) node[below left, color=black] {$+\bt(g,h,k)~$};
        \node[draw, fill=white, inner sep=1.5pt, text height=1.5ex] at (4*0.5, 4*0.866025){$v_{ghk,h^{-1}g^{-1}}$};
        \node[draw, fill=white, inner sep=1.5pt, text height=1.5ex] at (0.5, 0.866025){$v_{g,h}$};
        \node[draw, fill=white, inner sep=1.5pt, text height=1.5ex] at (7*0.5, 0.866025) {$v_{h^{-1},g^{-1}}$};
        \node[draw, fill=white, inner sep=1.5pt, text height=1.5ex] at (2*0.5, 2*0.866025) {$v_{gh,k}$};
    \end{tikzpicture}.
\end{equation}
We next reverse the orientation of the rightmost segment of~\eqref{fusiontreepostFmove}. This changes the fusion junction ${v_{h^{-1},g^{-1}}}$ to a splitting junction, which we denote by ${\bar{v}_{h^{-1},g^{-1}}\in \Hom_\G(gh, g\otimes h)}$. This causes~\eqref{fusiontreepostFmove} to become
\begin{equation}\label{fusiontreepostFmove2}
    \begin{tikzpicture}[decoration={markings, mark=at position 0.55 with {\arrow{>}}}, scale=1.5]
        \draw[postaction=decorate,color=black] (0, 0) -- (0.5, 0.866025) node[midway, left, color=black] {$g\,$}; 
        \draw[postaction=decorate,color=black] (1, 0) -- (0.5, 0.866025) node[midway, right, color=black] {$\,h$};
        \draw[postaction=decorate,color=black] (2, 0) -- (2*0.5, 2*0.866025) node[midway, above right, color=black] {$k$};
        \draw[postaction=decorate,color=black] (0.5, 0.866025) -- (2*0.5, 2*0.866025) node[midway, left, color=black] {$gh\,$};
        \draw[postaction=decorate,color=black] (7*0.5, 0.866025) -- (6*0.5, 0) node[midway, left, color=black] {$h\,$}; 
        \draw[postaction=decorate,color=black] (2*0.5, 2*0.866025) -- (4*0.5, 4*0.866025) node[midway, above left, color=black] {$ghk$}; 
        \draw[postaction=decorate,color=black] (7*0.5, 0.866025) -- (8*0.5, 0) node[midway, right, color=black] {$\,g$}; 
        \draw[postaction=decorate,color=black] (4*0.5, 4*0.866025) -- (7*0.5, 0.866025) node[midway, above right, color=black] {$gh$}; 
        \draw[postaction=decorate,color=black] (4*0.5, 4*0.866025) -- (3*0.5, 5*0.866025) node[midway, above right, color=black] {$k$}; 
        \fill[white] (0,3*0.866025) circle (1.75pt) node[above left, color=black] {$\bt(g,hk,h^{-1}) - \bt(ghk,h^{-1},g^{-1})~$};
        \fill[black] (0,3*0.866025) circle (1.75pt) node[below left, color=black] {$+\bt(g,h,k)~$};
        \node[draw, fill=white, inner sep=1.5pt, text height=1.5ex] at (4*0.5, 4*0.866025){$v_{ghk,(gh)^{-1}}$};
        \node[draw, fill=white, inner sep=1.5pt, text height=1.5ex] at (0.5, 0.866025){$v_{g,h}$};
        \node[draw, fill=white, inner sep=1.5pt, text height=1.5ex] at (7*0.5, 0.866025) {$\bar{v}_{h^{-1},g^{-1}}$};
        \node[draw, fill=white, inner sep=1.5pt, text height=1.5ex] at (2*0.5, 2*0.866025) {$v_{gh,k}$};
    \end{tikzpicture}.
\end{equation}

It remains to eliminate the leftmost and rightmost $g$ and $h$ lines. Since the lines are connected via the compact direction of the cylinder, we can do so using the relation
\begin{equation}\label{bubblepop}
    \begin{tikzpicture}[decoration={markings, mark=at position 0.55 with {\arrow{>}}}, scale=1.5]
        \draw[postaction=decorate, color=black]
        (0.5,0) to[bend left=80] node[midway, left, color=black] {$g$} (0.5,0.866025);
        \draw[postaction=decorate, color=black]
        (0.5,0) to[bend right=80] node[midway, right, color=black] {$h$} (0.5,0.866025);
        \draw[postaction=decorate,color=black] (0.5, 0.866025) -- (0.5, 2*0.866025) node[midway, above left, color=black] {$gh$};
        \draw[postaction=decorate,color=black] (0.5, -0.866025) -- (0.5, 0) node[midway, below left, color=black] {$gh$};
        \node[draw, fill=white, inner sep=1.5pt, text height=1.5ex] at (0.5, 0.866025){$v_{g,h}$};
        \node[draw, fill=white, inner sep=1.5pt, text height=2ex] at (0.5, 0){$v^{-1}_{g,h}$};
    \end{tikzpicture}
    \quad = \quad
    \begin{tikzpicture}[decoration={markings, mark=at position 0.55 with {\arrow{>}}}, scale=1.5] 
        \draw[postaction=decorate,color=black] (0.5, -0.866025) -- (0.5, 2*0.866025) node[midway, above left, color=black] {$gh$};
    \end{tikzpicture}~.
\end{equation}
However, it remains to find how the splitting junction $\bar{v}_{h^{-1},g^{-1}}$ is related to the splitting junction $v_{g,h}^{-1}$. 
Consider the injective map ${f\colon \Hom_\G(gh,g\otimes h)\to \Hom_\G(g\otimes h\otimes h^{-1}\otimes g^{-1}, 1)}$ defined graphically by
\begin{equation}
    f(s) = \begin{tikzpicture}[decoration={markings, mark=at position 0.55 with {\arrow{>}}}, scale=1.5]
        \draw[postaction=decorate,color=black] (0, 0) -- (0.5, 0.866025) node[midway, left, color=black] {$g\,$}; 
        \draw[postaction=decorate,color=black] (1, 0) -- (0.5, 0.866025) node[midway, right, color=black] {$\,h$};
        \draw[postaction=decorate,color=black] (0.5, 0.866025) -- (0.5, 2*0.866025) node[midway, left, color=black] {$gh\,$};
        \draw[postaction=decorate,color=black] (5*0.5, 0) -- (3*0.5, 3*0.866025) node[midway, left, color=black] {$h^{-1}$};
        \draw[postaction=decorate,color=black] (7*0.5, 0) -- (4*0.5, 4*0.866025) node[midway, above right, color=black] {$g^{-1}$};
        \draw[postaction=decorate,color=black] (0.5, 2*0.866025) -- (3*0.5, 3*0.866025) node[midway, below, color=black] {$h$};
        \draw[postaction=decorate,color=black] (3*0.5, 3*0.866025) -- (4*0.5, 4*0.866025)  node[midway, left, color=black] {$1\,$};
        \draw[postaction=decorate,color=black] (0.5, 2*0.866025) -- (3*0.5, 5*0.866025) node[midway, above left, color=black] {$g$};
        \draw[postaction=decorate,color=black] (4*0.5, 4*0.866025) -- (3*0.5, 5*0.866025) node[midway, right, color=black] {$\,g^{-1}$};
        \draw[postaction=decorate,color=black] (3*0.5, 5*0.866025) -- (3*0.5, 6*0.866025) node[midway, above right, color=black] {$1$};
        \node[draw, fill=white, inner sep=1.5pt, text height=1.5ex] at (0.5, 0.866025){$v_{g,h}$};
        \node[draw, fill=white, inner sep=1.5pt, text height=1.5ex] at (0.5, 2*0.866025){$\,s_{\,}$};
        \node[draw, fill=white, inner sep=1.5pt, text height=1.5ex] at (3*0.5, 3*0.866025){$v_{h,h^{-1}}$};
        \node[draw, fill=white, inner sep=1.5pt, text height=1.5ex] at (4*0.5, 4*0.866025){$v_{1,g^{-1}}$};
        \node[draw, fill=white, inner sep=1.5pt, text height=1.5ex] at (3*0.5, 5*0.866025){$v_{g,g^{-1}}$};
    \end{tikzpicture}.
\end{equation}
Note that $\bar{v}_{h^{-1}, g^{-1}}$ can be written as
\begin{equation}
    \begin{tikzpicture}[decoration={markings, mark=at position 0.55 with {\arrow{>}}}, scale=1.5]
        \draw[postaction=decorate,color=black] (0.5, 0) -- (0.5, 0.866025) node[midway, left, color=black] {$gh$};
        \draw[postaction=decorate,color=black] (0.5, 0.866025) -- (0, 2*0.866025) node[midway, left, color=black] {$g\,$};
        \draw[postaction=decorate,color=black] (0.5, 0.866025) -- (2*0.5, 2*0.866025) node[midway, right, color=black] {$\,h$};
        \node[draw, fill=white, inner sep=1.5pt, text height=1.5ex] at (0.5, 0.866025){$\bar{v}_{h^{-1}, g^{-1}}$};
    \end{tikzpicture}
    \quad = \quad
    \begin{tikzpicture}[decoration={markings, mark=at position 0.55 with {\arrow{>}}}, scale=1.5]
        \draw[postaction=decorate,color=black] (0, 0.9) -- (0, 3*0.866025) node[midway, above left, color=black] {$gh$};
        \draw[postaction=decorate, color=black]
        (0, 3*0.866025) to[bend left=90] node[midway, above, color=black] {$gh$} (4*0.5, 3*0.866025);
        \draw[postaction=decorate,color=black] (4*0.5, 3*0.866025) -- (3*0.5, 2*0.866025)  node[midway, left, color=black] {$h\,$};
        \draw[postaction=decorate,color=black] (4*0.5, 3*0.866025) -- (5*0.5, 2*0.866025)  node[midway, right, color=black] {$\,g$};
        \draw[postaction=decorate,color=black] (6*0.5, 2*0.866025) -- (6*0.5, 4*0.866025)  node[midway, right, color=black] {$\,g$};
        \draw[postaction=decorate,color=black] (8*0.5, 2*0.866025) -- (8*0.5, 4*0.866025)  node[midway, right, color=black] {$\,h$};
        \draw[postaction=decorate, color=black]
        (5*0.5, 2*0.866025) to[out=-60, in=-90, looseness=1.2](6*0.5, 2*0.866025);
        \draw[postaction=decorate, color=black]
        (3*0.5, 2*0.866025) to[out=-120, in=-90, looseness=1.2] (8*0.5, 2*0.866025);
        \node[draw, fill=white, inner sep=1.5pt, text height=1.5ex] at (4*0.5, 3*0.866025){$v_{h^{-1}, g^{-1}}$};
    \end{tikzpicture}
    .
\end{equation}
Using this, we can simplify $f(\bar{v}_{h^{-1}, g^{-1}})$ to
\begin{equation}\label{forienv}
    f(\bar{v}_{h^{-1}, g^{-1}}) = \begin{tikzpicture}[decoration={markings, mark=at position 0.55 with {\arrow{>}}}, scale=1.5]
        \draw[postaction=decorate,color=black] (1, 2*0.866025) -- (0.5, 3*0.866025) node[midway, right, color=black] {$\,h$};
        \draw[postaction=decorate,color=black] (4*0.5, 2*0.866025) -- (5*0.5, 3*0.866025) node[midway, left, color=black] {$h^{-1}$};
        \draw[postaction=decorate,color=black] (6*0.5, 2*0.866025) -- (5*0.5, 3*0.866025) node[midway, right, color=black] {$\,g^{-1}$};
        \draw[postaction=decorate,color=black] (0, 2*0.866025) -- (1*0.5, 3*0.866025) node[midway, left, color=black] {$g\,$};
        \draw[postaction=decorate,color=black] (5*0.5, 3*0.866025) -- (3*0.5, 5*0.866025) node[midway, above right, color=black] {$h^{-1}g^{-1}$};
        \draw[postaction=decorate,color=black] (1*0.5, 3*0.866025) -- (3*0.5, 5*0.866025) node[midway, above left, color=black] {$gh$};
        \draw[postaction=decorate,color=black] (3*0.5, 5*0.866025) -- (3*0.5, 6*0.866025) node[midway, above left, color=black] {$1$};
        \node[draw, fill=white, inner sep=1.5pt, text height=1.5ex] at (1*0.5, 3*0.866025){$v_{g,h}$};
        \node[draw, fill=white, inner sep=1.5pt, text height=1.5ex] at (5*0.5, 3*0.866025){$v_{h^{-1},g^{-1}}$};
        \node[draw, fill=white, inner sep=1.5pt, text height=1.5ex] at (3*0.5, 5*0.866025){$v_{gh,h^{-1}g^{-1}}$};
    \end{tikzpicture}.
\end{equation}
On the other hand, using that $v_{g,h}^{-1}$ is the inverse of $v_{g,h}$, $f(v_{g,h}^{-1})$ can be simplified to
\begin{equation}\label{finvv}
    f(v_{g,h}^{-1}) = \begin{tikzpicture}[decoration={markings, mark=at position 0.55 with {\arrow{>}}}, scale=1.5]
        \draw[postaction=decorate,color=black] (1, 2*0.866025) -- (3*0.5, 3*0.866025) node[midway, left, color=black] {$h\,$};
        \draw[postaction=decorate,color=black] (4*0.5, 2*0.866025) -- (3*0.5, 3*0.866025) node[midway, right, color=black] {$\,h^{-1}$};
        \draw[postaction=decorate,color=black] (6*0.5, 2*0.866025) -- (4*0.5, 4*0.866025) node[midway, above right, color=black] {$g^{-1}$};
        \draw[postaction=decorate,color=black] (3*0.5, 3*0.866025) -- (4*0.5, 4*0.866025)  node[midway, left, color=black] {$1\,$};
        \draw[postaction=decorate,color=black] (0, 2*0.866025) -- (3*0.5, 5*0.866025) node[midway, above left, color=black] {$g$};
        \draw[postaction=decorate,color=black] (4*0.5, 4*0.866025) -- (3*0.5, 5*0.866025) node[midway, right, color=black] {$\,g^{-1}$};
        \draw[postaction=decorate,color=black] (3*0.5, 5*0.866025) -- (3*0.5, 6*0.866025) node[midway, above right, color=black] {$1$};
        \node[draw, fill=white, inner sep=1.5pt, text height=1.5ex] at (3*0.5, 3*0.866025){$v_{h,h^{-1}}$};
        \node[draw, fill=white, inner sep=1.5pt, text height=1.5ex] at (4*0.5, 4*0.866025){$v_{1,g^{-1}}$};
        \node[draw, fill=white, inner sep=1.5pt, text height=1.5ex] at (3*0.5, 5*0.866025){$v_{g,g^{-1}}$};
    \end{tikzpicture}.
\end{equation}
The fusion trees~\eqref{forienv} and~\eqref{finvv} are related by two $F$-moves:
\begin{equation}
    \begin{tikzpicture}[decoration={markings, mark=at position 0.55 with {\arrow{>}}}, scale=1.5]
        \draw[postaction=decorate,color=black] (1, 2*0.866025) -- (0.5, 3*0.866025) node[midway, right, color=black] {$\,h$};
        \draw[postaction=decorate,color=black] (4*0.5, 2*0.866025) -- (5*0.5, 3*0.866025) node[midway, left, color=black] {$h^{-1}$};
        \draw[postaction=decorate,color=black] (6*0.5, 2*0.866025) -- (5*0.5, 3*0.866025) node[midway, right, color=black] {$\,g^{-1}$};
        \draw[postaction=decorate,color=black] (0, 2*0.866025) -- (1*0.5, 3*0.866025) node[midway, left, color=black] {$g\,$};
        \draw[postaction=decorate,color=black] (5*0.5, 3*0.866025) -- (3*0.5, 5*0.866025) node[midway, above right, color=black] {$h^{-1}g^{-1}$};
        \draw[postaction=decorate,color=black] (1*0.5, 3*0.866025) -- (3*0.5, 5*0.866025) node[midway, above left, color=black] {$gh$};
        \draw[postaction=decorate,color=black] (3*0.5, 5*0.866025) -- (3*0.5, 6*0.866025) node[midway, above left, color=black] {$1$};
        \node[draw, fill=white, inner sep=1.5pt, text height=1.5ex] at (1*0.5, 3*0.866025){$v_{g,h}$};
        \node[draw, fill=white, inner sep=1.5pt, text height=1.5ex] at (5*0.5, 3*0.866025){$v_{h^{-1},g^{-1}}$};
        \node[draw, fill=white, inner sep=1.5pt, text height=1.5ex] at (3*0.5, 5*0.866025){$v_{gh,h^{-1}g^{-1}}$};
    \end{tikzpicture}
    \quad = \quad
    \begin{tikzpicture}[decoration={markings, mark=at position 0.55 with {\arrow{>}}}, scale=1.5]
        \draw[postaction=decorate,color=black] (1, 2*0.866025) -- (3*0.5, 3*0.866025) node[midway, left, color=black] {$h\,$};
        \draw[postaction=decorate,color=black] (4*0.5, 2*0.866025) -- (3*0.5, 3*0.866025) node[midway, right, color=black] {$\,h^{-1}$};
        \draw[postaction=decorate,color=black] (6*0.5, 2*0.866025) -- (4*0.5, 4*0.866025) node[midway, above right, color=black] {$g^{-1}$};
        \draw[postaction=decorate,color=black] (3*0.5, 3*0.866025) -- (4*0.5, 4*0.866025)  node[midway, left, color=black] {$1\,$};
        \draw[postaction=decorate,color=black] (0, 2*0.866025) -- (3*0.5, 5*0.866025) node[midway, above left, color=black] {$g$};
        \draw[postaction=decorate,color=black] (4*0.5, 4*0.866025) -- (3*0.5, 5*0.866025) node[midway, right, color=black] {$\,g^{-1}$};
        \draw[postaction=decorate,color=black] (3*0.5, 5*0.866025) -- (3*0.5, 6*0.866025) node[midway, above right, color=black] {$1$};
        \node[draw, fill=white, inner sep=1.5pt, text height=1.5ex] at (3*0.5, 3*0.866025){$v_{h,h^{-1}}$};
        \node[draw, fill=white, inner sep=1.5pt, text height=1.5ex] at (4*0.5, 4*0.866025){$v_{1,g^{-1}}$};
        \node[draw, fill=white, inner sep=1.5pt, text height=1.5ex] at (3*0.5, 5*0.866025){$v_{g,g^{-1}}$};
        \fill[white] (0.1, 5*0.866025) circle (1.75pt) node[above left, color=black] {$\bt(h,h^{-1},g^{-1})$};
        \fill[black] (0.5, 5*0.866025) circle (1.75pt) node[below left, color=black] {$- \bt(g,h,h^{-1}g^{-1})~$};
    \end{tikzpicture}.
\end{equation}
Because $f$ is injective, this implies that
\begin{equation}\label{splitjunctrelation}
    \begin{tikzpicture}[decoration={markings, mark=at position 0.55 with {\arrow{>}}}, scale=1.5]
        \draw[postaction=decorate,color=black] (0.5, 0) -- (0.5, 0.866025) node[midway, left, color=black] {$gh$};
        \draw[postaction=decorate,color=black] (0.5, 0.866025) -- (0, 2*0.866025) node[midway, left, color=black] {$g\,$};
        \draw[postaction=decorate,color=black] (0.5, 0.866025) -- (2*0.5, 2*0.866025) node[midway, right, color=black] {$\,h$};
        \node[draw, fill=white, inner sep=1.5pt, text height=1.5ex] at (0.5, 0.866025){$\bar{v}_{h^{-1}, g^{-1}}$};
    \end{tikzpicture}
    \quad = \quad
    \begin{tikzpicture}[decoration={markings, mark=at position 0.55 with {\arrow{>}}}, scale=1.5]
        \draw[postaction=decorate,color=black] (0.5, 0) -- (0.5, 0.866025) node[midway, left, color=black] {$gh$};
        \draw[postaction=decorate,color=black] (0.5, 0.866025) -- (0, 2*0.866025) node[midway, left, color=black] {$g\,$};
        \draw[postaction=decorate,color=black] (0.5, 0.866025) -- (2*0.5, 2*0.866025) node[midway, right, color=black] {$\,h$};
        \node[draw, fill=white, inner sep=1.5pt, text height=2ex] at (0.5, 0.866025){$v^{-1}_{g,h}$};
        \fill[white] (-0.9, 0.866025) circle (1.75pt) node[above left, color=black] {$\bt(h,h^{-1},g^{-1})$};
        \fill[black] (-0.3, 0.866025) circle (1.75pt) node[below left, color=black] {$-\bt(g,h,h^{-1}g^{-1}) ~$};
    \end{tikzpicture}\,.
\end{equation}
Using the relation~\eqref{splitjunctrelation} and the identity~\eqref{bubblepop}, we can simplify the fusion tree~\eqref{fusiontreepostFmove2} to
\begin{equation}
    \begin{tikzpicture}[decoration={markings, mark=at position 0.55 with {\arrow{>}}}, scale=1.5]
        \draw[postaction=decorate,color=black] (3*0.5, 0.866025) -- (2*0.5, 2*0.866025) node[midway, above right, color=black] {$k$};
        \draw[postaction=decorate,color=black] (0.5, 0.866025) -- (2*0.5, 2*0.866025) node[midway, left, color=black] {$gh\,$};
        \draw[postaction=decorate,color=black] (2*0.5, 2*0.866025) -- (4*0.5, 4*0.866025) node[midway, above left, color=black] {$ghk$};  
        \draw[postaction=decorate,color=black] (4*0.5, 4*0.866025) -- (7*0.5, 0.866025) node[midway, above right, color=black] {$gh$}; 
        \draw[postaction=decorate,color=black] (4*0.5, 4*0.866025) -- (3*0.5, 5*0.866025) node[midway, above right, color=black] {$k$}; 
        \fill[black] (0.5,4*0.866025) circle (1.75pt) node[above left, color=black] {$\al_k(g,h)$};
        \node[draw, fill=white, inner sep=1.5pt, text height=1.5ex] at (4*0.5, 4*0.866025){$v_{ghk,(gh)^{-1}}$};
        \node[draw, fill=white, inner sep=1.5pt, text height=1.5ex] at (2*0.5, 2*0.866025) {$v_{gh,k}$};
    \end{tikzpicture}
    \quad \equiv \quad
    \begin{tikzpicture}[decoration={markings, mark=at position 0.55 with {\arrow{>}}}, scale=1.5]
        \draw[postaction=decorate,color=black] (0, 0) -- (0, 1.5*0.866025) node[midway, above left, color=black] {$k$};
        \draw[postaction=decorate,color=black] (0, 1.5*0.866025) -- (0, 3*0.866025) node[midway, above left, color=black] {$k$};
        \draw[postaction=decorate,color=black] (-1.5, 1.5*0.866025) -- (0, 1.5*0.866025) node[midway, above left, color=black] {$gh$}; 
        \draw[postaction=decorate,color=black] (0, 1.5*0.866025) -- (1.5, 1.5*0.866025) node[midway, above right, color=black] {$gh$}; 
        \fill[black] (-0.75,2.5*0.866025) circle (1.75pt) node[above left, color=black] {$\al_k(g,h)$};
    \end{tikzpicture}
    ~,
\end{equation}
where
\begin{equation}\label{alphakunsimplified}
    \al_k(g,h) = 
    \bt(g,h,k) -\bt(ghk,h^{-1},g^{-1})
    +\bt(g,hk,h^{-1}) 
    - \bt(g,h,h^{-1}g^{-1})
    + \bt(h,h^{-1},g^{-1}).
\end{equation}
This, of course, is the right-hand side of the equation~\eqref{Postnikovfromfusion}.

We next introduce the 1-cochain ${\mu_k\in\cC^1(G,A)}$ satisfying ${\mu_k(g) = -\bt(k, g, g^{-1})}$. Using the 3-cocycle condition for $\bt$,~\eqref{alphakunsimplified} can be simplified to
\begin{equation}
    \al_k(g,h) = \iota_k\bt(g,h) + \del \mu_k(g,h).
\end{equation}
Note that, since $\bt$ is a normalized 3-cocycle, ${\al_1(g,h) = 0}$, as expected.
Furthermore, the cohomology class
\begin{equation}
    [\al_k] = [\iota_k\bt],
\end{equation}
as claimed at the beginning of this appendix section.

\addcontentsline{toc}{section}{References}

\hypersetup{linkcolor=brn}

\bibliographystyle{ytphys}
\bibliography{local.bib} 

\providecommand{\href}[2]{#2}\begingroup\begin{thebibliography}{100}

\bibitem{GW14125148}
D.~Gaiotto, A.~Kapustin, N.~Seiberg, and B.~Willett, ``{Generalized Global Symmetries},'' \href{http://dx.doi.org/10.1007/JHEP02(2015)172}{{\em JHEP} {\bfseries 02} (2015) 172}, \href{http://arxiv.org/abs/1412.5148}{{\ttfamily arXiv:1412.5148 [hep-th]}}.

\bibitem{BT170402330}
L.~Bhardwaj and Y.~Tachikawa, ``{On finite symmetries and their gauging in two dimensions},'' \href{http://dx.doi.org/10.1007/JHEP03(2018)189}{{\em JHEP} {\bfseries 03} (2018) 189}, \href{http://arxiv.org/abs/1704.02330}{{\ttfamily arXiv:1704.02330 [hep-th]}}.

\bibitem{T171209542}
Y.~Tachikawa, ``{On gauging finite subgroups},'' \href{http://dx.doi.org/10.21468/SciPostPhys.8.1.015}{{\em SciPost Phys.} {\bfseries 8} no.~1, (2020) 015}, \href{http://arxiv.org/abs/1712.09542}{{\ttfamily arXiv:1712.09542 [hep-th]}}.

\bibitem{CLS180204445}
C.-M. Chang, Y.-H. Lin, S.-H. Shao, Y.~Wang, and X.~Yin, ``{Topological Defect Lines and Renormalization Group Flows in Two Dimensions},'' \href{http://dx.doi.org/10.1007/JHEP01(2019)026}{{\em JHEP} {\bfseries 01} (2019) 026}, \href{http://arxiv.org/abs/1802.04445}{{\ttfamily arXiv:1802.04445 [hep-th]}}.

\bibitem{TW191202817}
R.~Thorngren and Y.~Wang, ``{Fusion category symmetry. Part I. Anomaly in-flow and gapped phases},'' \href{http://dx.doi.org/10.1007/JHEP04(2024)132}{{\em JHEP} {\bfseries 04} (2024) 132}, \href{http://arxiv.org/abs/1912.02817}{{\ttfamily arXiv:1912.02817 [hep-th]}}.

\bibitem{KLW200514178}
L.~Kong, T.~Lan, X.-G. Wen, Z.-H. Zhang, and H.~Zheng, ``{Algebraic higher symmetry and categorical symmetry -- a holographic and entanglement view of symmetry},'' \href{http://dx.doi.org/10.1103/PhysRevResearch.2.043086}{{\em Phys. Rev. Res.} {\bfseries 2} no.~4, (2020) 043086}, \href{http://arxiv.org/abs/2005.14178}{{\ttfamily arXiv:2005.14178 [cond-mat.str-el]}}.

\bibitem{M220403045}
J.~McGreevy, ``{Generalized Symmetries in Condensed Matter},'' \href{http://dx.doi.org/10.1146/annurev-conmatphys-040721-021029}{{\em Ann. Rev. Condensed Matter Phys.} {\bfseries 14} (2023) 57--82}, \href{http://arxiv.org/abs/2204.03045}{{\ttfamily arXiv:2204.03045 [cond-mat.str-el]}}.

\bibitem{CDI220509545}
C.~C{\'o}rdova, T.~T. Dumitrescu, K.~Intriligator, and S.-H. Shao, ``{Snowmass White Paper: Generalized Symmetries in Quantum Field Theory and Beyond},'' in {\em {Snowmass 2021}}.
\newblock 5, 2022.
\newblock \href{http://arxiv.org/abs/2205.09545}{{\ttfamily arXiv:2205.09545 [hep-th]}}.

\bibitem{S230518296}
S.~Sch{\"a}fer-Nameki, ``{ICTP lectures on (non-)invertible generalized symmetries},'' \href{http://dx.doi.org/10.1016/j.physrep.2024.01.007}{{\em Phys. Rept.} {\bfseries 1063} (2024) 1--55}, \href{http://arxiv.org/abs/2305.18296}{{\ttfamily arXiv:2305.18296 [hep-th]}}.

\bibitem{S230800747}
S.-H. Shao, ``{What's Done Cannot Be Undone: TASI Lectures on Non-Invertible Symmetries},'' in {\em {Theoretical Advanced Study Institute in Elementary Particle Physics 2023}: {Aspects of Symmetry}}.
\newblock 8, 2023.
\newblock \href{http://arxiv.org/abs/2308.00747}{{\ttfamily arXiv:2308.00747 [hep-th]}}.

\bibitem{FTL0612341}
A.~Feiguin, S.~Trebst, A.~W.~W. Ludwig, M.~Troyer, A.~Kitaev, Z.~Wang, and M.~H. Freedman, ``{Interacting anyons in topological quantum liquids: The golden chain},'' \href{http://dx.doi.org/10.1103/PhysRevLett.98.160409}{{\em Phys. Rev. Lett.} {\bfseries 98} (2007) 160409}, \href{http://arxiv.org/abs/cond-mat/0612341}{{\ttfamily arXiv:cond-mat/0612341}}.

\bibitem{AMF160107185}
D.~Aasen, R.~S.~K. Mong, and P.~Fendley, ``{Topological Defects on the Lattice I: The Ising model},'' \href{http://dx.doi.org/10.1088/1751-8113/49/35/354001}{{\em J. Phys. A} {\bfseries 49} no.~35, (2016) 354001}, \href{http://arxiv.org/abs/1601.07185}{{\ttfamily arXiv:1601.07185 [cond-mat.stat-mech]}}.

\bibitem{BG170102800}
M.~Buican and A.~Gromov, ``{Anyonic Chains, Topological Defects, and Conformal Field Theory},'' \href{http://dx.doi.org/10.1007/s00220-017-2995-6}{{\em Commun. Math. Phys.} {\bfseries 356} no.~3, (2017) 1017--1056}, \href{http://arxiv.org/abs/1701.02800}{{\ttfamily arXiv:1701.02800 [hep-th]}}.

\bibitem{LDO211209091}
L.~Lootens, C.~Delcamp, G.~Ortiz, and F.~Verstraete, ``{Dualities in One-Dimensional Quantum Lattice Models: Symmetric Hamiltonians and Matrix Product Operator Intertwiners},'' \href{http://dx.doi.org/10.1103/PRXQuantum.4.020357}{{\em PRX Quantum} {\bfseries 4} no.~2, (2023) 020357}, \href{http://arxiv.org/abs/2112.09091}{{\ttfamily arXiv:2112.09091 [quant-ph]}}.

\bibitem{BBS240505964}
L.~Bhardwaj, L.~E. Bottini, S.~Sch{\"a}fer-Nameki, and A.~Tiwari, ``{Lattice models for phases and transitions with non-invertible symmetries},'' \href{http://dx.doi.org/10.21468/SciPostPhys.20.5.134}{{\em SciPost Phys.} {\bfseries 20} no.~5, (2026) 134}, \href{http://arxiv.org/abs/2405.05964}{{\ttfamily arXiv:2405.05964 [cond-mat.str-el]}}.

\bibitem{JSW241008884}
C.~Jones, K.~Schatz, and D.~J. Williamson, ``{Quantum Cellular Automata and Categorical Dualities of Spin Chains},'' \href{http://dx.doi.org/10.1007/s00220-026-05571-y}{{\em Commun. Math. Phys.} {\bfseries 407} no.~4, (2026) 66}, \href{http://arxiv.org/abs/2410.08884}{{\ttfamily arXiv:2410.08884 [math-ph]}}.

\bibitem{EJ250705185}
D.~E. Evans and C.~Jones, ``{An operator algebraic approach to fusion category symmetry on the lattice},'' \href{http://arxiv.org/abs/2507.05185}{{\ttfamily arXiv:2507.05185 [math-ph]}}.

\bibitem{I260212053}
K.~Inamura, ``{Remarks on non-invertible symmetries on a tensor product Hilbert space in 1+1 dimensions},'' \href{http://arxiv.org/abs/2602.12053}{{\ttfamily arXiv:2602.12053 [cond-mat.str-el]}}.

\bibitem{JY260309949}
C.~Jones and X.~Yang, ``{On the structure of categorical duality operators},'' \href{http://arxiv.org/abs/2603.09949}{{\ttfamily arXiv:2603.09949 [math.QA]}}.

\bibitem{WIS260515194}
R.~Wen, K.~Inamura, and S.~Sch{\"a}fer-Nameki, ``{Non-Invertible Symmetries on Tensor-Product Hilbert Spaces and Quantum Cellular Automata},'' \href{http://arxiv.org/abs/2605.15194}{{\ttfamily arXiv:2605.15194 [cond-mat.str-el]}}.

\bibitem{BJ260521327}
I.~Bunner and C.~Jones, ``{Universal fusion category symmetries on tensor products of infinite-dimensional Hilbert spaces},'' \href{http://arxiv.org/abs/2605.21327}{{\ttfamily arXiv:2605.21327 [math-ph]}}.

\bibitem{IO230505774}
K.~Inamura and K.~Ohmori, ``{Fusion surface models: 2+1d lattice models from fusion 2-categories},'' \href{http://dx.doi.org/10.21468/SciPostPhys.16.6.143}{{\em SciPost Phys.} {\bfseries 16} (2024) 143}, \href{http://arxiv.org/abs/2305.05774}{{\ttfamily arXiv:2305.05774 [cond-mat.str-el]}}.

\bibitem{EF240804006}
L.~Eck and P.~Fendley, ``{Generalizations of Kitaev{\textquoteright}s honeycomb model from braided fusion categories},'' \href{http://dx.doi.org/10.21468/SciPostPhys.18.6.170}{{\em SciPost Phys.} {\bfseries 18} no.~6, (2025) 170}, \href{http://arxiv.org/abs/2408.04006}{{\ttfamily arXiv:2408.04006 [cond-mat.str-el]}}.

\bibitem{E250114722}
L.~Eck, ``{Dualities between 2+1d fusion surface models from braided fusion categories},'' \href{http://dx.doi.org/10.21468/SciPostPhys.19.6.157}{{\em SciPost Phys.} {\bfseries 19} no.~6, (2025) 157}, \href{http://arxiv.org/abs/2501.14722}{{\ttfamily arXiv:2501.14722 [cond-mat.str-el]}}.

\bibitem{IHT250609177}
K.~Inamura, S.-J. Huang, A.~Tiwari, and S.~Sch{\"a}fer-Nameki, ``{(2+1)d lattice models and tensor networks for gapped phases with categorical symmetry},'' \href{http://dx.doi.org/10.21468/SciPostPhys.20.2.043}{{\em SciPost Phys.} {\bfseries 20} no.~2, (2026) 043}, \href{http://arxiv.org/abs/2506.09177}{{\ttfamily arXiv:2506.09177 [cond-mat.str-el]}}.

\bibitem{DR181211933}
C.~L. Douglas and D.~J. Reutter, ``{Fusion 2-categories and a state-sum invariant for 4-manifolds},'' \href{http://arxiv.org/abs/1812.11933}{{\ttfamily arXiv:1812.11933 [math.QA]}}.

\bibitem{DHJ241105907}
T.~D. D{\'e}coppet, P.~Huston, T.~Johnson-Freyd, D.~Nikshych, D.~Penneys, J.~Plavnik, D.~Reutter, and M.~Yu, ``{The Classification of Fusion 2-Categories},'' \href{http://arxiv.org/abs/2411.05907}{{\ttfamily arXiv:2411.05907 [math.CT]}}.

\bibitem{S190910544}
N.~Seiberg, ``{Field Theories With a Vector Global Symmetry},'' \href{http://dx.doi.org/10.21468/SciPostPhys.8.4.050}{{\em SciPost Phys.} {\bfseries 8} no.~4, (2020) 050}, \href{http://arxiv.org/abs/1909.10544}{{\ttfamily arXiv:1909.10544 [cond-mat.str-el]}}.

\bibitem{QRH201002254}
M.~Qi, L.~Radzihovsky, and M.~Hermele, ``{Fracton phases via exotic higher-form symmetry-breaking},'' \href{http://dx.doi.org/10.1016/j.aop.2020.168360}{{\em Annals Phys.} {\bfseries 424} (2021) 168360}, \href{http://arxiv.org/abs/2010.02254}{{\ttfamily arXiv:2010.02254 [cond-mat.str-el]}}.

\bibitem{OPH230104706}
Y.-T. Oh, S.~D. Pace, J.~H. Han, Y.~You, and H.-Y. Lee, ``{Aspects of $\mathbb{Z}_N$ rank-2 gauge theory in (2+1) dimensions: Construction schemes, holonomies, and sublattice one-form symmetries},'' \href{http://dx.doi.org/10.1103/PhysRevB.107.155151}{{\em Phys. Rev. B} {\bfseries 107} no.~15, (2023) 155151}, \href{http://arxiv.org/abs/2301.04706}{{\ttfamily arXiv:2301.04706 [cond-mat.str-el]}}.

\bibitem{Y150803468}
B.~Yoshida, ``{Topological phases with generalized global symmetries},'' \href{http://dx.doi.org/10.1103/PhysRevB.93.155131}{{\em Phys. Rev. B} {\bfseries 93} no.~15, (2016) 155131}, \href{http://arxiv.org/abs/1508.03468}{{\ttfamily arXiv:1508.03468 [cond-mat.str-el]}}.

\bibitem{KSK180505367}
R.~Kobayashi, K.~Shiozaki, Y.~Kikuchi, and S.~Ryu, ``{Lieb-Schultz-Mattis type theorem with higher-form symmetry and the quantum dimer models},'' \href{http://dx.doi.org/10.1103/PhysRevB.99.014402}{{\em Phys. Rev. B} {\bfseries 99} no.~1, (2019) 014402}, \href{http://arxiv.org/abs/1805.05367}{{\ttfamily arXiv:1805.05367 [cond-mat.stat-mech]}}.

\bibitem{W181202517}
X.-G. Wen, ``{Emergent anomalous higher symmetries from topological order and from dynamical electromagnetic field in condensed matter systems},'' \href{http://dx.doi.org/10.1103/PhysRevB.99.205139}{{\em Phys. Rev. B} {\bfseries 99} no.~20, (2019) 205139}, \href{http://arxiv.org/abs/1812.02517}{{\ttfamily arXiv:1812.02517 [cond-mat.str-el]}}.

\bibitem{PW230105261}
S.~D. Pace and X.-G. Wen, ``{Exact emergent higher-form symmetries in bosonic lattice models},'' \href{http://dx.doi.org/10.1103/PhysRevB.108.195147}{{\em Phys. Rev. B} {\bfseries 108} no.~19, (2023) 195147}, \href{http://arxiv.org/abs/2301.05261}{{\ttfamily arXiv:2301.05261 [cond-mat.str-el]}}.

\bibitem{SNH230404792}
C.~Stahl, R.~Nandkishore, and O.~Hart, ``{Topologically stable ergodicity breaking from emergent higher-form symmetries in generalized quantum loop models},'' \href{http://dx.doi.org/10.21468/SciPostPhys.16.3.068}{{\em SciPost Phys.} {\bfseries 16} no.~3, (2024) 068}, \href{http://arxiv.org/abs/2304.04792}{{\ttfamily arXiv:2304.04792 [cond-mat.stat-mech]}}.

\bibitem{HNK230507063}
J.~Huxford, D.~X. Nguyen, and Y.~B. Kim, ``{Gaining insights on anyon condensation and 1-form symmetry breaking across a topological phase transition in a deformed toric code model},'' \href{http://dx.doi.org/10.21468/SciPostPhys.15.6.253}{{\em SciPost Phys.} {\bfseries 15} no.~6, (2023) 253}, \href{http://arxiv.org/abs/2305.07063}{{\ttfamily arXiv:2305.07063 [cond-mat.str-el]}}.

\bibitem{TC230703180}
N.~Tantivasadakarn and X.~Chen, ``{String operators for Cheshire strings in topological phases},'' \href{http://dx.doi.org/10.1103/PhysRevB.109.165149}{{\em Phys. Rev. B} {\bfseries 109} no.~16, (2024) 165149}, \href{http://arxiv.org/abs/2307.03180}{{\ttfamily arXiv:2307.03180 [cond-mat.str-el]}}.

\bibitem{EHN231006701}
H.~Ebisu, M.~Honda, and T.~Nakanishi, ``{Foliated field theories and multipole symmetries},'' \href{http://dx.doi.org/10.1103/PhysRevB.109.165112}{{\em Phys. Rev. B} {\bfseries 109} no.~16, (2024) 165112}, \href{http://arxiv.org/abs/2310.06701}{{\ttfamily arXiv:2310.06701 [cond-mat.str-el]}}.

\bibitem{LLM231016839}
R.~Liu, H.~T. Lam, H.~Ma, and L.~Zou, ``{Symmetries and anomalies of Kitaev spin-S models: Identifying symmetry-enforced exotic quantum matter},'' \href{http://dx.doi.org/10.21468/SciPostPhys.16.4.100}{{\em SciPost Phys.} {\bfseries 16} no.~4, (2024) 100}, \href{http://arxiv.org/abs/2310.16839}{{\ttfamily arXiv:2310.16839 [cond-mat.str-el]}}.

\bibitem{XRK231116235}
W.-T. Xu, T.~Rakovszky, M.~Knap, and F.~Pollmann, ``{Entanglement Properties of Gauge Theories from Higher-Form Symmetries},'' \href{http://dx.doi.org/10.1103/PhysRevX.15.011001}{{\em Phys. Rev. X} {\bfseries 15} no.~1, (2025) 011001}, \href{http://arxiv.org/abs/2311.16235}{{\ttfamily arXiv:2311.16235 [cond-mat.str-el]}}.

\bibitem{XPK240200127}
W.-T. Xu, F.~Pollmann, and M.~Knap, ``{Critical behavior of Fredenhagen-Marcu string order parameters at topological phase transitions with emergent higher-form symmetries},'' \href{http://dx.doi.org/10.1038/s41534-025-01030-z}{{\em npj Quantum Inf.} {\bfseries 11} no.~1, (2025) 74}, \href{http://arxiv.org/abs/2402.00127}{{\ttfamily arXiv:2402.00127 [cond-mat.str-el]}}.

\bibitem{CSS240513105}
Y.~Choi, Y.~Sanghavi, S.-H. Shao, and Y.~Zheng, ``{Non-invertible and higher-form symmetries in 2+1d lattice gauge theories},'' \href{http://dx.doi.org/10.21468/SciPostPhys.18.1.008}{{\em SciPost Phys.} {\bfseries 18} no.~1, (2025) 008}, \href{http://arxiv.org/abs/2405.13105}{{\ttfamily arXiv:2405.13105 [cond-mat.str-el]}}.

\bibitem{LXP250217572}
Y.-J. Liu, W.-T. Xu, F.~Pollmann, and M.~Knap, ``{Information-theoretic principle of emergent 1-form symmetries},'' \href{http://arxiv.org/abs/2502.17572}{{\ttfamily arXiv:2502.17572 [quant-ph]}}.

\bibitem{HKP250410569}
P.-S. Hsin, R.~Kobayashi, and A.~Prem, ``{Higher-Form Anomalies Imply Intrinsic Long-Range Entanglement},'' \href{http://arxiv.org/abs/2504.10569}{{\ttfamily arXiv:2504.10569 [quant-ph]}}.

\bibitem{PAL250702036}
S.~D. Pace, {\"O}.~M. Aksoy, and H.~T. Lam, ``{Spacetime symmetry-enriched SymTFT: From LSM anomalies to modulated symmetries and beyond},'' \href{http://dx.doi.org/10.21468/SciPostPhys.20.1.007}{{\em SciPost Phys.} {\bfseries 20} no.~1, (2026) 007}, \href{http://arxiv.org/abs/2507.02036}{{\ttfamily arXiv:2507.02036 [cond-mat.str-el]}}.

\bibitem{FKCR250912304}
Y.~Feng, R.~Kobayashi, Y.-A. Chen, and S.~Ryu, ``{Higher-Form Anomalies on Lattices},'' \href{http://dx.doi.org/10.1103/2jz1-m1lb}{{\em Phys. Rev. Lett.} {\bfseries 136} no.~4, (2026) 046504}, \href{http://arxiv.org/abs/2509.12304}{{\ttfamily arXiv:2509.12304 [cond-mat.str-el]}}.

\bibitem{FCH251023701}
Y.~Feng, Y.-A. Chen, P.-S. Hsin, and R.~Kobayashi, ``{When Can Higher-Form Symmetries Be Made On Site},'' \href{http://dx.doi.org/10.1103/59dq-sl3m}{{\em Phys. Rev. Lett.} {\bfseries 137} no.~2, (2026) 026501}, \href{http://arxiv.org/abs/2510.23701}{{\ttfamily arXiv:2510.23701 [cond-mat.str-el]}}.

\bibitem{LTL260120935}
R.~Liu, P.~M. Tam, H.~T. Lam, and L.~Zou, ``{When does a lattice higher-form symmetry flow to a topological higher-form symmetry at low energies?},'' \href{http://arxiv.org/abs/2601.20935}{{\ttfamily arXiv:2601.20935 [cond-mat.str-el]}}.

\bibitem{HPC260512601}
D.~Hofmeier, G.~Pimenta, and W.~Cao, ``{Lattice Gauging Interfaces and Noninvertible Defects in Higher Dimensions},'' \href{http://arxiv.org/abs/2605.12601}{{\ttfamily arXiv:2605.12601 [cond-mat.str-el]}}.

\bibitem{KT13094721}
A.~Kapustin and R.~Thorngren, ``{Higher Symmetry and Gapped Phases of Gauge Theories},'' \href{http://dx.doi.org/10.1007/978-3-319-59939-7_5}{{\em Prog. Math.} {\bfseries 324} (2017) 177--202}, \href{http://arxiv.org/abs/1309.4721}{{\ttfamily arXiv:1309.4721 [hep-th]}}.

\bibitem{S150804770}
E.~Sharpe, ``{Notes on generalized global symmetries in QFT},'' \href{http://dx.doi.org/10.1002/prop.201500048}{{\em Fortsch. Phys.} {\bfseries 63} (2015) 659--682}, \href{http://arxiv.org/abs/1508.04770}{{\ttfamily arXiv:1508.04770 [hep-th]}}.

\bibitem{CI180204790}
C.~C{\'o}rdova, T.~T. Dumitrescu, and K.~Intriligator, ``{Exploring 2-Group Global Symmetries},'' \href{http://dx.doi.org/10.1007/JHEP02(2019)184}{{\em JHEP} {\bfseries 02} (2019) 184}, \href{http://arxiv.org/abs/1802.04790}{{\ttfamily arXiv:1802.04790 [hep-th]}}.

\bibitem{BH180309336}
F.~Benini, C.~C{\'o}rdova, and P.-S. Hsin, ``{On 2-Group Global Symmetries and their Anomalies},'' \href{http://dx.doi.org/10.1007/JHEP03(2019)118}{{\em JHEP} {\bfseries 03} (2019) 118}, \href{http://arxiv.org/abs/1803.09336}{{\ttfamily arXiv:1803.09336 [hep-th]}}.

\bibitem{BBCW14104540}
M.~Barkeshli, P.~Bonderson, M.~Cheng, and Z.~Wang, ``{Symmetry Fractionalization, Defects, and Gauging of Topological Phases},'' \href{http://dx.doi.org/10.1103/PhysRevB.100.115147}{{\em Phys. Rev. B} {\bfseries 100} no.~11, (2019) 115147}, \href{http://arxiv.org/abs/1410.4540}{{\ttfamily arXiv:1410.4540 [cond-mat.str-el]}}.

\bibitem{FV151101502}
L.~Fidkowski and A.~Vishwanath, ``{Realizing anomalous anyonic symmetries at the surfaces of three-dimensional gauge theories},'' \href{http://dx.doi.org/10.1103/PhysRevB.96.045131}{{\em Phys. Rev. B} {\bfseries 96} no.~4, (2017) 045131}, \href{http://arxiv.org/abs/1511.01502}{{\ttfamily arXiv:1511.01502 [cond-mat.str-el]}}.

\bibitem{BC170609464}
M.~Barkeshli and M.~Cheng, ``{Time-reversal and spatial-reflection symmetry localization anomalies in (2+1)-dimensional topological phases of matter},'' \href{http://dx.doi.org/10.1103/PhysRevB.98.115129}{{\em Phys. Rev. B} {\bfseries 98} no.~11, (2018) 115129}, \href{http://arxiv.org/abs/1706.09464}{{\ttfamily arXiv:1706.09464 [cond-mat.str-el]}}.

\bibitem{CET200805652}
Y.-A. Chen, T.~D. Ellison, and N.~Tantivasadakarn, ``{Disentangling supercohomology symmetry-protected topological phases in three spatial dimensions},'' \href{http://dx.doi.org/10.1103/PhysRevResearch.3.013056}{{\em Phys. Rev. Res.} {\bfseries 3} no.~1, (2021) 013056}, \href{http://arxiv.org/abs/2008.05652}{{\ttfamily arXiv:2008.05652 [cond-mat.str-el]}}.

\bibitem{BCH220807367}
M.~Barkeshli, Y.-A. Chen, S.-J. Huang, R.~Kobayashi, N.~Tantivasadakarn, and G.~Zhu, ``{Codimension-2 defects and higher symmetries in (3+1)D topological phases},'' \href{http://dx.doi.org/10.21468/SciPostPhys.14.4.065}{{\em SciPost Phys.} {\bfseries 14} no.~4, (2023) 065}, \href{http://arxiv.org/abs/2208.07367}{{\ttfamily arXiv:2208.07367 [cond-mat.str-el]}}.

\bibitem{BCHK221111764}
M.~Barkeshli, Y.-A. Chen, P.-S. Hsin, and R.~Kobayashi, ``{Higher-group symmetry in finite gauge theory and stabilizer codes},'' \href{http://dx.doi.org/10.21468/SciPostPhys.16.4.089}{{\em SciPost Phys.} {\bfseries 16} no.~4, (2024) 089}, \href{http://arxiv.org/abs/2211.11764}{{\ttfamily arXiv:2211.11764 [cond-mat.str-el]}}.

\bibitem{DT230101259}
C.~Delcamp and A.~Tiwari, ``{Higher categorical symmetries and gauging in two-dimensional spin systems},'' \href{http://dx.doi.org/10.21468/SciPostPhys.16.4.110}{{\em SciPost Phys.} {\bfseries 16} no.~4, (2024) 110}, \href{http://arxiv.org/abs/2301.01259}{{\ttfamily arXiv:2301.01259 [hep-th]}}.

\bibitem{BHK231105674}
M.~Barkeshli, P.-S. Hsin, and R.~Kobayashi, ``{Higher-group symmetry of (3+1)D fermionic $\mathbb{Z}_2$ gauge theory: Logical CCZ, CS, and T gates from higher symmetry},'' \href{http://dx.doi.org/10.21468/SciPostPhys.16.5.122}{{\em SciPost Phys.} {\bfseries 16} no.~5, (2024) 122}, \href{http://arxiv.org/abs/2311.05674}{{\ttfamily arXiv:2311.05674 [cond-mat.str-el]}}.

\bibitem{OE260402856}
T.~Oishi and H.~Ebisu, ``{Type-IV 't Hooft Anomalies on the Lattice: Emergent Higher-Categorical Symmetries and Applications to LSM Systems},'' \href{http://arxiv.org/abs/2604.02856}{{\ttfamily arXiv:2604.02856 [cond-mat.str-el]}}.

\bibitem{LSS260406307}
Z.~Lu, S.~Seifnashri, and S.-H. Shao, ``{Lattice chiral symmetry from bosons in 3+1D},'' \href{http://dx.doi.org/10.1103/pxvz-r2gf}{{\em Phys. Rev. D} {\bfseries 114} no.~3, (2026) 034518}, \href{http://arxiv.org/abs/2604.06307}{{\ttfamily arXiv:2604.06307 [hep-th]}}.

\bibitem{BL0307200}
J.~C. {Baez} and A.~D. {Lauda}, ``{Higher-dimensional algebra. V: 2-Groups},'' \href{http://eudml.org/doc/124217}{{\em Theory and Applications of Categories} {\bfseries 12} (2004) 423--491}, \href{http://arxiv.org/abs/math/0307200}{{\ttfamily arXiv:math/0307200}}.

\bibitem{PZB231008554}
S.~D. Pace, C.~Zhu, A.~Beaudry, and X.-G. Wen, ``{Generalized symmetries in singularity-free nonlinear {\ensuremath{\sigma}} models and their disordered phases},'' \href{http://dx.doi.org/10.1103/PhysRevB.110.195149}{{\em Phys. Rev. B} {\bfseries 110} no.~19, (2024) 195149}, \href{http://arxiv.org/abs/2310.08554}{{\ttfamily arXiv:2310.08554 [cond-mat.str-el]}}.

\bibitem{MP0608484}
J.~F. Martins and T.~Porter, ``{On Yetter's invariant and an extension of the Dijkgraaf-Witten invariant to categorical groups},'' {\em Theor. Appl. Categor.} {\bfseries 18} (2007) 118--150, \href{http://arxiv.org/abs/math/0608484}{{\ttfamily arXiv:math/0608484}}.

\bibitem{thorngrenThesis}
R.~G. Thorngren, {\em {Combinatorial Topology and Applications to Quantum Field Theory}}.
\newblock PhD thesis, UC, Berkeley (main), 2018.

\bibitem{DT180210104}
C.~Delcamp and A.~Tiwari, ``{From gauge to higher gauge models of topological phases},'' \href{http://dx.doi.org/10.1007/JHEP10(2018)049}{{\em JHEP} {\bfseries 10} (2018) 049}, \href{http://arxiv.org/abs/1802.10104}{{\ttfamily arXiv:1802.10104 [cond-mat.str-el]}}.

\bibitem{BSW220805973}
L.~Bhardwaj, S.~Sch{\"a}fer-Nameki, and J.~Wu, ``{Universal Non-Invertible Symmetries},'' \href{http://dx.doi.org/10.1002/prop.202200143}{{\em Fortsch. Phys.} {\bfseries 70} no.~11, (2022) 2200143}, \href{http://arxiv.org/abs/2208.05973}{{\ttfamily arXiv:2208.05973 [hep-th]}}.

\bibitem{BBFP220805993}
T.~Bartsch, M.~Bullimore, A.~E.~V. Ferrari, and J.~Pearson, ``{Non-invertible symmetries and higher representation theory I},'' \href{http://dx.doi.org/10.21468/SciPostPhys.17.1.015}{{\em SciPost Phys.} {\bfseries 17} no.~1, (2024) 015}, \href{http://arxiv.org/abs/2208.05993}{{\ttfamily arXiv:2208.05993 [hep-th]}}.

\bibitem{CT230700939}
S.~Chen and Y.~Tanizaki, ``{Solitonic symmetry as non-invertible symmetry: cohomology theories with TQFT coefficients},'' \href{http://arxiv.org/abs/2307.00939}{{\ttfamily arXiv:2307.00939 [hep-th]}}.

\bibitem{P230805730}
S.~D. Pace, ``{Emergent generalized symmetries in ordered phases and applications to quantum disordering},'' \href{http://dx.doi.org/10.21468/SciPostPhys.17.3.080}{{\em SciPost Phys.} {\bfseries 17} no.~3, (2024) 080}, \href{http://arxiv.org/abs/2308.05730}{{\ttfamily arXiv:2308.05730 [cond-mat.str-el]}}.

\bibitem{HJS260804248}
S.~Harder, T.~Jacobson, and Z.~Sun, ``{Tilts from 2-Groups},'' \href{http://arxiv.org/abs/2608.04248}{{\ttfamily arXiv:2608.04248 [hep-th]}}.

\bibitem{S230805151}
S.~Seifnashri, ``{Lieb-Schultz-Mattis anomalies as obstructions to gauging (non-on-site) symmetries},'' \href{http://dx.doi.org/10.21468/SciPostPhys.16.4.098}{{\em SciPost Phys.} {\bfseries 16} no.~4, (2024) 098}, \href{http://arxiv.org/abs/2308.05151}{{\ttfamily arXiv:2308.05151 [cond-mat.str-el]}}.

\bibitem{HKZ240520401}
P.-S. Hsin, R.~Kobayashi, and C.~Zhang, ``{Fractionalization of coset non-invertible symmetry and exotic Hall conductance},'' \href{http://dx.doi.org/10.21468/SciPostPhys.17.3.095}{{\em SciPost Phys.} {\bfseries 17} no.~3, (2024) 095}, \href{http://arxiv.org/abs/2405.20401}{{\ttfamily arXiv:2405.20401 [cond-mat.str-el]}}.

\bibitem{K9707021}
A.~Y. Kitaev, ``{Fault tolerant quantum computation by anyons},'' \href{http://dx.doi.org/10.1016/S0003-4916(02)00018-0}{{\em Annals Phys.} {\bfseries 303} (2003) 2--30}, \href{http://arxiv.org/abs/quant-ph/9707021}{{\ttfamily arXiv:quant-ph/9707021}}.

\bibitem{RSS220402407}
K.~Roumpedakis, S.~Seifnashri, and S.-H. Shao, ``{Higher Gauging and Non-invertible Condensation Defects},'' \href{http://dx.doi.org/10.1007/s00220-023-04706-9}{{\em Commun. Math. Phys.} {\bfseries 401} no.~3, (2023) 3043--3107}, \href{http://arxiv.org/abs/2204.02407}{{\ttfamily arXiv:2204.02407 [hep-th]}}.

\bibitem{PLA240918113}
S.~D. Pace, H.~T. Lam, and {\"O}.~M. Aksoy, ``{(SPT-)LSM theorems from projective non-invertible symmetries},'' \href{http://dx.doi.org/10.21468/SciPostPhys.18.1.028}{{\em SciPost Phys.} {\bfseries 18} no.~1, (2025) 028}, \href{http://arxiv.org/abs/2409.18113}{{\ttfamily arXiv:2409.18113 [cond-mat.str-el]}}.

\bibitem{FHH191007998}
M.~Freedman, J.~Haah, and M.~B. Hastings, ``{The Group Structure of Quantum Cellular Automata},'' \href{http://dx.doi.org/10.1007/s00220-022-04316-x}{{\em Commun. Math. Phys.} {\bfseries 389} no.~3, (2022) 1277--1302}, \href{http://arxiv.org/abs/1910.07998}{{\ttfamily arXiv:1910.07998 [quant-ph]}}.

\bibitem{ZLL241105004}
Z.~Zhang, Y.~Li, and T.-C. Lu, ``{Non-onsite symmetry breaking: Topological phase coexistence and criticality},'' \href{http://dx.doi.org/10.1103/rtk8-h9xz}{{\em Phys. Rev. B} {\bfseries 113} no.~12, (2026) 125123}, \href{http://arxiv.org/abs/2411.05004}{{\ttfamily arXiv:2411.05004 [cond-mat.str-el]}}.

\bibitem{CGL11064772}
X.~Chen, Z.-C. Gu, Z.-X. Liu, and X.-G. Wen, ``{Symmetry protected topological orders and the group cohomology of their symmetry group},'' \href{http://dx.doi.org/10.1103/PhysRevB.87.155114}{{\em Phys. Rev. B} {\bfseries 87} no.~15, (2013) 155114}, \href{http://arxiv.org/abs/1106.4772}{{\ttfamily arXiv:1106.4772 [cond-mat.str-el]}}.

\bibitem{WWW170506728}
J.~Wang, X.-G. Wen, and E.~Witten, ``{Symmetric Gapped Interfaces of SPT and SET States: Systematic Constructions},'' \href{http://dx.doi.org/10.1103/PhysRevX.8.031048}{{\em Phys. Rev. X} {\bfseries 8} no.~3, (2018) 031048}, \href{http://arxiv.org/abs/1705.06728}{{\ttfamily arXiv:1705.06728 [cond-mat.str-el]}}.

\bibitem{B190505790}
N.~Bultinck, ``{UV perspective on mixed anomalies at critical points between bosonic symmetry-protected phases},'' \href{http://dx.doi.org/10.1103/PhysRevB.100.165132}{{\em Phys. Rev. B} {\bfseries 100} no.~16, (2019) 165132}, \href{http://arxiv.org/abs/1905.05790}{{\ttfamily arXiv:1905.05790 [cond-mat.str-el]}}.

\bibitem{TTV211007599}
N.~Tantivasadakarn, R.~Thorngren, A.~Vishwanath, and R.~Verresen, ``{Pivot Hamiltonians as generators of symmetry and entanglement},'' \href{http://dx.doi.org/10.21468/SciPostPhys.14.2.012}{{\em SciPost Phys.} {\bfseries 14} no.~2, (2023) 012}, \href{http://arxiv.org/abs/2110.07599}{{\ttfamily arXiv:2110.07599 [cond-mat.str-el]}}.

\bibitem{SS240401369}
S.~Seifnashri and S.-H. Shao, ``{Cluster State as a Noninvertible Symmetry-Protected Topological Phase},'' \href{http://dx.doi.org/10.1103/PhysRevLett.133.116601}{{\em Phys. Rev. Lett.} {\bfseries 133} no.~11, (2024) 116601}, \href{http://arxiv.org/abs/2404.01369}{{\ttfamily arXiv:2404.01369 [cond-mat.str-el]}}.

\bibitem{PV250920431}
S.~Prembabu and R.~Verresen, ``{Multicriticality between Purely Gapless SPT Phases with Unitary Symmetry},'' \href{http://arxiv.org/abs/2509.20431}{{\ttfamily arXiv:2509.20431 [cond-mat.str-el]}}.

\bibitem{E0408120}
J.~Elgueta, ``{Representation theory of 2-groups on Kapranov and Voevodsky's 2-vector spaces},'' \href{http://dx.doi.org/https://doi.org/10.1016/j.aim.2006.11.010}{{\em Adv. Math.} {\bfseries 213} no.~1, (2007) 53--92}, \href{http://arxiv.org/abs/math/0408120}{{\ttfamily arXiv:math/0408120 [math.CT]}}.

\bibitem{HWW12113695}
Y.~Hu, Y.~Wan, and Y.-S. Wu, ``{Twisted quantum double model of topological phases in two dimensions},'' \href{http://dx.doi.org/10.1103/PhysRevB.87.125114}{{\em Phys. Rev. B} {\bfseries 87} no.~12, (2013) 125114}, \href{http://arxiv.org/abs/1211.3695}{{\ttfamily arXiv:1211.3695 [cond-mat.str-el]}}.

\bibitem{PRS220413708}
T.~Pantev, D.~G. Robbins, E.~Sharpe, and T.~Vandermeulen, ``{Orbifolds by 2-groups and decomposition},'' \href{http://dx.doi.org/10.1007/JHEP09(2022)036}{{\em JHEP} {\bfseries 09} (2022) 036}, \href{http://arxiv.org/abs/2204.13708}{{\ttfamily arXiv:2204.13708 [hep-th]}}.

\bibitem{PS230316220}
A.~Perez-Lona and E.~Sharpe, ``{Three-dimensional orbifolds by 2-groups},'' \href{http://dx.doi.org/10.1007/JHEP08(2023)138}{{\em JHEP} {\bfseries 08} (2023) 138}, \href{http://arxiv.org/abs/2303.16220}{{\ttfamily arXiv:2303.16220 [hep-th]}}.

\bibitem{BSTW250220440}
L.~Bhardwaj, S.~Schafer-Nameki, A.~Tiwari, and A.~Warman, ``{Gapped Phases in (2+1)d with Non-Invertible Symmetries: Part II},'' \href{http://arxiv.org/abs/2502.20440}{{\ttfamily arXiv:2502.20440 [hep-th]}}.

\bibitem{AFM200808598}
D.~Aasen, P.~Fendley, and R.~S.~K. Mong, ``{Topological Defects on the Lattice: Dualities and Degeneracies},'' \href{http://arxiv.org/abs/2008.08598}{{\ttfamily arXiv:2008.08598 [cond-mat.stat-mech]}}.

\bibitem{BBS240505302}
L.~Bhardwaj, L.~E. Bottini, S.~Sch{\"a}fer-Nameki, and A.~Tiwari, ``{Illustrating the categorical Landau paradigm in lattice models},'' \href{http://dx.doi.org/10.1103/PhysRevB.111.054432}{{\em Phys. Rev. B} {\bfseries 111} no.~5, (2025) 054432}, \href{http://arxiv.org/abs/2405.05302}{{\ttfamily arXiv:2405.05302 [cond-mat.str-el]}}.

\bibitem{CAW240505331}
A.~Chatterjee, {\"O}.~M. Aksoy, and X.-G. Wen, ``{Quantum phases and transitions in spin chains with non-invertible symmetries},'' \href{http://dx.doi.org/10.21468/SciPostPhys.17.4.115}{{\em SciPost Phys.} {\bfseries 17} no.~4, (2024) 115}, \href{http://arxiv.org/abs/2405.05331}{{\ttfamily arXiv:2405.05331 [cond-mat.str-el]}}.

\bibitem{CBN250811003}
K.~T.~K. Chung, U.~Borla, A.~H. Nevidomskyy, and S.~Moroz, ``{Spontaneously Broken Noninvertible Symmetries in Transverse-Field Ising Qudit Chains},'' \href{http://dx.doi.org/10.1103/ng8b-sdt4}{{\em Phys. Rev. Lett.} {\bfseries 136} no.~21, (2026) 216601}, \href{http://arxiv.org/abs/2508.11003}{{\ttfamily arXiv:2508.11003 [cond-mat.str-el]}}.

\bibitem{LCT260210183}
D.-C. Lu, A.~Chatterjee, and N.~Tantivasadakarn, ``{Generalized Kramers-Wannier Self-Duality in Hopf-Ising Models},'' \href{http://arxiv.org/abs/2602.10183}{{\ttfamily arXiv:2602.10183 [cond-mat.str-el]}}.

\bibitem{PKC250504684}
S.~D. Pace, M.~L. Kim, A.~Chatterjee, and S.-H. Shao, ``{Parity Anomaly from a Lieb-Schultz-Mattis Theorem: Exact Valley Symmetries on the Lattice},'' \href{http://dx.doi.org/10.1103/4x4w-pfgq}{{\em Phys. Rev. Lett.} {\bfseries 135} no.~23, (2025) 236501}, \href{http://arxiv.org/abs/2505.04684}{{\ttfamily arXiv:2505.04684 [cond-mat.str-el]}}.

\bibitem{KX250504719}
A.~Kapustin and S.~Xu, ``{Higher symmetries and anomalies in quantum lattice systems},'' \href{http://arxiv.org/abs/2505.04719}{{\ttfamily arXiv:2505.04719 [math-ph]}}.

\bibitem{KS250707430}
K.~Kawagoe and W.~Shirley, ``{Anomaly diagnosis via symmetry restriction in two-dimensional lattice systems},'' \href{http://arxiv.org/abs/2507.07430}{{\ttfamily arXiv:2507.07430 [cond-mat.str-el]}}.

\bibitem{GM250716475}
J.~Garre-Rubio and A.~Moln{\'a}r, ``{On two-dimensional tensor network group symmetries},'' \href{http://dx.doi.org/10.1088/1367-2630/ae124f}{{\em New J. Phys.} {\bfseries 27} no.~10, (2025) 104510}, \href{http://arxiv.org/abs/2507.16475}{{\ttfamily arXiv:2507.16475 [quant-ph]}}.

\bibitem{TLE250721209}
Y.-T. Tu, D.~M. Long, and D.~V. Else, ``{Anomalies of Global Symmetries on the Lattice},'' \href{http://dx.doi.org/10.1103/m188-w1ct}{{\em Phys. Rev. X} {\bfseries 16} no.~1, (2026) 011027}, \href{http://arxiv.org/abs/2507.21209}{{\ttfamily arXiv:2507.21209 [cond-mat.str-el]}}.

\bibitem{SZJ250721267}
W.~Shirley, C.~Zhang, W.~Ji, and M.~Levin, ``{Anomaly-Free Symmetries with Obstructions to Gauging and Onsiteability},'' \href{http://dx.doi.org/10.1103/zrs1-j2xd}{{\em Phys. Rev. Lett.} {\bfseries 136} no.~21, (2026) 216602}, \href{http://arxiv.org/abs/2507.21267}{{\ttfamily arXiv:2507.21267 [cond-mat.str-el]}}.

\bibitem{CGT251202105}
A.~M. Czajka, R.~Geiko, and R.~Thorngren, ``{Anomalies on the Lattice, Homotopy of Quantum Cellular Automata, and a Spectrum of Invertible States},'' \href{http://arxiv.org/abs/2512.02105}{{\ttfamily arXiv:2512.02105 [cond-mat.str-el]}}.

\bibitem{PB260211266}
S.~D. Pace and D.~Bulmash, ``{Lieb-Schultz-Mattis constraints from stratified anomalies of modulated symmetries},'' \href{http://arxiv.org/abs/2602.11266}{{\ttfamily arXiv:2602.11266 [cond-mat.str-el]}}.

\bibitem{KS250716966}
A.~Kapustin and L.~Spodyneiko, ``{Higher symmetries, anomalies, and crossed squares in lattice gauge theory},'' \href{http://arxiv.org/abs/2507.16966}{{\ttfamily arXiv:2507.16966 [hep-th]}}.

\bibitem{Vafa:1986wx}
C.~Vafa, ``{Modular Invariance and Discrete Torsion on Orbifolds},'' \href{http://dx.doi.org/10.1016/0550-3213(86)90379-2}{{\em Nucl. Phys. B} {\bfseries 273} (1986) 592--606}.

\bibitem{LSY240514939}
D.-C. Lu, Z.~Sun, and Y.-Z. You, ``{Realizing triality and $p$-ality by lattice twisted gauging in (1+1)d quantum spin systems},'' \href{http://dx.doi.org/10.21468/SciPostPhys.17.5.136}{{\em SciPost Phys.} {\bfseries 17} no.~5, (2024) 136}, \href{http://arxiv.org/abs/2405.14939}{{\ttfamily arXiv:2405.14939 [cond-mat.str-el]}}.

\bibitem{SSY250302925}
S.~Seifnashri, S.-H. Shao, and X.~Yang, ``{Gauging non-invertible symmetries on the lattice},'' \href{http://dx.doi.org/10.21468/SciPostPhys.19.2.063}{{\em SciPost Phys.} {\bfseries 19} no.~2, (2025) 063}, \href{http://arxiv.org/abs/2503.02925}{{\ttfamily arXiv:2503.02925 [cond-mat.str-el]}}.

\bibitem{VD250116301}
B.~Vancraeynest-De~Cuiper and C.~Delcamp, ``{Twisted gauging and topological sectors in (2+1)d Abelian lattice gauge theories},'' \href{http://dx.doi.org/10.21468/SciPostPhys.19.2.054}{{\em SciPost Phys.} {\bfseries 19} no.~2, (2025) 054}, \href{http://arxiv.org/abs/2501.16301}{{\ttfamily arXiv:2501.16301 [cond-mat.str-el]}}.

\bibitem{AA211112096}
V.~V. Albert, D.~Aasen, W.~Xu, W.~Ji, J.~Alicea, and J.~Preskill, ``{Spin chains, defects, and quantum wires for the quantum-double edge},'' \href{http://arxiv.org/abs/2111.12096}{{\ttfamily arXiv:2111.12096 [cond-mat.str-el]}}.

\bibitem{BS0412325}
J.~Baez and U.~Schreiber, ``{Higher gauge theory: 2-connections on 2-bundles},'' \href{http://arxiv.org/abs/hep-th/0412325}{{\ttfamily arXiv:hep-th/0412325}}.

\bibitem{BS0511710}
J.~C. Baez and U.~Schreiber, \href{http://dx.doi.org/10.1090/conm/431/08264}{``Higher gauge theory,''} in {\em Categories in Algebra, Geometry and Mathematical Physics}, A.~Davydov, M.~Batanin, M.~Johnson, S.~Lack, and A.~Neeman, eds., vol.~431 of {\em Contemporary Mathematics}, pp.~7--30.
\newblock American Mathematical Society, Providence, RI, 2007.
\newblock \href{http://arxiv.org/abs/math/0511710}{{\ttfamily arXiv:math/0511710 [math.DG]}}.

\bibitem{GST240612978}
P.~Gorantla, S.-H. Shao, and N.~Tantivasadakarn, ``{Tensor Networks for Noninvertible Symmetries in 3+1D and Beyond},'' \href{http://dx.doi.org/10.1103/p32z-v884}{{\em Phys. Rev. X} {\bfseries 15} no.~4, (2025) 041006}, \href{http://arxiv.org/abs/2406.12978}{{\ttfamily arXiv:2406.12978 [quant-ph]}}.

\bibitem{T170609769}
N.~Tantivasadakarn, ``{Dimensional Reduction and Topological Invariants of Symmetry-Protected Topological Phases},'' \href{http://dx.doi.org/10.1103/PhysRevB.96.195101}{{\em Phys. Rev. B} {\bfseries 96} no.~19, (2017) 195101}, \href{http://arxiv.org/abs/1706.09769}{{\ttfamily arXiv:1706.09769 [cond-mat.str-el]}}.

\end{thebibliography}\endgroup

\end{document}